\documentclass[12pt,a4paper]{book}
\usepackage{a4wide}
\usepackage{latexsym}
\usepackage{epsf}
\usepackage{amssymb}
\usepackage[spanish, activeacute]{babel}
\usepackage{lecciones}
\makeatletter
\@addtoreset{equation}{section}
\makeatother

\frontmatter

\begin{document}

\begin{center}

%title

\vspace{.7cm}

{\Huge \textbf{Agujeros negros cl'asicos}}

\vspace{.7cm}

{\Huge \textbf{y }}

\vspace{.7cm}

{\Huge \textbf{cu'anticos}}

\vspace{.7cm}

{\Huge \textbf{en}}

\vspace{.7cm}

{\Huge \textbf{Teor'ia de Cuerdas}}

\vspace{3cm}

%authors

{\bf\large Tom'as Ort'in}

\vspace{1cm}

{\it Instituto de F'isica Te'orica, C-XVI,
Universidad Aut'onoma de Madrid \\
E-28049-Madrid, Espa~na}

\vspace{.3cm}

{\it e}

\vspace{.3cm}

{\it I.M.A.F.F., C.S.I.C., Calle de Serrano 113 bis\\ 
E-28006-Madrid, Espa~na}

\vspace{1cm}

E-mail: {\tt tomas.ortin@cern.ch}

\newpage
~
\newpage
%%%%%%%%%%%%%%%%%%%%%%%%%%%%%%%%%%%%%%%%%%%%%%%%%%%%%%%%%%%%%%%%%%%%%%
%%%%%%%%%%%%%%%%%%%%%%%%%%%%%%%%%%%%%%%%%%%%%%%%%%%%%%%%%%%%%%%%%%%%%%
\pagestyle{plain}
%%%%%%%%%%%%%%%%%%%%%%%%%%%%%%%%%%%%%%%%%%%%%%%%%%%%%%%%%%%%%%%%%%%%%%
%%%%%%%%%%%%%%%%%%%%%%%%%%%%%%%%%%%%%%%%%%%%%%%%%%%%%%%%%%%%%%%%%%%%%%
%%%%%%%%%%%%%%%%%%%%%%%%%%%%%%%%%%%%%%%%%%%%%%%%%%%%%%%%%%%%%%%%%%%%%%
\underline{\underline{\Large\it Resumen}}

\end{center}

\begin{quotation}

\small

Tras una revisi'on de los resultados cl'asicos de termodin'amica de
los agujeros negros de Schwarzschild y de Reissner y Nordstr\"om,
estudiamos las soluciones de tipo agujero negro de las teor'ias de
supergravedad en cuatro dimensiones para, a continuaci'on estudiar
las teor'ias efectivas de cuerdas (supergravedades en diez y once
dimensiones), la construcci'on de soluciones de tipo agujero negro en
cuatro dimensiones a partir de soluciones {\it elementales} de estas
teor'ias en diez y la interpretaci'on microsc'opica de su
entrop'ia utilizando los grados de libertad de las teor'ias de
cuerdas.

\end{quotation}

\newpage
%%%%%%%%%%%%%%%%%%%%%%%%%%%%%%%%%%%%%%%%%%%%%%%%%%%%%%%%%%%%%%%%%%%%%%
%%%%%%%%%%%%%%%%%%%%%%%%%%%%%%%%%%%%%%%%%%%%%%%%%%%%%%%%%%%%%%%%%%%%%%
\tableofcontents
\chapter*{Introducci'on}

El siglo pasado ha visto el nacimiento y triunfo de dos
teor'ias-marco de la F'isica: la Relatividad Especial y la
Mec'anica Cu'antica, fruct'iferamente combinadas en las
Teor'ias de Campos Relativistas que describen tres de las cuatro
interacciones fundamentales de la Naturaleza con un detalle y una
precisi'on inconcebible para los propios pioneros. La cuarta fuerza,
la gravitaci'on, est'a descrita por lo que parece simplemente una
extensi'on de la Relatividad Especial: la Relatividad General en la
que se confunden las caracter'isticas de una teor'ia-marco a
cuyas reglas (covariancia general) deben adaptarse otras teor'ias
espec'ificas con las caracter'isticas de una teor'ia de la
interacci'on gravitatoria que afecta a todas las formas de
energ'ia. De este doble car'acter nacen la riqueza y todos los
problemas de interpretaci'on de esta teor'ia y su posible
versi'on cu'antica.

As'i, por ejemplo, es posible interpretar la Relatividad General
como una teor'ia que determina consistentemente el espacio-tiempo
en el que toda la F'isica se desarrolla ({\it geometrodin'amica})
y tambi'en como una teor'ia que describe la propagaci'on de un
campo que interact'ua d'ebilmente con toda la materia y consigo
mismo en un espacio-tiempo mincosquiano.  Este segundo punto de vista
surgi'o mucho despu'es del primero que es el originalmente adoptado
por Einstein al proponer su teor'ia y permitir'ia, en
principio, cuantizar la gravitaci'on siguiendo las pautas generales
de las Teor'ias de Campos Relativistas.

No es nuestra intenci'on revisar aqu'i la historia de los
diferentes intentos de llevar a cabo este programa que, aunque sin
'exito, no fueron infructuosos puesto que contribuyeron a poner los
fundamentos de la cuantizaci'on de las teor'ias con invariancias
locales. Lo que nos interesa es se~nalar que este camino lleva
indisolublemente asociada la idea del cuanto del campo gravitatorio:
el gravit'on, una part'icula sin masa de esp'in dos, que no
aparece en otras propuestas (gravitaci'on euclidiana, cuantizaci'on
por bucles etc.) sobre las que vamos a dar un salto en el tiempo y en
el espacio de teor'ias para llegar a las Teor'ias de Cuerdas,
en cuyo espectro s'i se encuentra. La importancia de este hecho es
que una part'icula sin masa de esp'in dos es siempre un
gravit'on, con acoplos que, a bajas energ'ias, son forzosamente
los predichos por la Relatividad General, por lo que la Teor'ia de
Cuerdas, originalmente utilizada para describir resonancias
hadr'onicas, es una teor'ia de gravitaci'on cu'antica, directa
heredera de los primeros intentos de cuantizar la gravitaci'on.

A finales de la d'ecada de los ochenta y principios de la de los
noventa, se hab'ia vuelto un lugar com'un el decir que la
Teor'ia de Cuerdas era la 'unica teor'ia de gravitaci'on
cu'antica consistente (ignorando problemas como la convergencia de
las series perturbativas). A'un cuando ofrec'ia ventajas como la
finitud de los diagramas, no resolv'ia ning'un problema (aparte
de la unificaci'on de las interacciones y part'iculas
elementales, y esto hasta cierto punto) ni hac'ia ninguna
predicci'on. Otras direcciones de investigaci'on basadas en el
aspecto geom'etrico de la gravitaci'on, sin embargo, hab'ian
dado como resultado que los agujeros negros tienen entrop'ia y
temperatura y emiten radiaci'on como si fuesen objetos negros que
tienen esa temperatura, pero sin poder ofrecer una interpretaci'on de
esa entrop'ia ni del mecanismo de radiaci'on y sin resolver el
problema de la aparente p'erdida de informaci'on en el Universo
implicada.

En la segunda mitad de los noventa, la comunidad de te'oricos de
cuerdas concentr'o sus energ'ias en la resoluci'on de estos
problemas elaborando modelos de agujero negro ``cuerd'istico'' en
los que los grados elementales de libertad dieran cuenta de la
entrop'ia y de la radiaci'on de los mismos.  Para ello hubo que
intensificar la investigaci'on en los aspectos geometrodin'amicos de
la teor'ia y hubo que hacer uso de nuevas herramientas: los
objetos extensos ({\it $p$-branas} etc.) y las dualidades. Hoy se
puede decir que la Teor'ia de Cuerdas ha conseguido su primer
'exito con la elaboraci'on de estos modelos pues, si bien los
modelos describen s'olo ciertos tipos de agujeros (extremos y
cuasi-extremos), no es menos cierto que ninguna otra teor'ia da
mejores modelos.

En estas lecciones se pretende revisar los problemas de la
entrop'ia y la informaci'on de los agujeros negros y c'omo son
resueltos en ciertos casos por la Teor'ia de Cuerdas, sin hacer
una innecesaria apolog'ia de la misma. El lector/oyente debe de
juzgar por s'i mismo si los resultados justifican los muchos
esfuerzos honestos que se han hecho para llegar a ellos o las
afirmaciones un tanto temerarias de algunos de los l'ideres de
este campo.

Este resumen no es, ni mucho menos, el 'unico que existe sobre este
tema y es obligado mencionar aqu'i a los m'as sobresalientes, a
los que debemos mucho. El m'as completo de todos, hasta la fecha, es
el de Peet Ref.~\cite{kn:Pee2}, que trata todos los temas de que vamos
a hablar. La tesis doctoral de J.~Maldacena Ref.~\cite{kn:M} es una
buena introducci'on pedag'ogica y el de Das y Mathur \cite{kn:DM} es
tambi'en razonablemente completo. Otros res'umenes interesantes por
su particular enfoque o como fuentes de bibliograf'ia son
Refs.~\cite{kn:Ho2,kn:Ho3,kn:M2,kn:Pee,kn:Ske,kn:BaKw,kn:Moh2,kn:Wad,kn:My}.

El plan general de estas lecciones es el siguiente: en la primera
estudiaremos las ideas b'asicas de la termodin'amica de los agujeros
negros, utilizando como ejemplos los agujeros negros de Schwarzschild
y de Reissner y Nordstr\"om. En la segunda lecci'on estudiaremos
estos mismos agujeros negros y otros m'as generales desde el punto de
vista de supersimetr'ia y supergravedad. En la tercera
comenzaremos el estudio de las teor'ias de cuerdas desde el punto
de vista de su acci'on efectiva y en la cuarta estudiaremos
soluciones fundamentales de estas acciones efectivas (que son acciones
de supergravedad) y c'omo construir con ellas soluciones de tipo
agujero negro.  Finalmente, en la quinta y 'ultima lecci'on veremos
c'omo, haciendo uso de las ideas desarrolladas en las lecciones
anteriores, con la teor'ia de cuerdas explicamos la entrop'ia
de una soluci'on concreta de agujero negro en cuatro dimensiones.

El ap'endice~\ref{sec-conventions} contiene nuestros convenios de
signatura, conexi'on, curvatura, 'indices etc.

\subsubsection{Agradecimientos}

Quisiera agradecer a los organizadores de esta escuela la oportunidad,
que con su invitaci'on me han dado, de participar en ella y tambi'en
manifestar mi agradecimiento a M.M.~Fern'andez por su ayuda y
est'imulo constantes.

Este trabajo ha sido financiado parcialmente por el proyecto espa~nol
FPA2000-1584.

\newpage
%%%%%%%%%%%%%%%%%%%%%%%%%%%%%%%%%%%%%%%%%%%%%%%%%%%%%%%%%%%%%%%%%%%%%%
%%%%%%%%%%%%%%%%%%%%%%%%%%%%%%%%%%%%%%%%%%%%%%%%%%%%%%%%%%%%%%%%%%%%%%

\mainmatter

%%%%%%%%%%%%%%%%%%%%%%%%%%%%%%%%%%%%%%%%%%%%%%%%%%%%%%%%%%%%%%%%%%%%%%
%%%%%%%%%%%%%%%%%%%%%%%%%%%%%%%%%%%%%%%%%%%%%%%%%%%%%%%%%%%%%%%%%%%%%%
\setcounter{page}{1}
\renewcommand{\thepage}{\arabic{page}}
\pagestyle{headings}
\chapter{Termodin'amica y acci'on euclidiana}

%%%%%%%%%%%%%%%%%%%%%%%%%%%%%%%%%%%%%%%%%%%%%%%%%%%%%%%%%%%%%%%%%%%%%%
\section*{Introducci'on}

En esta primera lecci'on vamos a repasar ideas y resultados bien
establecidos acerca de la termodin'amica de los agujeros negros de
Schwarzschild de Reissner y Nordstr\"om.  El agujero negro de
Schwarzschild es el arquetipo de agujero negro est'atico que se
esperar'ia encontrar en el Universo a escala macrosc'opica y uno
de los objetivos de cualquier candidato a teor'ia de gravitaci'on
cu'antica deber'ia de ser el reproducir estos resultados
cl'asicos a partir de sus grados de libertad y principios
fundamentales. El agujeros negro de Reissner y Nordstr\"om es el
arquetipo del agujero negro est'atico que est'a cargado con respecto
a otro campo (en este caso un campo electromagn'etico). Como tal, no
es relevante a escalas macrosc'opicas, pues los objetos
macrosc'opicos tienden a ser el'ectricamente neutros, pero puede
serlo a escalas microsc'opicas. Precisamente, la teor'ia de
cuerdas es capaz de dar una explicaci'on para el valor de la
entrop'ia de agujeros negros negros cargados en, o cerca de, el
{\it l'imite extremo} que los agujeros negros de Schwarzschild no
poseen.

Por razones de tiempo, y por ilustrar un m'etodo algo menos
convencional de calcular la temperatura y la entrop'ia de un
agujero negro, vamos a hacerlo {\it \`a la} Gibbons y
Hawking\footnote{Esta y otras referencias pueden encontrarse en
  Ref.~\cite{kn:GH5}.} \cite{kn:GH}. Para ver una introducci'on
pedag'ogica a la termodin'amica de los agujeros m'as convencional,
se pueden consultar las lecciones de E.~Verdager en esta escuela
\cite{kn:Ver}, los libros sobre Teor'ia Cu'antica de Campos en
espacios curvos \cite{kn:BiDa,kn:W10} el art'iculo \cite{kn:W8} y
el excelente libro de Novikov y Frolov \cite{kn:FN}. Tambi'en son
interesantes las lecciones de Townsend \cite{kn:Tow5}.

Para empezar, vamos a repasar las constantes y unidades relevantes
en los problemas que van a ocuparnos.

%%%%%%%%%%%%%%%%%%%%%%%%%%%%%%%%%%%%%%%%%%%%%%%%%%%%%%%%%%%%%%%%%%%%%%

\section{Preliminares}

En $d$ dimensiones, la acci'on de Einstein y Hilbert \cite{kn:H} para
el campo gravitacional acoplado a materia es \cite{kn:GH,kn:W}

\begin{equation}
\label{eq:EH1}
\begin{tabular}{|c|}
\hline \\  
$
S_{EH}[g] = \frac{c^{3}}{16\pi G_{N}^{(d)}} {\displaystyle\int_{\cal M}}
d^{d}x \sqrt{|g|}\ R(g) +(-1)^{d}\frac{c^{3}}{8\pi G_{N}^{(d)}}
{\displaystyle\int_{\partial {\cal M}}} d^{d-1}\Sigma\, {\cal K}
+S_{\rm materia}\, .
$
\\ \\ \hline
\end{tabular}
\end{equation}

\noindent $R(g)$ es el escalar de  Ricci de la m'etrica $g_{\mu\nu}$,
$G_{N}^{(d)}$ es la constante de Newton en $d$ dimensiones, ${\cal M}$
es la variedad $d$-dimensional sobre la que integramos y $\partial
{\cal M}$ es su frontera. ${\cal K}$ es la traza de la curvatura
extr'inseca de $\partial {\cal M}$ (v'ease el
ap'endice~\ref{sec-extrinsic}). Finalmente 

\begin{equation}
  \begin{array}{rcl}
d^{d-1}\Sigma & \equiv & n^{2} d^{d-1}\Sigma_{\rho} n^{\rho}\, ,  \\
& & \\
d^{d-1}\Sigma_{\rho} & = & {\textstyle\frac{1}{(d-1)!\sqrt{|g|}}}
\epsilon_{\rho\mu_{1}\cdots \mu_{d-1}} 
dx^{\mu_{1}}\wedge \ldots dx^{\mu_{d-1}}\, ,\\
\end{array}
\end{equation}

\noindent donde $n^{\mu}$ es el vector unitario normal a 
$\partial {\cal M}$.

Por definici'on la m'etrica $g_{\mu\nu}$ es adimensional y las
coordenadas tienen dimensiones de longitud. Las unidades de
$G_{N}^{(d)}$ en este sistema $c\neq 1$ son $M^{-1}L^{d-1}T^{-2}$.  El
factor convencional de $16\pi$ est'a asociado a unidades {\it
  racionalizadas} 'unicamente en $d=4$. Con estos convenios, la
fuerza gravitacional entre dos masas $m$ y $M$ en el l'imite
niutoniano es

\begin{equation}
\label{eq:gravforce}
\vec{F}= 
-\frac{8(d-3)\pi G_{N}^{(d)} mM}{(d-2)\omega_{d-2}} 
\frac{\vec{x}_{d-1}}{|\vec{x}_{d-1}|^{d-1}}\, ,
\end{equation}

\noindent donde $\omega_{d-2}$ es el volumen de la $(d-2)$ esfera
de radio unidad (v'ease el Ap'endice~\ref{sec-sph}).

En el exponente de la integral de Feynman 

\begin{equation}
{\cal Z} = \int Dg\ e^{+iS_{EH}/\hbar}\, ,
\end{equation}

\noindent tendr'iamos la combinaci'on adimensional

\begin{equation}
\frac{S_{EH}}{\hbar} =
\frac{2\pi}{\ell_{\rm Planck}^{d-2}}\int d^{d}x \ldots\, ,
\end{equation}

\noindent donde

\begin{equation}
\label{eq:Plancklength}
\frac{\ell_{\rm Planck}^{d-2}}{2\pi} =
\frac{16\pi G_{N}^{(d)} \hbar }{c^{3}}\, ,
\end{equation}

\noindent es la  {\it longitud de Planck} 
$d$-dimensional\footnote{A veces se usa la {\it longitud de Planck
    reducida}
\begin{equation}
\label{eq:reducedPlancklength}
{}^{-}\!\!\!\!\ell_{\rm Planck}=\frac{\ell_{\rm Planck}}{2\pi}\, .
\end{equation}
}. 'Esta es la 'unica combinaci'on de las
constantes $\hbar,c,G_{N}^{(d)}$ con dimensiones de longitud.
Sin embargo, si hay un objeto de masa $M$, hay dos combinaciones m'as
con dimensiones de longitud: la {\it longitud de onda Compton}
asociada al objeto

\begin{equation}
{}^{-}\!\!\!\!\lambda_{\rm Compton} = \frac{\hbar}{Mc}\, ,
\end{equation}

\noindent que es de naturaleza puramente mecano-cu'antica, y el
{\it radio de Schwarzschild} o {\it radio gravitacional}
$d$-dimensional

\begin{equation}
R_{s}=\left(\frac{16\pi MG_{N}^{(d)}c^{-2}}{(d-2)
\omega_{(d-2)}}\right)^{\frac{1}{d-3}}\, ,  
\end{equation}

\noindent de naturaleza puramente gravitacional y cl'asica.
${}^{-}\!\!\!\!\lambda_{\rm Compton}$ nos da una idea del ``tama~no
cu'antico'' y $R_{S}$ del ``tama~no gravitacional'' de un objeto de
masa $M$.

Con las constantes $\hbar,c,G_{N}^{(d)}$ tambi'en se puede construir
una combinaci'on con dimensiones de masa: la {\it masa de Planck}

\begin{equation}
M_{\rm Planck}   = 
\left(\frac{\hbar^{d-3}}{G_{N}^{(d)}c^{d-5}}\right)^{\frac{1}{d-2}}\, ,
\end{equation}

\noindent en t'erminos de la cual, el prefactor de la integral 
de Feynman es

\begin{equation}
\frac{c^{3}}{G_{N}^{(d)}\hbar} = 
\left(\frac{M_{\rm Planck}c}{\hbar}\right)^{d-2}\, .
\end{equation}

Evidentemente, en el sistema natural de unidades 
$\ell_{\rm Planck}=1/M_{\rm Planck}$.

Objetos cuya masa es del orden de la de Planck tienen una longitud de
onda Compton asociada que es del orden del radio de Schwarzschild del
objeto que, a su vez es del orden de la longitud de Planck:

\begin{equation}
M\sim M_{\rm Planck} \Rightarrow {}^{-}\!\!\!\!\lambda_{\rm Compton} 
\sim R_{s} \sim \ell_{\rm Planck}\, .
\end{equation}

Objetos con una masa mayor que la de Planck tienen una longitud de
onda Compton menor que su radio de Schwarzschild y se comportan como
agujeros negros bien descritos por la Relatividad General (RG),
mientras que si su masa es menor, no estar'an localizados dentro de
su radio de Schwarzschild y no se comportar'an como agujeros negros.
Adem'as a distancias menores que la longitud de Planck, los efectos
cu'anticos empiezan a ser importantes y la RG debe de ser reemplazada
por una teor'ia de gravitaci'on cu'antica.

Como veremos, en la Teor'ia de Cuerdas no hay una 'unica
constante que juega los papeles de constante de acoplo y escala de
longitud, como en la RG, sino que hay dos constantes: la constante de
acoplo de la cuerda $g$ y la longitud de la cuerda $\ell_{s}$ que se
combinan en la constante de Newton de acuerdo con
Ec.~(\ref{eq:GN10A}), y debemos comparar el radio de Schwarzschild de
los objetos con $\ell_{s}$, la escala a la que la Teor'ia de
Cuerdas como teor'ia de gravitaci'on cu'antica sustituye a la
RG.

%%%%%%%%%%%%%%%%%%%%%%%%%%%%%%%%%%%%%%%%%%%%%%%%%%%%%%%%%%%%%%%%%%%%%%
\section{El agujero negro de Schwarzschild}

Es una creencia firmemente establecida entre nuestra comunidad que los
agujeros negros macrosc'opicos (objeto de estudio de la
Astrof'isica) son el punto final del {\it colapso gravitacional} y
que, tras un periodo de tiempo m'as o menos largo, el colapso
gravitacional de una estrella neutra y sin momento angular, produce un
agujero negro de Schwarzschild est'atico y esf'ericamente
sim'etrico. Nosotros vamos a estar interesados en agujeros negros de
diversos tama~nos, sin hacer referencia a su posible origen
(primordial, cu'antico...).

La m'etrica del agujero negro de Schwarzschild es una soluci'on
cl'asica de las ecuaciones de Einstein en el vac'io

\begin{equation}
\label{eq:E}
R_{\mu\nu} -{\textstyle\frac{1}{2}} g_{\mu\nu} R = 0\, .
\Rightarrow R_{\mu\nu} = 0\, .
\end{equation}

Como es bien sabido, para resolver estas ecuaciones es necesario hacer
un {\it Ansatz} para la m'etrica que las simplifique. En este caso
queremos la m'etrica que describe el espacio-tiempo alrededor de un
objeto esf'ericamente sim'etrico y en reposo (en cierto sistema de
coordenadas) y podemos suponer que la m'etrica es
est'atica\footnote{Es decir: admite una vector de Killing temporal y
  el espacio-tiempo se puede foliar por hipersuperficies de car'acter
  espacial que son ortogonales a las 'orbitas del vector de Killing
  de forma que las hipersuperficies se pueden etiquetar por el
  par'ametro de estas 'orbitas.} y esf'ericamente sim'etrica. Una
m'etrica as'i, siempre puede escribirse de esta forma\footnote{En
  esta secci'on trabajamos en $d=4$.}:

\begin{equation}
\label{eq:statsphe}
\begin{array}{rcl}
ds^{2} & = & W (r) (dct)^{2} -W^{-1} (r) dr^{2}
-R^{2}(r)d\Omega^{2}_{(2)}\, , \\
& & \\
d\Omega^{2}_{(2)} & = & d\theta^{2} +\sin^{2}\theta d\varphi^{2}\, , \\
\end{array}
\end{equation}

\noindent donde $W(r)$ y $R(r)$ 
son dos funciones de la coordenada radial $r$ a determinar y donde
$d\Omega_{(2)}^{2}$ es la m'etrica de la 2-esfera unidad $S^{2}$
(v'ease el Ap'endice~\ref{sec-sph}).  Substituyendo este Ansatz en
las ecuaciones de Einstein, se encuentra inmediatemente una 'unica
soluci'on de $W$ y $R$ con dos constantes de integraci'on. Una de
ellas se fija imponiendo que el espacio-tiempo sea {\it
  asint'oticamente plano}, es decir, que la m'etrica se aproxime
cuanto queramos a la de Minkowski para valores suficientemente altos
de la coordenada radial. Esto es lo mismo que imponer que nuestra
soluci'on describa un sistema aislado en el que la fuente del campo
esta confinada en una regi'on finita. La otra constante de
integraci'on tiene dimensiones de longitud y la denotamos por
$\omega$. El resultado es la soluci'on de Schwarzschild
\cite{kn:Schw} en el {\it sistema de coordenadas de Schwarzschild}
$\{t,r,\theta,\varphi\}$

\begin{equation}
\label{Schwar}
\begin{tabular}{|c|}
\hline \\
${\displaystyle
ds^{2} = W (dct)^{2}-W^{-1}dr^{2} -r^{2} d\Omega^{2}_{(2)}\, ,
\hspace{1cm} W=1+\frac{\omega}{r}\, .}$
\\ \\ \hline
\end{tabular}
\end{equation}

Veamos ahora algunas de las propiedades de esta soluci'on.

%%%%%%%%%%%%%%%%%%%%%%%%%%%%%%%%%%%%%%%%%%%%%%%%%%%%%%%%%%%%%%%%%%%%%%

\subsection{Propiedades generales}

\begin{enumerate}
  
\item {\it La soluci'on de Schwarzschild es la 'unica soluci'on
    esf'ericamente sim'etrica (est'atica o no) de la ecuaci'on
    $R_{\mu\nu}=0$}. Este es el Teorema de Birkhoff \cite{kn:Birk5}.
  Demostraciones simples de este teorema se pueden encontrar en
  \cite{kn:MTW,kn:CW}.
  
\item La soluci'on de Schwarzschild es estable frente a peque~nas
  perturbaciones gravitacionales y de otros campos externos
  \cite{kn:chan}: las perturbaciones desaparecen con el tiempo, siendo
  transportadas por ondas (gravitacionales o de otro tipo) hacia el
  $r\rightarrow \infty$ o $r\rightarrow 0$.
  
\item La constante de integraci'on $\omega$ es, en principio,
  arbitraria. Su significado es el siguiente: para valores de $r$ muy
  grandes, donde el campo gravitacional es d'ebil, las trayectorias
  de las part'iculas de prueba (geod'esicas de este
  espacio-tiempo) son aproximadamente las 'orbitas keplerianas que
  describir'ian esas mismas part'iculas si estuviesen
  sometidas al campo gravitacional niutoniano producido por un objeto
  (puntual o esf'ericamente sim'etrico) de masa

  \begin{equation}
   M= -\frac{\omega c^{2}}{2G_{N}^{(4)}}\, ,\,\,\,
   \Rightarrow
   \omega= -R_{S}\, ,
  \end{equation}
  
  situado en el origen de coordenadas.  Por lo tanto, $M$ se puede
  interpretar como la masa del objeto descrito por la soluci'on de
  Schwarzschild. A veces recibe el nombre de {\it masa ADM}, porque en
  la formulaci'on can'onica de Arnowitt, Deser y Misner
  \cite{kn:ADM} de la RG, aparece como la
  energ'ia total, y se puede calcular usando la f'ormula ($c=1$)

\begin{equation}
M = \frac{1}{8\pi G_{N}^{(4)}} \int_{S^{2}_{\infty}} d^{2}S_{i} 
\left(\partial_{j}g_{ij} -\partial_{i} g_{jj}\right)\, ,
\end{equation}

\noindent donde la integral se hace sobre la 2-esfera en el infinito
definida por $t={\rm constante}\, ,\,\,\, r=\infty$ y los 'indices
$i,j=1,2,3$ corresponden a las tres coordenadas espaciales.

\item Como conclusi'on de lo anterior, podemos decir que la
  soluci'on de Schwarzschild describe el campo gravitacional creado
  por un objeto masivo, esf'ericamente sim'etrico tal y como es
  visto por un observador alejado de tal objeto (pues est'a en la
  regi'on de vac'io) y est'atico con respecto a tal objeto. A
  este observador est'an ligadas las coordenadas de Schwarzschild
  $\{t,r,\theta,\varphi\}$.
  
\item Normalmente la m'etrica de Schwarzschild se usa desde
  $r=\infty$ hasta un determinado valor de $r=r_{e}$ y all'i se
  contin'ua (``pega'') con otras m'etricas est'aticas y
  esf'ericamente sim'etricas que son soluciones de las ecuaciones de
  Einstein en presencia de materia

\begin{equation}
\label{eq:2}
R_{\mu\nu} -{\textstyle\frac{1}{2}} g_{\mu\nu} R = 
{\textstyle\frac{8\pi G_{N}^{(4)}}{c^{4}}}T_{{\rm materia}\, \mu\nu}\, .
\end{equation}

\noindent ({\it soluciones de Schwarzschild interiores}) que describen
el espacio-tiempo en el interior de diversas estrellas de radio
$r_{e}$ mientras que la soluci'on de Schwarzschild describe el
exterior de todas ellas (por el Teorema de Birkhoff)\footnote{Aunque
  esto es lo que uno esperar'ia siempre, es notoria la ausencia
  de soluciones interiores de Kerr.}

\item La m'etrica es singular (es decir ${\rm det}\ g_{\mu\nu}=0$ o
  ciertos componentes de la m'etrica divergen) en $r=0,R_{S}$.  La
  soluci'on de Schwarzschild es f'isicamente aceptable para
  valores grandes de $r$, pero no podemos tomarla en serio m'as
  all'a de la m'as externa de estas singularidades $r=R_{S}$.

Estas singularidades pueden ser f'isicas o debidas a una
elecci'on inapropiada de coordenadas (como es el caso de la
singularidad en el origen de la m'etrica euclidiana en
coordenadas esf'ericas). Para determinar la naturaleza de estas
singularidades, es necesario hacer un an'alisis de los invariantes de
curvatura y de las geod'esicas de la m'etrica en $r=0,R_{S}$.

\begin{itemize}
  
\item Obviamente, $R=0$ por doquier (pues la m'etrica es
  soluci'on de $R_{\mu\nu}=0$) excepto, quiz'a, en $r=0$, donde
  algunas funciones no son derivables.  Examinando invariantes de
  orden superior, que no son cero, se concluye que hay una
  singularidad de curvatura en $r=0$, pero no en $r=R_{S}$.
  
\item Si estudiamos el movimiento de observadores en ca'ida libre
  en la direcci'on radial, el resultado es que los observadores
  alcanzan el radio de Schwarzschild en un tiempo propio finito,
  aunque, en tiempo de Schwarzschild $t$ (el tiempo propio de un
  observador en reposo) tarde un tiempo infinito. Esto, junto a la
  finitud de las fuerzas de marea\footnote{Adem'as de ser finitas, son
    peque~nas para espacios-tiempos de Schwarzschild con $M$ grande.
    Sin embargo, 'este puede no ser un comportamiento universal
    \cite{kn:HoRo,kn:HoRo2}.} en $r=R_{S}$ sugiere que si
  utiliz'asemos el tiempo propio del observador en ca'ida libre
  como coordenada, la m'etrica resultante ser'ia regular
  ah'i.

\end{itemize}

Esta es esencialmente la idea en que se basan las coordenadas de
Eddington y Finkelstein \cite{kn:Edd,kn:Fink} $\{v,r,\theta,\varphi\}$
en las que la m'etrica de Schwarzschild toma la forma

\begin{equation}
ds^{2}= \left(1-\frac{R_{s}}{r} \right)dv^{2} 
-2dvdr -r^{2}d\Omega_{(2)}^{2}\, ,
\end{equation}
  
\noindent donde la coordenada $v$ viene dada en t'erminos 
de las $t$ y $r$ de Schwarzschild por

  \begin{equation}
  v=ct +r +R_{S}\log{|1-\frac{R_{s}}{r}|}  \, ,
  \end{equation}
  
\noindent  y es constante para geod'esicas radiales tipo luz (las 
trayectorias de fotones). La m'etrica es regular en la regi'on
$r>R_{S}$, pero tambi'en en $0<r\leq 0$ (la singularidad en $r=0$
permanece) y {\it extienden anal'iticamente} la soluci'on de
Schwarzschild a esta regi'on, permiti'endonos estudiar lo que pasa
en $r=R_{S}$.

Nosotros vamos a hacerlo usando las coordenadas de Kruskal y Szekeres
\cite{kn:Kr,kn:Sz} $\{T,X,\theta,\varphi\}$ que proporcionan la
m'axima extensi'on anal'itica del espacio-tiempo de
Schwarzschild (cuadrante I la Figura~\ref{fig:kruskal}), incluyendo
nuevas regiones (cuadrantes II, II y IV).  La m'etrica de
Schwarzschild en estas coordenadas es

\begin{equation}
\label{eq:kruskal}
ds^{2} = \frac{4 R_{S}^{3} e^{-r/R_{S}}}{r} \left[(d\ cT)^{2} -dX^{2}\right]
-r^{2}d\Omega^{2}_{(2)}\, ,
\end{equation}

\noindent donde $r$ es una funci'on de $T$ y  $X$ dada 
impl'icitamente por las transformaciones de coordenadas entre el par
$(t,r)$ y el  $(T,X)$:

\begin{eqnarray}
\left( \frac{r}{R_{S}} -1\right)e^{r/R_{S}} & = &
X^{2} -c^{2} T^{2}\, , \nonumber \\
& &  \label{eq:kruskaltrans} \\
\frac{ct}{R_{S}}  =  \ln \left( \frac{X +cT}{X -cT} \right) & = &
2\ {\rm arcth}\ (cT/X)\, .
\end{eqnarray}

\begin{figure}[!ht]
\begin{center}
\leavevmode
\epsfxsize= 10cm
\epsfysize= 8cm
\epsffile{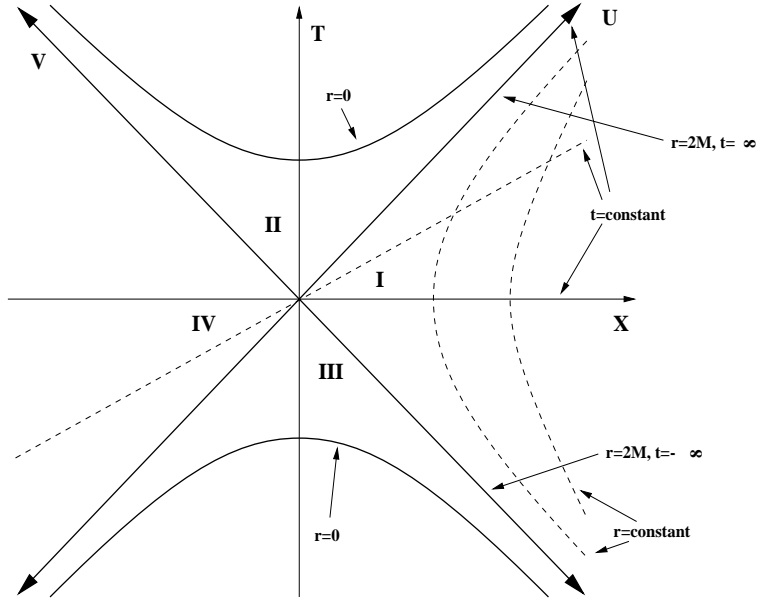}
\caption{El espacio-tiempo de  Schwarzschild en coordenadas 
  de Kruskal y Szekeres. Cada punto representa una 2-esfera de radio
  $r$.}
\label{fig:kruskal}
\end{center}
\end{figure}

En la Figura~\ref{fig:kruskal}) est'a representado el espacio-tiempo
de Schwarzschild en el plano $T,X$. Cada punto corresponde a una
esfera de radio $r(T,X)$. El tiempo de Schwarzschild $t$ es una
coordenada angular en este diagrama y las l'ineas de radio
constante se asemejan a hip'erbolas que se unen asint'oticamente a
las rectas $X=\pm T$. En este diagrama, los conos de luz son
exactamente como en el espacio de Minkowski, con las geod'esicas tipo
luz formando un 'angulo de $\pi/4$ con los ejes de coordenadas.

Hay tres puntos de particular inter'es para nosotros:

\begin{enumerate}
\item En $r=0$ (que en este diagrama es una hipersuperficie
  tridimensional de tipo espacio) seguimos encontrando la
  singularidad, como era de esperar.
\item La hipersuperficie $t=+\infty,r=R_{S}$ llamada {\it horizonte de
    eventos}, que es de tipo luz, s'olo se puede atravesar en un
  sentido (hacia el cuadrante II) y divide al espacio-tiempo de
  Schwarzschild en un {\it interior} en el que est'a la singularidad
  y del que no puede llegar ninguna se~nal al {\it exterior}. Por eso
  se le da al objeto descrito por la soluci'on completa de
  Schwarzschild (sin soluci'on interior que corresponda a una
  estrella) el nombre de {\it agujero negro}. Obs'ervese que la
  existencia del horizonte que nos separa de la singularidad depende
  de que la masa $M$ sea positiva.
\item La hipersuperficie $t=-\infty,r=R_{S}$ s'olo se puede atravesar
  en desde el cuadrante III al I. Un observador en ese cuadrante puede
  ver todo tipo de objetos provenientes de ese cuadrante y por ello se
  dice que esta parte del espacio-tiempo es un {\it agujero blanco}.
\end{enumerate}

\item Sabemos que en el Universo hay muchos objetos cuyo campo
  gravitacional externo est'a bien descrito por la parte $r> R_{S}$
  de la m'etrica de Schwarzschild, pero ?`qu'e tipo de objeto da
  lugar a la m'etrica incluyendo la regi'on $r\leq R_{S}$, es decir,
  la m'etrica del agujero negro?
  
  Para responder a esta pregunta nos vemos forzados a inventar un
  nuevo objeto: el agujero negro que es, por definici'on el objeto
  cuya m'etrica posee como caracter'istica principal un horizonte
  de eventos.
  
  ?`C'omo se originan los agujeros negros (si es que los hay) en el
  Universo? En el libro de Thorne \cite{kn:Th} se narra c'omo en un
  proceso que dur'o casi cincuenta a~nos, la comunidad
  cient'ifica lleg'o a la conclusi'on de que los agujeros negros
  pod'ian originarse en el {\it colapso gravitacional} de
  estrellas muy masivas y que, adem'as, este colapso es inevitable si
  la estrella tiene una masa varias veces la del Sol. Adem'as se ha
  considerado la posibilidad de que se formen en fen'omenos violentos
  como el Big Bang (agujeros negros {\it primordiales}).
  
  Evidentemente este espacio-tiempo completo no puede surgir del
  colapso gravitacional de ning'un objeto. Se dice que representa un
  {\it agujero negro eterno}. La Figura~\ref{fig:collap1} describe la
  formaci'on de un agujero negro de Schwarzschild por colapso
  gravitacional en coordenadas de tipo Kruskal-Szekeres. No hay
  agujero blanco ni las regiones III y IV. El agujero negro aparece
  cuando la estrella se contrae por debajo de su radio gravitacional.

\begin{figure}[!ht]
\begin{center}
\leavevmode
\epsfxsize= 5cm
\epsffile{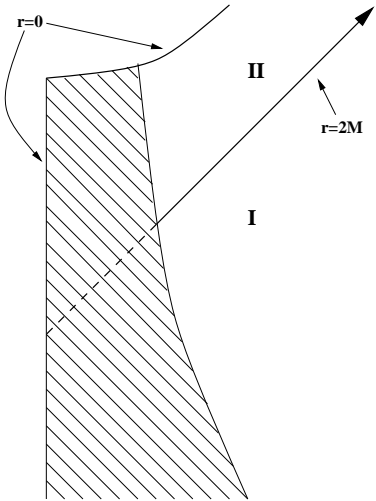}
\caption{Espacio-tiempo del colapso gravitacional de una estrella.}
\label{fig:collap1}
\end{center}
\end{figure}

\item Para estudiar las relaciones de causalidad en este
  espacio-tiempo s'olo se necesita la estructura de los conos de luz.
  Esta estructura es preservada por transformaciones conformes de la
  m'etrica. Se puede hacer una transformaci'on conforme que ``traiga
  el infinito a una distancia finita'' en la m'etrica transformada.
  El diagrama resultante (Figura~\ref{fig:penrose}) es un {\it diagrama
    de Penrose} y en 'el es f'acil identificar horizontes y ver
  qu'e pasa cuando prolongamos infinitamente las geod'esicas. Por
  ejemplo, vemos que cualquier geod'esica que atraviese el horizonte
  de eventos desde el cuadrante I cae inevitablemente en la
  singularidad.

\begin{figure}[!ht]
\begin{center}
\leavevmode
\epsfxsize= 7cm
\epsffile{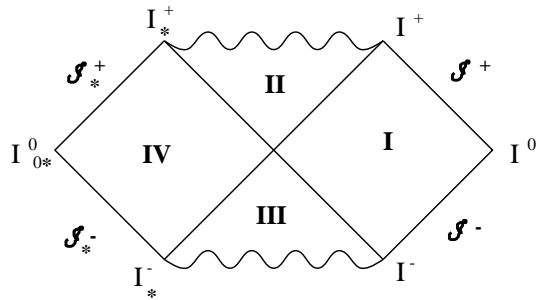}
\caption{Diagrama de Penrose del espacio-tiempo de Schwarzschild.}
\label{fig:penrose}
\end{center}
\end{figure}

\item Cuando $M$ es negativa, no hay horizonte que impida que la
  singularidad en $r=0$ sea ``vista'' por observadores externos (la
  singularidad est'a {\it desnuda}). El diagrama de Penrose
  correspondiente est'a en la Figura~\ref{fig:penronak}).
  
  Esto plantea numerosos problemas y para evitarlos podemos aducir que
  esta situaci'on nunca se va a originar en el colapso gravitacional
  de una estrella ordinaria (o de materia con propiedades de
  positividad de su tensor energ'ia-momento f'isicamente
  aceptables). Esta es la esencia de la {\it hip'otesis del censor
    c'osmico} de Penrose. Esta hip'otesis est'a fuertemente
  relacionada con la positividad de la energ'ia: dado que la
  energ'ia de ligadura gravitatoria es negativa, cuando una nube
  de materia empieza a comprimirse bajo el efecto de su propia
  gravitaci'on, su energ'ia total disminuye y podr'ia llegar
  a ser negativa. Antes de que esto ocurra, ha de formarse un
  horizonte de eventos.

\begin{figure}[!ht]
\begin{center}
\leavevmode
\epsfxsize= 4cm
\epsffile{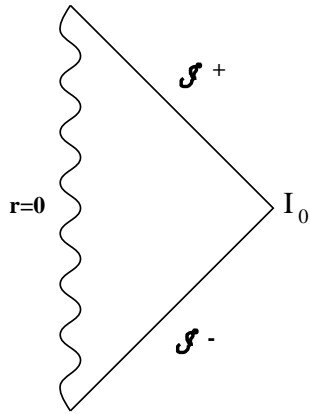}
\caption{El diagrama de Penrose del espacio-tiempo 
  t'ipico de una singularidad desnuda. Obs'ervese que la
  singularidad es de tipo tiempo.}
\label{fig:penronak}
\end{center}
\end{figure}

\item Adem'as del agujero negro de Schwarzschild, debe de haber otros
  tipos de agujeros negros: aquellos correspondientes a estadios
  intermedios del colapso gravitacional de una estrella, o aquellos
  que resultan de perturbar un agujero negro de Schwarzschild.
  Adem'as, dado que hay muchos posibles estados de una estrella,
  parece l'ogico pensar que su colapso gravitacional debe de dar
  lugar a agujeros negros distintos.
  
  El an'alisis de las perturbaciones del agujero negro de
  Schwarzschild \cite{kn:Pr1,kn:Pr2} demuestra, al contrario, que,
  tras un tiempo suficientemente largo, el agujero negro acaba siendo
  el de Schwarzschild\footnote{O el de Kerr, descrito 'unicamente por
    la masa $M$ y el momento angular $J$. Por simplicidad, vamos a
    ignorar en la mayor parte de nuestra discusi'on el momento
    angular.}, descrito 'unicamente por la masa $M$,
  independientemente del estado inicial del colapso gravitacional o de
  la perturbaci'on a que haya sido sometido. Todos los momentos
  multipolares del campo gravitacional\footnote{Cuadrupolar y
    superiores. El monopolar es la masa $M$ y el dipolar el momento
    angular $J$.} o del campo electromagn'etico\footnote{Dipolar y
    superiores, El monopolar es la carga el'ectrica $Q$.} son
  radiados al infinito de forma que el agujero negro resultante es
  siempre un agujero negro de Schwarzschild ($M\neq 0$, $Q,J=0$), Kerr
  ($M,J\neq 0$, $Q=0$) o de Kerr y Newmann ($M,J,Q\neq 0$). En el caso
  de un campo escalar, todos su momentos son radiados.
  
  Sin embargo, podr'ia haber soluciones de tipo agujero negro con
  estos momentos o con un campo escalar, aunque no se pudieran generar
  por colapso gravitacional. Pero se puede demostrar que no existen
  (teoremas de unicidad: dos referencias generales son
  \cite{kn:He,kn:He3}), y que el 'unico agujero negro sin momento
  angular u otros campos externos es el de Schwarzschild \cite{kn:I},
  sin momento angular pero con carga el'ectrica es el de Reissner y
  Nordstr\"om \cite{kn:I2} y con masa y momento angular es el de Kerr
  \cite{kn:Ca2,kn:W3}.  Adem'as, no hay agujeros negros con un campo
  escalar que no sea constante\footnote{Esta afirmaci'on ser'a
    matizada y precisada m'as adelante.}
  \cite{kn:Chas,kn:B4,kn:MB,kn:SZ}.
  
  Esto no quiere decir que no haya soluciones con momentos superiores
  de los campos gravitacionales o electromagn'eticos o que no haya
  soluciones con campos escalares no-triviales. Las hay, pero no son
  agujeros negros y tienen singularidades desnudas. Un ejemplo de
  familia de soluciones est'aticas y esf'ericamente sim'etricas con
  un campo escalar no-trivial es \cite{kn:JNW,kn:ALC}
  ($c=G_{N}^{(4)}=1$):

\begin{equation}
\label{eq:JNWALCsolutions}
\begin{tabular}{|c|}
\hline \\
$
\begin{array}{rcl}
ds^{2} & = & W^{\frac{2M}{\omega}-1}Wdt^{2} 
-W^{1-\frac{2M}{\omega}}\left[ W^{-1}dr^{2} 
+r^{2}d\Omega_{(2)}^{2}\right]\, ,  \\
& & \\
\varphi & = & \varphi_{0} +\frac{\Sigma}{\omega}\ln W\, ,\\
& & \\
W & = & 1+{\displaystyle\frac{\omega}{r}}\, ,
\hspace{1cm}
\omega  =  \pm 2\sqrt{M^{2} + \Sigma^{2}}\, .\\
\end{array}
$
\\ \\ \hline
\end{tabular}
\end{equation}

Las soluciones de esta familia vienen dadas por tres par'ametros
completamente independientes: la masa $M$, la carga escalar $\Sigma$ y
el valor del escalar en el infinito $\varphi_{0}$. Este 'ultimo no es
un par'ametro din'amico y no est'a excluido por los resultados
anteriores, pero $\Sigma$ s'i. Sin embargo, la 'unica soluci'on
de esta familia que es un agujero negro es la que tiene $\Sigma=0$
(Schwarzschild)\footnote{\label{foot:continuidad}Aqu'i tenemos
  tambi'en un primer ejemplo de un fen'omeno que os encontraremos a
  menudo: la familia de soluciones depende de par'ametros continuos,
  pero las propiedades f'isicas no son funciones continuas de esos
  par'ametros.}.

La conclusi'on es que no puede haber agujeros negros que tengan otras
caracter'isticas (``pelos'') distintas de $M,J,Q$ (y, en general,
cargas conservadas localmente). Aunque esto no ha sido completamente
demostrado para todos los casos \cite{kn:B2,kn:He2}, hay un consenso
general sobre que los agujeros negros estacionarios {\it no tienen
  pelos} \cite{kn:RW}. Nos gustar'ia hacer dos comentarios a esta
afirmaci'on:

\begin{enumerate}
\item Dado que la presencia de pelos est'a asociada a la ausencia de
  un horizonte de eventos, esta conjetura de que los agujeros negros
  no tiene pelos, est'a 'intimamente ligada a la del censor
  c'osmico: en el colapso gravitacional, para que se forme un
  horizonte, han de desaparecer por radiaci'on todos los momentos
  superiores de los campos presentes. La censura c'osmica est'a
  ligada a la positividad de la energ'ia y, de hecho, hay agujeros
  negros con pelo escalar si se permite que el campo escalar tenga
  energ'ia negativa. Agujeros negros no estacionarios con pelo
  escalar de energ'ia positiva existen \cite{kn:O}, pero las
  conjeturas sobre censura c'osmica y calvicie nos dicen que deben de
  evolucionar de forma que el pelo desaparezca por radiaci'on antes de
  llegar a un estado estacionario. Esto es posible porque la ``carga
  escalar'' no est'a sujeta a una ley de conservaci'on.
\item Dado que a trav'es del colapso gravitacional de muchos sistemas
  distintos se llega siempre a los mismos agujeros negros
  caracterizados por muy pocos par'ametros, cabe preguntarse qu'e ha
  pasado con toda la informaci'on sobre el estado original del
  sistema (el {\it problema de la informaci'on}) y cabe atribuir a
  los agujeros negros una entrop'ia muy grande que deber'iamos
  poder calcular si conoci'esemos los estados del agujero negro (el
  {\it problema de la entrop'ia}). Para resolver estos problemas
  necesitamos una teor'ia cu'antica de la gravitaci'on.
\end{enumerate}

\item El horizonte de eventos es una hipersuperficie de tipo luz cuyas
  secciones de $t$ constante tienen la topolog'ia de una 2-esfera.
  Esta es la 'unica topolog'ia permitida por los {\it teoremas de
    censura topol'ogica} Refs.\cite{kn:Haw7,kn:FSW} que, como
  siempre, utilizan como hip'otesis condiciones de positividad de la
  energ'ia. No es, por lo tanto, sorprendente que en presencia de
  una constante cosmol'ogica negativa se posible encontrar {\it
    agujeros negros topol'ogicos} cuyos horizontes pueden tener la
  topolog'ia de una superficie de Riemann compacta cualquiera
  \cite{kn:Va}.

\item El 'area de las secciones de $t$ constante 
del horizonte de eventos es

\begin{equation}
A = \int_{r=R_{S}} d\theta d\varphi\ r^{2} = 4 \pi R_{S}^{2}\, .
\end{equation}

Hawking demostr'o en Ref.~\cite{kn:Haw3} que las ecuaciones de
Einstein implican que $A$ nunca disminuye con el tiempo. Adem'as, si
dos agujeros negros se unen, el 'area del agujero negro resultante es
mayor que la suma de los de los dos iniciales.

Hay una clara analog'ia entre $A$ y la entrop'ia de un sistema
termodin'amico \cite{kn:Ch,kn:B,kn:B5,kn:B6,kn:B7}, aunque, de
momento, esto podr'ia ser s'olo una coincidencia.

\item La {\it gravedad superficial} $\kappa$ del horizonte es una
  cantidad que es constante sobre todo 'el (tambi'en en casos m'as
  generales \cite{kn:Haw,kn:Ca,kn:BCH}). En esto es similar a la
  temperatura de un sistema en equilibrio termodin'amico.
  F'isicamente es la fuerza que hace falta ejercer en el infinito
  para mantener una unidad de masa en reposo cuando $r\rightarrow
  R_{S}$ y tiene dimensiones de aceleraci'on $LT^{-2}$.  Se puede
  calcular usando la f'ormula

\begin{equation}
\label{eq:surgrav}
\kappa^{2} = - {\textstyle\frac{1}{2}}
\left. (\nabla^{\mu} k^{\nu})  (\nabla_{\mu} k_{\nu})
\right|_{\rm horizonte}\, ,
\end{equation}

\noindent donde $k^{\mu}$ es el vector de Killing temporal que 
es normal al horizonte (o, mejor, a sus secciones $t$ constante). En
el caso de m'etricas esf'ericamente sim'etricas, que se pueden
escribir de esta forma

\begin{equation}
\label{eq:genesph}
ds^{2} = g_{tt}(r) dt^{2} +g_{rr}(r) dr^{2} -r^{2} d\Omega^{2}_{(2)}\, ,
\end{equation}

\noindent $k^{\mu}=\delta^{\mu t}$ y

\begin{equation}
\label{eq:kappaesph}
\kappa = {\textstyle\frac{1}{2}}
\frac{\partial_{r} g_{tt}}{\sqrt{- g_{tt} g_{rr}}}  \, ,
\end{equation}

\noindent que, para Schwarzschild vale

\begin{equation}
\kappa = \frac{c^{4}}{4 G_{N}^{(4)}M}\, .
\end{equation}

\end{enumerate}

%%%%%%%%%%%%%%%%%%%%%%%%%%%%%%%%%%%%%%%%%%%%%%%%%%%%%%%%%%%%%%%%%%%%%%%%%%%%%

\subsection{Termodin'amica}
\label{sec-Schtherm}

En la secci'on anterior hemos visto que, de acuerdo con las
ecuaciones de Einstein, hay dos magnitudes, el 'area $A$ y la
gravedad superficial $\kappa$ del horizonte de un agujero negro que se
comportan en ciertos aspectos como la entrop'ia $S$ y la
temperatura $T$ de un sistema termodin'amico. Desde este punto de
vista, los teoremas de Hawking sobre el \cite{kn:Haw3} se pueden
interpretar como partes de la {\it segunda ley de la Termodin'amica
  de los agujeros negros}.

En un sistema termodin'amico $S,T$ y la energ'ia $E$ est'an
relacionadas por la {\it primera ley de la Termodin'amica}:

\begin{equation}
dE = TdS\, .
\end{equation}

Si hemos de tomar completamente en serio la analog'ia
termodin'amica es necesario demostrar que $\kappa$ y $A$ est'an
relacionados con el an'alogo de la energ'ia $E$ (que en un
agujero negro es, evidentemente $Mc^{2}$) de la misma forma:

\begin{equation} dM \sim {\textstyle\frac{1}{G_{N}^{(4)}}}\kappa dA\, .
\end{equation}

Sorprendentemente, esta relaci'on es correcta. El coeficiente de
proporcionalidad se puede calcular \cite{kn:Ch,kn:B,kn:Sm} y la {\it
  primera ley de la Termodin'amica de los agujeros negros} toma la
forma

\begin{equation}
\label{eq:1law}
dM = {\textstyle\frac{1}{8\pi G_{N}^{(4)}}} \kappa dA\, .
\end{equation}

Una versi'on integral de esta relaci'on de puede comprobar inmediatamente
para Schwarzschild: la {\it f'ormula de Smarr} \cite{kn:Sm}

\begin{equation}
\label{eq:smarr}
M = {\textstyle\frac{1}{4\pi G_{N}^{(4)}}} \kappa A\, .
\end{equation}

Estas dos relaciones, convenientemente generalizadas para tener en
cuenta otras cantidades conservadas (momento angular y carga
el'ectrica), siguen siendo ciertas bajo condiciones muy generales
\cite{kn:BCH} (v'eanse tambi'en \cite{kn:W2,kn:HS,kn:GKK}).

Este sorprendente conjunto de analog'ias sugiere la
identificaci'on de $A$ y $\kappa$ con la entrop'ia y la
temperatura del agujero negro. Estimulados por estas ideas, los
autores de Ref.~\cite{kn:BCH} aventuraron, dando algunos argumentos,
que tambi'en hay una {\it tercera ley de la Termodin'amica de los
  agujeros negros}: ``es imposible reducir $\kappa$ a cero a trav'es
de una secuencia finita de operaciones''. Varios ejemplos
espec'ificos fueron estudiados por Wald en Ref.~\cite{kn:W11}
Discutiremos esta ley m'as a fondo cuando estudiemos el agujero negro
de Reissner y Nordstr\"om.

Sin embargo, estas analog'ias no son suficiente para establecer
una identificaci'on completa. Los propios autores de
Ref.~\cite{kn:BCH} dicen

\begin{quote}
  
  {\it It can be seen that $\frac{\kappa}{8\pi}$ is analogous to the
    temperature in the same way that $A$ is analogous to the entropy.
    It should however be emphasized that $\frac{\kappa}{8\pi}$ and $A$
    are distinct from the temperature and entropy of the BH.
    
    In fact the effective temperature of a BHs is absolute zero. One
    way of seeing this is to note that a BH cannot be in equilibrium
    with black body radiation at any non-zero temperature, because no
    radiation could be emitted from the hole whereas some radiation
    would always cross the horizon into the BH.  }
\end{quote}

Por otro lado, en las identificaciones $A\sim S\, , \kappa\sim T$ hay
que determinar el valor de las constantes de proporcionalidad.

El descubrimiento de Hawking \cite{kn:Haw2} de que, la existencia de
un horizonte de eventos en un espacio-tiempo en el que hay campos
cu'anticos \footnote{El c'alculo original de Hawking es {\it
    semicl'asico}: la m'etrica es cl'asica y s'olo los otros
  campos son cuantizados. El horizonte (la m'etrica) da lugar a la
  radiaci'on, pero no se tiene en cuenta la reacci'on de la
  m'etrica frente a la radiaci'on. Por otro lado, es v'alido para
  todos los agujeros de la familia de Kerr y Newmann con $M,J,Q$.},
hace que los agujeros negros irradien cuantos de estos campos con el
espectro de energ'ias de un agujero negro con
temperatura\footnote{En nuestras unidades, la constante de Boltzmann
  $k_{B}=1$ y adimensional, con lo que $T$ tiene dimensiones de
  energ'ia y $S$ es adimensional.}

\begin{equation}
\label{eq:Tkappa}
T= {\displaystyle\frac{\hbar\kappa}{2\pi c}}\, ,
\end{equation}

\noindent cambi'o radicalmente esta situaci'on pues desaparece 
el obst'aculo mencionado en \cite{kn:BCH} y la constante de
proporcionalidad entre $\kappa$ y $T$ queda completamente determinada
y, por lo tanto

\begin{equation}
\label{eq:Sarea}
S = {\displaystyle\frac{Ac^{3}}{4\hbar G_{N}^{(4)}}}\, ,
\end{equation}

\noindent que se puede reescribir as'i:

\begin{equation}
S =\frac{1}{32\pi^{2}} {\displaystyle\frac{A}{\ell_{\rm Planck}^{2}}}\, ,
\end{equation}

\noindent esto es: esencialmente el 'area del horizonte medido en unidades
planquianas. Obs'ervese que la presencia de $\hbar$ en $T$ pone de
manifiesto su origen mecano-cu'antico. Este es un n'umero enorme
para agujeros negros astrof'isicos, como cab'ia esperar de los
argumentos que dimos cuando discutimos la pilosidad de los agujeros
negros.

\begin{figure}[!ht]
\begin{center}
\leavevmode
\epsfxsize= 4cm
\epsffile{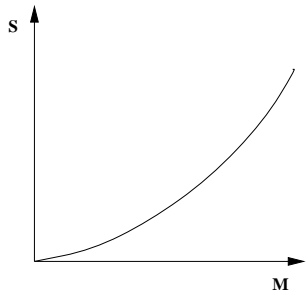}
\caption{$S$ frente a $M$ para el agujero negro de Schwarzschild.}
\label{fig:sschwarz}
\end{center}
\end{figure}

En particular, para el agujero de Schwarzschild tenemos (v'eanse las
Figuras \ref{fig:tschwarz} y \ref{fig:sschwarz})

\begin{equation}
T= {\displaystyle\frac{\hbar c^{3}}{8\pi G_{N}^{(4)}M}}\, ,
\hspace{1cm}
S = {\displaystyle\frac{4\pi G_{N}^{(4)} M^{2}}{\hbar c}}\, ,
\end{equation}

\noindent y la primera ley de la termodin'amica de los agujeros negros
y la f'ormula de Smarr toman la forma

\begin{equation}
dMc^{2} = TdS\, ,
\hspace{1cm}
Mc^{2} = 2TS\, .
\end{equation}

La termodin'amica de los agujeros negros presenta varios problemas
o peculiaridades:

\begin{figure}[!ht]
\begin{center}
\leavevmode
\epsfxsize= 4cm
\epsffile{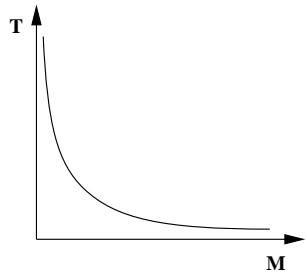}
\caption{$T$ como funci'on de  $M$ para el agujero negro de Schwarzschild.}
\label{fig:tschwarz}
\end{center}
\end{figure}

\begin{enumerate}
  
\item La temperatura del agujero negro de Schwarzschild BH (y de todos
  los agujeros negros conocidos, lejos del l'imite extremo, si lo
  hay) disminuye cuando la mas aumenta (Figura~\ref{fig:tschwarz}) y,
  por o tanto tiene un calor espec'ifico negativo
  (Figura~\ref{fig:cschwarz}):

\begin{equation}
C^{-1} = \frac{\partial T}{\partial M}=
\frac{-\hbar c^{3}}{8\pi G_{N}^{(4)} M^{2}} < 0\, ,
\end{equation}

\begin{figure}[!ht]
\begin{center}
\leavevmode
\epsfxsize= 4cm
\epsffile{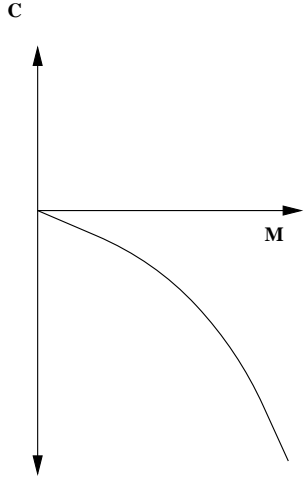}
\caption{El calor espec'ifico $C$ como funci'on de  $M$ 
para un agujero negro de Schwarzschild.}
\label{fig:cschwarz}
\end{center}
\end{figure}

\noindent y se enfr'ian al absorber masa. Por esto, un agujero 
negro no puede ponerse en equilibrio con una fuente de calor infinita.

\item La temperatura crece cuando la masa decrece (en la evaporaci'on
  por radiaci'on de Hawking, por ejemplo, lo que aumenta la
  radiaci'on) y diverge cuando la masa tiende a
  cero\footnote{\label{foot:continuidad2}Justo cuando el
    espacio-tiempo es el de Minkowski. Este es un segundo ejemplo del
    fen'omeno que comentamos en la nota \ref{foot:continuidad}.}. Las
  etapas finales de la evaporaci'on de Hawking de un agujero negro
  aislado ser'ian explosivas. Al mismo tiempo, cuando el $R_{S}$
  se acerca a su longitud de Compton (su masa es del orden de $M_{\rm
    Planck}$), los efectos de gravedad cu'antica empiezan a ser muy
  importantes y determinan (no sabemos c'omo) el destino final del
  agujero negro.

\item Si un agujero negro puede radiar, su entrop'ia puede
  disminuir.  Sin embargo la entrop'ia total (agujero m'as
  radiaci'on) siempre crece. Este resultado se conoce a veces como la
  {\it segunda ley de la termodin'amica de los agujeros negros
    generalizada}.
  
\item Volviendo al problema de la informaci'on en los agujeros
  negros, la radiaci'on de Hawking parece no tener m'as
  informaci'on que los datos $M,J,Q$ (como el agujero negro) pero
  cabe preguntarse si la contendr'ia en el caso de que fu'esemos
  capaces de hacer el c'alculo cu'antico completo del colapso
  gravitacional y la evaporaci'on. Esta es la creencia de 't Hooft,
  Susskind y otros que consideran un agujero negro es un sistema
  f'isico (cu'antico) m'as y que si entra informaci'on en el
  colapso gravitacional, 'esta vuelve a salir con la radiaci'on de
  Hawking, como en cualquier experimento de dispersi'on de los que se
  llevan a cabo en los aceleradores de part'iculas y que la
  teor'ia cu'antica de la gravitaci'on es una teor'ia
  unitaria.
  
  Si la radiaci'on de Hawking no contiene ninguna informaci'on,
  entonces, si el agujero negro se evapora indefinidamente, la
  informaci'on sobre el estado inicial que dio origen al agujero
  negro se pierde totalmente y la teor'ia cu'antica de la
  gravitaci'on es necesariamente no-unitaria, a diferencia de todas
  las dem'as teor'ias de l F'isica. 'Este es el punto de
  vista de Hawking.  Hay un tercer grupo minoritario que propone que
  la informaci'on permanece dentro del agujero negro y que la
  evaporaci'on deja un {\it remanente} que contiene esta
  informaci'on\footnote{En ciertos modelos el remanente sale de
    nuestro universo creando un universo beb'e.}.
  
  No hay ning'un resultado concluyente sobre el problema de la
  informaci'on en los agujeros negros. En los modelos basados en la
  Teor'ia de Cuerdas que vamos a explicar aqu'i, los agujeros
  negros son sistemas mecano-cu'anticos normales y la informaci'on
  se recupera (aunque sea tras un tiempo muy largo).
  
  Tambi'en debemos de mencionar una cuarta posibilidad poco explorada
  pero que, hasta cierto punto, concuerda con los resultados
  cl'asicos sobre la estabilidad de los agujeros negros: la
  informaci'on nunca entra en el agujero negro.
  
\item En cuanto al problema de la entrop'ia de los agujeros
  negros, en el conjunto microcan'onico, la entrop'ia de un
  sistema es

\begin{equation}
S(E) =\log \rho(E)\, , 
\end{equation}

\noindent donde $\rho(E)$ es la densidad de estados cuya energ'ia 
es $E$. Si un agujero negro es un sistema cu'antico m'as con $E=M$,
una buena teor'ia cu'antica de la gravitaci'on deber'ia de
permitirnos calcular $\rho(M)$, (y, por lo tanto, $S$) con el
conocimiento de los grados de libertad fundamentales de la
teor'ia. Para el agujero negro de Schwarzschild en particular
$S\sim M^{2}$ y

\begin{equation}
\rho(M) \sim e^{M^{2}}\, ,  
\end{equation}

\noindent que es un n'umero enorme de estados para un agujero negro 
de una masa solar: $10^{10^{78}}$.

\end{enumerate}

Como vamos a ver, la Teor'ia de Cuerdas nos va a permitir calcular
la entrop'ia y la radiaci'on de Hawking de ciertos agujeros
negros a partir del conocimiento de sus microestados asociados,
resolviendo (hasta cierto punto) los problemas de la informaci'on y
de la entrop'ia.  Estos agujeros negros son tratados como sistemas
cu'anticos ordinarios, unitarios, lo que apoya las tesis de 't Hooft.

%%%%%%%%%%%%%%%%%%%%%%%%%%%%%%%%%%%%%%%%%%%%%%%%%%%%%%%%%%%%%%%%%%%%%%%%%%%%%

\subsubsection{Gravitaci'on euclidiana}
\label{sec-Euclid}

Es posible calcular la temperatura y la entrop'ia de un agujero
negro por el m'etodo de la integral de Feynman euclidiana
propuesto por Gibbons y Hawking \cite{kn:GH,kn:Haw5}.

Para estudiar un sistema termodin'amico, primero se calcula un
potencial termodin'amico.  Si el sistema posee varias cargas
conservadas $C_{i}$ (cuyos potenciales asociados son $\mu_{i}$), en
conveniente trabajar en el conjunto can'onico grande, cuyo objeto
fundamental es la {\it gran funci'on de partici'on} ${\cal Z}$

\begin{equation}
{\cal Z}={\rm Tr}\, e^{-(H-\mu_{i}C_{i})/T} \, .
\end{equation}

\noindent El potencial termodin'amico $W$ 

\begin{equation}
W=E-TS-\mu_{i}C_{i} \, ,
\end{equation}

\noindent est'a relacionado con  ${\cal Z}$ por

\begin{equation}
e^{-\beta W}={\cal Z} \, .
\end{equation}

Todas las propiedades termodin'amicas del sistema se pueden obtener
de ${\cal Z}$. En particular, la entrop'ia 

\begin{equation}
\label{eq:entropy}
S  =  \frac{1}{T} (E-\mu_{i}C_{i})+\log {\cal Z}\, .
\end{equation}

Ahora queremos calcular la gran funci'on de partici'on  {\it
  t'ermica} de la gravitaci'on, calculando la integral de Feynman de
la acci'on de Einstein y Hilbert euclidiana $\tilde{S}_{EH}$ dada
en Ec.~(\ref{eq:EH1})
 
\begin{equation}
{\cal Z} = \int Dg\ e^{-\tilde{S}_{EH}/\hbar}\, ,
\end{equation}

\noindent donde hay que sumar sobre todas las m'etricas euclidianas
peri'odicas en una direcci'on (que interpretamos como el tiempo
euclidiano) con periodo $\beta=\hbar c/ T$. La 'unica modificaci'on
que hay que hacer a la acci'on de Einstein y Hilbert cl'asica
Ec.~(\ref{eq:EH1}) es la adici'on de un t'ermino de frontera que sirve
para normalizar la acci'on: la hace cero cuando se sustituye la
soluci'on de vac'io (el espacio euclidiano).  La acci'on completa es,
pues \cite{kn:GH}

\begin{equation}
\label{eq:EH4}
\begin{tabular}{|c|}
\hline \\ \\$
{\displaystyle
S_{EH}[g] = \frac{c^{3}}{16\pi G_{N}^{(4)}} \int_{\cal M} d^{4}x
\sqrt{|g|}\ R +\frac{c^{3}}{8\pi G_{N}^{(4)}}
\int_{\partial {\cal M}} ({\cal K}-{\cal K}_{0})\, ,}
$\\ \\ \hline
\end{tabular}
\end{equation}

\noindent donde  ${\cal K}_{0}$ se calcula sustituyendo 
la m'etrica de vac'io en la expresi'on de ${\cal K}$.

La integral de Feynman se calcula ahora semicl'asicamente utilizando
la aproximaci'on de ``punto de silla de montar'' (a partir de este
momento tomamos $\hbar=c=G_{N}^{(4)}=1$ por simplicidad)

\begin{equation}
\label{eq:z}
{\cal Z}= e^{-\tilde{S}_{EH}({\rm on-shell})}\, .
\end{equation}

Pasamos ahora a discutir la soluci'on euclidiano cl'asica con una
direcci'on peri'odica que debemos sustituir en la acci'on.

%%%%%%%%%%%%%%%%%%%%%%%%%%%%%%%%%%%%%%%%%%%%%%%%%%%%%%%%%%%%%%%%%%%%%%%%%%%%

\subsubsection{La soluci'on de Schwarzschild euclidiana}

La soluci'on de Schwarzschild euclidiana (signatura $(-,-,-,-)$ en
nuestro caso) que se obtiene haciendo una rotaci'on de Wick $\tau=it$
en la soluci'on lorenciana es real y resuelve las ecuaciones de
Einstein.  Si utilizamos las coordenadas de Kruskal y Szekeres (KS)
$\{T,X,\theta,\varphi\}$, tenemos que definir el tiempo KS
euclidiano ${\cal T}=iT$. La rotaci'on de Wick tiene importantes
efectos: la relaci'on entre la coordenada radial de Schwarzschild $r$
y $T,X$ es

\begin{equation}
\left(\frac{r}{R_{S}}-1 \right)e^{r/R_{S}} = X^{2} -T^{2}\, .
\end{equation}

EL lado izquierdo es mayo que $-1$ y es por ello que las coordenadas
$X,T$ tambi'en cubren el interior del horizonte. Sin embargo, en
t'erminos de ${\cal T}$

\begin{equation}
\left(\frac{r}{R_{S}}-1 \right)e^{r/R_{S}} = X^{2} +{\cal T}^{2}>0\, ,
\end{equation}

\noindent y el interior $r<R_{S}$ no est'a cubierto por 
las coordenadas KS euclidianas. Por otro lado, la relaci'on entre
el tiempo de Schwarzschild $t$ (que, recordemos, aparec'ia en el
diagrama de la Figura~\ref{fig:kruskal} como una coordenada angular) y
$X,{\cal T}$

\begin{equation}
\frac{X+T}{X-T} = e^{t/R_{S}}\, ,
\end{equation}

\noindent se transforma en

\begin{equation}
\frac{X-i{\cal T}}{X+{\cal T}}= e^{-2i\ {\rm Arg} (X+i{\cal T})}
= e^{-i\tau/R_{S}}\, .
\end{equation}

Dado que ${\rm Arg} (X+i{\cal T})\in[0,2\pi]$ (que debe ser tomado
como un c'irculo), por consistencia, para evitar singularidades
c'onicas, $\tau$ debe de ser una coordenada peri'odica con periodo
$8\pi M$. Este periodo puede interpretarse como el inverso de la
temperatura $\beta$, y coincide con el valor de la temperatura de
Hawking.

Por esta raz'on, podemos utilizar la soluci'on de Schwarzschild
euclidiana para calcular la funci'on de partici'on t'ermica.
Esta m'etrica cubre s'olo el exterior del agujero negro (la regi'on
I de la Figura~\ref{fig:kruskal}). La parte $X,{\cal T}$ de la
m'etrica describe un ``cigarro'' semi-infinito (veces una 2-esfera)
que va del horizonte al infinito. La topolog'ia de la soluci'on
es $\mathbb{R}^{2}\times S^{2}$

En la pr'actica no necesitamos usar realmente la m'etrica
euclidiana, sino tan s'olo la la informaci'on sobre la
periodicidad del tiempo euclidiano y sobre la regi'on de
integraci'on, de forma que sustituimos de nuevo $-\tilde{S}_{EH}(\rm
on-shell)$ por $+iS_{EH}({\rm on-shell})$ puesto que dan el mismo
resultado una vez tenidos en cuenta los datos anteriores.

%%%%%%%%%%%%%%%%%%%%%%%%%%%%%%%%%%%%%%%%%%%%%%%%%%%%%%%%%%%%%%%%%%%%%%%%%

\subsubsection{Los t'erminos de frontera}

Como la soluci'on de Schwarzschild euclidiana es una soluci'on de
vac'io, tiene $R=0$ y 'unicamente contribuyen a la acci'on los
t'erminos de frontera.

La 'unica frontera de esta soluci'on es $r\rightarrow \infty$, que
vamos a describir como la hipersuperficie $r=r_{0}$ cuando la
constante $r_{0}$ tiende a infinito. El vector normal a las
hipersuperficies $r=r_{0}$ es $n_{\mu} \sim
\partial_{\mu}(r-r_{0})=\delta_{\mu r}$, y, normalizado
($n_{\mu}n^{\mu}=-1$) y dotado del signo correcto para que apunte
hacia $r$ creciente es, para cualquier m'etrica est'atica y
esf'ericamente sim'etrica Ec.~(\ref{eq:genesph})

\begin{equation}
n_{\mu} = -\frac{\delta_{\mu r}}{\sqrt{-n^{2}}}
=-\sqrt{-g_{rr}}\delta_{\mu r}\, ,
\end{equation}

\noindent y la m'etrica inducida sobre las hipersuperficies
$r=r_{0}$ es para m'etricas est'aticas y esf'ericamente
sim'etricas generales Ec.~(\ref{eq:genesph})

\begin{equation}
ds^{2}_{(3)}=
\left.
h_{\mu\nu}dx^{\mu}dx^{\nu}= g_{tt}dt^{2} -r^{2}d\Omega^{2}_{(2)}
\right|_{r=r_{0}}\, .
\end{equation}

La derivada covariante de $n_{\mu}$ es

\begin{equation}
\nabla_{\mu}n_{\nu}=-\sqrt{-g_{rr}} \left\{\delta_{\mu r} \delta_{\nu r}
\partial_{r}\log{\sqrt{-g_{rr}}} -\Gamma_{\mu\nu}{}^{r}\right\}\, ,
\end{equation}

\noindent y la traza de la curvatura extr'inseca de las hipersuperficies
$r=r_{0}$  es

\begin{equation}
\label{eq:Kesph}
{\cal K}= 
\left.
h^{\mu\nu} \nabla_{\mu}n_{\nu} = \frac{1}{\sqrt{-g_{rr}}}
\left\{ {\textstyle\frac{1}{2}} \partial_{r}\log g_{tt}
+\frac{2}{r} \right\}
\right|_{r=r_{0}}\, .
\end{equation}

El regulador ${\cal K}_{0}$ es

\begin{equation}
\label{eq:K0esph}
{\cal K}_{0}= \left. \frac{2}{r}\right|_{r=r_{0}}\, .
\end{equation}

Por otro lado, para cualquier m'etrica est'atica, esf'ericamente
sim'etrica y asint'oticamente plana, para $r$ grande

\begin{equation}
g_{tt}\sim 1-\frac{2M}{r}\, ,
\hspace{1cm}
g_{rr}\sim -\left(1+\frac{2M}{r}\right)\, ,
\end{equation}

\noindent de forma que el integrando sobre la frontera 

\begin{equation}
\left.({\cal K}-{\cal K}_{0})\right|_{r=r_{0}}
\sim -\frac{M}{r_{0}^{2}}\, .
\end{equation}

Finalmente, tenemos

\begin{eqnarray}
\frac{i}{8\pi}\int_{r_{0}\rightarrow \infty} d^{3}x\sqrt{|h|}\, 
({\cal K}-{\cal K}_{0})
& = & \lim_{r_{0} \rightarrow \infty}
\frac{i}{8\pi}\int_{0}^{-i\beta} dt
\int_{S^{2}} d\Omega^{2} r_{0}^{2}\sqrt{g_{tt}(r_{0})}\, 
({\cal K}-{\cal K}_{0}) 
\nonumber \\
& & \nonumber \\
& = & \lim_{r_{0}\rightarrow \infty} \frac{\beta}{2}r_{0}^{2}\, 
({\cal K}-{\cal K}_{0}) = -\frac{\beta M}{2}\, .
\end{eqnarray}

Este resultado es v'alido para cualquier agujero negro est'atico,
esf'ericamente sim'etrico y asint'oticamente plano. Para
Schwarzschild $\beta=8\pi M$ y, usando las ecuaciones
(\ref{eq:entropy}) y (\ref{eq:z}) encontramos para la entrop'ia,
como antes

\begin{equation}
S = \beta M + \log {\cal Z} = \frac{\beta M}{2}= 4\pi M^{2}\, .
\end{equation}

%%%%%%%%%%%%%%%%%%%%%%%%%%%%%%%%%%%%%%%%%%%%%%%%%%%%%%%%%%%%%%%%%%%%%%
\section{El agujero negro de Reissner y Nordstr\"om}

%%%%%%%%%%%%%%%%%%%%%%%%%%%%%%%%%%%%%%%%%%%%%%%%%%%%%%%%%%%%%%%%%%%%%%
\subsection{El sistema de  Einstein y Maxwell}
\label{sec-EMsystem}

La acci'on que describe a la gravedad acoplada a un campo vectorial
abeliano $A_{\mu}$ es la de Einstein y Maxwell (EM)\footnote{En esta
  secci'on utilizamos el sistema de Heaviside en el que la fuerza
  entre dos cargas es
\begin{equation} 
{\textstyle\frac{1}{4\pi}}
\frac{q_{1}q_{2}}{r_{12}^{2}}\, . 
\end{equation}  
En el sistema de Gauss (que es un sistema racionalizado) se reemplaza
el prefactor $1/4c$ por $1/16\pi c$ y el factor $4\pi$ desaparece de
la ley de Coulomb. Despu'es introduciremos otro sistema en el que
trabajaremos usualmente, con $c=1$, reemplazando $1/4c$ por $1/64\pi
G_{N}^{(4)}$.}:

\begin{equation}
\begin{tabular}{|c|}
\hline \\ \\$
S_{EM}[g,A] = S_{EH}[g] +\frac{1}{c}{\displaystyle\int} d^{d}x \sqrt{|g|}\
\left[-\frac{1}{4}F^{2}\right]\, ,
$\\ \\ \hline
\end{tabular}
\end{equation}

\noindent donde

\begin{equation}
\label{eq:Fdef}
F_{\mu\nu}=2\partial_{[\mu}A_{\nu]}\, ,
\end{equation}

\noindent es el tensor intensidad de campo, invariante bajo 
las transformaciones gauge

\begin{equation}
A^{\prime}_{\mu} = A_{\mu} +\partial_{\mu} \Lambda\, ,
\end{equation}

\noindent y donde $F^{2}\equiv F_{\mu\nu}F^{\mu\nu}$. Obs'ervese que
no hay materia cargada en este sistema y por ello no hay constante de
acoplo ni unidad de carga definidas.

Las ecuaciones de movimiento son

\begin{eqnarray}
R_{\mu\nu} -{\textstyle\frac{1}{2}}g_{\mu\nu} R
-\frac{8\pi G_{N}^{(4)}}{c^{3}}T_{\mu\nu}
& = & 0\, , \\
& & \nonumber \\
\nabla_{\mu} F^{\mu\nu} & = & 0\, ,
\hspace{.5cm}
({\rm Ley\,\, de\,\, Gauss})
\label{eq:gaussdiff}
\end{eqnarray}

\noindent donde $T_{\mu\nu}$ es el tensor energ'ia-momento
del campo vector (sin traza en $d=4$)

\begin{equation}
T_{\mu\nu} = \frac{-2c}{\sqrt{|g|}} 
\frac{\delta S_{M}[A]}{\delta g^{\mu\nu}} = 
F_{\mu\rho} F_{\nu}{}^{\rho}-{\textstyle\frac{1}{4}}g_{\mu\nu}F^{2}\, .
\end{equation}

Las ecuaciones est'an escritas en t'erminos de $F$, pero la variable
fundamental es $A$ y tenemos que asegurarnos de que, dado $F$, existe
el $A$ del que procede a trav'es de Ec.~(\ref{eq:Fdef}). La
condici'on para que esto ocurra es la {\it identidad de
  Bianchi}\footnote{Dado un $F$ que satisface la identidad de Bianchi,
  un potencial $A$ se puede encontrar usando la f'ormula
\begin{equation}
\label{eq:AvsAdual}
A_{\mu}(x) = -\int_{0}^{1}d\lambda\lambda x^{\nu} F_{\mu\nu} 
(\lambda x)\, .  
\end{equation}
Obs'ervese que la relaci'on entre $F$ y $A$ no es local.
}

\begin{equation}
\label{eq:bianchidiff}
\nabla_{\mu}{}^{\star} F^{\mu\nu}  =  0\, .
\hspace{.5cm}
({\rm Identidad\,\, de\,\, Bianchi})
\end{equation}

En $d=4$, el par de ecuaciones (\ref{eq:gaussdiff}) y
(\ref{eq:bianchidiff}) es invariante bajo el intercambio de $F$ por
${}^{\star}F$ (${}^{\star\star}F=-F$), que tambi'en deja invariante
$T_{\mu\nu}$.  Esta es una transformaci'on de {\it dualidad}
el'ectrico-magn'etico que estudiaremos m'as tarde en
Sec.~\ref{sec-emdual}.

Para finalizar, en el lenguaje de formas diferenciales, los campos se
definen

\begin{equation}
A \equiv  A_{\mu} dx^{\mu}\, ,
\hspace{1cm}
F  = {\textstyle\frac{1}{2}}F_{\mu\nu}dx^{\mu}\wedge dx^{\nu} \equiv dA\, ,
\end{equation}

\noindent y la Ley de Gauss y la identidad de Bianchi se escriben

\begin{equation}
d{}^{\star}F=0\, ,
\hspace{1cm}
dF=0\, .
\hspace{.5cm}
\left( \partial_{[\alpha}F_{\beta\gamma]}=0\, \right)\, .
\end{equation}

\noindent y la invariancia de $F$ bajo
las transformaciones gauge $\delta A= d\Lambda$ es consecuencia de
$d^{\, 2}=0$.

%%%%%%%%%%%%%%%%%%%%%%%%%%%%%%%%%%%%%%%%%%%%%%%%%%%%%%%%%%%%%%%%%%%%%%

\subsubsection{La carga el'ectrica}

Para definir la carga el'ectrica, acoplamos el campo de Maxwell a una
fuente, descrita por la corriente electromagn'etica $j^{\mu}$. Su
acoplo al vector $A_{\mu}$ en la acci'on es

\begin{equation}
\label{eq:pointinter}
\frac{1}{c^{2}} \int d^{d}x \sqrt{|g|}\ \left[ -A_{\mu}j^{\mu} \right]\, ,
\end{equation}

\noindent t'ermino que rompe la invariancia gauge salvo que  $j^{\mu}$
tenga divergencia nula (o sea ``conservada''), es decir, satisfaga

\begin{equation}
\nabla_{\mu} j^{\mu}=0\, ,
\hspace{1cm}
(d {}^{\star}j=0)\, ,
\end{equation}

\noindent que implica la {\it ecuaci'on de continuidad} 

\begin{equation}
\partial_{\mu}\mathfrak{j}^{\mu} =0\, ,
\end{equation}

\noindent para la densidad $\mathfrak{j}^{\mu}\equiv
\sqrt{|g|}\, j^{\mu}$. La ecuaci'on de continuidad implica la
conservaci'on local de la carga el'ectrica que as'i aparece
como una consecuencia de la simetr'ia gauge.

En presencia de fuentes, la Ley de Gauss es

\begin{equation}
\label{eq:gauss2diff}
\nabla_{\mu}F^{\mu\nu} = {\textstyle\frac{1}{c}} j^{\nu}\, ,
\hspace{1cm}
(d{}^{\star}F={\textstyle\frac{1}{c}}{}^{\star}j)\, .
\end{equation}

Antes de definir carga el'ectrica, consideremos un ejemplo: la
corriente asociada a una part'icula con carga $q$ y l'inea del
Universo $\gamma$ parametrizada por $X^{\mu}(\xi)$, escrita en forma
manifiestamente covariante es, por definici'on

\begin{equation}
\label{eq:pointcurrent}
j^{\mu}(y) = qc\int_{\gamma} dX^{\mu} \frac{1}{\sqrt{|g|}}
\delta^{(4)}(y-X(\xi))\, ,
\end{equation}

\noindent donde $dX^{\mu}=d\xi dX^{\mu}/d\xi$. Tomando
 $\xi=X^{0}$ e integrado sobre $X^{0}$, encontramos

\begin{equation}
j^{\mu}(y^{0},\vec{y}) = qc \int dX^{0} \frac{dX^{\mu}}{dX^{0}}
 \frac{1}{\sqrt{|g|}}
\delta^{(3)}(\vec{y}-\vec{X})\delta (y^{0}-X^{0})
 =  q V^{\mu}
\frac{\delta^{(3)}(\vec{y}-\vec{X}(y^{0}))}{\sqrt{|g|}}\, ,
\end{equation}

\noindent y, si la part'icula est'a en reposo en el origen 
($V^{\mu}=c\delta^{\mu 0}$)

\begin{equation}
j^{\mu}(y^{0},\vec{y}) = qc\delta^{\mu 0}  
\frac{\delta^{(3)}(\vec{y})}{\sqrt{|g|}}\, .
\end{equation}

Para la corriente (\ref{eq:pointcurrent}), el t'ermino de
interacci'on (\ref{eq:pointinter}) es simplemente la integral; de la
1-forma $A$ sobre $\gamma$

\begin{equation}
\label{eq:WZ1}
-\frac{q}{c}\int_{\gamma(\xi)}
A_{\mu} \dot{x}^{\mu}d\xi =-\frac{q}{c}\int_{\gamma}A \, .
\end{equation}

La acci'on que rige el movimiento de una part'icula con masa $M$
y carga $q$ en campos gravitatorios y electromagn'eticos, incorpora
este t'ermino de interacci'on y es

\begin{equation}
\label{eq:electricNambuGotopar}
\begin{tabular}{|c|}
\hline \\
$
S_{M,q}[X^{\mu}(\xi)] = -Mc {\displaystyle\int} 
d\xi\ \sqrt{g_{\mu\nu}(X) \dot{X}^{\mu} 
\dot{X}^{\nu}}-{\displaystyle\frac{q}{c}\int} A_{\mu}\dot{X}^{\mu} \, ,
$
\\ \\ \hline
\end{tabular}
\end{equation}

\noindent Este tipo de  t'ermino (``topol'ogico'', pues no depende de 
la m'etrica) se llama {\it t'ermino de  Wess y Zumino (WZ)}. 

Pasemos ahora a definir la carga el'ectrica a trav'es de la
ecuaci'on de continuidad $d{}^{\star}j=0$, integr'andola sobre la
regi'on de espacio $d$-dimensional $V$ limitada por dos
hipersuperficies de tiempo constante $x^{0}=x^{0}_{1,2}$

\begin{equation}
0 = \int_{V}d\ {}^{\star}j\, .
\end{equation}

La frontera de $V$ consta de las dos hipersuperficies con
orientaciones opuestas. Utilizando el teorema de Stokes

\begin{equation}
\int_{V}d\ {}^{\star}j=\int_{x^{0}=x^{0}_{2}}{}^{\star}j 
-\int_{x^{0}=x^{0}_{1}}{}^{\star}j=0\, .
\end{equation}

\noindent Definiendo la carga el'ectrica por 

\begin{equation}
\label{eq:charge00}
q = {\textstyle\frac{1}{c}}  \int_{x^{0}={\rm constant}}{}^{\star}j\, ,
\end{equation}

\noindent vemos que la ecuaci'on anterior nos dice que es conservada.
Podemos introducir en esta expresi'on la corriente de una
part'icula puntual cargada Ec.~(\ref{eq:pointcurrent}) y ver que
obtenemos el mismo resultado.

Usando ahora Ec.~(\ref{eq:gauss2diff}) podemos reescribir esta
definici'on en t'erminos de $F$ y usar de nuevo el teorema de
Stokes.  Si la frontera de la hipersuperficie tiene la topolog'ia
de una $(d-2)$-esfera en el infinito, tenemos la definici'on
alternativa

\begin{equation}
\label{eq:charge}
\begin{tabular}{|c|}
\hline \\
$q ={\displaystyle\int}_{S^{2}_{\infty}}{}^{\star}F\, ,$
\\ \\ \hline
\end{tabular}
\end{equation}

\noindent que es f'acilmente generalizable y que tiene la ventaja de 
que no necesita del conocimiento expl'icito de la
fuente\footnote{Esta f'ormula no es sino una generalizaci'on de la
  Ley de Gauss de la electrost'atica.}.

Como veremos, en estas unidades $M$ aparece multiplicada por
$G_{N}^{(4)}$ (como en la soluci'on de Schwarzschild) pero $q$ no.
Algunas expresiones se simplifican  haciendo $c=1$ y reescribiendo 
la acci'on de Einstein y Maxwell en la forma

\begin{equation}
\label{eq:EMaction}
\begin{tabular}{|c|}
\hline \\ \\$
S_{EM}[g,A] = \frac{1}{16\pi G_{N}^{(d)}}{\displaystyle\int} 
d^{d}x \sqrt{|g|}\ \left[R -\frac{1}{4}F^{2}\right]\, .
$\\ \\ \hline
\end{tabular}
\end{equation}

\noindent EN estas unidades $A$ y $g$ son adimensionales. El factor 
$16\pi G_{N}^{(d)}$ desaparece de las ecuaciones de movimiento. Si no
ponemos ning'un factor de normalizaci'on en el t'ermino de WZ, la
carga el'ectrica es ahora

\begin{equation}
\label{eq:charge2}
\begin{tabular}{|c|}
\hline \\
$q =\frac{1}{16\pi G_{N}^{(d)}}
{\displaystyle\int}_{S^{d-2}_{\infty}}{}^{\star}F\, ,$
\\ \\ \hline
\end{tabular}
\end{equation}

\noindent y tiene dimensiones de masa. Finalmente, para cuna carga puntual,
para $r$ grande, esperamos

\begin{equation}
\label{eq:electricfield2}
E_{r}=F_{0r} \sim  \frac{4G_{N}^{(4)}q}{r^{2}}\, .
\end{equation}

%%%%%%%%%%%%%%%%%%%%%%%%%%%%%%%%%%%%%%%%%%%%%%%%%%%%%%%%%%%%%%%%%%%%%%

\subsection{La soluci'on el'ectrica de Reissner  y Nordstr\"om}
\label{sec-eRNsolution}

Estamos listos para buscar soluciones de tipo agujero negro de las
ecuaciones (\ref{eq:EMaction}) y (\ref{eq:charge2}),
(\ref{eq:electricfield2}).  Como en la secci'on anterior, hacemos el
Ansatz Ec.~(\ref{eq:statsphe}) para la m'etrica y, para el campo
electromagn'etico

\begin{equation}
F_{tr} \sim \pm\frac{1}{R^{2}(r)}\, ,
\end{equation}

\noindent que es apropiado para un objeto de simetr'ia 
esf'erica con carga el'ectrica y en reposo. La soluci'on que se
obtiene es la de Reissner y Nordstr\"om \cite{kn:Reiss,kn:No8}, que
depende de 3 constantes de integraci'on que fijamos imponiendo que la
soluci'on sea asint'oticamente plana e identificando la masa $M$ y
la carga $q$:

\begin{equation}
\label{eq:rnsolution}
\begin{tabular}{|c|}
\hline \\
$
\begin{array}{rcl}
ds^{2} & = & f(r)dt^{2}-f^{-1}(r)dr^{2} -r^{2}d\Omega_{(2)}^{2}\, , \\
& & \\
F_{tr} & = & -{\displaystyle\frac{4G_{N}^{(4)}q}{r^{2}}}\, , \\
& & \\
f(r) & = & r^{-2}(r-r_{+})(r-r_{-})\, , \\
& & \\
r_{\pm} & = & G_{N}^{(4)}M \pm r_{0}\, ,
\hspace{1cm}
r_{0} =G_{N}^{(4)}\left(M^{2}-4q^{2}\right)^{1/2}\, .\\
\end{array}
$
\\ \\ \hline
\end{tabular}
\end{equation}

Vamos a ver ahora algunas de las propiedades de esta soluci'on:

\begin{enumerate}

\item El campo vector asociado es 

\begin{equation}
A_{\mu}= \delta_{\mu t}\frac{-4G_{N}^{(4)}q}{r}\, .
\end{equation}

\item Se puede demostrar el an'alogo del teorema de Birkhoff para los
  agujeros negros de RN (ejercicio 32.1 de Ref.~\cite{kn:MTW}).

\item Esta m'etrica describe el campo gravitacional de un objeto
  esf'ericamente sim'etrico de masa $M$ y carga $aq$ visto desde
  lejos por un observador en reposo. La soluci'on de Schwarzschild
es el caso especial $q=0$.

\item La soluci'on de RN es v'alida para cualquier valor de 
  $M$ y $q$, por ello, de $r_{\pm}$, que pueden ser complejos.
  
\item La m'etrica es singular en $r=0,r_{\pm}$ (si $r_{\pm}$ son
  reales). De nuevo debemos analizar la naturaleza de estas
  singularidades. Como $T_{\mu}{}^{\mu}=0$ $R=0$, pero otros
  invariantes de curvatura nos dicen (como nos dice $F$) que hay una
  singularidad de curvatura en $r=0$, pero no en $r_{\pm}$. Si los
  $r_{\pm}$ son reales ($M^{2}\geq 4q^{2}$), entonces $r_{+}\geq
  r_{-}$ y un an'alisis similar al que hicimos para la soluci'on de
  Schwarzschild  nos dice que si $r_{+}>0$ ($M>0$) entonces hay un horizonte
  de eventos en $r_{+}$, cuyo 'area es

\begin{equation}
\label{eq:areahorizonteRN}
A=4\pi r_{+}^{2}\, ,
\end{equation}

\noindent mientras que $r_{-}$ es un {\it horizonte de Cauchy}
\footnote{Este horizonte parece ser inestable bajo peque~nas
  perturbaciones \cite{kn:Pen} que, se conjetura, deben transformarle
  en una singularidad de tipo espacio.}. Ambos horizontes existen si
$M>2|q|$ y entonces podemos decir que la soluci'on de RN describe un
agujero negro {\it no-extremo} (la Figura~\ref{fig:penrorn} es su
diagrama de Penrose). El caso $M=2|q|$ lo discutiremos m'as adelante.

\begin{figure}[!ht]
\begin{center}
\leavevmode
\epsfxsize= 8cm
\epsffile{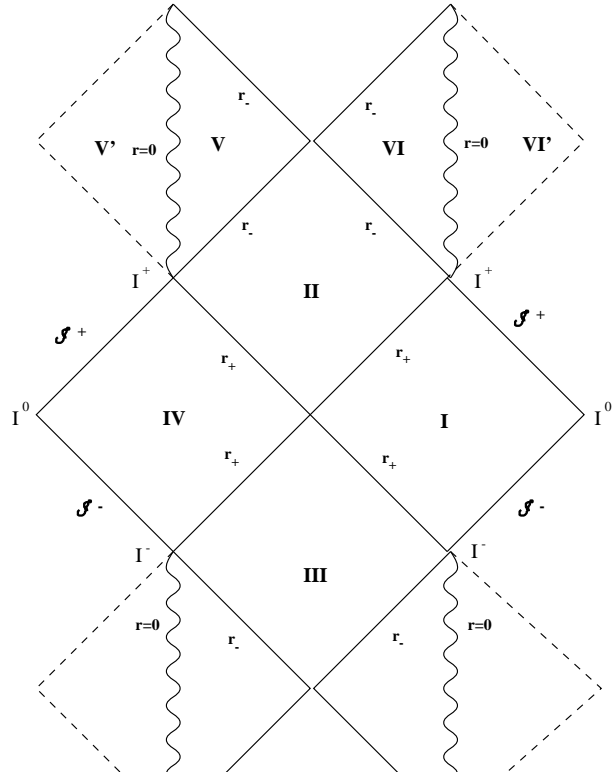}
\caption{Diagrama de Penrose de una agujero negro de Reissner y 
Nordstr\"om con $M>2|q|$.}
\label{fig:penrorn}
\end{center}
\end{figure}

\item Si $M<-2|q|$ no hay horizonte y tenemos una singularidad desnuda
  con diagrama de Penrose Fig.~\ref{fig:penronak}. Este caso debe de
  ser excluido invocando al censor c'osmico. Lo mismo ocurre en el
  intervalo $-2|q|<M<2|q|$, que incluye el caso $M=0$, un objeto
  cargado y sin masa, en reposo, bastante ex'otico. Si la
  energ'ia del campo electromagn'etico creado por la masa es
  positiva, entonces, debe de haber, intuitivamente, alguna densidad
  de energ'ia negativa que haga $M=0$, y por esto es razonable que
  sea c'osmicamente censurado. As'i pues, la censura c'osmica
  restringe $M$ al intervalo $M\geq 2|q|$.
  
\item El caso l'imite entre la singularidad desnuda y el agujero
  de RN regular $M =2|q|$ ({\it agujero de RN extremo (ERN)}) es muy
  especial.  Cuando $M=2|q|$ los dos horizontes coinciden $r_{+}
  =r_{-} =G_{N}^{(4)} M$ y tienen un 'area

\begin{equation}
\label{eq:areaERN}
A_{\rm extreme} = 4\pi r_{+}^{2} =4\pi \left(G_{N}^{(4)}M\right)^{2}\, . 
\end{equation}

\noindent Este objeto va a jugar un papel muy importante.
Algunas de sus propiedades son:

\begin{enumerate}

\item La distancia propia radial al horizonte a tiempo constante
%\begin{equation}
%\lim_{r_{2}\rightarrow r_{+}} \int_{r_{1}}^{r_{2}}ds
%=\lim_{r_{2}\rightarrow r_{+}} \int_{r_{1}}^{r_{2}}
%dr\left(1-\frac{r_{+}}{r}\right)^{-1}
%=\infty\, ,
%\end{equation}
diverge\footnote{Esto no pasa en direcciones temporales o tipo luz.}

\item El diagrama de Penrose est'a representado en la 
Figura~\ref{fig:penrextr}.

\begin{figure}[!ht]
\begin{center}
\leavevmode
\epsfxsize= 5cm
\epsffile{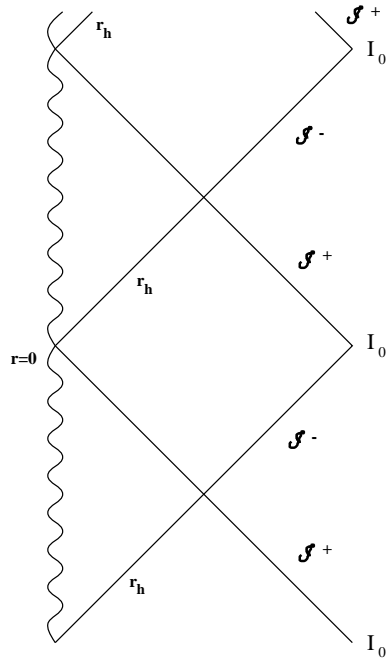}
\caption{Diagrama de Penrose de una agujero negro de Reissner y 
Nordstr\"om  extremo.}
\label{fig:penrextr}
\end{center}
\end{figure}

\item Si tenemos dos ERNs con $M_{1} =2|q_{1}|$, $M_{2} =2|q_{2}|$ y,
  {\it si ambas cargas tienen el mismo signo} y $r_{12}$ es la
  distancia entre ambos, tenemos que la fuerza no-relativista entre
  ambos

\begin{equation}
\label{eq:equilibrium}
F_{12} =-G_{N}^{(4)}\frac{M_{1}M_{2}}{r_{12}^{2}}
+4G_{N}^{(4)}\frac{q_{1}q_{2}}{r_{12}^{2}}=0\, ,
\end{equation}

\noindent  y ambos objetos {\it podr'ian} estar en equilibrio.
Esto sugiere que podr'ia haber soluciones est'aticas que
describan dos o m'as ERNs en equilibrio.

\item Si desplazamos la coordenada radial $r =\rho+G_{N}^{(4)}M$, y
  pasamos a coordenadas cartesianas $\vec{x}_{3}=(x^{1},x^{2},x^{3})$,
  con $|\vec{x}_{3}|=\rho$ y $d\vec{x}_{3}^{\, 2}=d\rho^{2}
  +\rho^{2}d\Omega_{(2)}^{2}$, tenemos una nueva forma de la
  soluci'on

\begin{equation}
\label{eq:ERN}
\begin{tabular}{|c|}
\hline \\
$
\begin{array}{rcl}
ds^{2} & = & H^{-2}dt^{2} -H^{2}d\vec{x}_{3}^{\, 2}\, , \\
& & \\
A_{\mu}& = & -2{\rm sign}(q)\left(H^{-1}-1\right)\delta_{\mu t}\, , \\
& & \\
H & = & 1+{\displaystyle\frac{G_{N}^{(4)}M}{|\vec{x}_{3}|}}\, .\\
\end{array}
$
\\ \\ \hline
\end{tabular}
\end{equation}

\noindent En estas coordenadas {\it is'otropas} el horizonte est'a
en $\rho=0$, donde son singulares pues todo el horizonte (que tiene
'area distinta de cero) aparece representado por un punto.

$H$ es una funci'on arm'onica en el espacio euclidiana tridimensional:

\begin{equation}
\partial_{\underline{i}}\partial_{\underline{i}} H=0\, .
\end{equation}

Este hecho puede parecer una mera coincidencia, pero, usando
Ec.~(\ref{eq:ERN}) como Ansatz con $H$ arbitrario, se ve que las
ecuaciones de movimiento se resuelven con un $H$ {\it cualquiera que
  satisfaga la ecuaci'on anterior}. As'i hemos obtenido la gran
familia de soluciones de Majumdar y Papapetrou (MP)
\cite{kn:Ma,kn:Pa}:

\begin{equation}
\label{eq:MP}
\begin{tabular}{|c|}
\hline \\
$
\begin{array}{rcl}
ds^{2} & = & H^{-2}dt^{2} -H^{2}d\vec{x}_{3}^{\, 2}\, , \\
& & \\
A_{\mu}& = & \delta_{\mu t}\alpha \left(H^{-1}-1\right)\, , 
\hspace{1cm} \alpha=\pm 2\\
& & \\
\partial_{\underline{i}}\partial_{\underline{i}} H & = & 0\, .\\
\end{array}
$
\\ \\ \hline
\end{tabular}
\end{equation}

Si queremos encontrar soluciones describiendo varios agujeros negros
ERN en equilibrio est'atico, dado que en coordenadas is'otropas el
horizonte es una singularidad puntual, podemos probar con una
funci'on arm'onica $H$ que tenga  varias ($N$):

\begin{equation}
H(\vec{x}_{3}) = 1 + \sum_{i=1}^{N}
\frac{2G_{N}^{(4)}|q_{i}|}{|\vec{x}_{3}-\vec{x}_{3, i}|}\, .
\end{equation}

$H$ est'a normalizada para que la m'etrica sea asint'oticamente
plana y los coeficientes de cada polo son positivos para que $H$ no se
anule y la m'etrica no sea singular.

Se puede demostrar \cite{kn:HaHa} que cada polo de $H$ es un
horizonte. La carga de cada agujero se puede calcular y el resultado
es ${\rm sign}(-\alpha)|q_{i}|$, es decir: todas las cargas tienen el
mismo signo. Pero es imposible calcular la masa de cada agujero porque
s'olo la masa total $M$ est'a bien definida. 'Esta resulta ser $M
=2\sum_{i=1}^{N}|q_{i}|$. Sin embargo, el equilibrio de fuerzas que
hay entre los agujeros negros sugiere que las densidades de
energ'ia de interacci'on electrost'aticas y gravitatorias se
cancelan mutuamente por doquier (una de las se~nales de que hay
supersimetr'ia, como veremos). As'i, las masas y cargas
estar'ian {\it localizadas} en las singularidades y podr'iamos
asignarlas masas $M_{i}=2|q_{i}|$. Esta es una idea muy atractiva,
pero no una prueba rigurosa. Sin embargo, veremos que para los ERNs (a
diferencia de Schwarzschild) es posible encontrar fuentes localizadas,
lo que apoya esta idea.

Si tomamos alguno de los coeficientes negativo, habr'a alguna masa
negativa que ser'a la causa de las singularidades desnudas. La
censura c'osmica deber'ia de prohibir estas situaciones.

\item En el l'imite de cercan'ia al horizonte $\rho\rightarrow
  0$ en la m'etrica Ecs.~(\ref{eq:ERN}), la constante $1$ se puede
  ignorar y encontramos otra soluci'on de MP con
  $H=R_{AdS}/\rho$ ($R_{AdS}=2G_{N}^{(4)}|q|$):

\begin{equation}
\label{eq:RB}
\begin{tabular}{|c|}
\hline \\
$
\begin{array}{rcl}
ds^{2} & = & 
{\displaystyle\frac{\rho^{2}}{R^{2}_{AdS}}}
dt^{2}
-R^{2}_{AdS} {\displaystyle\frac{d\rho^{2}}{\rho^{2}}}
-R^{2}_{AdS}d\Omega_{(2)}^{2}\, , \\
& &  \\
A_{t} & = & -{\displaystyle\frac{2\rho}{R_{AdS}}}\, ,
\hspace{1cm}
F_{\rho t} =-{\displaystyle\frac{2}{R_{AdS}}}\, . \\
\end{array}
$
\\ \\ \hline
\end{tabular}
\end{equation}

Esta es la soluci'on de Robinson y Bertotti (RB)
\cite{kn:Rob,kn:Bert}. No es asint'oticamente plana: es el producto
de dos espacios bidimensionales de curvatura constante anti-de Sitter
($AdS_{2}$) con ``radio'' $R_{AdS}=2G_{N}^{(4)}|q|$ y curvatura
$R^{(2)}=-2/R_{AdS}^{2}$ y una 2-esfera $S^{2}$ de radio $R_{AdS}$ y
curvatura $R^{(2)}=+2/R_{AdS}^{2}$.

$AdS_{2}$ es invariante bajo $SO(1,2)$ y $S^{2}$ bajo $SO(3)$.  El
grupo de isometr'ia de RB es mucho mayor que el de ERN
($SO(1,1)\times SO(3)$). En la pr'oxima lecci'on veremos que
tambi'en hay un incremento de supersimetr'ia, que es m'axima, por
lo que se podr'ia considerar a RB como un vac'io de l;a
teor'ia alternativo al de Minkowski. As'i podemos imaginar que
ERN interpola entre el vac'io de de Minkowski (en el infinito) y
el de RB (en el horizonte) y podemos interpretar ERN como un {\it
  solit'on gravitacional} \cite{kn:GWG}.

\end{enumerate}

\item Si desplazamos la coordenada radial de la soluci'on de RN
  $r=\rho +r_{-}$, podemos reescribirla de esta forma gen'erica:

\begin{equation}
\label{eq:RN3}
\begin{tabular}{|c|}
\hline \\
$
\begin{array}{rcl}
ds^{2} & = & H^{-2} W dt^{2} - H^{2}\left[W^{-1} d\rho^{2}
+\rho^{2}d\Omega_{(2)}^{2}\right]\, , \\
& & \\
A_{\mu} & = & \delta_{\mu t}\alpha \left(H^{-1}-1\right)\, , \\
& & \\
H & = & 1 + {\displaystyle\frac{h}{\rho}}\, , 
\hspace{1cm}
W = 1 + {\displaystyle\frac{\omega}{\rho}}\, ,
\hspace{1cm}
\omega=h\left[1-\left({\displaystyle\frac{\alpha}{2}}\right)^{2}\right]\, ,\\
\end{array}
$
\\ \\ \hline
\end{tabular}
\end{equation}

\noindent donde las constantes $h,\omega,\alpha$ est'an relacionadas
con las f'isicas por

\begin{eqnarray}
\alpha=-\frac{4G_{N}^{(4)}q}{r_{\pm}}\, ,
\hspace{.5cm}
h=r_{\pm}\, ,
\hspace{.5cm}
\omega =\pm 2r_{0}\, .
\end{eqnarray}

En esta forma la m'etrica parece la de un agujero de Schwarzschild de
masa $r_{0}/G^{(4)}_{N}$ ``vestido'' con ciertos factores (de $H$)
relacionados con el potencial gauge, o como la soluci'on de ERN
``vestida'' con factores tipo Schwarzschild ($W$). $W$ desaparece en el
l'imite extremo y $H$ cuando la carga es cero. Esta forma de la
soluci'on se puede generalizar a {\it $p$-branas cargadas}, como
veremos.

\item Consideremos, finalmente la acci'on 

\begin{equation}
\label{eq:EMactionN}
\begin{tabular}{|c|}
\hline \\ \\$
S[g,A^{I}] = \frac{1}{16\pi G_{N}^{(4)}}{\displaystyle\int} d^{4}x \sqrt{|g|}\
\left[R -\frac{1}{4}\sum_{I=1}^{I=N}\left(F^{I}\right)^{2}\right]\, .
$\\ \\ \hline
\end{tabular}
\end{equation}

\noindent Esta acci'on es invariante bajo rotaciones 
$O(N)$ de los $N$ vectores abelianos. Este es un ejemplo muy simple de
simetr'ia de dualidad\footnote{En general, efectos cu'anticos
  como la cuantizaci'on de la carga rompen el grupo de dualidad
  $O(N)$ al subgrupo discreto $O(N,\mathbb{Z}$).}.  Cualquier soluci'on
de EM es una soluci'on de 'esta con todos los vectores salvo uno
igual a cero, y, haciendo una rotaci'on $O(N)$ podemos generar
soluciones nuevas en las que los $N$ vectores sean no-nulos. Si la
soluci'on original ten'ia $q_{1}$, las nuevas soluciones
tendr'an $q_{i}$ con  $\sum_{i=1}^{N}q^{\prime\,
  2}_{i}=q^{2}_{1}$. Esta dualidad no act'ua sobre la m'etrica y,
por lo tanto, s'olo tenemos que sustituir $q_{1}^{2}$ por
$\sum_{i=1}^{N}q^{\prime\, 2}_{i}$ en ella. As'i, partiendo de la
soluci'on Ec.~(\ref{eq:RN3}) se obtiene

\begin{equation}
\label{eq:RN4}
\begin{tabular}{|c|}
\hline \\
$
\begin{array}{rcl}
ds^{2} & = & H^{-2} W dt^{2} - H^{2}\left[W^{-1} d\rho^{2}
+\rho^{2}d\Omega_{(2)}^{2}\right]\, , \\
& & \\
A_{\mu}^{i} & = & \delta_{\mu t}\alpha^{i} \left(H^{-1}-1\right)\, , \\
& & \\
H & = & 1 + {\displaystyle\frac{h}{\rho}}\, , 
\hspace{1cm}
W = 1 + {\displaystyle\frac{\omega}{\rho}}\, ,
\hspace{1cm}
\omega=h\left[1-\sum_{i=1}^{i=N}
\left({\displaystyle\frac{\alpha^{i}}{2}}\right)^{2}\right]\, ,\\
\end{array}
$
\\ \\ \hline
\end{tabular}
\end{equation}

\noindent con

\begin{eqnarray}
\alpha^{i}=-\frac{4G_{N}^{(4)}q^{i}}{r_{\pm}}\, ,
\hspace{.5cm}
h=r_{\pm}\, ,
\hspace{.5cm}
\omega =\pm 2r_{0}\, ,
\end{eqnarray}

\noindent  donde, ahora

\begin{equation}
r_{\pm}= G_{N}^{(4)}M \pm r_{0}\, ,
\hspace{1cm}
r_{0}=G_{N}^{(4)}\sqrt{M^{2}-4\sum_{i=1}^{i=N}q_{i}^{2}}\, .
\end{equation}

Este es el primer y m'as simple ejemplo de uso de simetr'ias de
dualidad para generar nuevas soluciones (el segundo lo encontraremos
en la Secci'on~\ref{sec-magneticdyonic}). El resultado es una familia
que, como tal, es invariante bajo cualquier transformaci'on de
dualidad adicional.Estas familias reflejan muchas de las
simetr'ias de la teor'ia y dependen de combinaciones de cargas
y {\it m'odulos} (que definiremos luego) que son invariantes bajo
dualidad. Sus propiedades f'isicas (temperatura, entrop'ia...)
tambi'en lo son.

\end{enumerate}

%%%%%%%%%%%%%%%%%%%%%%%%%%%%%%%%%%%%%%%%%%%%%%%%%%%%%%%%%%%%%%%%%%%%%%

\subsection{Las fuentes del agujero negro ERN}

La soluci'on de Schwarzschild satisface las ecuaciones de Einstein en
el ``vac'io'' excepto en $r=0$. El campo $A_{\mu}\sim Q/r
\delta_{\mu t}$ tambi'en satisface las ecuaciones de Maxwell en el
vac'io excepto en $r=0$. En este 'ultimo caso, sabemos que hay un
t'ermino de fuente correspondiente a una carga puntual situada en
$r=0$ $j^{\mu}\sim Q \delta^{(3)} (\vec{x}_{3}) \delta^{\mu t}$ tal
que el campo anterior es soluci'on de las ecuaciones de Maxwell con
esa fuente por doquier. En el primer caso, tal fuente no existe (o
no se conoce) y, en cualquier caso no corresponde a una part'icula
puntual por el problema de la localizaci'on de la energ'ia del
campo gravitacional.

En el caso general de RN no esperamos nada mejor, pero en el extremo,
de acuerdo con los argumentos que dimos, podr'ia haber una
localizaci'on de la densidad de energ'ia (igual que de la
densidad de carga el'ectrica). 

Con m'as precisi'on, queremos ver si ERN es parte de una soluci'on
completa de las ecuaciones de movimiento sistema $S[g,A,X]
=S_{EM}[g,A]+S_{M,q}[X^{\mu}]$, definidas en Ecs.~(\ref{eq:EMaction})
y~(\ref{eq:electricNambuGotopar}) que describe a una part'icula
puntual de masa $M$, carga $q$ y l'inea del Universo
$X^{\mu}(\xi)$. En la Secci'on~\ref{sec-fuentes} vamos a ver una
generalizaci'on de este sistema a dimensiones arbitrarias, con
potenciales que son $k$-formas y con un escalar que se acopla al
potencial de forma arbitraria. Entre las soluciones, est'a justamente
el ERN, lo que corrobora nuestras ideas.

Observemos que, si eliminamos la gravitaci'on del problema, no
tenemos una soluci'on de todas las ecuaciones: la ecuaci'on de
Maxwell con fuentes se resuelve como antes hemos indicado, pero la
ecuaci'on de movimiento de la part'icula no por el problema de la
fuerza infinita (o indefinida) que el campo creado por la
part'icula ejerce sobre la propia part'icula ({\it el}
problema de la electrodin'amica cl'asica). La gravitaci'on cura
este problema pues las fuerzas gravitatorias y electrost'aticas se
cancelan sobre la part'icula.

Esta sorprendente propiedad es una manifestaci'on m'as de la
supersimetr'ia del agujero negro ERN.

%%%%%%%%%%%%%%%%%%%%%%%%%%%%%%%%%%%%%%%%%%%%%%%%%%%%%%%%%%%%%%%%%%%%%%

\subsection{La termodin'amica de RN}

En esta secci'on vamos a usar unidades naturales $\hbar=c=1$
(adem'as de $k_{B}=1$).

Como ya dijimos, muchas de las propiedades termodin'amicas se aplican
a todos los agujeros negros conocidos, y, en particular, a los de RN.
Vamos repasar estas propiedades para un agujero de RN gen'erico y
m'as tarde veremos qu'e pasa en el caso de un ERN.

Para empezar, las leyes de la termodin'amica de los agujeros negros
se generalizan y aplican inmediatamente a los RN. La ley ``cero'' se
cumple en todos los agujeros estacionarios. La primera ley incluye un
t'ermino adicional que tiene en cuenta posibles variaciones de la
energ'ia debidas a variaciones de la carga:

\begin{equation}
dM = {\textstyle\frac{1}{8\pi G_{N}^{(4)}}}\kappa dA +\Phi d q\, ,  
\end{equation}

\noindent donde $\Phi=2q/r_{+}$ es el potencial electrost'atico 
en el horizonte y $\kappa$ ahora vale

\begin{equation}
\kappa =\frac{1}{G_{N}^{(4)}}
\frac{\sqrt{M^{2}-4q^{2}}}{\left(M+\sqrt{M^{2}-4q^{2}}\right)^{2}}\, .
\end{equation}

La segunda ley es universal. El 'area del horizonte de un RN est'a
dado en Ec.~(\ref{eq:areahorizonteRN}). Y la tercera ley se aplica,
como dijimos, a este tipo de agujeros que tienen un l'imite
``extremo'' $M=2|q|$ en el que $\kappa=0$ y nos dice que a este
l'imite no se puede llegar en un tiempo finito. Vamos a ver que,
en cualquier caso este l'imite es muy especial y cerca de 'el la
descripci'on termodin'amica del agujero negro deja de ser v'alida.

La relaci'on entre la temperatura de la radiaci'on de Hawking $T$ y
la gravedad superficial, es la misma que en el caso de Schwarzschild
Ec.~(\ref{eq:Tkappa}), lo que implica la misma relaci'on entre el
'area del horizonte y la entrop'ia Ec.~(\ref{eq:Sarea})

\begin{equation}
\begin{tabular}{|c|}
\hline \\
${\displaystyle
T 
%=\frac{r_{0}}{2\pi r_{+}^{2}}
=\frac{1}{2\pi G_{N}^{(4)}}
\frac{\sqrt{M^{2}-4q^{2}}}{\left(M+\sqrt{M^{2}-4q^{2}}\right)^{2}}\, ,
\hspace{.5cm}
S 
%= \frac{\pi r_{+}^{2}}{G_{N}^{(4)}}
=\pi G_{N}^{(4)}
\left(M+\sqrt{M^{2}-4q^{2}}\right)^{2}\, .
}
$
\\ \\ \hline
\end{tabular}
\end{equation}

\begin{figure}[!ht]
\begin{center}
\leavevmode
\epsfxsize= 5cm
\epsffile{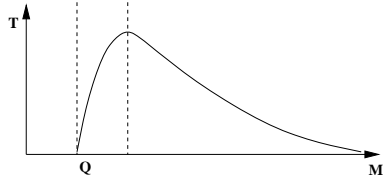}
\caption{La temperatura $T$ como funci'on de la masa $M$ para un agujero 
RN de carga $Q=2q$.}
\label{fig:trn}
\end{center}
\end{figure}

\begin{figure}[!ht]
\begin{center}
\leavevmode
\epsfxsize= 5cm
\epsffile{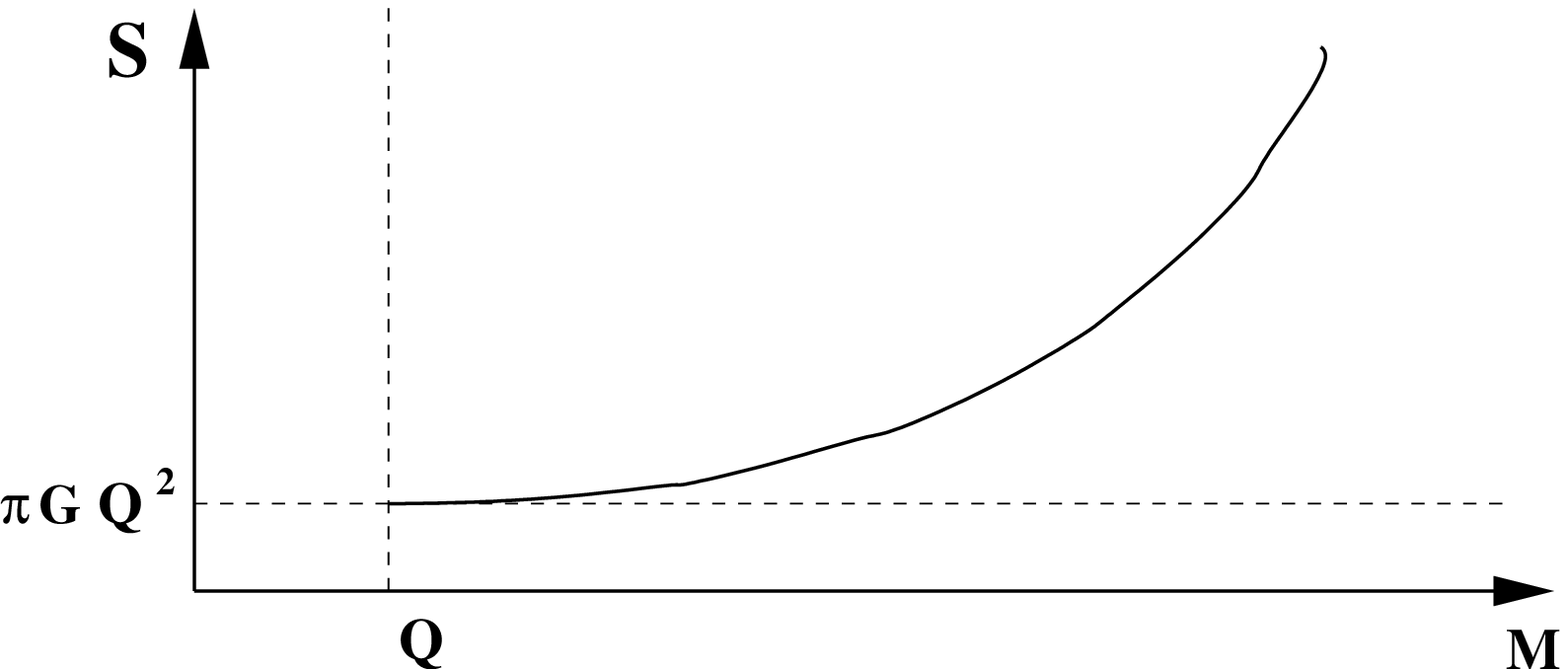}
\caption{La entrop'ia $S$ como funci'on de la masa $M$ para un agujero 
RN de carga $Q=2q$.}
\label{fig:srn}
\end{center}
\end{figure}

Las gr'aficas que representan a $T$ y $S$ como funciones de la masa
para un valor de la carga $q$ fija est'an representadas en las
Figuras~\ref{fig:trn} y~\ref{fig:srn}.

\begin{figure}[!ht]
\begin{center}
\leavevmode
\epsfxsize= 5cm
\epsffile{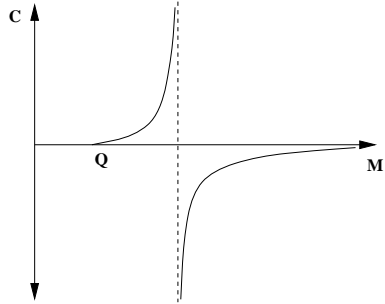}
\caption{El calor espec'ifico $C$ como funci'on de la masa $M$ para 
un agujero RN de carga $Q=2q$.}
\label{fig:crn}
\end{center}
\end{figure}

Observemos que la entrop'ia no tiende a cero cuando la
$M\rightarrow 2|q|$ y $T\rightarrow 0$. Tambi'en es interesante
observar la gr'afica del calor espec'ifico Figura~\ref{fig:crn}.
Como vemos, hay dos regiones bien diferenciadas en la termodin'amica
de RN: para valores de $M$ mucho m'as grandes que la carga, el
comportamiento es como el de Schwarzschild, con un calor
espec'ifico negativo y $S$ que aumenta con $M$ mientras que $T$
disminuye. Para $M$ cercana a $2|q|$, sin embargo, la termodin'amica
es como la de un sistema ordinario, con calor espec'ifico
positivo, lo que nos hace esperar que haya una descripci'on
estad'istica est'andar de su entrop'ia. La 'unica diferencia
es que cuando la temperatura tiende a cero, la entrop'ia tiende a
un valor distinto de cero, lo que no tendr'ia por qu'e
contradecir ninguna ley fundamental de la termodin'amica de acuerdo
con Wald \cite{kn:W12}.

Sin embargo, este es un punto conflictivo: ya hemos advertido varias
veces que las propiedades f'isicas de una familia de soluciones
que depende de varios par'ametros continuos no tienen por qu'e ser
ellas mismas funciones continuas de los par'ametros (ver las notas a
pie de p'agina \ref{foot:continuidad} y \ref{foot:continuidad2}).  En
el l'imite extremo no tenemos realmente control sobre lo que
ocurre\footnote{De hecho, la descripci'on termodin'amica del agujero
  negro deja de ser v'alida en su entorno \cite{kn:PSSTW}.} y es
necesario rehacer los c'alculos utilizando directamente la soluci'on
ERN, en vez de tomando el l'imite sobre resultados obtenidos
utilizando la RN general.

El c'alculo de la temperatura directamente sobre la soluci'on ERN no
da lugar a sorpresas: la temperatura es cero (mejor dicho: no tiene
sentido definirla, como en el espacio-tiempo de Minkowski). 

El c'alculo de la entrop'ia es m'as sutil: si la identificamos
ciegamente con el 'area del horizonte, su valor es el del l'imite
de la entrop'ia del RN gen'erico. Pero esta identificaci'on, en
un sistema de temperatura cero, es dudosa. Como alternativa, se puede
calcular por el m'etodo euclidiano explicado en la secci'on
anterior. El resultado es que la entrop'ia es id'enticamente cero
\cite{kn:GK,kn:HaHo}, resultado confirmado por otro m'etodo en
Ref.~\cite{kn:Te3}.

Este resultado es notable y debemos recordarlo porque la entrop'ia
que vamos a calcular utilizando un modelo microsc'opico basado en la
Teor'ia de Cuerdas es la de un agujero negro extremo, y
obtendremos justamente como resultado un cuarto del 'area del
horizonte, distinto de cero, el resultado euclidiana semicl'asico.
Siempre es posible pensar que la Teor'ia de Cuerdas funciona
ah'i donde la gravedad euclidiana falla. Se pueden encontrar
argumentos a favor de esta interpretaci'on en Ref.~\cite{kn:Ho2}.

Si esto fuese as'i, el agujero negro ERN ser'ia un buen
candidato a ``remanente'' en el que la informaci'on est'a acumulada
sin que pueda salir (ni cl'asicamente ni cu'anticamente, pues no hay
radiaci'on de Hawking), aunque es dudoso que todos los agujeros
negros acaben as'i.

Entre las dos regiones de calor espec'ifico positivo y negativo
que hemos descrito hay un punto en el que 'este diverge, lo cual
podr'ia ser interpretado como se~nal de un posible cambio de
fase. Esto ser'ia un problema a~nadido a la hora de extrapolar los
resultados obtenidos en el l'imite extremo o cuasi-extremo.

%%%%%%%%%%%%%%%%%%%%%%%%%%%%%%%%%%%%%%%%%%%%%%%%%%%%%%%%%%%%%%%%%%%%%%

\subsection{Dualidad el'ectrico-magn'etico}
\label{sec-emdual}

Como dijimos en la Secci'on~\ref{sec-EMsystem} el conjunto de las
ecuaciones de Maxwell sin fuentes (incluyendo la identidad de
Bianchi)\footnote{De momento ignoramos el acoplo a la gravedad y
  simplemente consideramos la acci'on de Maxwell en un espacio-tiempo
  curvo.}  es invariante bajo la sustituci'on de $F$ por
$\tilde{F}={}^{\star}F$. En el espacio-tiempo de Minkowski esta
transformaci'on corresponde a un intercambio de los campos
el'ectrico y magn'etico

\begin{equation}
\tilde{\vec{E}} = \vec{B}\, , 
\hspace{1cm}
\tilde{\vec{B}} = -\vec{E}\, ,
\end{equation}

\noindent por lo que esta simetr'ia $\mathbb{Z}_{2}$
de las ecuaciones de Maxwell se conoce como {\it dualidad
  electromagn'etica}. Este $\mathbb{Z}_{2}$ se extiende a un grupo
continuo\footnote{Como en el caso de $N$ vectores la simetr'ia
  $O(N)$, la cuantizaci'on de la carga romper'a esta simetr'ia a
  un subgrupo discreto.} $GL(2)$ de transformaciones:

\begin{equation}
\tilde{F}=aF+b{}^{\star}F\, , 
\,\,\,
\Rightarrow
{}^{\star}\tilde{F}= -bF +a{}^{\star}F\, ,
\,\,\,
a^{2}+b^{2}\neq 0\, .
\end{equation}

\noindent  Es conveniente definir el {\it vector de dualidad} $\vec{F}$

\begin{equation}
\vec{F}\equiv
\left(
\begin{array}{c}
F \\
{}^{\star} F \\
\end{array}
\right)\, ,
\hspace{1cm}
{}^{\star}\vec{F}= 
\left( 
\begin{array}{rr}
0 & 1 \\
-1 & 0 \\
\end{array}
\right)
\vec{F}\, ,
\end{equation}

\noindent puesto que con 'el las ecuaciones de Maxwell se escriben
en forma compacta

\begin{equation}
\nabla_{\mu}\vec{F}^{\mu\nu}=0\, .
\end{equation}

\noindent $\vec{F}$ se transforma en la representaci'on vectorial
del grupo de dualidad $GL(2)$

\begin{equation}
\tilde{\vec{F}}=M\vec{F}\, ,
\hspace{1cm}
M=
\left( 
\begin{array}{rr}
a & b \\
-b & a \\
\end{array}
\right)\, .
\end{equation}

Si integramos el dual de Hodge del vector de dualidad
${}^{\star}\vec{F}$ sobre una 2-esfera en el infinito, obtenemos un
vector cuya primer componente es $16\pi G_{N}^{(4)}q$, con nuestros
convenios. La segunda componente es, por definici'on, la carga
magn'etica $p$:

\begin{equation}
\int_{S^{2}_{\infty}} {}^{\star}\vec{F}=
\left(
\begin{array}{c}
16\pi G_{N}^{(4)}q \\
p\\
\end{array}
\right)
\equiv
16\pi G_{N}^{(4)} \vec{q}\, ,
\hspace{1cm}
\vec{q}=
\left(
\begin{array}{c}
q \\
p/16\pi G_{N}^{(4)}\\
\end{array}
\right)\, .
\end{equation}

Estas transformaciones son no-locales en t'erminos del potencial $A$:
usando Ec.~(\ref{eq:AvsAdual}) obtenemos la siguiente relaci'on entre
$A$ y $\tilde{A}$:

\begin{equation}
\tilde{A}_{\mu}(x) = -\int_{0}^{1}d\lambda\lambda x^{\nu} 
\frac{\epsilon_{\mu\nu}{}^{\rho\sigma}}{\sqrt{|g|}}
\partial_{\rho} A_{\sigma}(\lambda x)\, .  
\end{equation}

Esta no-localidad es, a la vez, lo que hace interesante esta dualidad
y el origen de muchos problemas. Para empezar, la acci'on de Maxwell
no es invariante bajo la simple sustituci'on de $F$ por ${}^{\star}F$
( $\left({}^{\star}F \right)^{2}=-F^{2}$). El procedimiento adecuado
para reescribir la acci'on en t'erminos de las variables duales se
llama {\it dualizaci'on de Poincar'e} y consiste en reescribir
primero la acci'on como un funcional de $F$ en vez de $A$.

Veamos ahora qu'e modificaciones produce el acoplo a la gravitaci'on.
Ahora hay una ecuaci'on m'as: la de Einstein, que podemos reescribir
as'i

\begin{equation}
G_{\mu\nu}
-\vec{F}^{\ T}_{\mu}{}^{\rho}\vec{F}_{\nu\rho}=0\, ,
\end{equation}

\noindent lo que pone de manifiesto que s'olo el subgrupo $O(2)$ 
la deja invariante y $SO(2)$ es el grupo de dualidad
el'ectrico-magn'etico del sistema de Einstein-Maxwell.

Como hemos hecho notar antes, 'esta es una teor'ia abeliana sin
materia y no hay constante de acoplo, pero podemos pensar que la
simetr'ia gauge $U(1)$ es parte de un un grupo no-abeliano que
est'a roto y entonces podemos introducir una constante de acoplo $e$
que aparece como un factor $1/e^{2}$ en frente del $F^{2}$ de la
acci'on. El vector de dualidad y la ecuaci'on de Einstein son ahora

\begin{equation}
\vec{F}\equiv
\left(
\begin{array}{c}
e^{-2}F \\
{}^{\star} F \\
\end{array}
\right)\, ,
\hspace{1cm}
G_{\mu\nu}
+\left(\vec{F}_{\mu}{}^{\rho}\right)^{T}
\left(
\begin{array}{rr}
0 & 1 \\
-1 & 0 \\
\end{array}
\right)
\vec{F}_{\nu\rho}=0\, ,
\end{equation}

\noindent y es invariante bajo $Sp(2,\mathbb{R})\sim
Sl(2,\mathbb{R})$. Sin embargo, de todo este grupo s'olo estas
transformaciones son consistentes con la ligadura del vector de
dualidad\footnote{Si hubi'esemos generalizado la teor'ia
  incluyendo un {\it 'angulo $\vartheta$}, todo el grupo
  $Sl(2,\mathbb{R})$ dejar'ia invariante la teor'ia. Este es
  el arquetipo de {\it grupo de dualidad~S} o el'ectrico-magn'etico.
  } (permitiendo cambios de escala de la constante de acoplo)

\begin{equation}
\begin{array}{rcl}
M & = & 
\left(
\begin{array}{rr}
a & 0 \\
0 & 1/a \\
\end{array}
\right)\, ,
\hspace{1cm}
e^{\prime} = a^{-1} e\, , \\
& & \\
M & = & 
\left(
\begin{array}{rr}
0 & 1 \\
-1 & 0 \\
\end{array}
\right)\, ,
\hspace{1cm}
e^{\prime} = {\displaystyle\frac{1}{e}}\, .\\
\end{array}
\end{equation}

\noindent Esta 'ultima transformaci'on es el intercambio de $F$ por
${}^{\star}F$ y como vemos invierte la constante de acoplo. 'Esta es
la propiedad que hace a la dualidad el'ectrico-magn'etico (o {\it
  dualidad~S}) tan interesante pues en las variables duales es
perturbativo lo que en las originales es no-perturbativo y
proporcionan una mejor descripci'on de la teor'ia.

Dualidades perturbativas como la rotaci'on $O(N)$ entre los $N$
campos vectoriales que vimos en la Secci'on~\ref{sec-eRNsolution} se
llaman {\it dualidades~T}, al menos en el contexto de Teor'ia de
Cuerdas.  En las Teor'ias de Cuerdas tipo~II, ambos tipos de
dualidades son parte de un grupo de dualidad mucho mayor (que no es
simplemente su producto directo) y que se conoce como el {\it grupo de
  dualidad~U} \cite{kn:HT}.

%%%%%%%%%%%%%%%%%%%%%%%%%%%%%%%%%%%%%%%%%%%%%%%%%%%%%%%%%%%%%%%%%%%%%%

\subsubsection{Agujeros negros RN magn'eticos y di'onicos}
\label{sec-magneticdyonic}

Como hicimos en el caso de dualidad $O(N)$, podemos usar
la invariancia de la teor'ia de Einstein y Maxwell bajo 

\begin{equation}
\left\{
\begin{array}{rcl}
\tilde{F} & = & \cos{\xi}F +\sin{\xi}{}^{\star}F\, ,\\
& & \\
{}^{\star}\tilde{F} & = & -\sin{\xi}F +\cos{\xi}{}^{\star}F\, ,\\
\end{array}
\right.
\end{equation}

\noindent para generar soluciones nuevas, partiendo de RN o de las MP.
Tomemos la soluci'on el'ectrica de RN Ec.~(\ref{eq:rnsolution}).
Inmediatamente obtenemos una soluci'on con la misma
m'etrica\footnote{M'as adelante veremos dualidades que generan
  m'etricas distintas a la original.} y con el campo
electromagn'etico

\begin{equation}
\left\{
\begin{array}{rcl}
\tilde{F}_{tr} & = & {\displaystyle\frac{-4G_{N}^{(4)}\cos{\xi}q}{r^{2}}}\, ,\\
& & \\
\tilde{F}_{\theta\varphi} & = & 4G_{N}^{(4)}\sin{\xi}q\sin{\theta}\, .\\
\end{array}
\right.
\end{equation}

\noindent Ahora hay que expresar la nueva soluci'on en t'erminos de los
nuevos par'ametros f'isicos $\tilde{q},\tilde{p}$ que est'an
relacionados con los antiguos por

\begin{equation}
\tilde{q}= \cos{\xi}q\, ,
\hspace{.5cm}
\tilde{p}= -16G_{N}^{(4)}\sin{\xi}q\, ,
\, \,
\Rightarrow
\, \,
\tilde{q}^{2}+\left({\displaystyle\frac{\tilde{p}}{16\pi G_{N}^{(4)}}}
\right)=q^{2}\, ,
\end{equation}

\noindent eliminando el par'ametro de dualidad $\xi$, que no lo es.
La 'ultima ecuaci'on se debe a que $SO(2)$ deja invariante
$|\vec{q}|$. Suprimiendo las tildes, el resultado es un agujero negro
RN con carga el'ectrica y magn'etica ({\it di'onico})

\begin{equation}
\label{eq:dyonicRN}
\begin{tabular}{|c|}
\hline \\
$
\begin{array}{rcl}
ds^{2} & = & f(r)dt^{2}-f^{-1}(r)dr^{2} -r^{2}d\Omega_{(2)}^{2}\, , \\
& & \\
F_{tr} & = & -{\displaystyle\frac{4G_{N}^{(4)}q}{r^{2}}}\, , 
\hspace{1cm}
F_{\theta\varphi} = -{\displaystyle\frac{p}{4\pi}}\sin{\theta}\, ,\\
& & \\
f(r) & = & r^{-2}(r-r_{+})(r-r_{-})\, , \\
& & \\
r_{\pm} & = & G_{N}^{(4)}M \pm r_{0}\, ,
\hspace{1cm}
r_{0} =G_{N}^{(4)}\left\{M^{2}-4\left[q^{2}
+\left({\displaystyle\frac{p}{16\pi G_{N}^{(4)}}}\right)^{2}\right]\right\}^{1/2}\, .\\
\end{array}
$
\\ \\ \hline
\end{tabular}
\end{equation}

Todas las propiedades que dependen de la m'etrica (entre ellas
temperatura y entrop'ia) son id'enticas a las del agujero negro
RN puramente el'ectrico (con $q^{2}$ reemplazado por el invariante
$|\vec{q}|$), pero hay sutilezas con las propiedades cu'anticas y con
la versi'on euclidiana de la soluci'on y el c'alculo
euclidiano de la entrop'ia Refs.~\cite{kn:HR,kn:DHT,kn:De}.

El hecho de que las propiedades f'isicas dependan de las cargas a
trav'es de invariantes bajo dualidad (aqu'i $|\vec{q}|$) es un
hecho muy importante que se repetir'a en casos m'as complicados.

\newpage
\chapter{Agujeros negros en Supergravedad}

%%%%%%%%%%%%%%%%%%%%%%%%%%%%%%%%%%%%%%%%%%%%%%%%%%%%%%%%%%%%%%%%%%%%%%
\section{Introducci'on}

Uno de los avances m'as interesantes que se han producido en las
'ultimas d'ecadas ha sido la invenci'on de la supersimetr'ia y
su aplicaci'on en la teor'ia de las part'iculas e
interacciones fundamentales. Esta simetr'ia que relaciona bosones
y fermiones se puede entender como una generalizaci'on del grupo de
Poincar'e que es la simetr'ia del espacio de Minkowski a un {\it
  supergrupo} que lo contiene y que es la simetr'ia de un {\it
  superespacio} que est'a descrito por coordenadas bos'onicas
$x^{\mu}$ y fermi'onicas $\theta^{\alpha}$
(anticonmutantes)\footnote{Aqu'i $\alpha,\beta,\ldots$ son
  'indices de una representaci'on espinorial del grupo de Lorentz
  en la dimensi'on de que se trate.}. Las transformaciones de
supersimetr'ia relacionan estas coordenadas.

Dado que se puede entender en cierto sentido la RG como la teor'ia
gauge del grupo de Poincar'e\footnote{Una referencia muy pedag'ogica
  sobre 'este y otros temas es \cite{kn:dSG}} , es natural estudiar
su generalizaci'on basada en el supergrupo correspondiente. Esta
generalizaci'on llamada {\it Supergravedad} (SUGRA) no es 'unica
porque dado un espacio-tiempo bos'onico podemos extenderlo a un
superespacio a~nadiendo uno o $N$ conjuntos de coordenadas
fermi'onicas $\theta^{i\, \alpha}$ $i=1,\ldots,N$. Las SUGRAs basadas
en estos superespacios extendidos tienen $N$ supersimetr'ias y se
las llama {\it Supergravedades Extendidas} (SUEGRAs). Sin embargo, no
hay un n'umero infinito de SUEGRAs porque para gaugear supergrupos
con $N>8$ o con $N=1$ en $d>11$ hacen falta o varios gravitones o
part'iculas de esp'in superior que no sabemos como tratar
consistentemente.

En esta lecci'on vamos a estudiar muy someramente las SUEGRAs m'as
sencillas en $d=4$ y sus soluciones de tipo agujero negro. Definiremos
el concepto de {\it supersimetr'ia residual}, caracteriz'andola en
el super'algebra y a trav'es de los {\it espinores de Killing} y
veremos qu'e relaci'on hay entre esta propiedad y las propiedades
termodin'amicas de las soluciones.

Hay muchas referencias generales excelentes sobre Supersimetr'ia y
Supergravedad. Algunas de ellas son el art'iculo de van
Nieuwenhuizen \cite{kn:vN}, los libros de Wess y Bagger \cite{kn:WB},
de West \cite{kn:Wes2} y de Freund \cite{kn:Fr} y los art'iculos
m'as recientes de van Proeyen \cite{kn:vP} y Bilal \cite{kn:Bi2}. En
el libro de Salam y Sezgin \cite{kn:SaSe} est'an reproducidos muchos
de los art'iculos originales sobre Supergravedad y contiene una
extensa bibliograf'ia.

%%%%%%%%%%%%%%%%%%%%%%%%%%%%%%%%%%%%%%%%%%%%%%%%%%%%%%%%%%%%%%%%%%%%%%
\section{Supersimetr'ia y Supergravedad}

%%%%%%%%%%%%%%%%%%%%%%%%%%%%%%%%%%%%%%%%%%%%%%%%%%%%%%%%%%%%%%%%%%%%%%
\subsection{El super'algebra de Poincar'e $N=1,d=4$}

El supergrupo de Poincar'e $N=1,d=4$ se puede construir por
exponenciaci'on del super'algebra de Poincar'e $N=1,d=4$, que es
una extensi'on del 'algebra de Poincar'e con generadores
bos'onicos $P_{a},M_{ab}$ con generadores fermi'onicos $Q^{\alpha}$
(los {\it generadores o cargas de supersimetr'ia}) que se
transforman como espinores de Majorana bajo transformaciones de
Lorentz, por lo que tienen 4 componentes reales y

\begin{equation}
[Q^{\alpha},M_{ab}]= 
\Gamma_{s}\left(M_{ab} \right)^{\alpha}{}_{\beta} Q^{\beta}\, ,
\end{equation}

\noindent donde $\Gamma_{s}\left(M_{ab} \right)$ es el generador del 
grupo de Lorentz $M_{ab}$ en la representaci'on espinorial, mientras
que el conmutador con los $P_{a}$s es cero.  El super'algebra se
completa con el {\it anticonmutador} de las $Q^{\alpha}$s

\begin{equation}
\{Q^{\alpha},Q^{\beta}\}= 
i\left(\gamma^{a}{\cal C}^{-1}\right)^{\alpha\beta}P_{a}\, ,
\end{equation}

\noindent de forma que las relaciones de conmutaci'on no-nulas completas
del super'algebra de Poincar'e $N=1,d=4$ son

\begin{equation}
\begin{tabular}{|c|}
\hline \\ \\ 
$
\begin{array}{rcl}
[M_{ab},M_{cd}] & = & -M_{eb}\Gamma_{v}\left(M_{cd} \right)^{e}{}_{a}
-M_{ae}\Gamma_{v}\left(M_{cd} \right)^{e}{}_{b}\, ,\\
& & \\
\left[ P_{a} , M_{bc} \right] & = & 
-P_{e}\Gamma_{v}\left(M_{bc} \right)^{e}{}_{a}\, ,\\
& & \\
\left[ Q^{\alpha},M_{ab} \right] & = &  
\Gamma_{s}\left(M_{ab} \right)^{\alpha}{}_{\beta} Q^{\beta}\, ,\\
& & \\
\{Q^{\alpha},Q^{\beta}\} & = &  
i\left(\gamma^{a}{\cal C}^{-1}\right)^{\alpha\beta}P_{a}\, .\\  
\end{array}
$
\\ \\ \hline
\end{tabular}
\end{equation}

Para construir la teor'ia de SUGRA $N=1,d=4$, se ``gaugea'' este
'algebra: introducimos el potencial gauge $A_{\mu}$

\begin{equation}
A_{\mu}= e^{a}{}_{\mu}P_{a} +{\textstyle\frac{1}{2}}\omega_{\mu}{}^{ab}M_{ab}  
+\bar{\psi}_{\mu\, \alpha }Q^{\alpha}\, ,
\end{equation}

\noindent donde $e^{a}{}_{\mu}$ es la t'etrada (que va a representar
gravitones) $\omega_{\mu}{}^{ab}$ es la conexi'on de esp'in (que
ser'a una funci'on de los otros campos) y $\psi_{\mu\, \alpha}$ es
el campo de Rarita-Schwinger (que va a representar {\it gravitinos},
part'iculas de esp'in $3/2$, relacionadas con los gravitones
por supersimetr'ia), y los par'ametros infinitesimales de
transformaci'on

\begin{equation}
\Lambda =  \sigma^{a}P_{a} +{\textstyle\frac{1}{2}}\sigma^{ab}M_{ab}  
+\bar{\epsilon}_{\alpha }Q^{\alpha} \, ,
\end{equation}

\noindent donde $\sigma^{a}$ genera translaciones locales, $\sigma^{ab}$
rotaciones de Lorentz locales y $\epsilon$ transformaciones de
supersimetr'ia locales cuya acci'on en el potencial gauge es, por
definici'on

\begin{equation}
\delta A_{\mu} =\partial_{\mu}\Lambda +{\Lambda,A_{\mu}} \equiv 
\mathbb{D}_{\mu}\Lambda\, ,  
\end{equation}

\noindent la derivada supercovariante.

A partir de aqu'i los pasos son los habituales: definici'on de
curvatura y construcci'on de una acci'on que dependa de la curvatura
y que tenga las simetr'ias apropiadas (ver, por ejemplo,
\cite{kn:Fr}). El resultado es la acci'on de SUGRA $N=1,d=4$

\begin{equation}
\label{eq:n1d4psugraaction2}
\begin{tabular}{|c|}
\hline \\ \\ 
$
S[e^{a}{}_{\mu},\omega_{\mu}{}^{ab},\psi_{\mu}] =
{\displaystyle\int} d^{4}x\, e \left[ R(e,\omega) 
+2 e^{-1}\epsilon^{\mu\nu\rho\sigma} \bar{\psi}_{\mu} \gamma_{5}
\gamma_{\nu} \nabla_{\rho}\psi_{\sigma}
\right]\, ,
$
\\ \\ \hline
\end{tabular}
\end{equation}

\noindent donde

\begin{equation}
R(e,\omega) = e_{a}{}^{\mu} e_{b}{}^{\nu} R_{\mu\nu}{}^{ab} (\omega)\, ,
\end{equation}

\noindent y $R_{\mu\nu}{}^{ab} (\omega)$ es la curvatura asociada a
la conexi'on de esp'in $\omega_{\mu}{}^{ab}$ 
por Ec.~(\ref{eq:curvatures}). 

En esta acci'on se sobreentiende que se ha resuelto la ecuaci'on de
movimiento de la conexi'on de esp'in en t'erminos de
$e^{a}{}_{\mu}$ y $\psi_{\mu}$ y se ha sustituido en ella la
soluci'on\footnote{Esto se conoce como {\it formalismo de orden
    1.5}.}

\begin{equation}
\label{eq:n1d4psuperconnection}
\begin{array}{rcl}
\omega_{abc} & = & -\Omega_{abc} +\Omega_{bca} -\Omega_{cab}\, ,\\
& & \\
\Omega_{\mu\nu}{}^{a} & = & \Omega_{\mu\nu}{}^{a} (e) 
+\frac{1}{2} T_{\mu\nu}{}^{a}\, ,\\
& & \\
\Omega_{\mu\nu}{}^{a} (e) & = & \partial_{[\mu}e^{a}{}_{\nu]}\, ,
\hspace{1cm}
T_{\mu\nu}{}^{a} =  i\bar{\psi}_{\mu}\gamma^{a}\psi_{\nu}\, .\\
\end{array}
\end{equation}

La acci'on Ec.~(\ref{eq:n1d4psugraaction2}) es invariante bajo

\noindent {\bf Transformaciones generales de coordenadas}

\begin{equation}
\left\{
\begin{array}{rcl}
\delta_{\xi}x^{\mu} & = & \xi^{\mu}\, ,\\
& & \\
\delta_{\xi} e^{a}{}_{\mu} & = & 
-\xi^{\nu}\partial_{\nu} e^{a}{}_{\mu} 
-\partial_{\mu}\xi^{\nu} e^{a}{}_{\nu}\, ,\\
& & \\
\delta_{\xi} \psi_{\mu} & = & 
-\xi^{\nu}\partial_{\nu} \psi_{\mu} 
-\partial_{\mu}\xi^{\nu} \psi_{\nu}\, ,\\
\end{array}
\right.
\end{equation}

\noindent {\bf Transformaciones de Lorentz locales}

\begin{equation}
\left\{
\begin{array}{rcl}
\delta_{\sigma} e^{a}{}_{\mu} & = & 
\sigma^{a}{}_{b}e^{b}{}_{\mu}\, ,\\
& & \\
\delta_{\sigma} \psi_{\mu} & = & 
\frac{1}{2}\sigma^{ab}\gamma_{ab}\psi_{\mu}\, ,\\
\end{array}
\right.
\end{equation}

\noindent {\bf Transformaciones de supersimetr'ia locales $N=1$ }

\begin{equation}
\label{eq:n1d4psusyrules}
\left\{
\begin{array}{rcl}
\delta_{\epsilon} e^{a}{}_{\mu} & = & 
-i\bar{\epsilon}\gamma\psi_{\mu}\, ,\\
& & \\
\delta_{\epsilon} \psi_{\mu} & = & 
\nabla_{\mu}\epsilon\, .\\
\end{array}
\right.
\end{equation}

De acuerdo con este resultado, SUGRA $N=1$
no es m'as que RG acoplada a un tipo especial de materia. Esto es
cierto en todas las SUGRAs. S'olo los campos bos'onicos tienen un
l'imite cl'asico y, si buscamos soluciones cl'asicas, s'olo hace
falta considerarles a ellos\footnote{Las soluciones de la acci'on que
  resulta de poner los fermiones a cero son siempre soluciones de la
  teor'ia completa.}. En este caso, el sector bos'onico es RG en
el vac'io y todas las soluciones de las ecuaciones de Einstein en
el vac'io ({\it vgr.} Schwarzschild) son autom'aticamente
soluciones de SUGRA $N=1$.

%%%%%%%%%%%%%%%%%%%%%%%%%%%%%%%%%%%%%%%%%%%%%%%%%%%%%%%%%%%%%%%%%%%%%%
\subsection{Supersimetr'ia extendida y cargas centrales}

Si nuestro superespacio tiene $N$ conjuntos de coordenadas
fermi'onicas, el super'algebra (extendida) correspondiente tendr'a
$N$ conjuntos de generadores de supersimetr'ia que denotamos con
$i=1,\ldots N$, $Q^{i\, \alpha}$. Estas super'algebras no son muy
diferentes de las $N=1$, salvo en el anticonmutador de dos
supercargas, que ahora puede incluir {\it cargas centrales}
el'ectricas $Q^{ij}$ o magn'eticas $P^{ij}$ que, por definici'on,
conmutan con todos los dem'as generadores\footnote{Si utilizamos
  espinores de Weyl en vez de Majorana, las cargas el'ectricas y
  magn'eticas se combinan en una 'unica matriz de cargas centrales
  compleja.}. En $d=4$ el super'algebra tipo Poincar'e m'as general
que da teor'ias con invariancia Poincar'e, es, de acuerdo con el
teorema de Haag-Lopuszanski-Sohnius \cite{kn:HLS}

\begin{equation}
\begin{tabular}{|c|}
\hline \\ \\ 
$
\begin{array}{rcl}
[M_{ab},M_{cd}] & = & -M_{eb}\Gamma_{v}\left(M_{cd} \right)^{e}{}_{a}
-M_{ae}\Gamma_{v}\left(M_{cd} \right)^{e}{}_{b}\, ,\\
& & \\
\left[ P_{a} , M_{bc} \right] & = & 
-P_{e}\Gamma_{v}\left(M_{bc} \right)^{e}{}_{a}\, ,\\
& & \\
\left[ Q^{\alpha\, i},M_{ab} \right] & = &  
\Gamma_{s}\left(M_{ab} \right)^{\alpha}{}_{\beta} Q^{\beta\, i}\, ,\\
& & \\
  \{Q^{\alpha\, i},Q^{\beta\, j}\} & = & 
i\delta^{ij} \left(\gamma^{a}{\cal C}^{-1}\right)^{\alpha\beta}P_{a}
-i \left({\cal C}^{-1}\right)^{\alpha\beta}Q^{ij}
-\gamma_{5} \left({\cal C}^{-1}\right)^{\alpha\beta}P^{ij}\, .\\
\end{array}
$
\\ \\ \hline
\end{tabular}
\end{equation}

La construcci'on de SUEGRAs a partir de super'algebras extendidas es
m'as compleja y hay que utilizar otros m'etodos como el reon'omico
\cite{kn:CDAF}, pero a'un as'i podemos aprender muchas cosas
``leyendo'' el super'algebra:

\begin{enumerate}
\item El superpotencial gauge ser'ia\footnote{No se pueden gaugear
    simult'aneamente las cargas centrales el'ectricas y
    magn'eticas.}

\begin{equation}
A_{\mu}= e^{a}{}_{\mu}P_{a} +{\textstyle\frac{1}{2}}\omega_{\mu}{}^{ab}M_{ab}  
+{\textstyle\frac{1}{2}} A^{ij}{}_{\mu}Q^{ij}
+\bar{\psi}^{i}{}_{\mu\, \alpha }Q^{i\, \alpha}\, ,
\end{equation}

\noindent de lo que deducimos que la SUEGRA correspondiente tendr'a $N$
gravitinos adem'as de $N(N-1)/2$ potenciales gauge abelianos. 

\item La teor'ia ser'a invariante bajo rotaciones $SO(N)$ de estos
  vectores y gravitinos (el super'algebra lo es).
  
\item Los gravitinos no est'an cargados con respecto a los
  potenciales gauge\footnote{Pero se puede hacer que lo est'en,
    gaugeando la simetr'ia global $SO(N)$. La teor'ia contiene
    siempre el n'umero preciso de vectores.}.
  
\item El super'algebra es invariante bajo transformaciones {\it
    quirales-duales} que intercambian las cargas el'ectricas y
  magn'eticas. La SUEGRA correspondiente tendr'a como simetr'ia
  de las ecuaciones del movimiento transformaciones {\it
    quirales-duales} en las que los potenciales gauge son sustituidos
  por sus duales el'ectricos-magn'eticos.

\end{enumerate}

Como veremos los posibles estados de las SUEGRAS van a estar
caracterizados tanto por sus propiedades de transformaci'on con
respecto a los generadores Poincar'e (masa, momento, esp'in) como
por sus propiedades de transformaci'on con respecto a las cargas
centrales (estar'an cargados el'ectrica o magn'eticamente con
respecto a ciertos potenciales). Nosotros vamos a estar interesados en
soluciones cl'asicas que podamos asociar a estos estados
interpret'andolas como los campos de largo alcance generados por esos
estados.

%%%%%%%%%%%%%%%%%%%%%%%%%%%%%%%%%%%%%%%%%%%%%%%%%%%%%%%%%%%%%%%%%%%%%%

\subsubsection{Extensiones cuasi-centrales}

Cabe ahora preguntarse cu'al es el super'algebra de tipo
Poincar'e m'as general si no mantenemos la invariancia Poincar'e.
El resultado \cite{kn:vHvP} es que el super'algebra anterior se puede
generalizar incluyendo ``cargas centrales'' con $n$ 'indices
Lorentz antisim'etricos y dos 'indices $SO(N)$ $Z_{a_{1}\cdots
  a_{n}}^{ij}$ que aparecen en el anticonmutador de dos supercargas
gen'ericamente as'i

\begin{equation}
{\textstyle\frac{1}{n!}}
\left(\gamma^{a_{1}\cdots a_{n}}{\cal C}^{-1}\right)^{\alpha\beta}
Z_{a_{1}\cdots a_{n}}^{ij}\, .
\end{equation}

\noindent Estas cargas no son centrales en sentido estricto (de ah'i 
el adjetivo de {\it cuasi-centrales}) pues no conmutan con los
generadores Lorentz, sino que

\begin{equation}
\left[Z^{kl}_{c_{1}\cdots c_{n}},
M_{ab} \right]  =-n\Gamma_{v}\! \left(M_{ab}\right)\!{}^{e}
{}_{[c_{1}} Z^{kl}_{|e|c_{2}\cdots c_{n}]}\, .
\end{equation}

\noindent Por otro lado, son sim'etricas o antisim'etricas
en los 'indices $SO(N)$ dependiendo de si
$\left(\gamma^{a_{1}\cdots a_{n}}{\cal C}^{-1}\right)^{\alpha\beta}$
es sim'etrico o antisim'etrico en los 'indices espinoriales
$\alpha\beta$. En $d=4$ (y, de forma similar, en cualquier otra dimensi'on)
es f'acil determinar la simetr'ia de los posibles t'erminos:

\begin{equation}
{\cal C}^{-1}\, ,\,\,\,\,
\gamma_{5}{\cal C}^{-1}\, ,\,\,\,\,
\gamma_{5}\gamma_{a}{\cal C}^{-1}\, ,\,\,\,\,  
\gamma_{abc}{\cal C}^{-1}\, ,\,\,\,\,
\gamma_{abcd}{\cal C}^{-1}\, ,\,\,\,\,
\end{equation}

\noindent son antisim'etricos, pero el primero y el quinto y el segundo
 y el cuarto est'an relacionados por Ec.~(\ref{eq:dualgammaidentityind4}).
Por otro lado,

\begin{equation}
\gamma_{a}{\cal C}^{-1}\, ,\,\,\,\,  
\gamma_{ab}{\cal C}^{-1}\, ,\,\,\,\,
\gamma_{5}\gamma_{ab}{\cal C}^{-1}\, ,\,\,\,\,
\gamma_{5}\gamma_{abc}{\cal C}^{-1}\, ,\,\,\,\,
\end{equation}

\noindent son sim'etricas. La primera y la cuarta y la segunda y la 
tercera est'an relacionadas por Ec.~(\ref{eq:dualgammaidentityind4}).

As'i, el anticonmutador m'as general de dos supercargas en $d=4$ es

\begin{equation}
\begin{array}{rcl}
\{Q^{\alpha\, i},Q^{\beta\, j}\} & = & 
i\delta^{ij} \left(\gamma^{a}{\cal C}^{-1}\right)^{\alpha\beta}P_{a}
+i \left({\cal C}^{-1}\right)^{\alpha\beta}Z^{[ij]}
+\gamma_{5} \left({\cal C}^{-1}\right)^{\alpha\beta}\tilde{Z}^{[ij]}\\
& & \\
& & 
+\left(\gamma^{a}{\cal C}^{-1}\right)^{\alpha\beta} Z_{a}^{(ij)}
+i\left(\gamma_{5}\gamma^{a}{\cal C}^{-1}\right)^{\alpha\beta} Z_{a}^{[ij]}\\
& & \\
& & 
+i\left(\gamma^{ab}{\cal C}^{-1}\right)^{\alpha\beta} Z_{ab}^{(ij)}
+\left(\gamma_{5}\gamma^{ab}{\cal C}^{-1}\right)^{\alpha\beta}
\tilde{Z}_{ab}^{(ij)}\, .\\
\end{array}  
\end{equation}

?`Qu'e tipo de estado puede portar estas cargas? Veremos que de forma
natural son los objetos extensos (no-puntuales, como cuerdas,
membranas y sus generalizaciones con m'as dimensiones ) los que las
portan. Estos objetos rompen la invariancia Poincar'e en las
direcciones transversas a su volumen, y esa rotura est'a asociada a
los 'indices Poincar'e de las cargas.

%%%%%%%%%%%%%%%%%%%%%%%%%%%%%%%%%%%%%%%%%%%%%%%%%%%%%%%%%%%%%%%%%%%%%%
\subsection{Supersimetr'ia residual}

Antes de definir supersimetr'ia residual, es oportuno hacer
algunas observaciones de car'acter general.

Las soluciones de una teor'ia dada normalmente rompen la
mayor'ia (o todas) las simetr'ias de la misma. A veces
preservan alguna de 'estas, que recibe el nombre de {\it
  simetr'ia residual}. Las dem'as simetr'ias de la
teor'ia de pueden utilizar para generar nuevas soluciones. Veamos
dos ejemplos:

\begin{description}
\item[Mec'anica:] El lagrangiano de una part'icula libre es
  invariante bajo todo el grupo de Poincar'e, pero cualquier
  soluci'on es una recta, invariante tan s'olo bajo translaciones a
  lo largo de ella misma y rotaciones en las que es tomada como eje.
  Estas son las simetr'ias residuales de las soluciones. Las otras
  transformaciones de Poincar'e mueven la recta y generan otras
  soluciones.
  
\item[Teor'ia de Campos:] Las ecuaciones de Einstein son
  invariantes bajo difeomorfismos arbitrarios, pero las
  transformaciones que dejan invariante una soluci'on determinada
  forman su {\it grupo de isometr'ias}, que es de dimensi'on
  finita. Los generadores de estas transformaciones se hallan
  resolviendo la {\it ecuaci'on de Killing}

  \begin{equation}
    \delta g_{\mu\nu} =-2\nabla_{(\mu}k_{\nu)}=0\, .
  \end{equation}
  
  Las dem'as transformaciones generan m'etricas que son equivalentes
  (si las transformaciones dejan invariantes las condiciones de
  contorno) o no (si no lo hacen).

\end{description}

El segundo ejemplo es evidentemente el m'as interesante. En 'el la
existencia de simetr'ias residuales tiene consecuencias m'as
transcendentales: las part'iculas que se mueven en un espacio-tiempo
con isometr'ias tienen cantidades conservadas asociadas a 'estas
({\it vgr.}~en Minkowski, cuyo grupo de isometr'ias es el de
Poincar'e, el momento y el momento angular de las part'iculas
est'an conservados). Si hacemos Teor'ia de Campos es ese
espacio-tiempo el grupo de isometr'ia ser'a el grupo de
simetr'ia de esa Teor'ia de Campos.

En este contexto, los vac'ios de las teor'ias de campos se
suelen identificar con las soluciones cl'asicas que tienen un grupo
de simetr'ias residuales m'aximo. As'i el espacio-tiempo de
Minkowski es el vac'io de la RG sin constante cosmol'ogica porque
tiene un grupo de isometr'ias m'aximo (10-dimensional) y el
espacio de Sitter (anti-de Sitter) es el vac'io de la RG con
constante cosmol'ogica positiva (negativa) porque su grupo de
isometr'ia es tambi'en m'aximo: $SO(1,4)$ ($SO(2,3)$).

El concepto de supersimetr'ia residual (conservada o preservada)
no es m'as que la aplicaci'on directa del concepto general de
simetr'ia residual a las soluciones cl'asicas (es decir:
puramente bos'onicas) de las SUGRAs: una soluci'on tiene
supersimetr'ias residuales (o es supersim'etrica o BPS) si es
invariante bajo alguna transformaci'on de supersimetr'ia local.
Si denotamos por $B$ a los bosones y $F$ a los fermiones, buscamos
entonces soluciones tales que

\begin{equation}
  \begin{array}{rcl}
\delta_{\epsilon} B & \sim & \epsilon F =0\, ,\\
& & \\
\delta_{\epsilon} F & \sim & 
\left\{
\begin{array}{l}
\partial\epsilon +B\epsilon\\
\partial B \epsilon +B\epsilon\\
\end{array}
\right\}
=0\, .\\
  \end{array}
\end{equation}

\noindent Las primeras ecuaciones se cumplen siempre ($F=0$). Las segundas
s'olo en ciertos casos y se llaman {\it ecuaciones de los espinores
  de Killing}.

Por ejemplo, en SUGRA $N=1,d=4$ la ecuaci'on del espinor de Killing
es, de acuerdo con Ecs.~(\ref{eq:n1d4psusyrules})

\begin{equation}
\delta_{\epsilon} \psi_{\mu} = \nabla_{\mu}\epsilon=0\, ,  
\end{equation}

\noindent y s'olo admite soluciones en dos casos: Minkowski, con $\epsilon$
constante (4 supersimetr'ias residuales, una por cada componente
arbitraria de $\epsilon$) que es m'aximamente supersim'etrica y las
ondas planas para las que $\epsilon$ ha de satisfacer la ligadura
algebraica

\begin{equation}
\label{eq:contraintmasslessparticle}
(1-\gamma^{0}\gamma^{1})\epsilon=0\, ,
\end{equation}

\noindent que s'olo deja dos componentes ($1/2$ del total) arbitrarias,
por lo que s'olo preserva $1/2$ de las supersimetr'ias.

En 'esta y las pr'oximas lecciones vamos a ver muchos m'as ejemplos
de ecuaciones de espinores de Killing y de espacio-tiempos en los que
tienen soluciones.

Las soluciones supersim'etricas tienen interesantes propiedades:

\begin{enumerate}
  
\item La existencia de supersimetr'ias residuales implica la de
  isometr'ias puesto que el anticonmutador de dos transformaciones
  de supersimetr'ia es esencialmente un difeomorfismo. El vector
  de Killing $k^{\mu}$ est'a dado por una expresi'on de este tipo:

  \begin{equation}
  k^{\mu}\sim\bar{\epsilon}\gamma^{\mu}\epsilon\, ,
  \end{equation}

\noindent donde $\epsilon$ es un espinor de Killing.

\item Las supersimetr'ias residuales y las isometr'ias forman
  un supergrupo. 
  
\item Si hacemos Teor'ia de Campos en una soluci'on
  supersim'etrica, tal teor'ia tendr'a el anterior supergrupo
  como grupo supersimetr'ia global.
  
\item Las soluciones supersim'etricas en general no tienen
  correcciones cu'anticas (o 'estas son limitadas, depende de la
  cantidad de supersimetr'ia preservada). Tambi'en son estables
  frente a perturbaciones cl'asicas. Las razones para esto se pueden
  explicar mejor en el lenguaje del super'algebra.

\end{enumerate}

%%%%%%%%%%%%%%%%%%%%%%%%%%%%%%%%%%%%%%%%%%%%%%%%%%%%%%%%%%%%%%%%%%%%%%
\subsubsection{Supersimetr'ias residuales en el super'algebra}

Hemos dicho que estamos particularmente interesados en soluciones que
se pueden interpretar como configuraciones de campos debidas a ciertos
estados de nuestra teor'ia. Hay pues una relaci'on entre la
acci'on de las transformaciones de simetr'ia sobre las soluciones
y la acci'on de los elementos del super'algebra sobre los estados
asociados. Si la soluci'on es invariante bajo cierta
transformaci'on, entonces el estado correspondiente ser'a aniquilado
por cierto elemento del super'algebra. Si hay supersimetr'ias
residuales, esperamos que la carga 

\begin{equation}
\delta_{\epsilon}|s>\sim\bar{\epsilon}^{i}_{\alpha}Q^{i\, \alpha}|s>=0\, .  
\end{equation}

El anticonmutador de esta supercarga consigo misma nos da una
expresi'on del tipo (en $d=4$)

\begin{equation}
  \begin{array}{rcl}
\bar{\epsilon}\mathfrak{M} \epsilon & = &0\, ,  \\
& & \\
\mathfrak{M} & \equiv & i\delta^{ij} \gamma^{a}P_{a}
+i Z^{[ij]}
+\gamma_{5}\tilde{Z}^{[ij]}
+\gamma^{a} Z_{a}^{(ij)}
+i\gamma_{5}\gamma^{a} Z_{a}^{[ij]}
+i\gamma^{ab} Z_{ab}^{(ij)}
+\gamma_{5}\gamma^{ab}\tilde{Z}_{ab}^{(ij)}\, ,\\
\end{array}
\end{equation}

\noindent donde hemos sustituido los generadores por su valor sobre 
el estado $|s>$. Esta ecuaci'on tiene soluciones si el determinante
de la matriz $\mathfrak{M}$ es cero. Resolverla completamente es un
problema muy complicado, pero afortunadamente hay soluciones simples
con interpretaci'on f'isica simple. Consideremos, por ejemplo, el
estado correspondiente a una part'icula puntual sin carga y sin
masa, cuyo momento es de tipo luz $P^{2}=0$. En el sistema de
referencia en el que se mueve en la direcci'on 1, $(P^{\mu})=(p,\pm
p,0,\ldots,0)$ y (por simplicidad, en el caso $N=1$) la matriz $M$ es

\begin{equation}
\mathfrak{M}= ip\gamma^{0}\left(1\pm \gamma^{0}\gamma^{1} \right)\, .  
\end{equation}

Es f'acil ver que $M$ es singular porque la mitad de los autovalores
de $\gamma^{0}\gamma^{1}$ son $+1$ y la otra mitad $-1$, con lo que la
mitad de los autovalores de $\mathfrak{M}$ son $0$. Las
supersimetr'ias que dejan invariantes el estado de esa
part'icula son generadas por $\epsilon$s que satisfacen
$\mathfrak{M}\epsilon=0$, que coincide precisamente con la ligadura
Ec.~(\ref{eq:contraintmasslessparticle}). De esto concluimos que
podemos asociar las soluciones que describen ondas gravitacionales
planas con estado de part'iculas sin masa movi'endose a la
velocidad de la luz y que ambos preservan la mitad de las
supersimetr'ias. Soluciones de este tipo hay en todas las SUEGRAS.

El siguiente caso en simplicidad es el de una part'icula de masa
$M$ y carga el'ectrica $Q$ en SUEGRA $N=2,d=4$
($Q^{ij}=Q\epsilon^{ij}$). En el sistema de referencia en el que la
part'icula est'a en reposo $(P^{\mu})=(M,,0,\ldots,0)$ y

\begin{equation}
\mathfrak{M}=i\gamma^{0}M \left(
\delta^{ij}+\frac{Q}{M} \gamma^{0}\epsilon^{ij}
\right)\, .  
\end{equation}

Esta matriz $8\times 8$ es singular s'olo cuando $Q=\pm M$, en cuyo
caso habr'ia supersimetr'ias residuales generadas por cargas
de supersimetr'ia $\bar{\epsilon}^{i}Q^{i}$ donde
$\bar{\epsilon}^{i}$ satisface la ligadura

\begin{equation}
(\delta^{ij}\pm \gamma^{0}\epsilon^{ij})\epsilon^{j}=0\, .
\end{equation}

Esta ecuaci'on tiene 4 soluciones independientes de las 8 posibles y
as'i el estado de que hablamos preserva la mitad de las
supersimetr'ias. ?`A qu'e soluci'on de la SUEGRA $N=2,d=4$
corresponde este estado? Ha de corresponder a un objeto
esf'ericamente sim'etrico (lo m'as parecido a un objeto puntual),
est'atico, de carga $M=|Q|=2|q|$ con la normalizaci'on correcta:
!`un agujero ERN con carga el'ectrica\footnote{Con carga magn'etica
  $P=\pm M$ tendr'iamos el proyector $(\delta^{ij}\pm
  \gamma_{5}\gamma^{0}\epsilon^{ij})$ y en el caso di'onico
  $P^{2}+Q^{2}=M^{2}$ el proyector ser'ia $[\delta^{ij}\pm
  (\cos{\xi}+i\sin{\xi}\gamma_{5})\gamma^{0}\epsilon^{ij}]=
  [\delta^{ij}\pm e^{i\gamma_{5}}\gamma^{0}\epsilon^{ij}]$. }! Esto
debemos demostrarlo encontrando los espinores de Killing. Veremos que
efectivamente esto es as'i en la siguiente secci'on.

?`Qu'e pasa en el caso general? Centr'emonos primero en $d=4$ con
cargas estrictamente centrales. Para atacar el problema general es
conveniente usar una base de Weyl en vez de una base Majorana, de
forma que las matrices de cargas el'ectricas y magn'eticas se
combinan en una sola matriz antisim'etrica ay compleja de cargas
centrales $\mathbf{Z}^{ij}$, que tiene $[N/2]$ autovalores $Z_{i}$. De
acuerdo con los autores de la Ref.~\cite{kn:FSZ}, en el sistema de
referencia en el que la part'icula masiva est'a en reposo, la
matriz $\mathfrak{M}$ es singular cuando el m'odulo de uno o varios
de estos autovalores es igual a la masa $M=|Z_{i}|$. La cantidad de
supersimetr'ia preservada depende del n'umero de m'odulos de
autovalores que sean iguales. Si todos son iguales e igual a $M$,
entonces se preservan la mitad de las supersimetr'ias y en los
dem'as casos, menos. Veamos qu'e pasa en los casos que nos interesan
$N=1,2,4,8,d=4$:

\begin{description}
\item[N=1] No hay estados masivos supersim'etricos.
\item[N=2] $1/2$ de la supersimetr'ia se preserva
  cuando\footnote{En $N=2$ el 'unico autovalor es $Z=Q+iP$.}
  \begin{displaymath}
    M=|Z|\, ,
  \end{displaymath}
\item[N=4] $1/2$ de las supersimetr'ias se preserva si
  \begin{displaymath}
    M=|Z_{1}|=|Z_{2}|\, ,
  \end{displaymath}
  y $1/4$ si 
  \begin{displaymath}
    M=|Z_{1}|\neq |Z_{2}|\, .
  \end{displaymath}
\item[N=8] $1/2$ de las supersimetr'ias se preserva si
  \begin{displaymath}
    M=|Z_{1}|=|Z_{2}|=|Z_{3}|=|Z_{4}|\, ,
  \end{displaymath}
$1/4$ si
  \begin{displaymath}
    M=|Z_{1}|=|Z_{2}|\neq |Z_{3,4}|\, ,
  \end{displaymath}
  y $1/8$ si 
  \begin{displaymath}
    M=|Z_{1}|\neq |Z_{2,3,4}|\, .
  \end{displaymath}
\end{description}

En todos los casos, los proyectores sobre los espinores son de la
forma

\begin{equation}
(\delta^{ij}+ \gamma^{0}\alpha^{ij})\, ,
\end{equation}

\noindent donde $\alpha^{ij}$ es una matriz que depende de la SUEGRA
concreta con que estemos trabajando. Estos proyectores deben de
asociarse, pues a estados de part'iculas puntuales masivas. Hay un
proyector por cada factor de $1/2$ de supersimetr'ia y deben
conmutar entre s'i para que haya supersimetr'ia.

?`A qu'e tipo de soluciones van a estar asociados estos estados?
Claramente, a generalizaciones de ERN (soluciones esf'ericamente
sim'etricas, est'aticas, cargadas) que depender'an de los campos
que haya en la SUEGRA. Las m'as estudiadas y conocidas son las de
$N=4,d=4$ que veremos en la Secci'on~\ref{sec-n4d4solutions}. Sobre
$N=8,d=4$ diremos algo en las lecciones siguientes.

Pasemos ahora a estudiar la situaci'on cuando hay una carga
cuasi-central (real) $Z^{(p)}_{a_{1}\cdots a_{p}}$. Utilizando
identidades de las matrices gamma y el sistema de referencia apropiado,
siempre es posible escribir 

\begin{equation}
\label{eq:proyectorpbrana}
\mathfrak{M}=i\gamma^{0}M \left(
\delta^{ij}+\frac{Z^{(p)}}{M} \gamma^{0}\gamma^{1}\cdots\gamma^{p}\alpha^{ij}
\right)\, ,  
\end{equation}

\noindent que es singular si $M=Z^{(p)}$. Aqu'i el proyector no puede
ser interpretado como el de una part'icula puntual. De hecho, las
cargas con $p$ 'indices Lorentz est'an asociadas en la SUEGRA a
potenciales que son $(p+1)$-formas diferenciales
$A^{(p+1)}{}_{\mu_{1}\cdots\mu_{p+1}}$ que, al igual que una 1-forma
(potencial vector) se acopla a la l'inea del universo de una
part'icula a trav'es de un t'ermino de Wess-Zumino, se acopla al
{\it volumen del Universo} de un objeto de $p$ dimensiones espaciales
o {\it p-brana}. As'i 'este es el proyector asociado a una
$p$-brana supersim'etrica que preserva $1/2$ de las
supersimetr'ias.

En la Secci'on~\ref{sec-intersections} estudiaremos el caso en que
hay varias cargas de este tipo distintas de cero a un tiempo, que
interpretaremos como intersecciones de $p$-branas. En estos casos, el
espinor $\epsilon$ ha de satisfacer varias ligaduras
simult'aneamente. Esto sugiere la interpretaci'on de los agujeros
negros con varias cargas como superposiciones de agujeros negros m'as
simples. Veremos en la Secci'on~\ref{sec-n8d4solutions} que esta
interpretaci'on es posible en ciertos casos.

%%%%%%%%%%%%%%%%%%%%%%%%%%%%%%%%%%%%%%%%%%%%%%%%%%%%%%%%%%%%%%%%%%%%%%
\subsubsection{Estabilidad y cotas BPS}

La matriz $\mathfrak{M}$ tiene otra propiedad importante: es
esencialmente el cuadrado de las supercargas y sus autovalores deben
de tener propiedades de positividad. De hecho, se puede demostrar que
el jamiltoniano de las teor'ias supersim'etricas es siempre
no-negativo \cite{kn:DT2} o, lo que es lo mismo, que $M\geq 0$.  A'un
m'as: en teor'ias con supersimetr'ia extendida, la masa
est'a acotada por debajo por los autovalores de la matriz de cargas
centrales $\mathbf{Z}^{ij}$ \cite{kn:WO,kn:FSZ}

\begin{equation}
M\geq |Z_{i}|\, ,
\hspace{1cm}
i=1,\ldots,[N/2]\, ,  
\end{equation}

\noindent lo que se conoce como {\it cotas de Bogomol\'\ nyi} o {\it cotas BPS}.
Es cuando alguna de estas cotas se satura, cuando el estado tiene
supersimetr'ias residuales (es ``BPS'').

Estas cotas juegan un papel crucial en la estabilidad de los estados y
de las teor'ias. Los estados BPS tienen los valores m'inimos
de la energ'ia para valores dados de las cargas centrales. Estos
'ultimos no pueden variar porque en las teor'ias no hay campos
cargados con respecto a ellas (por eso a veces se las califica como
cargas topol'ogicas) y por lo tanto los estados BPS son estables.

Despu'es de ver el lado algebraico/cu'antico\footnote{Todo el
  razonamiento y los resultados anteriores se aplican s'olo a
  teor'ias cu'anticas en las que los estados han de estar en
  representaciones unitarias de los grupos de simetr'ia (T. de
  Wigner). En teor'ias cl'asicas de campos, estos resultados no
  tienen aplicaci'on directa y se necesitan condiciones adicionales:
  comportamiento asint'otico y condiciones sobre la energ'ia. A
  estos resultados nos vamos a referir ahora.} de las cotas, veamos el
lado ``supergravitatorio''.  Como hemos visto, SUGRA $N=1$ no es m'as
que RG acoplada a un campo fermi'onico. Esto implica que la
energ'ia (masa) en RG tambi'en ha de ser no-negativa
\cite{kn:Gri}. La demostraci'on rigurosa de este teorema fue hecha
por Schoen y Yau en Ref.~\cite{kn:SY} con t'ecnicas distintas, pero
esta demostraci'on fue seguida inmediatamente por otra versi'on de
Witten \cite{kn:Wi5} basada de nuevo en SUGRA. Esta t'ecnica fue
perfeccionada por Nester e Israel Refs.~\cite{kn:Ne,kn:IN} y
generalizada por Gibbons, Hull y otros \cite{kn:GHu,kn:ILPT}. Los
resultados de la t'ecnica de Witten, Israel y Nester (WIN) son la
versi'on en SUEGRA de las cotas BPS. La relaci'on est'a explicada
en el caso $N=1$ en Ref.~\cite{kn:Hu5}.

Las estados BPS de SUEGRA tienen cargas y masas que saturan cotas BPS
y admiten espinores de Killing. Dentro de la teor'ia de SUEGRA
concreta de que se trate, gozan de estabilidad. En el caso concreto de
ERN, esta estabilidad se manifiesta en la ausencia de radiaci'on de
Hawking ($T=0$). Aqu'i aparece un nexo entre las propiedades
termodin'amicas y la supersimetr'ia del que vamos a encontrar
m'as ejemplos.

Pasamos ahora a ver ejemplos concretos de todos estos resultados
generales.

%%%%%%%%%%%%%%%%%%%%%%%%%%%%%%%%%%%%%%%%%%%%%%%%%%%%%%%%%%%%%%%%%%%%%%
\section{Agujeros negros en Supergravedad $N=2,d=4$}

%%%%%%%%%%%%%%%%%%%%%%%%%%%%%%%%%%%%%%%%%%%%%%%%%%%%%%%%%%%%%%%%%%%%%%
\subsection{SUEGRA $N=2,d=4$}

SUEGRA $N=2,d=4$ fue construida originalmente por Ferrara y van
Nieuwenhuizen \cite{kn:FPVN} acoplando un supermultiplete vectorial
(compuesto por un vector y un gravitino) a SUGRA $N=1,d=4$ y
observando que la teor'ia resultante era invariante bajo una
transformaci'on de supersimetr'ia adicional\footnote{Esto es
  similar a la construcci'on de SUGRA $N=1,d=4$ a partir de $N=0$ (RG)
  acoplando un gravitino.}. As'i, el supermultiplete de SUEGRA
$N=2,d=4$ consta de una t'etrada, una pareja de gravitinos reales y
un campo vectorial

\begin{equation}
\{e^{a}{}_{\mu},\psi_{\mu}=
\left(
\begin{array}{c}
\psi^{1}_{\mu}\\
\psi^{2}_{\mu}\\
\end{array}
\right),A_{\mu}\}\, .
\end{equation}

\noindent Los 'indices  $SO(2)$  $i=1,2$ no los escribiremos 
excepto cuando sea completamente necesario. La acci'on en el
formalismo de orden 1.5 es

\begin{equation}
\label{eq:n2d4psugraaction}
\begin{tabular}{|c|}
\hline \\ \\
$
S={\displaystyle\int} d^{4}x\, e \left\{ R(e,\omega) 
+2 e^{-1}\epsilon^{\mu\nu\rho\sigma}\bar{\psi}_{\mu}\gamma_{5}\gamma_{\nu}
\nabla_{\rho}\psi_{\sigma}
-{\cal F}^{2} +{\cal J}_{(m)}{}^{\mu\nu}
({\cal J}_{(e) \mu\nu} +{\cal J}_{(m) \mu\nu}) \right\}\, ,
$
\\ \\ \hline
\end{tabular}
\end{equation}

\noindent donde

\begin{equation}
\left\{
\begin{array}{rcl}
{\cal F}_{\mu\nu} & = &  \tilde{F}_{\mu\nu} +{\cal J}_{(m) \mu\nu}\, , \\
& & \\
\tilde{F}_{\mu\nu} & = & F_{\mu\nu} +{\cal J}_{(e) \mu\nu}\, ,\\
& & \\
F_{\mu\nu} & = & 2\partial_{[\mu} A_{\nu]}\, ,\\
\end{array}
\right.
\end{equation}

\noindent y

\begin{equation}
\left\{
\begin{array}{rcl}
{\cal J}_{(e) \mu\nu} & = & i\bar{\psi}_{\mu}\sigma^{2}\psi_{\nu}\, ,\\
& & \\
{\cal J}_{(m) \mu\nu} & = & -\frac{1}{2e}\epsilon^{\mu\nu\rho\sigma}
\bar{\psi}_{\rho}\gamma_{5}\sigma^{2}\psi_{\sigma}\, .\\
\end{array}
\right.
\end{equation}

La ecuaci'on de $\omega_{\mu}{}^{ab}$ es la misma que en el caso
$N=1$ y la soluci'on es tambi'en Ec.~(\ref{eq:n1d4psuperconnection})
pero con la torsi'on dada por

\begin{equation}
T_{\mu\nu}{}^{a}  =  i\bar{\psi}_{\mu}\gamma^{a}\psi_{\nu}
\,\,\, (\equiv i\bar{\psi}_{j\, \mu}\gamma^{a}\psi^{j}_{\nu})\, .
\end{equation}

Esta acci'on es invariante bajo 1. difeomorfismos, 2. transformaciones
Lorentz locales, 3. transformaciones gauge del potencial vector,
4. rotaciones globales $SO(2)$ de los gravitinos y 5. transformaciones
locales de supersimetr'ia $N=2$

\begin{equation}
\label{eq:susyt}
\left\{
\begin{array}{rcl}
\delta_{\epsilon} e^{a}{}_{\mu} & = & 
-i\bar{\epsilon}\gamma^{a}\psi_{\mu}\, ,\\
& & \\
\delta_{\epsilon} A_{\mu} & = & 
-i\bar{\epsilon}\sigma^{2}\psi_{\mu}\, ,\\
& & \\
\delta_{\epsilon} \psi_{\mu} & = & 
\tilde\nabla_{\mu}\epsilon\, ,\\
\end{array}
\right.
\end{equation}

\noindent donde

\begin{equation}
\label{eq:supercovariantderivativeungauged}
\tilde{\nabla}_{\mu}= \nabla_{\mu}
+{\textstyle\frac{1}{4}}\not\!\tilde{F}\gamma_{\mu}\sigma^{2}\, ,
\end{equation}

\noindent es la {\it derivada supercovariante} sobre $\epsilon$.

Las ecuaciones de movimiento, pero no la acci'on, son invariantes
bajo un grupo global $SO(2)$ de transformaciones quirales-duales 

\begin{equation}
\label{eq:chiral-dual}
\left\{
\begin{array}{rcl}
\tilde{F}_{\mu\nu}^{\prime} & = & 
\cos{\theta}\, \tilde{F}_{\mu\nu} 
+\sin{\theta}\, {}^{\star}\tilde{F}_{\mu\nu}\, ,\\
& & \\
\psi_{\mu}^{\prime} & = & e^{\frac{i}{2}\theta\gamma_{5}}\psi_{\mu}\, .\\ 
\end{array}
\right.
\end{equation}

La correspondencia de todas estas simetr'ias con propiedades del
super'algebra correspondiente es obvia.

%%%%%%%%%%%%%%%%%%%%%%%%%%%%%%%%%%%%%%%%%%%%%%%%%%%%%%%%%%%%%%%%%%%%%%
\subsection{Soluciones}

Para nuestros prop'ositos, lo m'as interesante de la SUEGRA
$N=2,d=4$ es que su sector bos'onico no es m'as que la teor'ia
de Einstein y Maxwell que ya vimos y todas las soluciones de 'esta
lo son de la primera, en particular RN y MP.

Si RN describe un estado (posiblemente solit'onico, pues no hay
materia cargada en esta SUEGRA) de masa $M$ y carga $|Z|=|Q+iP|$,
cu'anticamente $M\geq |Z|$, justamente la condici'on de censura
c'osmica. Esto no es sorprendente, dada la relaci'on en la que hemos
insistido varias veces entre censura c'osmica y positividad de la
energ'ia. Cl'asicamente se puede probar \cite{kn:GHu} que bajo
condiciones energ'eticas usuales y en espacios asint'oticamente
planos, esa relaci'on se tiene que cumplir siempre.

La saturaci'on de la cota BPS coincide con el l'imite de
extremalidad de RN, la desaparici'on de la radiaci'on de Hawking y
la aparici'on de soluciones est'aticas con varios ERNs en equilibrio
(MP). Todas las soluciones de la familia MP admiten espinores de
Killing \cite{kn:GHu}, pero hay a'un m'as soluciones que los admiten
como demostr'o Tod en Ref.~\cite{kn:To}: la familia de soluciones de
Israel, Wilson y Perj'es (IWP) \cite{kn:IW,kn:Pe} y una
generalizaci'on de las ondas gravitatorias planas que incluyen campos
electromagn'eticos. Vamos a describir la familia IWP en la siguiente
secci'on. Aqu'i basta decir que en ella hay soluciones con
momento angular y carga NUT, pero que las 'unicas soluciones
regulares sin curvas cerradas de tipo tiempo son las de la subfamilia
de MP \cite{kn:HaHa}. En particular, no hay agujeros negros
supersim'etricos regulares con momento angular en SUGRA $N=2,d=4$.
Este resultado es v'alido para $N=4,8, d=4$ tambi'en, pero no en
$d=5$.

En cuanto al n'umero de supersimetr'ias residuales,
gen'ericamente las soluciones de IWP tiene $1/2$. Las dos excepciones
conocidas son Minkowski y RB que tienen el m'aximo n'umero de
supersimetr'ias preservadas y se pueden considerar vac'ios de
la teor'ia.

%%%%%%%%%%%%%%%%%%%%%%%%%%%%%%%%%%%%%%%%%%%%%%%%%%%%%%%%%%%%%%%%%%%%%%
\section{Agujeros negros en Supergravedad $N=4,d=4$}

%%%%%%%%%%%%%%%%%%%%%%%%%%%%%%%%%%%%%%%%%%%%%%%%%%%%%%%%%%%%%%%%%%%%%%
\subsection{SUEGRA $N=4,d=4$}

El supermultiplete de SUEGRA $N=4,d=4$ \cite{kn:CSF} consta de la
t'etrada, seis vectores abelianos, un escalar ({\it dilat'on}) un
pseudoescalar ({\it axi'on}) cuatro gravitinos reales y cuatro {\it
  dilatinos} reales (esp'in $1/2$)

\begin{equation}
\{e^{a}{}_{\mu},A^{(n)}{}_{\mu},\phi,a,\psi^{i}_{\mu},\lambda^{i}\}\, ,
\end{equation}

\noindent respectivamente. La acci'on de los campos bos'onicos es

\begin{equation}
\label{eq:n4d4bosonicaction}
S  =\int d^{4}x \sqrt{|g|}\, \left\{ R 
+2(\partial\phi)^{2} + {\textstyle\frac{1}{2}}e^{4\phi}(\partial a)^{2}
-e^{-2\phi}\sum_{n=1}^{6} F^{(n)}F^{(n)} 
+a \sum_{n=1}^{6} F^{(n)}\ ^{\star}F^{(n)} \right\}\, .
\end{equation}

Adem'as de la invariancia bajo difeomorfismos, transformaciones de
Lorentz locales, transformaciones gauge de los seis vectores y
transformaciones de supersimetr'ia local $N=4$, rotaciones
globales $SO(4)$ de vectores y fermiones y rotaciones quirales-duales,
el sector bos'onico de esta teor'ia es invariante bajo rotaciones
globales $SO(6)$ de los seis vectores\footnote{Esta simetr'ia
  est'a relacionada con la de rotaciones $SO(4)$ que esperamos a
  partir del super'algebra que se ve aumentada hasta $SU(4)$ por las
  rotaciones quirales-duales.} y (s'olo las ecuaciones del
movimiento) bajo transformaciones de dualidad $SL(2,\mathbb{Z})$ que
act'uan sobre los escalares y los vectores. Para describir esta
simetr'ia es conveniente combinar los escalares en un escalar
complejo que a veces es llamado {\it axidilat'on}

\begin{equation}
\tau =a+ie^{-2\phi}\, ,
\end{equation}

\noindent y definir los duales $SL(2,\mathbb{Z})$ de los vectores

\begin{equation}
\tilde{F}_{\mu\nu}^{(n)}=\partial_{\mu}\tilde{A}_{\nu}^{(n)}-
\partial_{\nu}\tilde{A}_{\mu}^{(n)}\, .
\end{equation}

Si $\Lambda$ es una matriz $SL(2,\mathbb{R})$ 
 
\begin{equation}
\Lambda = \left( \begin{array}{cc} 
             a & b \\ 
             c & d \\
             \end{array}\right)\, , 
\hspace{1cm}
ad-bc=1\, ,
\end{equation}

\noindent entonces los vectores y sus duales se transforman en dobletes 
(lo que antes llamamos vector de dualidad)

\begin{equation}
\left( 
\begin{array}{c} 
\tilde{F}^{(n)}{}_{\mu\nu} \\
F^{(n)}{}_{\mu\nu} \\ 
\end{array} 
\right) 
\longrightarrow 
\Lambda
\left( 
\begin{array}{c} 
\tilde{F}^{(n)}{}_{\mu\nu} \\
F^{(n)}{}_{\mu\nu} \\ 
\end{array} 
\right) \, ,
\label{Transformation-A}
\end{equation}

\noindent y $\tau$ se transforma no-linealmente

\begin{equation}
\label{Transformation-l}
\tau \longrightarrow
\frac{a\tau+b}{c\tau+d}\, .
\end{equation}

Aqu'i $e^{\phi}$ juega el mismo papel que la constante de acoplo
$e$ en la dualidad el'ectrico-magn'etico y $a$ juega el papel de un
'angulo $\theta$ local. De hecho, desde el punto de vista de la
Teor'ia de Cuerdas $e^{\phi}$ es realmente la constante de acoplo.

El grupo de dualidad de esta teor'ia es, pues el producto directo
de la dualidad $T$ ($SO(6)$) y la $S$ ($SL(2,\mathbb{R})$).

En $N=4$ hay dos cotas BPS $M^{2}-|Z_{1,2}|^{2}\geq 0$.  Ninguna de
ellas es separadamente invariante bajo dualidad, que las intercambia.
Para escribir una cota invariante ({\it cota BPS generalizada})
tomamos el producto de ambas cotas y dividimos por $M^{2}$ 

\begin{equation}
\label{eq:genBbound}
M^{2} +\frac{|Z_{1}Z_{2}|^{2}}{M^{2}} -|Z_{1}|^{2} -|Z_{2}|^{2}\geq 0\, ,
\end{equation}

\noindent que es la que debe de aparecer en la m'etrica, que es 
invariante bajo dualidad. El t'ermino $|Z_{1}Z_{2}|^{2}M^{-2}$ se
puede interpretar con las cargas de los escalares en los agujeros
negros regulares. Como esperamos por el teorema de ``no-pelo'', no son
independientes de las otras cargas.

%%%%%%%%%%%%%%%%%%%%%%%%%%%%%%%%%%%%%%%%%%%%%%%%%%%%%%%%%%%%%%%%%%%%%%
\subsection{Soluciones}
\label{sec-n4d4solutions}

Las soluciones de tipo agujero negro m'as generales de esta
teor'ia se conocen desde hace poco Ref.~\cite{kn:L-TO}.  Su
construcci'on ha sido la culminaci'on del esfuerzo de muchos
autores: las primeras soluciones fueron obtenidas por Gibbons en la
Ref.~\cite{kn:G} y la supersimetr'ia de las extremas fue
demostrada en la Ref.~\cite{kn:KLOPP}. En la Ref.~\cite{kn:STW} se
utiliz'o por primera la dualidad~S para generar soluciones
di'onicas, posteriormente generalizadas en la Ref.~\cite{kn:O1},
donde se observ'o por primera vez que las transformaciones de
dualidad~S preservan las propiedades de supersimetr'ia de las
soluciones originales. Las soluciones m'as generales sin momento
angular ni carga NUT, fueron obtenidas en la Ref.~\cite{kn:KO} y la
adici'on de carga NUT fue estudiada en la Ref.~\cite{kn:KKOT}.  La
soluciones supersim'etricas m'as generales, incluyendo momento
angular y carga NUT fueron halladas por Tod en la Ref.~ \cite{kn:To2}
y poco despu'es, independientemente en la Ref.~\cite{kn:BKO3}.

%%%%%%%%%%%%%%%%%%%%%%%%%%%%%%%%%%%%%%%%%%%%%%%%%%%%%%%%%%%%%%%%%%%%

Por simplicidad vamos a presentar 'unicamente las soluciones
supersim'etricas, llamadas SWIP en la Ref.~\cite{kn:BKO3}. Todas las
funciones que aparecen en los distintos campos de las soluci'on se
pueden expresar en t'erminos de dos funciones arm'onicas complejas
completamente arbitrarias
 ${\cal H}_{1,2}(\vec{x})$

\begin{equation}
\partial_{\underline{i}}\partial_{\underline{i}} {\cal H}_{1}=
\partial_{\underline{i}}\partial_{\underline{i}} {\cal H}_{2}= 0\, ,
\end{equation}

\noindent y un conjunto de constante complejas $k^{(n)}$ que satisfacen
las relaciones

\begin{equation}
\sum_{n=1}^{N} (k^{(n)})^{2}=0\, ,
\hspace{1cm}
\sum_{n=1}^{N} |k^{(n)}|^{2}={\textstyle\frac{1}{2}}\, ,
\end{equation}

\noindent en el caso general\footnote{Si una de las funciones 
arm'onicas es constante, entonces s'olo hace falta la segunda relaci'on.}.

Las funciones arm'onicas aparecen en la m'etrica a trav'es de las
siguientes combinaciones:

\begin{eqnarray}
e^{-2U} & = & 2\ \Im{\rm m}\ ({\cal H}_{1}\overline{\cal H}_{2})\, ,\\
& & \nonumber \\
\partial_{[\underline{i}}\
\omega_{\underline{j}]} & = &
\epsilon_{ijk}\
\Re{\rm e}\
\left( {\cal H}_{1} \partial_{\underline{k}}\overline{\cal H}_{2}
-\overline{\cal H}_{2} \partial_{\underline{k}}{\cal H}_{1} \right)\, .
\end{eqnarray}

Los campos de la soluci'on son

\begin{eqnarray}
ds^{2}  & = &
e^{2U} (dt^{2}+\omega_{\underline{i}}dx^{\underline{i}})^{2}
-e^{-2U}d\vec{x}^{2}\, ,
\\
& & \nonumber\\
\lambda & = &
\frac{{\cal H}_{1}}{{\cal H}_{2}}\, ,
\\
& & \nonumber \\
A_{t}^{(n)} & = &
2e^{2U}\Re {\rm e}\left( k^{(n)}{\cal H}_{2} \right)\, ,
\\
& & \nonumber \\
\tilde{A}^{(n)}_{t} & = &
-2e^{2U}\Re {\rm e} \left( k^{(n)}{\cal H}_{1} \right)\, .
\end{eqnarray}

%%%%%%%%%%%%%%%%%%%%%%%%%%%%%%%%%%%%%%%%%%%%%%%%%%%%%%%%%%%%%%%%%%%%%%

Estas soluciones generales se reducen a las m'etricas IWP de las que
hablamos en la Secci'on anterior cuando 

\begin{equation}
{\cal H}_{1}=i{\cal H}_{2}={\textstyle\frac{1}{\sqrt{2}}} V^{-1}\, ,
\end{equation}

\noindent donde  $V^{-1}$ es una funci'on arm'onica compleja. 
Todas las soluciones de esta subfamilia tienen $1/2$ de las
supersimetr'ias preservadas.

%%%%%%%%%%%%%%%%%%%%%%%%%%%%%%%%%%%%%%%%%%%%%%%%%%%%%%%%%%%%%%%%%%%%%%
\subsubsection{Propiedades de dualidad}

La familia de soluciones es invariante, como familia, bajo todas las
dualidades de la teor'ia, que no sirven ya para generar ninguna
soluci'on nueva. En particular, ${\cal H}_{1,2}$ se transforman como
un doblete (linealmente) bajo $SL(2,\mathbb{R})$ mientras que las $k^{(n)}$s
son invariantes y 'estas se transforman como un vector de $SO(6)$
mientras que las ${\cal H}_{1,2}$s son invariantes. 

Estas propiedades de transformaci'on de los componentes funcionales
de la soluci'on tiene una interesante interpretaci'on desde el punto
de vista de la {\it geometr'ia especial} de las SUGRAS $N=2,d=4$
(con materia en general) \cite{kn:FKS}. Esta interpretaci'on ha sido
usada para construir soluciones de tipo agujero negro extremos en
SUGRAS $N=2,d=4$ \cite{kn:BLS} lo que tiene gran importancia porque
las SUEGRAS $N=4,8,d=4$ se pueden interpretar como SUEGRA $N=2,d=4$
con una materia supersim'etrica de tipo particular. Sin embargo, esta
direcci'on de trabajo no ha sido capaz de proporcionar soluciones tan
completas como las SWIP en esos casos \cite{kn:Moh,kn:DAF}, sino que
en general s'olo ha dado como resultado ``soluciones generatrices''
con las que se podr'ia (pero nadie lo ha hecho de forma
expl'icita) generar la soluci'on m'as general con
transformaciones de dualidad. 

%%%%%%%%%%%%%%%%%%%%%%%%%%%%%%%%%%%%%%%%%%%%%%%%%%%%%%%%%%%%%%%%%%%%%%
\subsubsection{Propiedades de supersimetr'ia}

Si elegimos adecuadamente las funciones arm'onicas, las soluciones
describen objetos puntuales con masa $M$, cargas el'ectricas,
magn'eticas $\Gamma^{(n)}$, escalares $\Upsilon$ que son funci'on de las anteriores

\begin{equation}
\Upsilon=-2\frac{\sum_{n}\overline{\Gamma^{(n)}}^{2}}{M}\, ,  
\end{equation}

\noindent y momento angular
$J$. Estas cargas siempre satisfacen la identidad

\begin{equation}
M^{2} + |\Upsilon|^{2}-4\sum_{n}|\Gamma^{(n)}|^{2}=0\, ,
\end{equation}

\noindent que tiene la forma exacta de la cota BPS generalizada. De hecho,
podemos identificar los autovalores de las cargas centrales 

\begin{equation}
{\textstyle\frac{1}{2}}|Z_{1,2}|^{2} =
\sum_{n}|\Gamma^{(n)}|^{2}
\pm \left[\left(\sum_{n} |\Gamma^{(n)}|^{2}\right)^{2}
-|\sum_{n}\Gamma^{(n)2}|^{2}\right]^{\frac{1}{2}}\, .
\end{equation}

El 'area del horizonte es

\begin{equation}
A=4\pi(| M|^{2}-|\Upsilon|^{2})= 4\pi || Z_{1}|^{2} -|Z_{2}|^{2}|\, ,
\end{equation}

\noindent que se hace cero cuando las dos cargas son iguales. Este es el caso
en el que las dos cotas BPS se saturan simult'aneamente y tenemos
$1/2$ de las supersimetr'ias preservadas. El resultado es que {\it
  s'olo con dos cargas centrales distintas y $1/4$ de las
  supersimetr'ias preservadas los agujeros negros de SUEGRA
  $N=4,d=4$ tienen un 'area finita}.

Si se buscan los espinores de Killing se encuentra que siempre hay
$1/4$ de las supersimetr'ias preservadas, como indica el hecho de
que la cota BPS generalizada est'e autom'aticamente saturada en
estas soluciones. En el caso $|Z_{1}|=|Z_{2}|$, como esperamos, $1/2$
de las supersimetr'ias est'an preservadas \cite{kn:BKO3}.

%%%%%%%%%%%%%%%%%%%%%%%%%%%%%%%%%%%%%%%%%%%%%%%%%%%%%%%%%%%%%%%%%%%%%%

Finalmente, observamos que si reescribimos la f'ormula del 'area del
horizonte en t'erminos de las cargas el'ectricas y magn'eticas
normalizadas de forma que al ser cuantizadas tomen valores enteros,
$\tilde{q}^{(n)},\tilde{p}^{(n)}$, obtenemos una expresi'on que es
independiente de todos los {\it m'odulos} de la soluci'on, como los
valores asint'oticos de los escalares

\begin{equation}
A = 8\pi \sqrt{{\rm det} \left[ 
\left( 
\begin{array}{c}
\vec{\tilde{p}}^{\hspace{2pt}t} \\
\vec{\tilde{q}}^{\hspace{2pt}t} \\
\end{array}
\right) 
\left( 
\vec{\tilde{p}} \,\,
\vec{\tilde{q}} 
\right)
\right]}\, , 
\end{equation}

\noindent donde la acci'on del grupo de dualidad en el vector de carga
 $\left(\begin{array}{c}\vec{\tilde{p}} \\
\vec{\tilde{q}} \\ \end{array} \right)$ es

\begin{equation}
\left(
\begin{array}{c}
\vec{\tilde{p}} \\ 
\vec{\tilde{q}} \\ 
\end{array} 
\right)^{\prime}
=
R\otimes S
\left(
\begin{array}{c}
\vec{\tilde{p}} \\ 
\vec{\tilde{q}} \\ 
\end{array} 
\right)\, ,
\end{equation}

\noindent donde  $R\in SO(6)$ y $S\in SL(2,\mathbb{R})$.

Esta es una propiedad fundamental de la entrop'ia de los agujeros
negros extremos, que vamos a utilizar en la 'ultima lecci'on.

%%%%%%%%%%%%%%%%%%%%%%%%%%%%%%%%%%%%%%%%%%%%%%%%%%%%%%%%%%%%%%%%%%%%%%
\subsection{SUEGRA $N=8,d=4$}

%%%%%%%%%%%%%%%%%%%%%%%%%%%%%%%%%%%%%%%%%%%%%%%%%%%%%%%%%%%%%%%%%%%%%%
\subsection{Soluciones: agujeros negros compuestos}
\label{sec-n8d4solutions}

Como hemos dicho, no existen soluciones similares a las SWIP de
$N=4,d=4$ para el caso $N=8,d=4$ que es bastante complicado de
describir. Nos interesa m'as en este momento describir las
propiedades generales de las soluciones supersim'etricas.

Primeramente, es posible ver que s'olo si las cuatro cargas centrales
son distintas entre s'i y la soluci'on tiene $1/8$ de las
supersimetr'ias preservadas, las soluciones tienen un horizonte
regular. El 'area del horizonte se puede expresar tambi'en de forma
manifiestamente invariante bajo dualidad (aqu'i llamada
dualidad~U, correspondiente al grupo $E_{7,7}$) \cite{kn:KK}. La
soluciones m'as simples est'an descritas en \cite{kn:KhO1}.

En $N=8,d=4$ aparecen dos fen'omenos nuevos: hay combinaciones de las
cargas que nos permiten tener agujeros negros ``sin masa''
supersim'etricos \cite{kn:Ra,kn:O3}. Por otro lado, hay agujeros
negros extremos supersim'etricos que pierden la supersimetr'ia
cuando se cambia el signo de una carga (lo que deja invariante la
m'etrica y los escalares). Este es un hecho de dif'icil
interpretaci'on \cite{kn:O7}.

\newpage
\chapter{Teor'ias efectivas de cuerdas: acciones}

%%%%%%%%%%%%%%%%%%%%%%%%%%%%%%%%%%%%%%%%%%%%%%%%%%%%%%%%%%%%%%%%%%%%%%
\section{Introducci'on}

En esta lecci'on vamos a iniciar el estudio de los agujeros negros
desde el punto de vista de la Teor'ia de Cuerdas. Nuestro punto de
vista ser'a el de las acciones efectivas que describen la din'amica
de los campos sin masa de la Teor'ia de Cuerdas a bajas
energ'ias. Estas acciones efectivas son en general acciones de
SUEGRAS en diversas dimensiones, al menos cuando el espectro de la
teor'ia de cuerdas correspondiente es supersim'etrico y contiene
un gravit'on. De ah'i nuestro inter'es en SUEGRAS en la
lecci'on anterior.  Las referencias esenciales de introducci'on a la
Teor'ia de Cuerdas son los libros de Green, Schwarz y Witten
\cite{kn:GSW}, L\"ust y Theysen \cite{kn:LT} y el de Polchinski
\cite{kn:P3}, m'as reciente. Menos extensas, pero igualmente
interesantes son \cite{kn:Ki2,kn:EM,kn:J}.

\begin{figure}[!ht]
\begin{center}
\leavevmode
\epsfxsize= 7cm
\epsffile{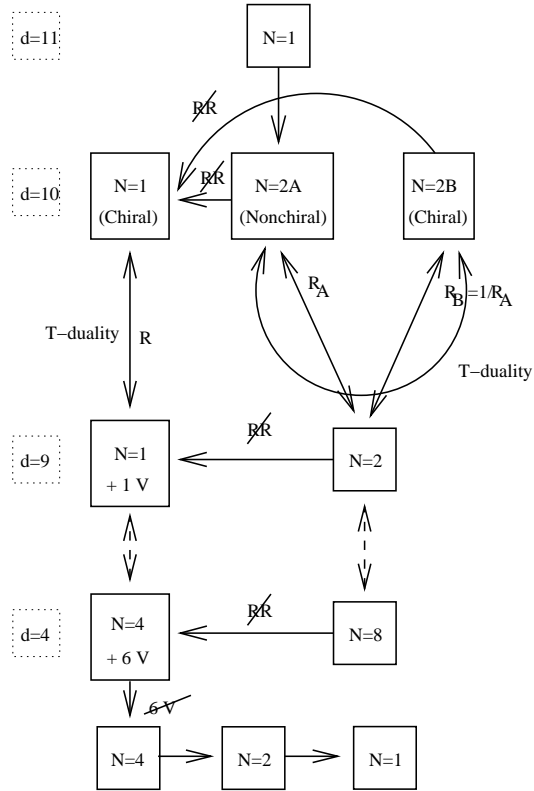}
\caption{Directorio de SUGRAs. La relaci'on entre distintas teor'ias 
  compactificadas en c'irculos (l'ineas con dos puntas de
  flecha) y truncaci'on (l'ineas con una 'unica punta de flecha)
  est'a representada aqu'i esquem'aticamente}
\label{fig:sugras}
\end{center}
\end{figure}

En este momento los te'oricos de cuerdas creen que existe la
denominada {\it Teor'ia~M} que se manifiesta en diversos
l'imites y contextos como lo que antes se cre'ia eran
diferentes teor'ias de cuerdas y supergravedad. Estas diferentes
teor'ias est'an conectadas por una red de dualidades que las
relacionan y que, se supone, ponen de manifiesto la relaci'on de
todas ellas con la Teor'ia~M. En un cierto l'imite del que
para nosotros es conveniente partir, la Teor'ia~M se manifiesta
como una teor'ia que a bajas energ'ias es SUGRA $N=1,d=11$.
Cuando esta teor'ia es compactificada en un c'irculo se
obtiene una teor'ia que a bajas energ'ias no es sino SUGRA
$N=2A,d=10$. Esta teor'ia era conocida como el l'imite a bajas
energ'ias de la Teor'ia de Cuerdas tipo~IIA y, de hecho,
cuando incluimos en ella todos los modos masivos de Kaluza-Klein que
aparecen en la compactificaci'on desde $d=11$, se obtiene todo el
espectro de esta teor'ia de cuerdas. A'un m'as interesante, el
dilat'on de la Teor'ia de Cuerdas tipo~IIA (que es su
``constante'' de acoplo) es el radio de la dimensi'on compactificada,
lo que permite visualizar SUGRA $N=1,d=11$ como el l'imite de
acoplo fuerte de la dicha teor'ia de cuerdas \cite{kn:Wi}. 'Esta
y otras relaciones que vamos a ver est'an representadas en la
Figura~\ref{fig:sugras}.

En $d=10$ hay otra SUEGRA conocida: $N=2B,d=10$, que es quiral, y es
el l'imite de baja energ'ia de la Teor'ia de Cuerdas
tipo~IIB. Resulta que cuando esta teor'ia de cuerdas y la tipo~IIA
son compactificadas en c'irculos de radios inversos, tienen
exactamente el mismo espectro y las mismas interacciones
\cite{kn:DHS,kn:DLP}. Esta {\it dualidad~T} entre ambas teor'ias
se manifiesta en sus l'imites de baja energ'ia en que las
SUEGRAS $N=2A,d=10$ y $N=2B,d=10$ son id'enticas cuando se las somete
a compactificaciones duales \cite{kn:BHO,kn:BRGPT}. Esta dualidad~T
que relaciona radios grandes y peque~nos es una de las propiedades
m'as interesantes y especiales de la Teor'ia de
Cuerdas\footnote{Una monograf'ia interesante sobre este tema es
  \cite{kn:GPR}, aunque no contiene algunos de los resultados m'as
  interesantes sobre dualidad en cuerdas abiertas y {\it D-branas}.}
que se aplica no s'olo a vac'ios simples con una dimensi'on
compacta como el producto directo de Minkowski por un c'irculo
sino a variedades m'as generales con el resultado de que
geometr'ias muy diferentes son indistinguibles desde el punto de
vista de la Teor'ia de Cuerdas \cite{kn:Bu}.

Si el l'imite de acoplo fuerte de la Teor'ia de Cuerdas
tipo~IIA es SUGRA $N=1,d=11$, ?`Cu'al es del de la tipo~IIB? Esto es
muy dif'icil de decir desde el punto de vista de la teor'ia de
cuerdas, que est'a definida perturbativamente (es decir: para constante
de acoplo peque~na), pero su teor'ia efectiva de bajas
energ'ias nos da una pista: las ecuaciones del movimiento de la
SUEGRA $N=2B,d=10$ son invariantes baso una {\it dualidad~S} que
relaciona acoplos fuertes y d'ebiles. As'i, cuando el acoplo es
fuerte, la teor'ia se puede reescribir en t'erminos de {\it otra}
Teor'ia de Cuerdas tipo~IIB con constante de acoplo peque~na.

Conforme vamos compactificando m'as dimensiones de estas teor'ias
de cuerdas, aparecen otras dualidades que se manifiestan (no siempre)
como simetr'ias globales de las SUEGRAS correspondientes. El grupo
de todas estas dualidades de las teor'ias tipo~II compactificadas
en toros es el {\it grupo de dualidad~U} \cite{kn:HT}.

Hasta este momento s'olo hemos hablado de teor'ias con dos
supersimetr'ias en $d=10$ ($N=1$ en $d=11$ $N=8$ en $d=4$), pero
una de las teor'ias de cuerdas m'as interesantes, la
heter'otica, tiene la mitad. Para hablar de 'esta y otras
teor'ias de cuerdas con $N=1,d=10$ tenemos que mencionar la
segunda idea que ha modificado nuestra forma de entender la
Teor'ia de Cuerdas: que 'esta describe no s'olo objetos
unidimensionales (cuerdas) sino objetos de m'as dimensiones {\it
  $p$-branas} y {\it $p$-orientiplanos} y que se pueden construir
nuevos vac'ios de la teor'ia no s'olo por el procedimiento de
compactificar en ciertas variedades, sino poniendo $p$-branas o
$p$-orientiplanos, cuyo efecto es el de romper parcialmente las
supersimetr'ias, hacer que aparezcan nuevos estados sin masa (en
particular campos de Yang-Mills con grupos gauge no-Abelianos) y
truncar las teor'ias. Existen varios tipos de estos objetos,
algunos de ellos no muy bien conocidos, y todos ellos est'an
relacionados con las cargas cuasi-centrales de los super'algebras de
las SUEGRAS correspondientes.

As'i, si para SUGRA $N=1,d=10$ elegimos un vac'io de la forma
$S^{1}/\mathbb{Z}_{2}$ y ponemos dos 9-branas (``M9-branas'') en los
puntos fijos, en vez de obtener la Teor'ia de Cuerdas tipo~IIA
obtenemos la Teor'ia Heter'otica $E_{8}\times E_{8}$
\cite{kn:HorWit} cuyo l'imite de baja energ'ia est'a descrito
por SUGRA $N=1,d=10$ acoplado a los correspondientes supermultipletes
vectoriales. Y si en el vac'io de la Teor'ia de Cuerdas
tipo~IIB colocamos el n'umero adecuado de 9-branas de tipo Dirichlet
(``D9-branas'') y 9-orientiplanos (``O9-planos'') obtenemos la
Teor'ia de Cuerdas tipo~I $SO(32)$ \cite{kn:Da2} cuyo l'imite
de baja energ'ia est'a tambi'en descrito por SUGRA $N=1,d=10$
acoplado a los correspondientes supermultipletes vectoriales. En el
l'imite de acoplo fuerte esta teor'ia es S~dual a la
Teor'ia Heter'otica $SO(32)$ \cite{kn:Hu6,kn:Da3,kn:PW}. 

Para interpretar esta 'ultima teor'ia hay que introducir un
importante aspecto de las $p$-branas y $p$-orientiplanos: que
tambi'en se transforman bajo las dualidades~T y~S. En particular,
bajo dualidad~S, las D9-branas se transforman en S9-branas (tambi'en
llamadas NSNS9-branas) y los O9-planos en otros objetos de forma que
la Teor'ia Heter'otica $SO(32)$ se puede considerar como la
Teor'ia de Cuerdas tipo~IIB en acoplamiento fuerte (por la
dualidad~S) con S9-branas y los duales~S de los O9-planos
\cite{kn:Hu7,kn:BEHvdSHL}.

Hora que conocemos las reglas fundamentales de este juego, podemos
empezar a hacernos preguntas como ?`cu'al es el T~dual de la
Teor'ia de Cuerdas tipo~I $SO(32)$? (Es decir: si la
compactificamos en un c'irculo, hay otra teor'ia de cuerdas
compactificada en un c'irculo de radio inverso que sea
equivalente? De acuerdo con lo que sabemos, esta teor'ia ha de ser
el resultado de poner en la Teor'ia de Cuerdas tipo~IIA los
T~duales de las D9s y los O9s (D8s y O8s) con la dimensi'on que no
est'a dentro de 'estas compactificada en un c'irculo de radio
inverso, que recibe el nombre de Teor'ia de Cuerdas tipo~I'.

As'i se pueden generar muchas nuevas teor'ias. En esta
lecci'on vamos a estudiar las teor'ias efectivas a bajas
energ'ias de las teor'ias de cuerdas m'as relevantes para
nosotros: $N=1,d=11$, $N=2A,d=10$ y $N=2B,d=10$, c'omo se relacionan
los campos y constantes que aparecen en sus acciones con los de las
teor'ias de cuerdas y c'omo se reflejan en sus relaciones mutuas
las dualidades entre las correspondientes teor'ias de cuerdas.

%%%%%%%%%%%%%%%%%%%%%%%%%%%%%%%%%%%%%%%%%%%%%%%%%%%%%%%%%%%%%%%%%%%%%%
\section{Acciones b'asicas y dualidades}

%%%%%%%%%%%%%%%%%%%%%%%%%%%%%%%%%%%%%%%%%%%%%%%%%%%%%%%%%%%%%%%%%%%%%%
\subsection{SUGRA $N=1,d=11$}

%%%%%%%%%%%%%%%%%%%%%%%%%%%%%%%%%%%%%%%%%%%%%%%%%%%%%%%%%%%%%%%%%%%%%%
\subsubsection{El super'algebra}

El super'algebra $N=1,d=11$ con todas las posibles extensiones
cuasi-centrales posibles tiene el siguiente anticonmutador de
supercargas es

\begin{equation}
  \begin{array}{rcl}
\left\{\hat{\hat{Q}}{}^{\hat{\hat{\alpha}}},
\hat{\hat{Q}}{}^{\hat{\hat{\beta}}}\right\} & = & 
i \left(\hat{\hat{\Gamma}}{}^{\hat{\hat{a}}}
\hat{\hat{\cal C}}{}^{-1}\right)
{}^{\hat{\hat{\alpha}}\hat{\hat{\beta}}}\hat{\hat{P}}_{\hat{\hat{a}}}
+{\textstyle\frac{1}{2}}
\left(\hat{\hat{\Gamma}}{}^{\hat{\hat{a}}_{1}\hat{\hat{a}}_{2}}
\hat{\hat{\cal C}}{}^{-1}\right){}^{\hat{\hat{\alpha}}\hat{\hat{\beta}}}
\hat{\hat{\cal Z}}{}^{(2)}_{\hat{\hat{a}}_{1}\hat{\hat{a}}_{2}}
+{\textstyle\frac{i}{5!}}
\left(\hat{\hat{\Gamma}}{}^{\hat{\hat{a}}_{1}\cdots \hat{\hat{a}}_{5}}
\hat{\hat{\cal C}}{}^{-1}\right){}^{\hat{\hat{\alpha}}\hat{\hat{\beta}}}
\hat{\hat{\cal Z}}{}^{(5)}_{\hat{\hat{a}}_{1}\cdots \hat{\hat{a}}_{5}}\\
& & \\
& & 
+{\textstyle\frac{1}{6!}}
\left(\hat{\hat{\Gamma}}{}^{\hat{\hat{a}}_{1}\cdots \hat{\hat{a}}_{6}}
\hat{\hat{\cal C}}{}^{-1}\right){}^{\hat{\hat{\alpha}}\hat{\hat{\beta}}}
\hat{\hat{\cal Z}}{}^{(6)}_{\hat{\hat{a}}_{1}\cdots \hat{\hat{a}}_{6}}
+{\textstyle\frac{i}{9!}}
\left(\hat{\hat{\Gamma}}{}^{\hat{\hat{a}}_{1}\cdots \hat{\hat{a}}_{9}}
\hat{\hat{\cal C}{}}{}^{-1}\right){}^{\hat{\hat{\alpha}}\hat{\hat{\beta}}}
{\cal Z}^{(9)}_{a_{1}\cdots a_{9}}\\
& & \\
& & 
+{\textstyle\frac{1}{10!}}
\left(\hat{\hat{\Gamma}}{}^{\hat{\hat{a}}_{1}\cdots \hat{\hat{a}}_{10}}
\hat{\hat{\cal C}}{}^{-1}\right){}^{\hat{\hat{\alpha}}\hat{\hat{\beta}}}
\hat{\hat{\cal Z}}{}^{(10)}_{\hat{\hat{a}}_{1}\cdots \hat{\hat{a}}_{10}}\, .\\
\end{array}
\end{equation}

No todas las cargas cuasi-centrales pueden ser asociadas a potenciales
de una SUGRA. Este es el caso de $\hat{\hat{\cal Z}}{}^{(10)}$ cuyo
potencial asociado (una 11-forma) no es campos din'amico y no
describe ning'un grado de libertad continuo. Otras cargas
cuasi-centrales simplemente no aparecen (y los potenciales asociados
tampoco). Este es el caso de $\hat{\hat{\cal Z}}{}^{(6)}$ y
$\hat{\hat{\cal Z}}{}^{(9)}$. Las dos cargas centrales que quedan
$\hat{\hat{\cal Z}}{}^{(2)}, \hat{\hat{\cal Z}}{}^{(5)}$est'an
asociadas a una 3-forma y una 6-forma, que son duales de Hodge una de
la otra.  As'i esperamos que SUGRA $N=1,d=11$ tenga m'etrica,
gravitino (representado por un vector-espinor de Majorana con 32
componentes reales) y una 3-forma\footnote{O una 6-forma, pero no se
  sabe c'omo formular la teor'ia 'unicamente en funci'on de la
  6-forma.}:

\begin{equation}
\left\{\hat{\hat{e}}_{\hat{\hat{\mu}}}{}^{\hat{\hat{a}}},
\hat{\hat{C}}_{\hat{\hat{\mu}}\hat{\hat{\nu}}\hat{\hat{\rho}}},
\hat{\hat{\psi}}_{\hat{\hat{\mu}}}
\right\}\, .
\end{equation}

\noindent  Los estados/soluciones asociados a los campos bos'onicos 
y a las cargas del super'algebra son la onda gravitatoria (MW), la
membrana (M2-brana) y la 5-brana (M5-brana).

Cuando la teor'ia est'a compactificada en un c'irculo,
aparecen nuevos estados en la teor'ia: el monopolo de Kaluza-Klein
(KK7) y la 9-brana (KK9 o M9-brana para otros autores) y las dos
cargas cuasi-centrales $\hat{\hat{\cal Z}}{}^{(6)}$ y $\hat{\hat{\cal
    Z}}{}^{(9)}$ aparecen. En la teor'ia en $d=10$ que se obtiene
por reducci'on dimensional s'i se puede definir un potencial
7-forma asociado a la $\hat{\hat{\cal Z}}{}^{(6)}$ y un potencial
9-forma asociado a la $\hat{\hat{\cal Z}}{}^{(9)}$ (uno de sus
'indices tiene que ser siempre la direcci'on compacta y s'olo
los 8 restantes juegan alg'un papel).

%%%%%%%%%%%%%%%%%%%%%%%%%%%%%%%%%%%%%%%%%%%%%%%%%%%%%%%%%%%%%%%%%%%%%%
\subsubsection{La acci'on}

La acci'on para los campos bos'onicos de SUGRA $N=1,d=11$ es
\cite{kn:CJS}

\begin{equation}
\begin{tabular}{|c|}
\hline
\\
$
\hat{\hat{S}}= \frac{1}{16\pi G_{N}^{(11)}}
{\displaystyle\int} d^{11}\hat{\hat{x}}\
\sqrt{|\hat{\hat{g}}|}\ \left[\hat{\hat{R}}
-{\textstyle\frac{1}{2\cdot 4!}}\hat{\hat{G}}{}^{\, 2}
-{\textstyle\frac{1}{(144)^{2}}} \frac{1}{\sqrt{|\hat{\hat{g}}}|}
\hat{\hat{\epsilon}} \hat{\hat{G}} \hat{\hat{G}} \hat{\hat{C}}
\right]\, ,
$
\\
\\ 
\hline
\end{tabular}
\label{eq:11daction}
\end{equation}

\noindent donde 

\begin{equation}
\hat{\hat{G}} = 4\partial \hat{\hat{C}}\, ,
\end{equation}

\noindent es la intensidad de campo de la 3-forma y es invariante
bajo las transformaciones gauge

\begin{equation}
\label{eq:3formgauge}
\delta_{\hat{\hat{\chi}}} \hat{\hat{C}}= 3\partial\hat{\hat{\chi}}\, .
\end{equation}

La acci'on es invariante bajo las transformaciones de supersimetr'ia
locales

\begin{equation}
\begin{tabular}{|c|}
\hline \\
$
\left\{
\begin{array}{rcl}
\delta_{\hat{\hat{\epsilon}}} 
\hat{\hat{e}}_{\hat{\hat{\mu}}}{}^{\hat{\hat{a}}}
& = & 
-\frac{i}{2} \bar{\hat{\hat{\epsilon}}}\,
\hat{\hat{\Gamma}}{}^{\hat{\hat{a}}}\,
\hat{\hat{\psi}}_{\hat{\hat{\mu}}}\, , \\
& & \\
\delta_{\hat{\hat{\epsilon}}} 
\hat{\hat{\psi}}_{\hat{\hat{\mu}}} & = & 
2\nabla_{\hat{\hat{\mu}}} \hat{\hat{\epsilon}}
+\frac{i}{144}
\left(
\hat{\hat{\Gamma}}{}^{\hat{\hat{\alpha}}\hat{\hat{\beta}}
\hat{\hat{\gamma}}\hat{\hat{\delta}}}{}_{\hat{\hat{\mu}}}
-8 \hat{\hat{\Gamma}}{}^{\hat{\hat{\beta}}
\hat{\hat{\gamma}}\hat{\hat{\delta}}}
\hat{\hat{\eta}}_{\hat{\hat{\mu}}}{}^{\hat{\hat{\alpha}}}
\right)
\hat{\hat{\epsilon}}\,
\hat{\hat{G}}_{\hat{\hat{\alpha}}\hat{\hat{\beta}}
\hat{\hat{\gamma}}\hat{\hat{\delta}}} \, , \\
& & \\
\delta_{\hat{\hat{\epsilon}}} 
\hat{\hat{C}}_{\hat{\hat{\mu}}\hat{\hat{\nu}}\hat{\hat{\rho}}}
& = & 
\frac{3}{2} \bar{\hat{\hat{\epsilon}}}\,
\hat{\hat{\Gamma}}_{[\hat{\hat{\mu}}\hat{\hat{\nu}}}\,
\hat{\hat{\psi}}_{\hat{\hat{\rho}}]}\, .
\end{array}
\right.
$
\\ \\ \hline
\end{tabular}
\label{eq:d11susyrules}
\end{equation}

Esta teor'ia s'olo tiene una constante de acoplo: $G^{(11)}_{N}$
o, equivalentemente, la longitud de Planck $\ell^{(11)}_{\rm Planck}$.

Es interesante tambi'en observar la presencia de un {\it t'ermino de
  Chern-Simons} $\hat{\hat{\epsilon}} \hat{\hat{G}} \hat{\hat{G}}
\hat{\hat{C}}$ en la acci'on. Este t'ermino topol'ogico (no depende
de la m'etrica) modifica la ecuaci'on del movimiento de
$\hat{\hat{C}}$ pero no la de Einstein, y, de acuerdo con Townsend
\cite{kn:Tow7} rige las posibles intersecciones de $M2$- y
$M5$-branas, como veremos en la Secci'on~\ref{sec-intersections}.

%%%%%%%%%%%%%%%%%%%%%%%%%%%%%%%%%%%%%%%%%%%%%%%%%%%%%%%%%%%%%%%%%%%%%%
\subsubsection{El potencial magn'etico}

La ecuaci'on de movimiento de la 3-forma se puede escribir as'i:

\begin{equation}
\partial\left( {}^{\star}\hat{\hat{G}} 
+{\textstyle\frac{35}{2}}\hat{\hat{C}}\hat{\hat{G}}  \right)=0\, ,
\end{equation}

\noindent como una identidad de Bianchi. As'i podemos identificar
la expresi'on que est'a entre par'entesis con
$7\partial\hat{\hat{\tilde{C}}}$ donde $\hat{\hat{\tilde{C}}}$ es, por
definici'on, el potencial 6-forma dual. Esto implica que la
intensidad de campo de la 6-forma dual $\hat{\hat{\tilde{C}}}$ es
\cite{kn:A,kn:BRO,kn:BLO}:

\begin{equation}
{}^{\star}\hat{\hat{G}} = 
7\left(\partial\hat{\hat{\tilde{C}}} 
-10\hat{\hat{C}}\partial\hat{\hat{C}} \right)
\equiv \hat{\hat{\tilde{G}}}\, .
\end{equation}

$\hat{\hat{\tilde{G}}}$ es evidentemente invariante bajo las transformaciones
gauge

\begin{equation}
\delta_{\hat{\hat{\tilde{\chi}}}}  \hat{\hat{\tilde{C}}}
=6\partial \hat{\hat{\tilde{\chi}}}\, , 
\end{equation}

\noindent donde $\hat{\hat{\tilde{\chi}}}$ es una 5-forma.
Sin embargo, contiene expl'icitamente la 3-forma y, para que sea
invariante bajo las transformaciones gauge de 'esta
(\ref{eq:3formgauge}) $\hat{\hat{\tilde{C}}}$ tiene que transformarse
as'i:

\begin{equation}
\delta_{\hat{\hat{\chi}}}  \hat{\hat{\tilde{C}}}
=-30\partial \hat{\hat{\chi}} \hat{\hat{C}}\, .
\end{equation}

Este procedimiento para definir el dual de $\hat{\hat{C}}$ en el que
$\hat{\hat{C}}$ no es completamente eliminado recibe el nombre de {\it
  dualizaci'on ``on-shell''}.

%%%%%%%%%%%%%%%%%%%%%%%%%%%%%%%%%%%%%%%%%%%%%%%%%%%%%%%%%%%%%%%%%%%%%%
\subsection{Reducci'on a $d=10$: la teor'ia 
tipo~IIA}

%%%%%%%%%%%%%%%%%%%%%%%%%%%%%%%%%%%%%%%%%%%%%%%%%%%%%%%%%%%%%%%%%%%%%%
\subsubsection{El super'algebra}

Como hemos dicho, al definir supersimetr'ia en el espacio de
Minkowski 10-dimensional veces un c'irculo, aparecen las cargas
$\hat{\hat{\cal Z}}{}^{(6)}$ y $\hat{\hat{\cal Z}}{}^{(9)}$ y hay que
tenerlas en cuenta para hallar todas las cargas del super'algebra
10-dimensional.

La reducci'on dimensional del super'algebra es sencilla: los
espinores\footnote{Aunque los espinores son los mismos, en $d=10$ son
  reducibles en dos espinores de Majorana-Weyl con quiralidades
  opuestas. De ah'i que pasemos de $N=1$ a $N=2$.} (y las
supercargas) en $d=10$ son los mismos que en $d=11$, de forma que
s'olo hace falta quitar una de las tildes, las matrices gamma en
$d=11$ y $d=10$ se relacionan entre s'i de acuerdo con
Ec.~(\ref{eq:11vs10gamma}) de forma que $\hat{\hat{\cal C}} =\hat{\cal
  C}$ y las cargas bos'onicas se descomponen

\begin{equation}
\hat{\hat{P}}_{\hat{\hat{a}}} =(\hat{P}_{\hat{a}},\hat{Z}^{(0)})\, ,
\hspace{.5cm}
\hat{\hat{Z}}{}^{(2)}_{\hat{\hat{a}}\hat{\hat{b}}} 
=(\hat{Z}^{(2)}_{\hat{a}\hat{b}},\hat{Z}^{(1)}_{\hat{a}})\, ,
\hspace{.5cm}
\hat{\hat{Z}}{}^{(5)}_{\hat{\hat{a}}_{1}\cdots\hat{\hat{a}}_{5}} 
=(\hat{Z}^{(5)}_{\hat{a}_{1}\cdots\hat{a}_{5}},
\hat{Z}^{(4)}_{\hat{a}_{1}\cdots\hat{a}_{4}})\, ,
\end{equation}

\noindent mientras que

\begin{equation}
\hat{\hat{Z}}{}^{(6)}_{\hat{\hat{a}}_{1}\cdots\hat{\hat{a}}_{6}} 
=(\hat{Z}^{(6)}_{\hat{a}_{1}\cdots\hat{a}_{6}},\cdot)\, ,
\hspace{.5cm}
\hat{\hat{Z}}{}^{(9)}_{\hat{\hat{a}}_{1}\cdots\hat{\hat{a}}_{9}} 
=(\cdot,\hat{Z}^{(8)}_{\hat{a}_{1}\cdots\hat{a}_{8}})\, ,
\end{equation}

\noindent con lo que se obtiene el super'algebra $N=2A,d=10$

\begin{equation}
\begin{array}{rcl}
\{\hat{Q}^{\hat{\alpha}},\hat{Q}^{\hat{\beta}}\} & = & 
i\left(\hat{\Gamma}^{\hat{a}}\hat{\cal C}^{-1} \right){}^{\hat{\alpha}\hat{\beta}}
\hat{P}_{\hat{a}} +\sum_{n=0,1,4,8}\frac{c_{n}}{n!}
\left(\hat{\Gamma}^{\hat{a}_{1}\cdots\hat{a}_{n}}\hat{\Gamma}_{11}
\hat{\cal C}^{-1} \right){}^{\hat{\alpha}\hat{\beta}}
\hat{Z}^{(n)}_{\hat{a}_{1}\cdots\hat{a}_{n}}\\
& & \\
& & 
+\sum_{n=2,5,6}\frac{c_{n}}{n!}
\left(\hat{\Gamma}^{\hat{a}_{1}\cdots\hat{a}_{n}}\hat{\Gamma}_{11}
\hat{\cal C}^{-1} \right){}^{\hat{\alpha}\hat{\beta}}
\hat{Z}^{(n)}_{\hat{a}_{1}\cdots\hat{a}_{n}}\, .\\
\end{array}
\end{equation}

\noindent Las cargas est'an asociadas a un gravitino, una m'etrica y 
potenciales que son $(p+1)$-formas con $p=0,1,2,4,5,6,8$. De 'estos
los $p=0,6$ y $p=2,4$ est'an relacionados por dualidad de Hodge y
s'olo uno de ellos (el de rango menor) aparece en la SUGRA. En vez de
la 9-forma lo que aparece en la supergravedad es un par'ametro
constante con dimensiones de masa ({\it SUGRA masiva de Romans}
\cite{kn:Ro2}), pero aqu'i vamos a ignorar esta complicaci'on de
forma que s'olo consideraremos los campos\footnote{Adem'as de los
  campos asociados a las cargas que aparecen en el 'algebra, hay un
  dilat'on y un dilatino.}

\begin{equation}
\left\{
\hat{g}_{\hat{\mu}\hat{\nu}},
\hat{B}_{\hat{\mu}\hat{\nu}},
\hat{\phi},
\hat{C}^{(3)}{}_{\hat{\mu}\hat{\nu}\hat{\rho}},
\hat{C}^{(1)}{}_{\hat{\mu}},
\right\}\, .
\end{equation}

\noindent Los estados/soluciones asociados son la onda plana 
gravitacional, las D$p$-branas con $p=0,2,4,6,8$, la {\it cuerda
  fundamental} (F1A o 1-brana NSNS) y la {\it 5-brana solit'onica}
(S5A).

De forma general, es posible introducir m'as cargas cuasi-centrales en
este super'algebra. En particular, podemos introducir una de 5
'indices distinta (aunque relacionada por dualidad) y una de 9:

\begin{equation}
{\textstyle\frac{c_{5}}{5!}}
\left(\hat{\Gamma}^{\hat{a}_{1}\cdots\hat{a}_{5}}\hat{\Gamma}_{11}
\hat{\cal C}^{-1} \right){}^{\hat{\alpha}\hat{\beta}}
\hat{Z}^{(5)}_{\hat{a}_{1}\cdots\hat{a}_{5}}
+{\textstyle\frac{c_{9}}{9!}}
\left(\hat{\Gamma}^{\hat{a}_{1}\cdots\hat{a}_{9}}
\hat{\cal C}^{-1} \right){}^{\hat{\alpha}\hat{\beta}}
\hat{Z}^{(9)}_{\hat{a}_{1}\cdots\hat{a}_{9}}\, ,
\end{equation}

\noindent que hemos de interpretar de nuevo que s'olo aparecen cuando 
la teor'ia est'a compactificada en un c'irculo. Los
estados/soluciones asociados son el monopolo de Kaluza-Klein (KK6A) y
la 9-brana NSNS (KK9A) \cite{kn:BEHvdSHL}. Adem'as hay que incluir
una segunda carga de 8 'indices uno de los cuales est'a siempre
en la direcci'on compacta \cite{kn:Hu3}.

%%%%%%%%%%%%%%%%%%%%%%%%%%%%%%%%%%%%%%%%%%%%%%%%%%%%%%%%%%%%%%%%%%%%%%
\subsubsection{La acci'on bos'onica}

Para obtener la acci'on para los modos sin masa que se obtienen al
compactificar SUGRA $N=1,d=11$ en un c'irculo seguimos el
procedimiento general de Scherk y Schwarz \cite{kn:SS}. Suponemos que
todos los campos son independientes de la coordenada espacial $z
=x^{\underline{10}}\in [0,2\pi {}^{-}\!\!\!\!\ell_{\rm
  Planck}^{(11)}]$ y reescribimos la teor'ia en forma
10-dimensional.  La reducci'on de la m'etrica 11-dimensional da
lugar a la m'etrica 10-dimensional, un vector y un escalar (el {\it
  dilat'on}) mientras que la 3-forma da lugar a una 3- y una 2-forma.
La m'etrica, la 2-forma y el dilat'on son campos del sector
Neveu-Schwarz-Neveu-Schwarz (NSNS) en el espectro de la Teor'ia de
Cuerdas tipo~IIA y la 3-forma y el vector pertenecen al sector
Ramond-Ramond (RR).

Los campos 11-dimensionales se descomponen en campos 10-dimensionales
as'i:

\begin{equation}
\label{eq:11vs10fields}
\begin{array}{rclrcl}
\hat{\hat{g}}_{\hat{\mu}\hat{\nu}}
&
=
&
e^{-\frac{2}{3}\hat{\phi}}\hat{g}_{\hat{\mu}\hat{\nu}}
- e^{\frac{4}{3}\hat{\phi}}\hat{C}^{(1)}{}_{\hat{\mu}}
\hat{C}^{(1)}{}_{\hat{\nu}}\, ,\hspace{1cm}
&
\hat{\hat{C}}_{\hat{\mu}\hat{\nu}\hat{\rho}}
&
=
&
\hat{C}^{(3)}{}_{\hat{\mu}\hat{\nu}\hat{\rho}}\, ,
\\
& & & & &
\\
\hat{\hat{g}}_{\hat{\mu}\underline{z}}
&
=
&
-e^{\frac{4}{3}\hat{\phi}}\hat{C}^{(1)}{}_{\hat{\mu}}\, ,
&
\hat{\hat{C}}_{\hat{\mu}\hat{\nu}\underline{z}}
&
=
&
\hat{B}_{\hat{\mu}\hat{\nu}}\, ,
\\
& & & & &
\\
\hat{\hat{g}}_{\underline{z}\underline{z}}
&
=
&
-e^{\frac{4}{3}\hat{\phi}}\, .
& & &
\\
\end{array}
\end{equation}

\noindent Los {\it Elfbein} se descomponen as'i:

\begin{equation}
\begin{array}{rcl}
\left( \hat{\hat{e}}_{\hat{\hat{\mu}}}{}^{\hat{\hat{a}}} \right) & = &
\left(
\begin{array}{cc}
e^{-\frac{1}{3}\hat{\phi}} \hat{e}_{\hat{\mu}}{}^{\hat{a}}
&
e^{\frac{2}{3}\hat{\phi}} \hat{C}^{(1)}{}_{\hat{\mu}}
\\
&
\\
0
&
e^{\frac{2}{3}\hat{\phi}}
\\
\end{array}
\right)
\, , \\
& & \\
\left( \hat{\hat{e}}_{\hat{\hat{a}}}{}^{\hat{\hat{\mu}}} \right) & = &
\left(
\begin{array}{cc}
e^{\frac{1}{3}\hat{\phi}} \hat{e}_{\hat{a}}{}^{\hat{\mu}}
&
-e^{\frac{1}{3}\hat{\phi}} \hat{C}^{(1)}{}_{\hat{a}}
\\
&
\\
0
&
e^{-\frac{2}{3}\hat{\phi}}
\\
\end{array}
\right)\, . \\
\end{array}
\label{eq:basis}
\end{equation}

La acci'on para estos campos es la de la parte bos'onica de SUEGRA
$N=2A,d=10$ \cite{kn:HW} pero escrita en el {\it sistema de referencia
  conforme de la cuerda}\footnote{Por definici'on, el {\it sistema de
    referencia conforme de la cuerda} es aqu'el en el que la
  m'etrica es la m'etrica que aparece en el modelo sigma de la
  cuerda. El {\it sistema de referencia conforme de Einstein} es
  aqu'el en el que no hay ning'un factor extra multiplicando al
  escalar de Ricci en la acci'on de Einstein-Hilbert. En el sistema
  de referencia conforme de la cuerda, siempre aparece el prefactor
  $e^{-2\hat{\phi}}$. Ambos sistemas de referencia conforme est'an
  relacionados por una transformaci'on conforme de la m'etrica, como
  veremos.}

\begin{equation}
\label{eq:IIAaction1}
\begin{array}{rcl}
\hat{S} & = & 
\frac{2\pi{}^{-}\!\!\!\!\ell_{\rm Planck}^{(11)}}{16\pi G_{N}^{(11)}} 
{\displaystyle\int} d^{10}\hat{x}\,
\sqrt{|\hat{g}|} \left\{ e^{-2\hat{\phi}}
\left[ \hat{R} -4\left( \partial\hat{\phi} \right)^{2}
+{\textstyle\frac{1}{2\cdot 3!}} \hat{H}^{2}\right] \right.\\
& & \\
& & 
-\left[
{\textstyle\frac{1}{4}} \left( \hat{G}^{(2)} \right)^{2}
+{\textstyle\frac{1}{2\cdot 4!}}\left(\hat{G}^{(4)}\right)^{2}
\right]
-{\textstyle\frac{1}{144}} \frac{1}{\sqrt{|\hat{g}|}}\
\hat{\epsilon}\partial\hat{C}^{(3)}\partial\hat{C}^{(3)}\hat{B}
\biggr \}\, .\\
\end{array}
\end{equation}

Como vemos, los campos del sector NSNS vienen afectados por un
prefactor com'un $e^{-2\hat{\phi}}$, mientras que los del sector RR
no llevan ning'un prefactor.

Esta acci'on tiene un problema: la m'etrica no puede ser
asint'oticamente plana en $d=10$ y $d=11$ simult'aneamente, puesto
que, al ser $z$ compacta $\hat{\hat{g}}_{\underline{z}\underline{z}}$
(el dilat'on) no tiene por qu'e ser $-1$ en el infinito en las
direcciones no-compactas, sino que, esencialmente, da la dimensi'on
asint'otica de la direcci'on compacta, que es un nuevo par'ametro
que hemos introducido impl'icitamente con la compactificaci'on.
Si la m'etrica 11-dimensional  es asint'oticamente plana y
denotamos por $\hat{\phi}_{0}$ el valor asint'otico del dilat'on,
entonces la m'etrica que hemos definido se comporta en el infinito
as'i:

\begin{equation}
\hat{g}_{\hat{\mu}\hat{\nu}} \rightarrow 
e^{\frac{2}{3}\hat{\phi}_{0}} \hat{\eta}_{\hat{\mu}\hat{\nu}}\, .
\end{equation}

Para obtener la m'etrica de la cuerda ``correcta'' hemos de
multiplicar 'esta, y todos los campos, por un factor num'erico, de
acuerdo con

\begin{equation}
\label{eq:rescalings10-11}
\begin{array}{rclrcl}
\hat{g}_{\hat{\mu}\hat{\nu}} & \rightarrow & 
e^{\frac{2}{3}\hat{\phi}_{0}}\hat{g}_{\hat{\mu}\hat{\nu}}\, ,\hspace{2cm} &
\hat{C}^{(1)}{}_{\hat{\mu}} & \rightarrow & 
e^{\frac{1}{3}\hat{\phi}_{0}}\hat{C}^{(1)}{}_{\hat{\mu}}\, ,\\
& & & & & \\
\hat{B}_{\hat{\mu}\hat{\nu}} & \rightarrow & 
e^{\frac{2}{3}\hat{\phi}_{0}}\hat{B}_{\hat{\mu}\hat{\nu}}\, , &
\hat{C}^{(3)}{}_{\hat{\mu}\hat{\nu}\hat{\rho}} & \rightarrow & 
e^{\hat{\phi}_{0}}\hat{C}^{(3)}{}_{\hat{\mu}\hat{\nu}\hat{\rho}}\, ,\\
\end{array}
\end{equation}

\noindent de forma que la acci'on queda as'i:

\begin{equation}
  \begin{tabular}{|c|}
\hline \\ \\
$\begin{array}{rcl}
\hat{S} & = & 
\frac{g_{A}^{2}}{16\pi G_{N\, A}^{(10)}} 
{\displaystyle\int} d^{10}\hat{x}\,
\sqrt{|\hat{g}|} \left\{ e^{-2\hat{\phi}}
\left[ \hat{R} -4\left( \partial\hat{\phi} \right)^{2}
+{\textstyle\frac{1}{2\cdot 3!}} \hat{H}^{2}\right] \right.\\
& & \\
& & 
-\left[
{\textstyle\frac{1}{4}} \left( \hat{G}^{(2)} \right)^{2}
+{\textstyle\frac{1}{2\cdot 4!}}\left(\hat{G}^{(4)}\right)^{2}
\right]
-{\textstyle\frac{1}{144}} \frac{1}{\sqrt{|\hat{g}|}}\
\hat{\epsilon}\partial\hat{C}^{(3)}\partial\hat{C}^{(3)}\hat{B}
\biggr \}\, .\\
\end{array}
$\\ \\ \hline
\end{tabular}
\label{eq:IIAaction}
\end{equation}

Aqu'i hemos hecho las siguientes identificaciones de constantes.
Primero, hemos identificado la {\it constante de acoplo de la cuerda}
$g_{A}$, que cuenta los lazos en las amplitudes de cuerdas con el
exponencial del valor asint'otico del dilat'on

\begin{equation}
g_{A} = e^{\hat{\phi}_{0}}\, .
\end{equation}

\noindent y, seguidamente, hemos identificado la constante de Newton 
10-dimensional as'i\footnote{El factor $g_{A}^{2}$ absorbe el
  valor asint'otico del dilat'on en la acci'on.}:

\begin{equation}
\frac{2\pi{}^{-}\!\!\!\!\ell_{\rm Planck}^{(11)}
e^{\frac{8}{3}\hat{\phi}_{0}}}{16\pi G_{N}^{(11)}} 
= 
\frac{g_{A}^{2}}{16\pi G_{N\, A}^{(10)}}\, .
\end{equation}

\noindent  Esto implica la relaci'on

\begin{equation}
G^{(10)}_{N}
=\frac{ G^{(11)}_{N}}{2\pi {}^{-}\!\!\!\!\ell_{\rm Planck}^{(11)} 
g_{A}^{2/3}}\, .  
\end{equation}

Es 'util introducir el radio de la direcci'on compacta $R_{11}$ (que no es
${}^{-}\!\!\!\!\ell^{(11)}_{\rm Planck}$)

\begin{equation} 
\label{eq:radio11}
R_{11} = \frac{1}{2\pi}\lim_{r\rightarrow\infty} \int
\sqrt{|\hat{\hat{g}}_{\underline{z}\underline{z}}|}dz 
= \ell_{\rm Planck}^{(11)}e^{\frac{2}{3}\hat{\phi}_{0}}
= \ell_{\rm Planck}^{(11)} g_{A}^{2/3}\, . 
\end{equation}

\noindent con el que la relaci'on entre las constantes de Newton en $d=10$
y $d=11$ se escribe 

\begin{equation}
G^{(10)}_{N\, A}=\frac{ G^{(11)}_{N}}{2\pi R_{11}}
=\frac{ G^{(11)}_{N}}{V_{11}}\, ,  
\end{equation}

\noindent como esperamos en teor'ias de Kaluza-Klein ($V_{11}$
es el volumen del espacio compacto).

Utilizando la definici'on de la longitud de Planck
Ec.~(\ref{eq:Plancklength}) tenemos

\begin{equation}
G^{(10)}_{N\, A} 
=\frac{(\ell_{\rm Planck}^{(11)})^{8}}{32\pi^{2}g_{A}^{2/3}}\, .
\end{equation}

Este resultado ha de ser comparado con el valor de $G^{(10)}_{N}$ que
se obtiene, con argumentos puramente cuerd'isticos en t'erminos
de las variables $\ell_{s},g_{A}$

\begin{equation}
\label{eq:GN10A}
\begin{tabular}{|c|}
\hline \\
$
G_{N\, A}^{(10)} =8\pi^{6}g_{A}^{2}\ell_{s}^{8}\, ,
$
\\ \\ \hline
\end{tabular}
\end{equation}

\noindent donde $\ell_{s}=\sqrt{\alpha^{\prime}}$ es la {\it longitud
de la cuerda}. Ambos resultados son consistentes si 

\begin{equation}
\label{eq:10-11constants}
\begin{tabular}{|c|}
\hline \\
$
\begin{array}{rcl}
\ell_{\rm Planck}^{(11)} & = & 2\pi \ell_{s}g_{A}^{1/3}\, ,\\
& & \\
R_{11} & = & \ell_{s} g_{A}\, ,\\
\end{array}
$
\\ \\ \hline
\end{tabular}
\end{equation}

\noindent que son las relaciones entre constantes 11- y 10-dimensionales
m'as importantes.

Es f'acil ver que el l'imite de acoplo fuerte de la teor'ia
tipo~IIA ($g_{A}\rightarrow \infty$) coincide con el l'imite de
{\it descompactificaci'on} ($R_{11}\rightarrow \infty$) en el que una
nueva dimensi'on se vuelve macrosc'opica.

Esta teor'ia hereda de la 11-dimensional un t'ermino de
Chern-Simons que determina las intersecciones posibles de los objetos
extensos de la misma \cite{kn:Tow7} como veremos en la
Secci'on~\ref{sec-intersections}.

%%%%%%%%%%%%%%%%%%%%%%%%%%%%%%%%%%%%%%%%%%%%%%%%%%%%%%%%%%%%%%%%%%%%%%
\subsubsection{Potenciales magn'eticos}

Los potenciales duales de la SUEGRA $N=2A,d=10$ s'olo se pueden
introducir por dualizaci'on {\it on-shell}. Es posible relacionar las
intensidades de campo ``el'ectricas'' y ``magn'eticas'' de acuerdo
con la relaci'on general

\begin{equation}
\label{eq:RRduals}
\hat{G}^{(10-k)} = (-1)^{\left[k/2\right]} \,{}^{\star} \hat{G}^{(k)}\, , 
\end{equation}

\noindent de forma que todas las intensidades de campo RR 
se escriben as'i\footnote{Estamos utilizando el lenguaje de formas
  diferenciales y la notaci'on en la que cada letra representa la
  suma formal de las formas diferenciales de todos los rangos $\hat{C}
  = \hat{C}^{(0)} + \hat{C}^{(1)} + \hat{C}^{(2)} +\ldots $ etc.}

\begin{equation}
\label{eq:RRfieldstrengths}
\hat{G} =d\hat{C} -\hat{H}\wedge \hat{C}\, ,  
\end{equation}

\noindent y las identidades de Bianchi y las ecuaciones
de movimiento as'i:

\begin{equation}
\label{eq:RReom&Bianchis}
d\hat{G} -\hat{H}\wedge \hat{G} =0\, ,
\hspace{1cm}
d{}^{\star}\hat{G} +\hat{H}\wedge {}^{\star}\hat{G} =0\, .
\end{equation}

Para la 2-forma NSNS, la definici'on de la intensidad de campo dual es

\begin{equation}
\hat{H}^{(7)}  =  e^{-2\hat{\phi}}\,{}^{\star}\hat{H}\, ,  
\end{equation}

\noindent y la identidad de Bianchi de la 3-forma se vuelve la ecuaci'on
 de movimiento de la 7-forma dual

 \begin{equation}
d H = 0\, ,
\hspace{1cm}
d\left( e^{2\phi} \,{}^{\star} \hat{H}^{(7)}  \right) = 0\, ,
 \end{equation}

\noindent y viceversa

\begin{equation}
d \left(e^{-2\phi} \,{}^{\star} \hat{H} \right) 
+{\textstyle\frac{1}{2}} \,{}^{\star} \hat{G} \wedge \hat{G} = 0\, ,
\hspace{1cm}
d\hat{H}^{(7)} +{\textstyle\frac{1}{2}}
 \,{}^{\star} \hat{G} \wedge \hat{G} = 0\, .
\end{equation}

Una definici'on posible es

\begin{equation}
\hat{H}^{(7)}  =  d\hat{B}^{(6)}
-{\textstyle\frac{1}{2}} \sum_{n=1}^{n=4} \,{}^{\star} 
\hat{G}^{(2n+2)}\wedge \hat{C}^{(2n-1)}\, .
\end{equation}

%%%%%%%%%%%%%%%%%%%%%%%%%%%%%%%%%%%%%%%%%%%%%%%%%%%%%%%%%%%%%%%%%%%%%%%%%%%
\subsubsection{Fermiones y reglas de supersimetr'ia}
\label{sec-susyd11d10}

Los espinores 11-dimensionales se expresan en t'erminos de los
10-dimensionales (gravitino $\hat{\psi}_{\hat{\mu}}$ y dilatino
$\hat{\lambda}$ y el par'ametro de transformaci'on de supersimetr'ia
$\hat{\epsilon}$) as'i:

\begin{equation}
\left\{
\begin{array}{rcl}
\hat{\hat{\epsilon}} & = &e^{-\frac{1}{6}(\hat{\phi}-\hat{\phi}_{0})}\, 
\hat{\epsilon}\, , \\
& & \\
\hat{\hat{\psi}}_{\hat{a}} & = &
e^{\frac{1}{6}(\hat{\phi}-\hat{\phi}_{0})}
\left(2\hat{\psi}_{\hat{a}} -{\textstyle\frac{1}{3}}
\hat{\Gamma}_{\hat{a}} \hat{\lambda}\right)\, ,
\\
& & \\
\hat{\hat{\psi}}_{z} & = & 
{\textstyle\frac{2i}{3}} e^{\frac{1}{6}(\hat{\phi}-\hat{\phi}_{0})}
\hat{\Gamma}_{11} \hat{\lambda}\, . \\
\end{array}
\right.
\end{equation}

Obs'ervese que con estas definiciones el gravitino
$\hat{\psi}_{\hat{\mu}}$ es real pero el dilatino $\hat{\lambda}$ es
imaginario puro.

Las reglas de transformaci'on de supersimetr'ia (al orden m'as
bajo en fermiones y s'olo con los potenciales NSNS y RR
``el'ectricos'') son:

\begin{equation}
\label{eq:IIAsusyrules1}
\begin{tabular}{|c|}
\hline \\
$
\begin{array}{rcl}
\delta_{\hat{\epsilon}} \hat{e}_{\hat{\mu}}{}^{\hat{a}} & = & 
-i\bar{\hat{\epsilon}}\hat{\Gamma}^{\hat{a}} \hat{\psi}_{\hat{\mu}}\, ,\\
& & \\
\delta_{\hat{\epsilon}} \hat{\psi}_{\hat{\mu}} & = & 
\left\{ 
\partial_{\hat{\mu}} -\frac{1}{4}
\left(\not\!\hat{\omega}_{\hat{\mu}} 
+\frac{1}{2}\Gamma_{11}\not\!\! \hat{H}_{\mu}\right)
\right\} \hat{\epsilon}
+\frac{i}{8} e^{\hat{\phi}} \Sigma_{n=1,2} \frac{1}{(2n)!}
\not\! \hat{G}^{(2n)} \hat{\Gamma}_{\hat{\mu}} 
\left( -\hat{\Gamma}_{11} \right)^{n}\hat{\epsilon}\, , \\
& & \\
\delta_{\hat{\epsilon}} \hat{B}_{\hat{\mu}\hat{\nu}} & = & 
-2i\bar{\hat{\epsilon}}\hat{\Gamma}_{[\hat{\mu}}
\hat{\Gamma}_{11}\hat{\psi}_{\hat{\nu}]}\, , \\
& & \\
\delta_{\hat{\epsilon}} \hat{C}^{(1)}{}_{\hat{\mu}} & = & 
-e^{\hat{\phi}}\bar{\hat{\epsilon}} \hat{\Gamma}_{11}
\left(\hat{\psi}_{\hat{\mu}} 
-\frac{1}{2}\hat{\Gamma}_{\hat{\mu}}\hat{\lambda} \right)\, ,\\
& & \\
\delta_{\hat{\epsilon}} \hat{C}^{(3)}{}_{\hat{\mu}\hat{\nu}\hat{\rho}}
& = & 
3 e^{\hat{\phi}} \bar{\epsilon} \hat{\Gamma}_{\hat{\mu}\hat{\nu}}
\left(\hat{\psi}_{\hat{\rho}]} 
-\frac{1}{3!}\hat{\Gamma}_{\hat{\rho}]}\hat{\lambda} \right) 
+3\hat{C}^{(1)}{}_{[\hat{\mu}}\delta_{\hat{\epsilon}}
\hat{B}_{\hat{\mu}\hat{\nu}]}\, ,\\
& & \\
\delta_{\hat{\epsilon}}\hat{\lambda} & = &   
\left(\not\!\partial\hat{\phi}
+\frac{1}{12}\hat{\Gamma}_{11}\not\!\! \hat{H}\right)\hat{\epsilon}
+ \frac{i}{4} e^{\hat{\phi}} \sum_{n=1,2} \frac{5-2n}{(2n)!} 
\not\! \hat{G}^{(2n)} \left(-\hat{\Gamma}_{11} \right)^{n} 
\hat{\epsilon}\, ,\\
& & \\
\delta_{\hat{\epsilon}}\hat{\phi} & = & 
-\frac{i}{2} \bar{\hat{\epsilon}}\hat{\lambda}\, .\\
\end{array}
$
\\ \\ \hline
\end{tabular}
\end{equation}

%%%%%%%%%%%%%%%%%%%%%%%%%%%%%%%%%%%%%%%%%%%%%%%%%%%%%%%%%%%%%%%%%%%%%%
\subsection{La teor'ia tipo~IIB. Dualidad~S}

%%%%%%%%%%%%%%%%%%%%%%%%%%%%%%%%%%%%%%%%%%%%%%%%%%%%%%%%%%%%%%%%%%%%%%
\subsubsection{El super'algebra}

Las dos supercargas del super'algebra $N=2B,d=10$ tienen la misma
quiralidad (positiva, por simplicidad) y no es posible combinarlas en
una sola, como en el caso anterior. Por ello hemos de introducir
'indices $i,j=1,2$ de $SO(2)$ para distinguirlas. De acuerdo con
los principios generales, el anticonmutador m'as general posible de
las supercargas es

\begin{equation}
\begin{array}{rcl}
\{\hat{Q}^{i\, \hat{\alpha}},\hat{Q}^{j\, \hat{\beta}}\} & = & 
i\delta^{ij}
\left(\hat{\Gamma}^{\hat{a}}\hat{\cal C}^{-1} \right)
{}^{\hat{\alpha}\hat{\beta}}
\hat{P}_{\hat{a}} +\left(\hat{\Gamma}^{\hat{a}}\hat{\cal C}^{-1} 
\right){}^{\hat{\alpha}\hat{\beta}}\hat{Z}^{(1)\, (ij)}_{\hat{a}}
+\frac{i}{3!} \left(\hat{\Gamma}^{\hat{a}_{1}\hat{a}_{2}\hat{a}_{3}}
\hat{\cal C}^{-1} \right){}^{\hat{\alpha}\hat{\beta}}
\hat{Z}^{(3)\, [ij]}_{\hat{a}_{1}\hat{a}_{2}\hat{a}_{3}}\\
& & \\
& & 
+\frac{i}{5!} \left(\hat{\Gamma}^{\hat{a}_{1}\cdots \hat{a}_{5}}
\hat{\cal C}^{-1} \right){}^{\hat{\alpha}\hat{\beta}}
\hat{Z}^{(5)\, (ij)}_{\hat{a}_{1}\cdots\hat{a}_{5}}
+\frac{i}{7!} \left(\hat{\Gamma}^{\hat{a}_{1}\cdots \hat{a}_{7}}
\hat{\cal C}^{-1} \right){}^{\hat{\alpha}\hat{\beta}}
\hat{Z}^{(7)\, [ij]}_{\hat{a}_{1}\cdots\hat{a}_{7}}\\
& & \\
& & 
+\frac{i}{9!} \left(\hat{\Gamma}^{\hat{a}_{1}\cdots \hat{a}_{9}}
\hat{\cal C}^{-1} \right){}^{\hat{\alpha}\hat{\beta}}
\hat{Z}^{(9)\, (ij)}_{\hat{a}_{1}\cdots\hat{a}_{9}}\, .\\
\end{array}
\end{equation}

Las cargas sim'etricas en el par de 'indices $ij$ se descomponen
en las 3 posibilidades independientes:

\begin{equation}
\hat{Z}^{(1)\, (ij)}_{\hat{a}} = \hat{Z}^{(1)\, 0}_{\hat{a}} \delta^{ij} 
\hat{Z}^{(1)\, 1}_{\hat{a}}\sigma^{1} 
+\hat{Z}^{(1)\, 3}_{\hat{a}} \sigma^{3}\, ,
\end{equation}

\noindent etc. El primer t'ermino es un singlete de $SO(2)$ y los 
otros dos forman un doblete. En el caso de $\hat{Z}^{(1)}$ no existe
un singlete independiente del momento y en los de $\hat{Z}^{(5)}$ y
$\hat{Z}^{(9)}$ tampoco existe un singlete si no se compactifica en un
c'irculo.

Las cargas asim'etricas son singletes de $SO(2)$. Por ejemplo:

\begin{equation}
\hat{Z}^{(3)\, [ij]}_{\hat{a}_{1}\hat{a}_{2}\hat{a}_{3}}
=\hat{Z}^{(3)}_{\hat{a}_{1}\hat{a}_{2}\hat{a}_{3}} i\sigma^{2}\, .
\end{equation}

Todas estas cargas est'an asociadas a dos gravitinos, una m'etrica y
potenciales que son $(p+1)$-formas con $p=1,3,5,7,9$. Los casos
$p=1,5$ est'an relacionados por dualidad de Hodge y s'olo los de
$p=1$ aparecen en la acci'on, formando un doblete de dualidad~S (un
potencial NSNS y uno RR, como veremos) que est'a relacionada con la
simetr'ia $SO(2)$ del super'algebra. En el caso $p=3$ s'olo hay
un potencial que es autodual (de hecho su intensidad de campo es una
5-forma autodual). En el caso $p=7$ s'olo hay un potencial: un
escalar que de hecho es el dual de la 8-forma correspondiente. Este
potencial no es invariante bajo dualidad~S, pero no parece formar un
doblete con ning'un otro campo. Argumentos de dualidad~T entre las
teor'ias~IIA y~IIB parecen sugerir que se debe de introducir una
segunda carga $\hat{Z}^{(7)\prime}$. As'i los campos bos'onicos
son\footnote{Como en el caso~IIA, hay dilat'on y dilatino que no
  aparecen asociados directamente a ninguna carga del 'algebra.}

\begin{equation} 
\{\hat{\jmath}_{\hat{\mu}\hat{\nu}}, 
\hat{\cal B}_{\hat{\mu}\hat{\nu}}, \hat{\varphi}\}\, , 
\end{equation}

\noindent en el sector NSNS y 

\begin{equation}
\{\hat{C^{(0)}}, \hat{C^{(2)}}{}_{\hat{\mu}\hat{\nu}},
\hat{C^{(4)}}{}_{\hat{\mu}\hat{\nu}\hat{\rho}\hat{\sigma}}\}\, ,
\end{equation}

\noindent en el sector RR.

Los estados/soluciones asociados son la onda gravitacional, las
D$p$-branas con $p=1,3,5,7,9$ y, en el sector NSNS, la cuerda
fundamental $F1B$, la 5-brana solit'onica $S5B$ y la 9-brana
solit'onica $S9B$ (a veces llamada NSNS9). Dada la no-invariancia de
la D7-brana bajo dualidad~S, es tentador introducir un objeto dual: la
S7-brana que adem'as parece ser requerida por dualidad~T
\cite{kn:MO,kn:L-TO2}, pero el {\it status} de este objeto no est'a
completamente claro.

Al compactificar la teor'ia en un c'irculo hay que considerar
una carga con 5 'indices singlete asociada al monopolo de
Kaluza-Klein (KK6B).

%%%%%%%%%%%%%%%%%%%%%%%%%%%%%%%%%%%%%%%%%%%%%%%%%%%%%%%%%%%%%%%%%%%%%%
\subsubsection{La acci'on de la teor'ia tipo~IIB}

Las intensidades de campo de los potenciales RR de esta teor'ia se
pueden escribir exactamente igual que en el caso~IIA, de acuerdo con
las Ecs.~(\ref{eq:RRfieldstrengths}),(\ref{eq:RReom&Bianchis}).
As'i tenemos

\begin{equation}
\left\{
\begin{array}{rcl}
\hat{\cal H} & = & 3\partial {\cal B}\, , \\
& & \\
\hat{G}^{(1)} & = & \partial \hat{C}^{(0)}\, ,\\
& & \\
\hat{G}^{(3)} & = & 3\left(\partial \hat{C}^{(2)}
-\partial \hat{\cal B} \hat{C}^{(0)}\right)\, ,\\
& & \\
\hat{G}^{(5)} & = & 5\left(\partial \hat{C}^{(4)}
-6\partial \hat{\cal B} \hat{C}^{(2)}\right)\, .\\
\end{array}
\right.
\end{equation}

Una de las ecuaciones de esta teor'ia es la condici'on de
auto-dualidad de la 5-forma RR \cite{kn:JHS}

\begin{equation}
\hat{G}^{(5)}= +{}^{\star} \hat{G}^{(5)}\, .  
\end{equation}

Es imposible escribir una acci'on covariante que d'e esta ecuaci'on
del movimiento, al menos sin introducir campos auxiliares,
b'asicamente porque la autodualidad implica

\begin{equation}
\left( \hat{G}^{(5)}  \right)^{2} = \left( {}^{\star}\hat{G}^{(5)}  \right)^{2}
=-\left( \hat{G}^{(5)}  \right)^{2} \Rightarrow =0\, .
\end{equation}

Si prescindimos de esta condici'on, s'i que podemos escribir una
acci'on ({\it acci'on NSD}) \cite{kn:BBO} de la cual derivamos
ecuaciones de movimiento que simplemente debemos de complementar con
la condici'on de autodualidad. Esta acci'on es en el sistema de
referencia conforme de la cuerda

\begin{equation}
\begin{tabular}{|c|}
\hline \\
$
\begin{array}{rcl}
S_{\rm NSD} & = &
\frac{g_{B}^{2}}{16\pi G_{N\, B}^{(10)}}
{\displaystyle\int} d^{10}\hat{x}\
\sqrt{|\hat{\jmath}|}\ \left\{ e^{-2\hat{\varphi}}
\left[ \hat{R}(\hat{\jmath}) -4\left( \partial\hat{\varphi} \right)^{2}
+{\textstyle\frac{1}{2\cdot 5!}} \hat{\cal H}^{2}\right]\right.\\
& & \\
& & 
\hspace{1cm}+{\textstyle\frac{1}{2}} \left( \hat{G}^{(0)} \right)^{2}
+{\textstyle\frac{1}{2\cdot 3!}}\left(\hat{G}^{(3)}\right)^{2}
+{\textstyle\frac{1}{4\cdot 3!}}\left(\hat{G}^{(5)}\right)^{2} \\
& & \\
& &
\hspace{1cm}\left.-{\textstyle\frac{1}{192}} \frac{1}{\sqrt{|\hat{\jmath}|}}\
\epsilon\  \partial \hat{C}^{(4)}\partial \hat{C}^{(2)} \hat{\cal B}
\right\}\, ,\\
\end{array}
$
\\ \\ \hline
\end{tabular}
\label{eq:IIBaction}
\end{equation}

Obs'ervese que el t'ermino cin'etico de la 4-forma tiene un factor
extra de $1/2$ que, en cierto sentido, tiene en cuenta que la 4-forma
no-autodual describe el doble de grados de libertad de los que debe.
Obs'ervese tambi'en que hemos introducido, como en el caso~IIA un
prefactor  $g_{B}^{2}$ para absorber el valor asint'oicos del dilat'on,
usando la definici'on

\begin{equation}
g_{B} = e^{\hat{\varphi}_{0}}\, .  
\end{equation}

%%%%%%%%%%%%%%%%%%%%%%%%%%%%%%%%%%%%%%%%%%%%%%%%%%%%%%%%%%%%%%%%%%%%%%

\subsubsection{Potenciales magn'eticos}

Los potenciales magn'eticos se introducen tambi'en usando la misma
relaci'on que en el caso~IIA Ec.~(\ref{eq:RRduals}), consistentemente
con las Ecs.~(\ref{eq:RRfieldstrengths}),(\ref{eq:RReom&Bianchis}).

%%%%%%%%%%%%%%%%%%%%%%%%%%%%%%%%%%%%%%%%%%%%%%%%%%%%%%%%%%%%%%%%%%%%%%

\subsubsection{Las reglas de transformaci'on bajo  supersimetr'ia}

Son

\begin{equation}
\label{eq:susiibstringgeneral}
\begin{tabular}{|c|}
\hline \\
$
\begin{array}{rcl}
\delta_{\hat{\varepsilon}} \hat{e}_{\hat{\mu}}{}^{\hat{a}} & = & 
-i\bar{\hat{\varepsilon}}\hat{\Gamma}^{\hat{a}} \hat{\zeta}_{\hat{\mu}}\, ,\\
& & \\
\delta_{\hat{\varepsilon}} \hat{\zeta}_{\hat{\mu}} & = & 
\nabla_{\hat{\mu}} \hat{\varepsilon}
-\frac{1}{8}\not\!\! \hat{\cal H}_{\hat{\mu}}\sigma_{3} \hat{\varepsilon} 
+\frac{1}{8}e^{\hat{\varphi}} \sum_{n=1,2,3} 
\frac{1}{(2n-1)!} \not\!\hat{G}^{(2n-1)} 
\hat{\Gamma}_{\hat{\mu}} {\cal P}_{n}\hat{\varepsilon}\, , \\
& & \\
\delta_{\hat{\varepsilon}} \hat{\cal B}_{\hat{\mu}\hat{\nu}} & = & 
-2i\bar{\hat{\varepsilon}} \sigma^{3} \hat{\Gamma}_{[\hat{\mu}} 
\hat{\zeta}_{\hat{\nu}]}\, , \\
& & \\
\delta_{\hat{\varepsilon}} 
\hat{C}^{(2n-2)}{}_{\hat{\mu}_{1}\cdots\hat{\mu}_{2n-2}} & = &
i(2n-2) e^{-\hat{\varphi}} \bar{\hat{\varepsilon}} {\cal P}_{n} 
\hat{\Gamma}_{[\hat{\mu}_{1}\cdots\hat{\mu}_{2n-3}} 
\left(\hat{\zeta}_{\hat{\mu}_{2n-2}]} -\frac{1}{2(2n-2)}
\hat{\Gamma}_{\hat{\mu}_{2n-2}]}\hat{\chi}
\right)\\
& & \\
& &  + \frac{1}{2}(2n-2)(2n-3) 
\hat{C}^{(2n-4)}{}_{[\hat{\mu}_{1}\cdots\hat{\mu}_{2n-4}}
\delta_{\hat{\varepsilon}} 
\hat{\cal B}_{\hat{\mu}_{2n-3}\hat{\mu}_{2n-4}}\, ,\\
& & \\
\delta_{\hat{\varepsilon}}\hat{\chi} & = & 
\left(\not\!\partial\hat{\varphi}
-\frac{1}{12}  \not\!\! \hat{\cal H}\sigma^{3}\right) \hat{\varepsilon}
+\frac{1}{2}e^{\hat{\varphi}} \sum_{n=1,2,3} 
\frac{(n-3)}{(2n-1)!}\not\! \hat{G}^{(2n-1)} 
{\cal P}_{n}\varepsilon\, , \\
& & \\
\delta_{\hat{\varepsilon}}\hat{\varphi} & = & 
-\frac{i}{2}\bar{\hat{\varepsilon}}\hat{\chi}\, ,\\
\end{array}
$
\\ \\ \hline
\end{tabular}
\end{equation}

\noindent donde

\begin{equation}
{\cal P}_{n} =
\left\{
\begin{array}{l}
\sigma^{1}\, ,\hspace{.5cm} n\,\, {\rm even}\, , \\
\\
i\sigma^{2}\, ,\hspace{.5cm} n\,\, {\rm odd}\, .\\
\end{array}
\right. 
\end{equation}

%%%%%%%%%%%%%%%%%%%%%%%%%%%%%%%%%%%%%%%%%%%%%%%%%%%%%%%%%%%%%%%%%%%%%%

\subsubsection{Dualidad~S en la teor'ia tipo~IIB}

Hemos mencionado que la teor'ia tipo~IIB es invariante bajo una
dualidad~S. Como la mayor parte de las dualidades, 'esta s'olo es
manifiesta en el sistema de referencia conforme de Einstein. La
raz'on es que si las dualidades involucran a los escalares y 'estos
aparecen como prefactores de la acci'on, s'olo transformando la
m'etrica vamos a ver la dualidad, lo cual puede ser muy complicado.

Por lo tanto, para empezar, reescaleamos la m'etrica con el dilat'on
para reescribir la acci'on en el sistema de referencia conforme de
Einstein

\begin{equation}
\hat{\jmath}_{E\, \mu\nu}=e^{-\varphi/2}\jmath_{\mu\nu}\, .
\end{equation}

Esto no es suficiente para hacer la simetr'ia manifiesta. Hemos de
redefinir los potenciales, puesto que los que estamos utilizando son
adecuados para describir dualidad~T, pero no dualidad~S. Los nuevos
potenciales son

\begin{equation}
\left\{
\begin{array}{rcl}
\hat{\vec{\cal B}} & = & 
\left(
\begin{array}{c}
\hat{C}^{(2)}  \\
\hat{\cal B}   \\
\end{array}
\right)\, , \\
& & \\
\hat{D} & = & \hat{C}^{(4)} -3\hat{\cal B}\hat{C}^{(2)}\, ,\\
\end{array}
\right.
\end{equation}

\noindent  y sus intensidades de campo son

\begin{equation}
\label{eq:fieldstrengths10B}
\left\{
\begin{array}{rcl}
\hat{\vec{\cal H}} & = & 3\partial \hat{\vec{\cal B}}\, ,\\
& & \\
\hat{F} & = & \hat{G}^{(5)}=+\ {}^{\star} \hat{F} \\
& & \\
& = & 5\left(\partial \hat{D} 
-\hat{\vec{\cal B}}{}^{\ T}\eta\ \hat{\vec{\cal H}}\right)\, ,\\
\end{array}
\right.
\end{equation}

\noindent donde $\eta$ es la matriz $2\times 2$ 

\begin{equation}
\eta= i\sigma^{2} =
\left(
\begin{array}{rc}
0  & 1 \\
-1 & 0 \\
\end{array}
\right)
=
-\eta^{-1}
=
-\eta^{T}\, ,
\end{equation}

\noindent que, dado el isomorfismo $SL(2,\mathbb{R})\sim Sp(2,\mathbb{R})$,
puede ser identificada con una m'etrica invariante

\begin{equation}
\label{eq:propertyeta}
\Lambda \eta \Lambda^{T}=\eta\, ,
\,\,
\Rightarrow
\,\,
\eta\Lambda\eta^{T} = (\Lambda^{-1})^{T}\, ,
\hspace{.5cm}
\Lambda \in SL(2,\mathbb{R})\, .
\end{equation}

Finalmente, definimos la matriz de escalares $2\times 2$ $\hat{\cal M}_{ij}$

\begin{equation}
\hat{\cal M}
=e^{\hat{\varphi}}
\left(
\begin{array}{cc}
|\hat{\tau}|^{2} &    \hat{C}^{(0)}  \\
& \\
\hat{C}^{(0)}       &  1          \\
\end{array}
\right)\, ,
\hspace{1cm}
\hat{\cal M}^{-1}
=e^{\hat{\varphi}}
\left(
\begin{array}{cc}
1                     &  -\hat{C}^{(0)}      \\
& \\
-\hat{C}^{(0)}        &  |\hat{\tau}|^{2} \\
\end{array}
\right)\, ,
\end{equation}

\noindent donde $\hat{\tau}$ es el escalar complejo

\begin{equation}
\hat{\tau} =\hat{C}^{(0)} +ie^{-\hat{\varphi}}\, .
\end{equation}

Bajo $\Lambda\in SL(2,\mathbb{R})$ los diferentes campos que hemos
definido se transforman as'i:

\begin{equation}
\begin{array}{rcl}
\hat{\cal M}^{\prime} & =  & \Lambda \hat{\cal M} \Lambda^{T}\, ,\\
& & \\
\hat{\vec{\cal B}}{}^{\prime}
 & =  & 
\Lambda \hat{\vec{\cal B}}\, ,\\
\end{array}
\end{equation}

\noindent mientras que la 4-forma y la m'etrica de Einstein son invariantes.
La regla de transformaci'on de $\hat{\cal M}$ implica, para $\hat{\tau}$

\begin{equation}
\hat{\tau}^{\prime} = \frac{a\hat{\tau} +b}{c\hat{\tau} +d}\, .  
\end{equation}

En t'erminos de los nuevos campos, la acci'on NSD es manifiestamente
invariante bajo dualidad~S:

\begin{equation}
\begin{tabular}{|c|}
\hline \\ \\
$
\begin{array}{rcl}
\hat{S}_{\rm NSD} & = &
\frac{g^{2}_{B}}{16\pi G_{N}^{(10)}} {\displaystyle\int} d^{10}\hat{x}\
\sqrt{|\hat{\jmath}_{E}|} \left\{ \hat{R}(\hat{\jmath}_{E})
+{\textstyle\frac{1}{4}}
{\rm Tr}\left(\partial\hat{\cal M} \hat{\cal M}^{-1}\right)^{2}
\right.\\
& & \\
& & 
\left.
+{\textstyle\frac{1}{2\cdot 3!}} 
\hat{\vec{\cal H}}{}^{\ T}\hat{\cal M}^{-1}\hat{\vec{\cal H}}
+{\textstyle\frac{1}{4\cdot 5!}}\hat{F}^{2}
-{\textstyle\frac{1}{2^{7}\cdot 3^{3}}} 
\frac{1}{\sqrt{|\hat{\jmath}_{E}|}}\
\epsilon\ \hat{D}\ \hat{\vec{\cal H}}{}^{\ T}\eta\ \hat{\vec{\cal H}}
\right\}\, , \\
\end{array}
$
\\ \\ \hline
\end{tabular}
\label{eq:SinvariantIIBaction}
\end{equation}

Obs'ervese que el factor $\frac{g^{2}_{B}}{16\pi G_{N}^{(10)}}$ es
invariante bajo dualidad~S porque no depende de $g_{B}$ (v'ease la
Ec.~(\ref{eq:GN10A})). S'olo en este sistema de referencia conforme
la m'etrica es invariante bajo dualidad~S. Sin embargo, 'este no es
en realidad el sistema de referencia conforme de Einstein porque, si
la m'etrica de la que partimos era asint'oticamente plana, 'esta no
lo es\footnote{En la literatura, sin embargo, se le denomina sistema
  de referencia de Einstein y, al que deber'ia denominarse de
  Einstein se le llama {\it sistema de referencia conforme de Einstein
    modificado} \cite{kn:M}.}. Hay que volver a reescalearla con el
valor asint'otico del dilat'on para llegar a la aut'entica
m'etrica de Einstein ({\it m'etrica de Einstein modificada})que, por
lo tanto, no es invariante bajo dualidad~S. Tampoco lo ser'an las
masas medidas en este sistema.  'Este es un dato importante que
utilizaremos en la siguiente lecci'on.

Es costumbre llamar transformaci'on de dualidad~S a la producida por
la matriz de $SL(2,\mathbb{R})$ $\Lambda=\eta$. Esta transformaci'on
intercambia las 2-formas NSNS y RR e invierte el escalar complejo
$\hat{\tau}^{\prime}= -1/\hat{\tau}$. En ausencia de 0-forma RR, esta
transformaci'on invierte el dilat'on y, por lo tanto la constante de
acoplo de la cuerda

\begin{equation}
\label{eq:dualcouplingB}
g^{\prime}_{B}=1/g_{B}\, .  
\end{equation}

La m'etrica de la cuerda se transforma bajo $SL(2,\mathbb{R})$
as'i:

\begin{equation}
\hat{\jmath}^{\prime} = |c\hat{\lambda} +d| \hat{\jmath}\, ,  
\end{equation}

\noindent y, bajo la transformaci'on de dualidad~S anterior, en ausencia de 0-forma RR

\begin{equation}
\hat{\jmath}^{\prime} = e^{-\hat{\varphi}}\hat{\jmath}\, ,  
\end{equation}

\noindent lo que implica para los radios de las dimensiones
compactas medidos en este sistema de referencia

\begin{equation}
\label{eq:dualradiiB}
R^{\prime}=R/g_{B}   
\end{equation}

%%%%%%%%%%%%%%%%%%%%%%%%%%%%%%%%%%%%%%%%%%%%%%%%%%%%%%%%%%%%%%%%%%%%%%
\subsection{Dualidad~T entre las teor'ias tipo~II}

En la introducci'on hemos dicho que las Teor'ias de Cuerdas
tipo~IIA y~IIB son equivalentes cuando se las compactifica en
c'irculos de radios rec'iprocamente inversos (son duales~T).
Esto permite relacionar todos los grados de libertad de las
teor'ias. Hay dos relaciones principales:

\begin{enumerate}
\item La relaci'on entre modos de momento (Kaluza-Klein) y modos de
  enrollamiento ({\it winding}): los estados que en una teor'ia
  corresponden a cuerdas (cerradas) movi'endose en la dimensi'on
  compacta, en la otra corresponden a cuerdas que est'an enrolladas
  en la dimensi'on compacta dual, de radio dual. En los espectros
  respectivos, ambos tipos de modos aparecen con la misma masa.
\item La relaci'on entre D$p$-branas de una teor'ia con una de
  sus dimensiones enrollada en la dimensi'on compacta del
  espacio-tiempo y D$(p-1)$-branas de la otra teor'ia ortogonales
  a la dimensi'on compacta dual.
\end{enumerate}

La dualidad~T es la m'as interesante de las propiedades de la
Teor'ia de Cuerdas. Una de sus implicaciones es que no podemos
hacer desaparecer una dimensi'on compactific'andola en un
c'irculo y haciendo su radio tender a cero, porque (a diferencia
de lo que pasa en las teor'ias de Kaluza-Klein) esta teor'ia
es equivalente a otra con un radio que tiende a infinito. Por otro
lado sugiere que con las cuerdas (objetos de tama~no finito) no se
pueden medir distancias peque~nas, lo que implicar'ia una
modificaci'on del Principio de Incertidumbre de Heisenberg.

En las acciones efectivas la dualidad~T se manifiesta relaci'on entre
los campos 10-dimensionales de ambas teor'ias que permite
transformar cualquier soluci'on de una de ellas que no dependa de la
coordenada compacta en una soluci'on de la otra que no depende de la
otra coordenada compacta. Las reglas que permiten hacer esta
transformaci'on son las {\it reglas de Buscher}. Para obtenerlas hay
que realizar la reducci'on dimensional de las acciones de las SUEGRAS
$N=2A,d=10$ y $N=2B,d=10$ de tal manera que den la misma acci'on en
$d=9$.

En esta Secci'on vamos simplemente a contar c'omo se reducen ambas
teor'ias y dar las reglas de Buscher resultantes, que
necesitaremos en la pr'oxima lecci'on para relacionar soluciones.
Tambi'en podemos aprender de ellas la relaci'on entre los grados de
libertad de las dos teor'ias tipo~II, comprobando lo dicho m'as
arriba.

Para empezar y como preparaci'on, vamos a reducir las super'algebras
de estas teor'ias. De esta reducci'on podemos aprender tambi'en
c'omo se relacionan los grados de libertad solit'onicos de estas dos
teor'ias a trav'es de las relaciones entre las cargas
cuasi-centrales.

%%%%%%%%%%%%%%%%%%%%%%%%%%%%%%%%%%%%%%%%%%%%%%%%%%%%%%%%%%%%%%%%%%%%%%
\subsubsection{Reducci'on del super'algebra $N=2A,d=10$ a $d=9$}

Los espinores en $d=10$ se descomponen en dos espinores en $d=9$, que
indicamos con un 'indice espinorial y un 'indice de $SO(2)$:
$\hat{\alpha}=(\alpha,i)$, $i=1,2$. Las matrices gamma
10-dimensionales se descomponen en el producto tensorial de matrices
gamma 9-dimensionales y matrices de Pauli como viene indicado en el
Ap'endice~\ref{sec-d9gammas}. Incluyendo todas las cargas que
aparecen cuando una de las dimensiones es compacta, se obtiene el
siguiente super'algebra $N=2,d=9$

\begin{equation}
\begin{array}{rcl}
\{Q^{i\, \alpha},Q^{j\, \beta}\} & = & 
i\left(\Gamma^{a}{\cal C}^{-1} \right)^{\alpha\beta}
\left(\delta^{ij}P_{a} +\sigma^{1\, ij}Z^{(1)\, 1}_{a}
+\sigma^{3\, ij}Z^{(1)\, 3}_{a} \right)\\
& & \\
& & 
+\left({\cal C}^{-1} \right)^{\alpha\beta}
\left(\delta^{ij}Z^{(0)\, 0} +\sigma^{1\, ij}Z^{(0)\, 1}
+\sigma^{3\, ij}Z^{(0)\, 3} \right)\\
& & \\
& & 
+\frac{i}{2!}\left(\Gamma^{a_{1}a_{2}}{\cal C}^{-1} \right)^{\alpha\beta}
\sigma^{2\, ij}Z^{(2)}_{a_{1}a_{2}}
+\frac{1}{3!}\left(\Gamma^{a_{1}a_{2}a_{3}}{\cal C}^{-1} \right)^{\alpha\beta}
\sigma^{2\, ij}Z^{(3)}_{a_{1}a_{2}a_{3}}\\
& & \\
& & 
+\frac{1}{4!}\left(\Gamma^{a_{1}\cdot a_{4}}{\cal C}^{-1} \right)^{\alpha\beta}
\left(\sigma^{1\, ij}Z^{(4)\, 1}_{a_{1}\cdot a_{4}}
+\sigma^{3\, ij}Z^{(4)\, 3}_{a_{1}\cdot a_{4}} \right)\\
& & \\
& & 
+\frac{i}{5!}\left(\Gamma^{a_{1}\cdot a_{5}}{\cal C}^{-1} \right)^{\alpha\beta}
\left(\delta^{ij}Z^{(5)\, 0}_{a_{1}\cdot a_{5}}
+\sigma^{1\, ij}Z^{(5)\, 1}_{a_{1}\cdot a_{5}}
+\sigma^{3\, ij}Z^{(5)\, 3}_{a_{1}\cdot a_{5}} \right)\\
& & \\
& & 
+\frac{i}{6!}\left(\Gamma^{a_{1}\cdots a_{6}}
{\cal C}^{-1} \right)^{\alpha\beta}
\sigma^{2\, ij}\left( Z^{(6)}_{a_{1}\cdots a_{6}}
+ Z^{(6)\prime}_{a_{1}\cdots a_{6}}\right)\\
& & \\
& & 
+\frac{1}{7!}\left(\Gamma^{a_{1}\cdots a_{7}}
{\cal C}^{-1} \right)^{\alpha\beta}
\sigma^{2\, ij}\left( Z^{(7)}_{a_{1}\cdots a_{7}}
+ Z^{(7)\prime}_{a_{1}\cdots a_{7}}\right) \\
& & \\
& & 
+\frac{1}{8!}\left(\Gamma^{a_{1}\cdot a_{8}}{\cal C}^{-1} \right)^{\alpha\beta}
\left(\sigma^{1\, ij}Z^{(8)\, 1}_{a_{1}\cdot a_{8}}
+\sigma^{3\, ij}Z^{(8)\, 3}_{a_{1}\cdot a_{8}} \right)\, .\\
\end{array}
\end{equation}

La invariancia (o covariancia) $SO(2)$ de este super'algebra es
herencia de su origen 11-dimensional a trav'es de la
compactificaci'on en un 2-toro. El super'algebra $N=2B,d=10$ tiene
la misma estructura $SO(2)$, pero no relacionada con
compactificaciones, sino con la dualidad~S. Esto sugiere una
relaci'on entre la dualidad~S de la teor'ia tipo~IIB y la
simetr'ia del 2-toro\footnote{El grupo modular del toro es
  precisamente $PSL(2,\mathbb{Z})$.}.

%%%%%%%%%%%%%%%%%%%%%%%%%%%%%%%%%%%%%%%%%%%%%%%%%%%%%%%%%%%%%%%%%%%%%%
\subsubsection{Reducci'on del super'algebra $N=2B,d=10$ a $d=9$}

La reducci'on de este super'algebra es muy sencilla y da el mismo
super'algebra en $d=9$. Esto nos permite relacionar las componentes
de las cargas de las dos super'algebras 10-dimensionales, conformando
las dos reglas generales de T~dualidad.

%%%%%%%%%%%%%%%%%%%%%%%%%%%%%%%%%%%%%%%%%%%%%%%%%%%%%%%%%%%%%%%%%%%%%%
\subsubsection{Reducci'on de la tipo~IIA a $d=9$}

Para establecer dualidad~T entre las acciones efectivas de las SUEGRAS
$N=2A,d=10$ y $N=2B,d=10$ tenemos que hacer la reducci'on
dimensional de ambas acciones, utilizando variables duales. Por
razones de espacio, no podemos explicar en detalle todo el proceso y
simplemente escribimos c'omo los campos de las dos teor'ias
10-dimensionales se descomponen en t'erminos de los campos de la
'unica teor'ia 9-dimensional.

Campos del sector NSNS:

\begin{equation}
\begin{array}{ll}
\begin{array}{lcl}
\hat{g}_{\mu\nu} & = & g_{\mu\nu} -k^{2}A^{(1)}{}_{\mu}A^{(1)}{}_{\nu}\, ,\\
& & \\
\hat{B}_{\mu\nu} & = & B_{\mu\nu} +A^{(1)}{}_{[\mu}A^{(2)}{}_{\nu]}\, ,\\& & \\
\hat{\phi} & = & \phi + \frac{1}{2} \log{k}\, ,\\
& & \\
\hat{g}_{\mu\underline{x}} & = & -k^{2}A^{(1)}{}_{\mu}\, ,\\
& & \\
\hat{B}_{\mu\underline{x}} & = & -A^{(2)}{}_{\mu}\, ,\\
& & \\
\hat{g}_{\underline{x}\underline{x}} & = & -k^{2}\, ,\\
\end{array}
\hspace{.5cm}
&
\begin{array}{lcl}
g_{\mu\nu} & = & \hat{g}_{\mu\nu} 
-\hat{g}_{\mu\underline{x}}\hat{g}_{\nu\underline{x}}/
\hat{g}_{\underline{x}\underline{x}}\, , \\
& & \\
B_{\mu\nu} & = & \hat{B}_{\mu\nu} 
+\hat{g}_{[\mu|\underline{x}|}\hat{B}_{\nu]\underline{x}}/
\hat{g}_{\underline{x}\underline{x}}\, , \\
& & \\
\phi & = & \hat{\phi} 
-\frac{1}{4} \log{|\hat{g}_{\underline{x}\underline{x}}|}\, ,\\
& & \\
A^{(1)}{}_{\mu} & = & 
\hat{g}_{\mu\underline{x}}/\hat{g}_{\underline{x}\underline{x}}\, ,\\
& & \\
A^{(2)}{}_{\mu} & = & -\hat{B}_{\mu\underline{x}}\, ,\\
& & \\
k & = & |\hat{g}_{\underline{x}\underline{x}}|^{1/2}\, .\\
\end{array}
\\
\end{array}
\end{equation}

Campos del sector RR:

\begin{equation}
\begin{array}{lcl}
\hat{C}^{(2n-1)}{}_{\mu_{1}\cdots\mu_{2n-1}} & = & C^{(2n-1)}{}_{\mu_{1}\cdots\mu_{2n-1}}
+(2n-1)A^{(1){}_{[\mu_{1}}}C^{(2n-2)}{}_{\mu_{2}\cdots\mu_{2n-1}]}\, ,\\
& & \\
\hat{C}^{(2n+1)}{}_{\mu_{1}\cdots\mu_{2n}\underline{x}} & = & 
C^{(2n)}{}_{\mu_{1}\cdots\mu_{2n}}\, ,\\ 
& & \\
C^{(2n-1)}{}_{\mu_{1}\cdots\mu_{2n-1}} & = &
\hat{C}^{(2n-1)}{}_{\mu_{1}\cdots\mu_{2n-1}} 
-(2n-1)\hat{g}_{[\mu_{1}|\underline{x}|} 
\hat{C}^{(2n-1)}{}_{\mu_{2}\cdots\mu_{2n-1}]\underline{x}}/
\hat{g}_{\underline{x}\underline{x}}\, ,\\
& & \\
C^{(2n)}{}_{\mu_{1}\cdots\mu_{2n}} & = & 
\hat{C}^{(2n+1)}{}_{\mu_{1}\cdots\mu_{2n}\underline{x}}\, .\\
\end{array}
\end{equation}

%%%%%%%%%%%%%%%%%%%%%%%%%%%%%%%%%%%%%%%%%%%%%%%%%%%%%%%%%%%%%%%%%%%%%%
\subsubsection{Reducci'on de la tipo~IIB a $d=9$}

Campos del sector NSNS:

\begin{equation}
\begin{array}{ll}
\begin{array}{lcl}
\hat{\jmath}_{\mu\nu} & = & 
g_{\mu\nu} -k^{-2}A^{(2)}{}_{\mu}A^{(2)}{}_{\nu}\, ,\\
& & \\
\hat{\cal B}_{\mu\nu} & = & 
B_{\mu\nu} +A^{(1)}{}_{[\mu}A^{(2)}{}_{\nu]}\, ,\\
& & \\
\hat{\varphi} & = & \phi -\frac{1}{2} \log{k}\, ,\\
& & \\
\hat{\jmath}_{\mu\underline{y}} & = & -k^{-2}A^{(2)}{}_{\mu}\, ,\\
& & \\
\hat{\cal B}_{\mu\underline{y}} & = & A^{(1)}{}_{\mu}\, ,\\
& & \\
\hat{\jmath}_{\underline{y}\underline{y}} & = & -k^{-2}\, ,\\
\end{array}
\hspace{.5cm}
&
\begin{array}{lcl}
g_{\mu\nu} & = & \hat{\jmath}_{\mu\nu} 
-\hat{\jmath}_{\mu\underline{y}}\hat{\jmath}_{\nu\underline{y}}/
\hat{\jmath}_{\underline{y}\underline{y}}\, , \\
& & \\
B_{\mu\nu} & = & \hat{\cal B}_{\mu\nu} 
+\hat{\jmath}_{[\mu|\underline{y}|}\hat{\cal B}_{\nu]\underline{y}}/
\hat{\jmath}_{\underline{y}\underline{y}}\, , \\
& & \\
\phi & = & \hat{\varphi} 
-\frac{1}{4} \log{|\hat{\jmath}_{\underline{y}\underline{y}}|}\, ,\\
& & \\
A^{(1)}{}_{\mu} & = & \hat{\cal B}_{\mu\underline{y}}\, ,\\
& & \\
A^{(2)}{}_{\mu} & = & 
\hat{\jmath}_{\mu\underline{y}}
/\hat{\jmath}_{\underline{y}\underline{y}}\, ,\\
& & \\
k & = & |\hat{\jmath}_{\underline{y}\underline{y}}|^{-1/2}\, .\\
\end{array}
\\
\end{array}
\end{equation}

Campos del sector RR:

\begin{equation}
\begin{array}{lcl}
\hat{C}^{(2n)}{}_{\mu_{1}\cdots\mu_{2n}} & = & 
C^{(2n)}{}_{\mu_{1}\cdots\mu_{2n}}
-(2n)A^{(2)}{}_{[\mu_{1}}C^{(2n-1)}{}_{\mu_{2}\cdots\mu_{2n}]}\, ,\\
& & \\
\hat{C}^{(2n)}{}_{\mu_{1}\cdots\mu_{2n-1}\underline{y}} & = & 
-C^{(2n-1)}{}_{\mu_{1}\cdots\mu_{2n-1}}\, ,\\ 
& & \\
C^{(2n)}{}_{\mu_{1}\cdots\mu_{2n}} & = &
\hat{C}^{(2n)}{}_{\mu_{1}\cdots\mu_{2n}} 
+(2n)\hat{\jmath}_{[\mu_{1}|\underline{y}|} 
\hat{C}^{(2n)}{}_{\mu_{2}\cdots\mu_{2n}]\underline{y}}/
\hat{\jmath}_{\underline{y}\underline{y}}\, ,\\
& & \\
C^{(2n-1)}{}_{\mu_{1}\cdots\mu_{2n-1}} & = & 
-\hat{C}^{(2n)}{}_{\mu_{1}\cdots\mu_{2n-1}\underline{y}}\, .\\
\end{array}
\end{equation}

%%%%%%%%%%%%%%%%%%%%%%%%%%%%%%%%%%%%%%%%%%%%%%%%%%%%%%%%%%%%%%%%%%%%%%
\subsubsection{Reglas de Buscher tipo~II}

Para hallar las reglas de Buscher no tenemos m'as que utilizar las
relaciones anteriores \cite{kn:BHO,kn:MO}:

\textbf{De IIA a IIB:}

\begin{equation}
\begin{tabular}{|c|}
\hline \\ \\ 
$
\begin{array}{l}
\begin{array}{lcllcl}
\hat{\jmath}_{\mu\nu} & = & \hat{g}_{\mu\nu} 
-\left(\hat{g}_{\mu\underline{x}}\hat{g}_{\nu\underline{x}} 
-\hat{B}_{\mu\underline{x}}\hat{B}_{\nu\underline{x}}\right)/
\hat{g}_{\underline{x}\underline{x}}\, , 
\hspace{.5cm}&
\hat{\jmath}_{\mu\underline{y}} & = & \hat{B}_{\mu\underline{x}}/
\hat{g}_{\underline{x}\underline{x}}\, , \\
& & & & & \\
\hat{\cal B}_{\mu\nu} & = & \hat{B}_{\mu\nu} 
+2\hat{g}_{[\mu|\underline{x}}\hat{B}_{\nu]\underline{x}}/
\hat{g}_{\underline{x}\underline{x}}\, , &
\hat{\cal B}_{\mu\underline{y}} & = & \hat{g}_{\mu\underline{x}}/
\hat{g}_{\underline{x}\underline{x}}\, , \\
& & & & & \\
\hat{\varphi} & = & \hat{\phi} 
-\frac{1}{2}\log{|\hat{g}_{\underline{x}\underline{x}}|}\, , &
\hat{\jmath}_{\underline{y}\underline{y}} & = & 
1/\hat{g}_{\underline{x}\underline{x}}\, ,\\
\end{array}
\\
\\
\begin{array}{lcl}
\hat{C}^{(2n)}{}_{\mu_{1}\ldots\mu_{2n}} & = & 
\hat{C}^{(2n+1)}{}_{\mu_{1}\ldots\mu_{2n}\underline{x}}
+2n \hat{B}_{[\mu_{1}|\underline{x}|}
\hat{C}^{(2n-1)}{}_{\mu_{2}\ldots\mu_{2n}]}\\
& & \\
& &
-2n (2n -1) \hat{B}_{[\mu_{1}|\underline{x}|}\hat{g}_{\mu_{2}|\underline{x}|}
\hat{C}^{(2n-1)}{}_{\mu_{3}\ldots\mu_{2n}]\underline{x}}/
\hat{g}_{\underline{x}\underline{x}}\, , \\
& & \\
\hat{C}^{(2n)}{}_{\mu_{1}\ldots\mu_{2n-1}\underline{y}} & = & 
-\hat{C}^{(2n-1)}{}_{\mu_{1}\ldots\mu_{2n-1}} \\
& & \\
& &
+(2n -1) \hat{g}_{[\mu_{1}|\underline{x}|}
\hat{C}^{(2n-1)}{}_{\mu_{2}\ldots\mu_{2n-1}]\underline{x}}/
\hat{g}_{\underline{x}\underline{x}}\, .\\
\end{array}\\
\end{array}
$
\\
\\
\hline
\end{tabular}
\label{eq:reglasBuscherIIAtoIIB}
\end{equation}

\textbf{De IIB a IIA:}

\begin{equation}
\begin{tabular}{|c|}
\hline \\ \\ 
$
\begin{array}{l}
\begin{array}{lcllcl}
\hat{g}_{\mu\nu} & = & \hat{\jmath}_{\mu\nu} 
-\left(\hat{\jmath}_{\mu\underline{y}}\hat{\jmath}_{\nu\underline{y}} 
-\hat{\cal B}_{\mu\underline{y}}\hat{\cal B}_{\nu\underline{y}}\right)/
\hat{\jmath}_{\underline{y}\underline{y}}\, , 
\hspace{.5cm}&
\hat{g}_{\mu\underline{x}} & = & \hat{\cal B}_{\mu\underline{y}}/
\hat{\jmath}_{\underline{y}\underline{y}}\, , \\
& & & & & \\
\hat{B}_{\mu\nu} & = & \hat{\cal B}_{\mu\nu} 
+2\hat{\jmath}_{[\mu|\underline{y}}\hat{\cal B}_{\nu]\underline{y}}/
\hat{\jmath}_{\underline{y}\underline{y}}\, , &
\hat{B}_{\mu\underline{x}} & = & \hat{\jmath}_{\mu\underline{y}}/
\hat{\jmath}_{\underline{y}\underline{y}}\, , \\
& & & & & \\
\hat{\phi} & = & \hat{\varphi} 
-\frac{1}{2}\log{|\hat{\jmath}_{\underline{y}\underline{y}}|}\, , &
\hat{g}_{\underline{x}\underline{x}} & = & 
1/\hat{\jmath}_{\underline{y}\underline{y}}\, ,\\
\end{array}
\\
\\
\begin{array}{lcl}
\hat{C}^{(2n+1)}{}_{\mu_{1}\ldots\mu_{2n+1}} & = & 
-\hat{C}^{(2n+2)}{}_{\mu_{1}\ldots\mu_{2n+1}\underline{y}}
+(2n+1) \hat{\cal B}_{[\mu_{1}|\underline{y}|}
\hat{C}^{(2n)}{}_{\mu_{2}\ldots\mu_{2n+1}]}\\
& & \\
& &
-2n (2n +1) \hat{\cal B}_{[\mu_{1}|\underline{y}|}
\hat{\jmath}_{\mu_{2}|\underline{y}|}
\hat{C}^{(2n)}{}_{\mu_{3}\ldots\mu_{2n+1}]\underline{y}}/
\hat{\jmath}_{\underline{y}\underline{y}}\, , \\
& & \\
\hat{C}^{(2n+1)}{}_{\mu_{1}\ldots\mu_{2n}\underline{x}} & = & 
\hat{C}^{(2n)}{}_{\mu_{1}\ldots\mu_{2n}} \\
& & \\
& &
+2n \hat{\jmath}_{[\mu_{1}|\underline{y}|}
\hat{C}^{(2n)}{}_{\mu_{2}\ldots\mu_{2n}]\underline{y}}/
\hat{\jmath}_{\underline{y}\underline{y}}\, .\\
\end{array}\\
\end{array}
$
\\
\\
\hline
\end{tabular}
\label{eq:reglasBuscherIIBtoIIA}
\end{equation}

%%%%%%%%%%%%%%%%%%%%%%%%%%%%%%%%%%%%%%%%%%%%%%%%%%%%%%%%%%%%%%%%%%%%%%
\subsubsection{Comentarios}

Podemos ahora ver c'omo reflejan estas reglas de dualidad~T las
relaciones generales que anunciamos al principio de esta secci'on.

Para empezar, dado que las componentes de la m'etrica en la
direcci'on compacta nos dan su medida (como en la
Ec.~(\ref{eq:radio11})), la regla

\begin{equation}
\hat{\jmath}_{\underline{y}\underline{y}}  =  
1/\hat{g}_{\underline{x}\underline{x}}\, ,  
\hspace{1cm}
\hat{g}_{\underline{x}\underline{x}} =
1/\hat{\jmath}_{\underline{y}\underline{y}}\, ,
\end{equation}

\noindent nos dice que los radios de estas dos teor'ias son
rec'iprocamente inversos. Si las coordenadas que parametrizan las
dimensiones compactas toman valores entre $0$ y $2\pi\ell_{s}$, entonces,
asint'oticamente

\begin{equation}
\label{eq:asintoticxx}
\hat{g}_{\underline{x}\underline{x}} \rightarrow 
\left(R_{A}/\ell_{s}  \right)^{2}\, ,
\hspace{1cm}
\hat{\jmath}_{\underline{y}\underline{y}} \rightarrow 
\left(R_{B}/\ell_{s}  \right)^{2}
\end{equation}

\noindent de donde deducimos 

\begin{equation}
\label{eq:dualradiiAB}
R_{A,B}=\ell_{s}^{2}/R_{B,A}\, .  
\end{equation}

Seguidamente vemos que el vector de Kaluza-Klein que viene de la
m'etrica de una teor'ia y se acopla a sus modos de Kaluza-Klein
se intercambia con el vector que viene de la 2-forma NSNS, que se
acopla a los modos de enrollamiento de las cuerdas.

Tambi'en vemos que las $(p+1)$-formas RR de una teor'ia que
tienen una componente en la dimensi'on compacta (lo que asociamos a
enrollamiento de la D$p$-brana correspondiente) se transforman en
$p$-formas RR de la otra, asociadas a D$(p-1)$-branas.

Finalmente observamos que los dilatones se transforman bajo
dualidad~T, lo que implica que las constantes de acoplamiento de las
teor'ias de cuerdas lo hacen. Utilizando
Ec.~(\ref{eq:asintoticxx}) es inmediato obtener

\begin{equation}
\label{eq:dualcouplingAB}
 g_{A,B}=g_{B,A}/R_{B,A}\, . 
\end{equation}

%%%%%%%%%%%%%%%%%%%%%%%%%%%%%%%%%%%%%%%%%%%%%%%%%%%%%%%%%%%%%%%%%%%%%%%
%\subsubsection{Truncaciones a SUGRAS $N=1,d=10$}?????

%%%%%%%%%%%%%%%%%%%%%%%%%%%%%%%%%%%%%%%%%%%%%%%%%%%%%%%%%%%%%%%%%%%%%%
\section{Consideraciones finales}

El orden que hemos escogido para presentar las acciones efectivas de
las teor'ias de cuerdas es quiz'a el m'as eficaz para entender
las relaciones de dualidad que existen entre ellas, pero no para
entender c'omo se obtienen estas acciones a partir de la definici'on
perturbativa de las teor'ias de cuerdas y cu'ales son sus
l'imites de validez.

La acciones efectivas aparecen de dos modos que en principio no tienen
nada que ver. El primer modo es el que corresponde a su definici'on:
se trata de una acci'on con la propiedad de reproducir, en un cierto
l'imite, las amplitudes de dispersi'on de los modos sin masa de
las teor'ias de cuerdas. Las amplitudes de dispersi'on se definen
como expansiones perturbativas con el dilat'on (la constante de
acoplo de la cuerda) como par'ametro peque~no. Estas amplitudes
dependen de la longitud de la cuerda $\ell_{s}$ y en el l'imite
$\ell_{s}\rightarrow 0$ (que podemos entender como el l'imite en
el que la cuerda se vuelve un punto) s'olo los modos sin masa
aparecen en ellas.  'Este l'imite coincide con el l'imite de
bajas energ'ias. La teor'ia de campos efectiva que describe
la din'amica de estos modos sin masa se define como un doble
desarrollo perturbativo en $\ell_{s}$ y en la constante de acoplo de
la cuerda $g$. Las acciones de SUEGRA $N=2A,2B, d=10$ son los
t'erminos m'as bajos en el desarrollo en $\ell_{s}$ pero contienen
contribuciones de dos tipos de diagramas de cuerdas distintos, y por
ello una parte de las mismas (el sector NSNS) tiene el factor global
$e^{-2\hat{\phi}}$ mientras que la otra (el sector RR) no contiene
ning'un factor con dilat'on.  

As'i, principio, estas acciones describen bien la teor'ia de
cuerdas correspondiente si

\begin{equation}
g <<1\, ,\hspace{1cm} \ell_{s} <<1\, .  
\end{equation}

La primera condici'on afecta a un par'ametro sin dimensiones,que es
lo correcto. La segunda no: hay que comparar $\ell_{s}$ con la escala
de longitud relevante\footnote{De otra forma, podr'iamos
  satisfacer esa condici'on con un simple cambio de unidades.} que no
puede ser otra que el radio de curvatura del espacio-tiempo $R_{c}$,
de forma que debemos de reescribir la condici'on as'i

\begin{equation}
\ell_{s}/R_{c}<<1\, .
\end{equation}

Ambas condiciones son en realidad locales: la primera sobre el
dilat'on y la segunda sobre los invariantes de curvatura del
espacio-tiempo, y pueden satisfacerse en unas regiones del
espacio-tiempo y violarse en otras. Si el dilat'on diverge, entonces
la teor'ia perturbativa de cuerdas no ser'ia v'alida
ah'i. Si el radio de curvatura es m'as peque~no que $\ell_{s}$,
el l'imite de teor'ia de campos $\ell_{s}\rightarrow 0$ deja
de ser v'alido ah'i y los efectos {\it cuerd'isticos}
empiezan a ser importantes.

La segunda forma de llegar a las anteriores acciones efectivas es
m'as sutil. Uno de los principios fundamentales de la Teor'ia de
Cuerdas es la invariancia conforme bidimensional, que es una propiedad
de la acci'on cl'asica de una cuerda propag'andose en el espacio de
Minkowski. Al cuantizar hay una anomal'ia que s'olo se anula si
el espacio tiene la dimensi'on cr'itica $d=10$ para supercuerdas.
En espacios m'as generales, con configuraciones no triviales de los
campos sin masa de la cuerda (m'etrica, 2-forma NSNS, dilat'on etc.)
la invariancia conforme se mantiene cu'anticamente si los campos
obedecen ciertas ecuaciones. Estas ecuaciones son precisamente las
mismas que se derivan de las acciones efectivas. La cuantizaci'on de
la cuerda se hace perturbativamente en $\ell_{s}$ con $\ell_{s}/R_{c}$
como par'ametro adimensional perturbativo, lo que nos lleva a la
segunda condici'on para que las ecuaciones garanticen la invariancia
cu'antica conforme al orden m'as bajo en $\ell_{s}$. Las condiciones
sobre el dilat'on est'an impl'icitas en la topolog'ia del
espacio bidimensional en el que definimos la acci'on de la cuerda.

Es realmente sorprendente que por caminos tan dispares se llegue a una
misma acci'on efectiva, lo que nos da confianza en la consistencia de
la Teor'ia de Cuerdas. Hay, adem'as, una tercera forma de llegar
a la misma acci'on en los casos supersim'etricos (que son los que
nos ocupan). Tiene que ver con la invariancia de la acci'on
bidimensional de la supercuerda en la formulaci'on de Green y Schwarz
bajo la {\it simetr'ia $\kappa$}, que s'olo se mantiene si los
campos obedecen las ecuaciones de la teor'ia de supergravedad
correspondiente. Esta misteriosa simetr'ia est'a presente en las
acciones de la mayor parte de los objetos extensos que vamos a
estudiar en la pr'oxima lecci'on e implica siempre las ecuaciones de
supergravedad (!`lo que incluye la ecuaci'on de Einstein!), mientras
que la invariancia conforme no lo est'a y no se sabe c'omo cuantizar
estas acciones.

\newpage
\chapter{Teor'ias efectivas de cuerdas: Soluciones}

%%%%%%%%%%%%%%%%%%%%%%%%%%%%%%%%%%%%%%%%%%%%%%%%%%%%%%%%%%%%%%%%%%%%%%
\section{Introducci'on}

En la 'ultima lecci'on hemos visto las acciones efectivas de las
teor'ias de cuerdas que m'as nos interesan. Tambi'en hemos
estudiado sus super'algebras y hemos anticipado qu'e
estados/soluciones esperamos en estas teor'ias. En esta lecci'on
nos vamos a dedicar a buscar y estudiar estos estados/soluciones.

Primeramente vamos a estudiar de forma gen'erica los objetos extensos
($p$-branas etc.), las acciones efectivas que dictan su din'amica,
sus masas y sus cargas, lo que nos ayudar'a a hacer una
clasificaci'on preliminar. Seguidamente, y usando los resultados
parciales de las lecciones anteriores, vamos a estudiar los objetos
extensos que de hecho aparecen en las teor'ias de cuerdas
10-dimensionales (y en SUGRA $N=1,d=11$, a la que nos referiremos sin
precisar demasiado como ``Teor'ia~M'') que nos interesan,
determinando sus masas y cargas, relaciones de dualidad etc.

A continuaci'on vamos a elaborar modelos gen'ericos de los que sacar
familias de soluciones de las que luego extraeremos las que
corresponden a los objetos que buscamos. Exhibiremos estas 'ultimas
con todo lujo de detalles y, cuando corresponda, daremos sus espinores
de Killing.

Una vez que tengamos las soluciones ``elementales'' podremos
combinarlas en soluciones m'as complejas en forma de intersecciones
({\it estados ligados ``en el umbral''}) o estados ligados.

Algunas referencias generales sobre objetos extensos gravitantes en RG
o en teor'ias de supergravedad y supercuerdas son
\cite{kn:Du4,kn:Tow2,kn:Tow3,kn:DKL,kn:CvSo,kn:Sc,kn:Tow,kn:Ste,kn:Ga2,kn:Y,kn:Ca3}

%%%%%%%%%%%%%%%%%%%%%%%%%%%%%%%%%%%%%%%%%%%%%%%%%%%%%%%%%%%%%%%%%%%%%%
\section{Objetos extensos: acciones, masas y cargas}
\label{sec-objetosextensos}

Un objeto extenso con $p$ dimensiones espaciales barre un {\it volumen
  del universo} que es un espacio-tiempo $(p+1)$-dimensional que se
puede describir intr'insecamente con coordenadas $\xi^{i}$,
$i=0,1,\ldots,p$. La posici'on de cada punto de este volumen en el
espacio-tiempo ambiente $d$-dimensional de coordenadas $x^{\mu}$ viene
dada por las $d$ funciones $X^{\mu}(\xi)$. La acci'on b'asica para
un objeto masivo de este tipo es la {\it acci'on de Nambu y Goto}
\cite{kn:Na,kn:Go}, que da el volumen del volumen del universo

\begin{equation}
S_{NG}^{(p)}[X^{\mu}(\xi)] = -T_{(p)} \int d^{p+1}\xi\ 
\sqrt{|g_{ij}|}\, ,
\end{equation}

\noindent donde 

\begin{equation}
g_{ij} =g_{\mu\nu}(X) \partial_{i}X^{\mu}\partial_{j}X^{\nu}\, ,
\end{equation}

\noindent es la m'etrica inducida sobre la $p$-brana y $|g_{ij}|$ 
denota al valor absoluto de su determinante. $T_{(p)}$ es la {\it
  tensi'on} de la $p$-brana y tiene dimensiones de masa por unidad de
volumen $p$-dimensional\footnote{De hecho, si compactificamos $p$
  dimensiones espaciales del espacio-tiempo y enrollamos en ellas la
  $p$-brana, congelando los grados de libertad que describen sus
  fluctuaciones, la acci'on de Nambu y Goto se transforma en la
  acci'on de una part'icula puntual con masa $T_{(p)}V_{(p)}$.},
o sea, $L^{p+1}$.  La acci'on anterior es una teor'ia de campos
para $d$ campos escalares, altamente no-lineal.  A veces es m'as
conveniente usar la {\it acci'on de Polyakov} \cite{kn:BDH} que
contiene una m'etrica auxiliar $\gamma_{ij}(\xi)$ definida sobre el
volumen del mundo:

\begin{equation}
S_{P}^{(p)}[X^{\mu},\gamma_{ij}] =  -{\textstyle\frac{T_{(p)}}{2}}
  \int d^{p+1}\xi
\sqrt{|\gamma|} \left[\gamma^{ij}\partial_{i}X^{\mu}\partial_{j}X^{\nu}
g_{\mu\nu} + (1-p) \right]\, .
\end{equation}

La ecuaci'on de $\gamma_{ij}(\xi)$ tiene la soluci'on

\begin{equation}
\gamma_{ij}=g_{ij}\, ,
\end{equation}

\noindent que, sustituida en la acci'on de Polyakov, nos da la
de Nambu y Goto. En el caso $p=1$ (cuerda) hay muchas soluciones,
todas de la forma

\begin{equation}
\gamma_{ij}=\Omega(\xi) g_{ij}\, ,
\end{equation}

\noindent debido a la invariancia conforme de la acci'on de Polyakov.
En otros contextos, la acci'on de Polyakov es tambi'en conocida como
{\it modelo sigma}.

Estas dos acciones son invariantes bajo reparametrizaciones del
volumen del mundo y covariantes (la m'etrica del espacio-tiempo
cambia) bajo reparametrizaciones del espacio-tiempo ambiente.

Estas dos acciones describen 'unicamente el acoplo de la $p$-brana a
la gravitaci'on del espacio-tiempo ambiente, pero hay dos tipos de
campos m'as del espacio ambiente a los que una $p$-brana se puede
acoplar: escalares, que juegan el papel de constantes de acoplo
locales (como el dilat'on) y $(p+1)$-formas diferenciales, a trav'es
de t'erminos de Wess-Zumino.  As'i, la acci'on de una $p$-brana
gen'erica con los acoplos m'as generales es

\begin{equation}
S_{NG}^{(p)}[X^{\mu}(\xi)] = -\frac{T_{(p)}}{K^{\alpha}_{0}} 
\int d^{p+1}\xi\ K(X)^{\alpha}
\sqrt{|g_{ij}|} -\frac{\mu}{K^{\alpha}_{0} (p+1)!} \int d^{p+1}\xi\ 
\epsilon^{i_{1}\cdots i_{p+1}} A_{(p+1)\, i_{1}\cdots i_{p+1}}\, ,
\end{equation}

\noindent donde $K(X)$ es un cierto escalar, $a$ un cierto n'umero real

\begin{equation}
A_{(p+1)\, i_{1}\cdots i_{p+1}} = A_{(p+1)\, \mu_{1}\cdots \mu_{p+1}}(X)
\partial_{i_{1}}X^{\mu_{1}}\ldots\partial_{i_{p+1}}X^{\mu_{p+1}}\, ,
\end{equation}

\noindent y $\mu$ la carga de la $p$-brana con respecto a este 
campo\footnote{Cuando las intensidades de campo de los potenciales
  contienen t'erminos de Chern-Simons, como es el caso en las SUGRAS
  que nos ocupan, se pueden definir otros dos tipos de carga distintos
  de la {\it carga de la brana} que acabamos de definir \cite{kn:Mar}.
  La conservaci'on ( o no conservaci'on) de estas tres cargas est'a
  relacionada con la posibilidad de que las branas acaben en otras
  \cite{kn:Tow7}.}, que suele ser igual a la tensi'on en los casos
que nos ocupan.  Hemos introducido el factor $K^{-\alpha}_{0}$, donde
$K_{0}$ es el valor asint'otico del escalar $K$, de forma que
$T_{(p)}$ sigue siendo la tensi'on f'isica de la $p$-brana. El
t'ermino a~nadido tiene las anteriores invariancias y, adem'as, es
invariante bajo transformaciones gauge de la $(p+1)$-forma

\begin{equation}
\delta A_{(p+1)} = (p+1)\partial \Lambda_{(p)}\, .  
\end{equation}

No hay m'as acoplos posibles, pero s'i dos generalizaciones de
esta acci'on: la introducci'on de m'as campos que viven en el
volumen del mundo, adem'as de los escalares $X^{\mu}(\xi)$, y la
eliminaci'on de ciertos grados de libertad a trav'es del ``gaugeo''
de simetr'ias globales de la acci'on asociadas a isometr'ias
de la m'etrica del espacio-ambiente. De esta 'ultima
generalizaci'on no vamos a decir m'as. De la primera, hay que decir
que el campo de volumen del mundo que m'as a menudo aparece es el
{\it campo vectorial de Born e Infeld} (BI) $V_{i}$, y lo hace de una
forma muy caracter'istica (acci'on de Born e Infeld):

\begin{equation}
S_{NG}^{(p)}[X^{\mu}(\xi)] = -\frac{T_{(p)}}{K^{\alpha}_{0}} 
\int d^{p+1}\xi\ K^{\alpha}(X)
\sqrt{|g_{ij}+F_{ij}|}+ \ldots\, ,
\hspace{1cm}
F_{ij}=2\partial_{[i}V_{j]}\, .
\end{equation}

Utilizando la m'etrica inducida $g_{ij}$ para subir y bajar
'indices, podemos expandir la acci'on de Born e Infeld as'i:

\begin{equation}
S_{NG}^{(p)}[X^{\mu}(\xi)] = -\frac{T_{(p)}}{K^{\alpha}_{0}} 
\int d^{p+1}\xi\ K^{\alpha}(X)
\sqrt{|g_{ij}|}\{1 -{\textstyle\frac{1}{2}}{\cal F}^{2}+\ldots\} \, ,
\end{equation}

\noindent en donde se ve un t'ermino cin'etico  est'andar 
(cuadr'atico) para un vector abeliano con correcciones de orden
superior que le dan un car'acter no-lineal.

?`Cu'al es la acci'on de los campos del espacio-tiempo ambiente a
los que se acopla la $p$-brana? Suele ser de esta forma:

\begin{equation}
\label{eq:acciongenerica}
S = {\textstyle\frac{1}{16\pi G^{(d)}_{N}}}\int d^{d}x\, \sqrt{|g|}\, 
[R + 2(\partial\log K)^{2} +{\textstyle\frac{(-1)^{p+1}}{2\cdot (p+2)!}}
K^{\beta}F^{2}_{(p+2)}]\, ,  
\end{equation}

\noindent donde 

\begin{equation}
F_{(p+2)}= (p+2)\partial A_{(p+1)}\, ,   
\end{equation}

\noindent es la intensidad de campo de $A_{(p+1)}$, invariante gauge.
Esta acci'on gen'erica suele corresponder a una truncaci'on de una
teor'ia de SUGRA como las que hemos visto en la lecci'on
anterior.  La consistencia del acoplo de la $p$-brana a los campos de
una teor'ia de SUGRA se consigue s'olo si la acci'on de la
$p$-brana es invariante bajo la simetr'ia $\kappa$, que juega un
papel crucial. Por otro lado, la simetr'ia $\kappa$ exige que los
campos satisfagan las ecuaciones de movimiento de la SUGRA, cerrando
el c'irculo de relaciones. 

%%%%%%%%%%%%%%%%%%%%%%%%%%%%%%%%%%%%%%%%%%%%%%%%%%%%%%%%%%%%%%%%%%%%%%
\subsection{Los objetos extensos de las teor'ias de cuerdas tipo~II:
 acciones efectivas y masas}

El acoplo de la $p$-brana al escalar $K$ depende evidentemente del
sistema de referencia conforme en el que estemos trabajando. Siempre
hay un sistema de referencia conforme en el que $\alpha=0$: {\it el
  sistema de referencia conforme de la $p$-brana}, en el que 'esta es
``fundamental'' (explicaremos esto m'as tarde). Fijado este sistema
de referencia conforme, podemos clasificar las dem'as $p$-branas con
respecto a 'esta por c'omo se acoplan a $K$. En Teor'ia de
Cuerdas, el sistema de referencia can'onico es el de la cuerda, y el
escalar es el dilat'on, la ``constante'' de acoplo de la cuerda.  En
el sistema de referencia conforme de la cuerda, las cuerdas son los
objetos fundamentales, elementales, que dan lugar a los estados
perturbativos de la teor'ia, y todos los objetos cuyas acciones en
este sistema de referencia no se acoplen al dilat'on (con tensiones
independientes de $g$), son tambi'en {\it $p$-branas fundamentales}.
Pr'acticamente la 'unica es la

%%%%%%%%%%%%%%%%%%%%%%%%%%%%%%%%%%%%%%%%%%%%%%%%%%%%%%%%%%%%%%%%%%%%%%
\subsubsection{Cuerda fundamental}

'Esta se acopla de forma natural a la 2-forma NSNS as'i:

\begin{equation}
S = -T\int d^{2}\xi\ \sqrt{|\hat{g}_{ij}|}
-{\textstyle\frac{T}{2}}\int d^{2}\xi\ \epsilon^{ij} \hat{B}_{ij} \, .
\end{equation}

\noindent La tensi'on $T$ de la cuerda es $T=1/\alpha^{\prime}$, donde
$\alpha^{\prime}=\ell_{s}^{2}$ es la {\it pendiente de Regge} y
$\ell_{s}$ es la {\it longitud de la cuerda}.

Los objetos no-perturbativos de una teor'ia de campos est'andar
tienen masas proporcionales a $g^{-2}$. Por esta raz'on, objetos con
acciones de la forma

\begin{equation}
S^{(p)} = -T_{(p)}e^{2\phi_{0}}\int d^{p+1}\xi\ e^{-2\phi} \sqrt{|g_{ij}|}
+\ldots\, ,
\end{equation}

\noindent son {\it $p$-branas solit'onicas}. Las principales en las 
teor'ias tipo~II en $d=10$ son las dos

%%%%%%%%%%%%%%%%%%%%%%%%%%%%%%%%%%%%%%%%%%%%%%%%%%%%%%%%%%%%%%%%%%%%%%
\subsubsection{5-branas solit'onicas}

($S5A$ y $S5B$), que se acoplan a las 6-formas de NSNS $B^{(6)}$ que
los las duales de Hodge de las 2-formas a las que se acoplan las
cuerdas fundamentales:

\begin{equation}
S = -T_{S5}e^{2\hat{\phi}_{0}}\int d^{6}\xi\ 
e^{-2\hat{\phi}} \sqrt{|\hat{g}_{ij}|}
-{\textstyle\frac{T_{S5} e^{2\hat{\phi}_{0}}}{6!}}\int d^{6}\xi\ 
\epsilon^{i_{1}\cdots i_{6}} \hat{B}^{(6)}_{i_{1}\cdots i_{6}}\, .
\end{equation}

En las teor'ias de cuerdas tipo~II, hay adem'as otros objetos que
est'an a medio camino entre los perturbativos y los no-perturbativos:
las\footnote{Hay bastantes referencias sobre las D$p$-branas. Adem'as
  del resumen de Johnson \cite{kn:J} son interesantes las lecciones de
  Bachas \cite{kn:Ba,kn:Ba2}, muy pedag'ogicas, Douglas
  \cite{kn:Dou}, el propio Polchinski \cite{kn:P2}, Vancea
  \cite{kn:Van} y F\"orste \cite{kn:For}.}

%%%%%%%%%%%%%%%%%%%%%%%%%%%%%%%%%%%%%%%%%%%%%%%%%%%%%%%%%%%%%%%%%%%%%%
\subsubsection{D$p$-branas}

($p$ par en la~IIA, $p$~impar en la~IIB), con tensiones proporcionales
a $g^{-1}$ y que se acoplan a las $(p+1)$-formas RR y a la 2-forma
NSNS:

\begin{equation}
S = -T_{Dp}e^{\hat{\phi}_{0}}\int d^{p+1}\xi\ 
e^{-\hat{\phi}} \sqrt{|\hat{g}_{ij}+2\pi\alpha^{\prime} {\cal F}_{ij}|}
-{\textstyle\frac{T_{SDp} e^{\hat{\phi}_{0}}}{6!}}\int d^{6}\xi\ 
\epsilon^{i_{1}\cdots i_{p+1}} \hat{C}^{(p+1)}_{i_{1}\cdots i_{p+1}}\, ,
\end{equation}

\noindent donde $V_{i}$ es el vector de Born e Infeld (BI) con intensidad 
de campo

\begin{equation}
{\cal F}_{ij}=F_{ij}+{\textstyle\frac{1}{2\pi\alpha^{\prime}}}\hat{B}_{ij}\, ,
\hspace{1cm}
F_{ij} = 2\partial_{[i}V_{j]}\, .
\end{equation}

La existencia de D$p$-branas en la Teor'ia de Cuerdas y sus
propiedades fundamentales fueron descubiertas por Polchinski en
Ref.~\cite{kn:P}.  Representan hipersuperficies sobre las que los
extremos de las cuerdas abiertas pueden moverse sin salir de ellas.
Estos extremos parecen puntos (desde el punto de vista del volumen del
universo de la brana) el'ectricamente cargados con respecto al vector
de BI $V_{i}$.

La din'amica de las cuerdas abiertas ligadas a las D$p$-branas
est'a, pues, descrita por el campo de BI, cuya existencia es lo que
las hace interesantes: si hay una 'unica D$p$-brana, s'olo puede
haber cuerdas abiertas con los dos extremos sobre la misma. En el
espectro de la cuerda con estas condiciones de contorno, hay un modo
sin masa asociado que es precisamente el vector de BI. Si hay $N$
D$p$-branas paralelas\footnote{Que esto es posible se debe a que son
  objetos supersim'etricos que pueden estar en equilibrio como los
  agujeros negros de Reissner y Nordstr\"om extremos \cite{kn:P}.},
adem'as de los $N$ vectores de BI asociados a cuerdas con los dos
extremos en la misma brana, habr'a cuerdas abiertas con extremos en
branas distintas. En general, el modo m'as ligero de 'estas tiene
una masa proporcional a la separaci'on entre ambas D$p$-branas y se
hace cero si las branas se superponen. As'i, si las $N$ branas se
superponen, hay $N^{2}$ vectores de BI sin masa que corresponden a la
representaci'on adjunta de $U(N)$. Este mecanismo nos permite tener
teor'ias de (super-) Yang-Mills $U(N)$ en las $p+1$ dimensiones
del volumen del mundo de las D$p$-branas.

No se conoce una acci'on que describa la din'amica de $N$
D$p$-branas superpuestas, aunque su expansi'on es, al orden m'as
bajo en $\alpha^{\prime}$, una teor'ia de super-Yang-Mills en
$p+1$ dimensiones. Una referencia reciente sobre el status de la
generalizaci'on no-abeliana de la acci'on de Born e Infeld es
\cite{kn:KS}.

La presencia del vector de BI y su interpretaci'on sugieren la
posible presencia de otros campos en los vol'umenes del mundo de los
objetos extensos y su asociaci'on a los extremos o fronteras que
otros objetos puedan tener en los primeros. Que esta interpretaci'on
es correcta fue explicado por Townsend en \cite{kn:Tow7}. En cuanto a
la presencia de otros campos, los m'as sencillos son los duales
el'ectrico-magn'eticos de los BI en las D$p$-branas: $(p-2)$-formas
diferenciales que viven en los vol'umenes del mundo de las
D$p$-branas y est'an asociadas, como veremos, a intersecciones
$(p-3)$-dimensionales con otros objetos.  A trav'es de dualidades es
posible ver que el los vol'umenes del mundo de las 5-branas
solit'onicas hay otras formas diferenciales con interpretaciones
an'alogas: un vector en la S5B, que representa la intersecci'on de
D1-branas con la 5-brana (por dualidad~S con la intersecci'on de
cuerdas fundamentales con la D5-brana) y una 2-forma cuya intensidad
de campo es autodual en la S5A.  La autodualidad conlleva el problema
de la ausencia de acci'on covariante que se puede resolver por el
m'etodo utilizado en el caso de SUEGRA $N=2B,d=10$ \cite{kn:BdRO} o
por el m'etodo alternativo de introducir campos auxiliares
\cite{kn:BLNPST}.

%%%%%%%%%%%%%%%%%%%%%%%%%%%%%%%%%%%%%%%%%%%%%%%%%%%%%%%%%%%%%%%%%%%%%%
\subsection{Relaciones de dualidad y masas}

\begin{figure}[!ht]
\begin{center}
  \leavevmode \epsfxsize= 13cm \epsfysize= 17cm
  \epsffile{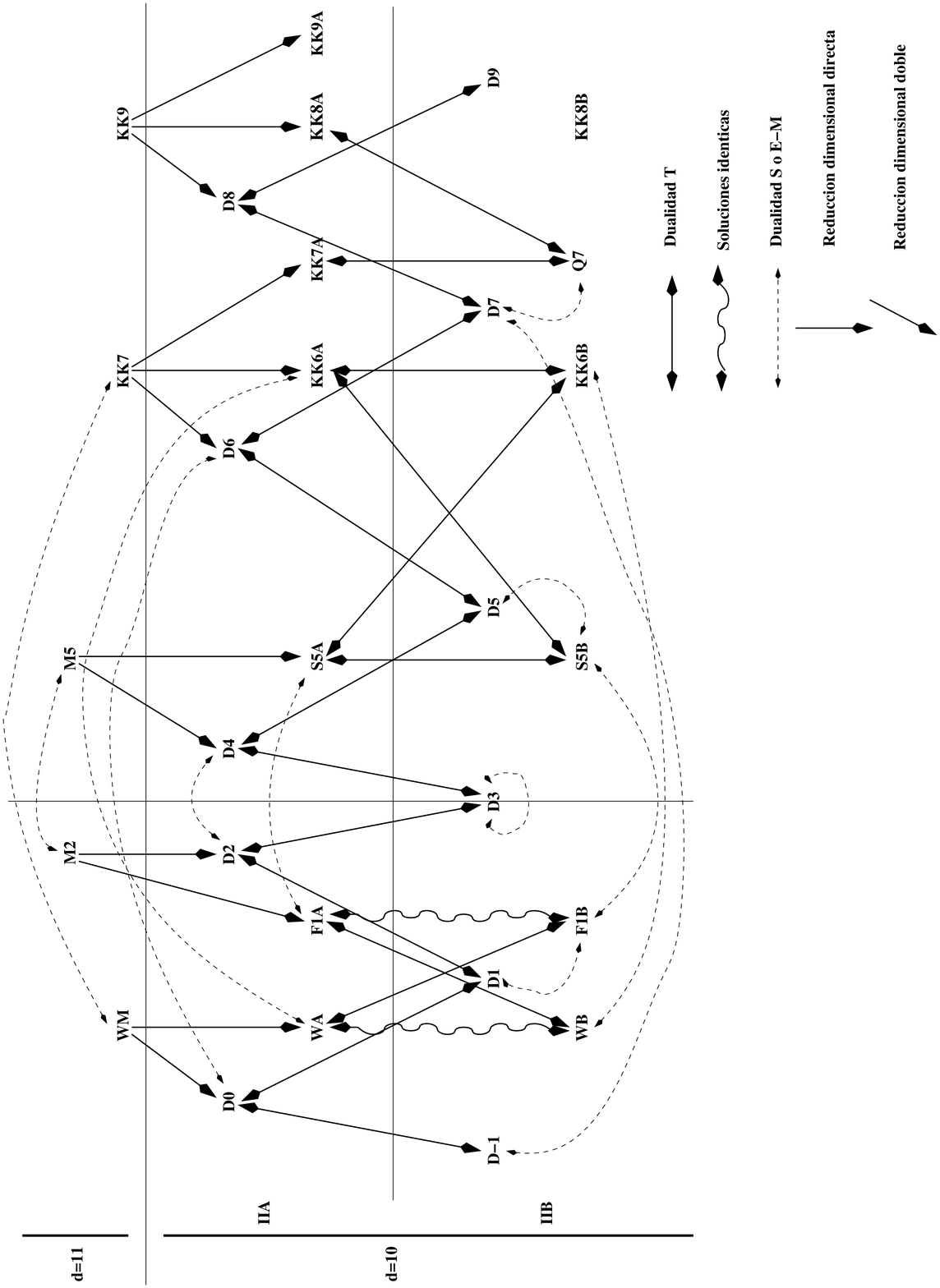}
\caption{\footnotesize Relaciones de dualidad entre soluciones 
  cl'asicas de las SUGRAS en $d=10,11$: cuerdas fundamentales $F1$,
  5-branas solit'onicas $S5$, M$p$-branas, D$p$-branas, ondas
  gravitacionales $W$ y monopolo de Kaluza-Klein $KK$. Las l'ineas
  con dos flechas se~nalan dualidad~T y las discontinuas,
  dualidad~S. Las l'ineas con una sola flecha indican relaciones
  por compactificaci'on en c'irculos: vertical (el objeto no
  est'a enrollado en la dimensi'on compacta) o diagonal (s'i lo
  est'a.\normalsize}
\label{fig:dualbran}
\end{center}
\end{figure}

Vamos ahora a explorar las relaciones de dualidad que hay entre estos
objetos (y alg'un otro que aparecer'a por el camino) que ya hemos
discutido anteriormente, representadas en el diagrama de la
Figura~\ref{fig:dualbran}.  

Las relaciones de dualidad entre los objetos extensos se reflejan en
sus acciones efectivas, aunque no es sencillo verlo en el espacio y
tiempo de que disponemos. Por ello, vamos a explorar las relaciones de
dualidad de las masas de estos objetos cuando est'an enrollados en
$p$-toros. Las reglas fundamentales de este juego vienen dadas por las
Ecs.~(\ref{eq:dualradiiAB}) y (\ref{eq:dualcouplingAB}), que
reescribimos aqu'i por comodidad

\begin{eqnarray}
\label{eq:tdual}
\begin{tabular}{|c|}
\hline \\
$
\begin{array}{rcl}
R_{A,B} & = & \ell_{s}^{2}/R_{B,A}\, , \\
& & \\
g_{A,B} & = & g_{B} \ell_{s}/R_{B,A}\, .\\
\end{array}
$
\\ \\ \hline
\end{tabular}
\end{eqnarray}

\noindent  para dualidad~T, y las Ecs.~(\ref{eq:dualcouplingB}) y 
(\ref{eq:dualradiiB}) que tambi'en reescribimos aqu'i por
comodidad

\begin{equation}
\label{eq:sdual}
\begin{tabular}{|c|}
\hline \\
$
\begin{array}{rcl}
g^{\prime} & = & 1/g\, , \\
& & \\
R_{i}^{\prime} & = & R_{i}/\sqrt{g}\, ,\\
\end{array}
$
\\ \\ \hline
\end{tabular}
\end{equation}

\noindent para la dualidad~S en la tipo~IIB. Estas reglas han de ser 
complementadas por la reglas de transformaci'on de las masas medidas
en el sistema de referencia conforme de Einstein ``modificado'' (es
decir, el correcto):

\begin{equation}
\label{eq:sdual2}
\begin{tabular}{|c|}
\hline \\
$
M^{\prime} =  g^{1/2} M\, .
$
\\ \\ \hline
\end{tabular}
\end{equation}

El punto de partida, es la masa de una cuerda fundamental enrollada
sobre una dimensi'on compacta $x^{9}$ de radio $R_{9}$. Este dato
hemos de tomarlo de la cuantizaci'on de la Teor'ia de Cuerdas:

\begin{equation}
M_{F1}= \frac{R_{9}}{\ell_{s}^{2}}\, . 
\end{equation}

Si 'esta es la cuerda fundamental de la teor'ia~IIB, entonces
est'a relacionada por dualidad~S con la D-cuerda (D1-brana).
Utilizando las reglas Ecs.~(\ref{eq:sdual}) y (\ref{eq:sdual2}),
tenemos

\begin{equation}
M_{D1}=M^{\prime}_{F1}= g^{1/2}M_{F1}
= g^{1/2} \frac{R_{9}}{\ell_{s}^{2}} 
=\frac{R^{\prime}_{9}}{g^{\prime}\ell_{s}^{2}}\, .
\end{equation}

Ahora podemos hacer una transformaci'on de dualidad~T en la
direcci'on $x^{9}$, con lo que se obtiene una D0-brana de la~IIA de masa 

\begin{equation}
M_{D0}=M^{\prime}_{D1}= \frac{R_{9}}{g\ell_{s}^{2}}=
\frac{\ell_{s}^{2}/R^{\prime}_{9}}
{g^{\prime} \ell_{s}/R^{\prime}_{9} \ell_{s}^{2}}
=\frac{1}{g^{\prime}\ell_{s}}\, .
\end{equation}
 
Si hacemos dualidad~T en una direcci'on ortogonal a la D-cuerda
($x^{8}$), tenemos la D-membrana (D2-brana) de la teor'ia~IIB:

\begin{equation}
M_{D0}=M^{\prime}_{D1}= \frac{R_{9}}{g\ell_{s}^{2}}=
\frac{R_{9}}{g^{\prime} \ell_{s}/R^{\prime}_{8} \ell_{s}^{2}}
=\frac{R_{8}R_{9}}{g^{\prime}\ell^{3}_{s}}\, .
\end{equation}

Repitiendo suficientes veces este procedimiento, obtenemos la masa de
una D$p$-brana enrollada en un $p$-toro (quitando las primas):

\begin{equation}
\label{eq:mdp}
M_{Dp} = \frac{R_{10-p}\ldots R_{9}}{g\ell_{s}^{p+1}}\, .
\end{equation}

La D5-brana est'a relacionada por dualidad~S con la 5-brana
solit'onica de la~IIB:

\begin{equation}
M_{S5}=g^{1/2} M_{D5}^{\prime}=g^{1/2}\frac{R_{5}\ldots R_{9}}{g\ell_{s}^{6}}=
g^{\prime -1/2}\frac{R^{\prime}_{5}/g^{\prime 1/2}\ldots 
R^{\prime}_{9}/g^{\prime 1/2}}{g^{\prime -1}\ell_{s}^{6}}=
\frac{R_{5}\ldots R_{9}}{g^{2}\ell_{s}^{6}}\, ,
\end{equation}

\noindent lo que confirma sus car'acter solit'onico. Los resultados 
para 'estos y otros objetos extensos de las teor'ias tipo~II
est'an reunidos en las Tablas~\ref{tab-IIAmasses}
y~\ref{tab-IIBmasses}.

Las tensiones que aparecen en las acciones de estos objetos se pueden
calcular simplemente dividiendo las masas por el volumen espacial en
el que est'an compactificadas (el producto de las
longitudes de los c'irculos).

\begin{table}
\footnotesize
\begin{center}
\begin{tabular}{||c||c|c||c||}
\hline\hline
& & & \\
Objeto IIA  & Masa en const. $d=10$  &  Masa en cons. $d=11$  & Objeto $d11$  \\
\hline\hline
& & & \\
F1m & $R_{9}^{-1}$ &  & \\
& & & \\
\hline
& & & \\
D0 & $g_{A}^{-1}\ell_{s}^{-1}$ & $R_{10}^{-1}$ & WM$(+,-^{10})$ \\
& & & \\
\hline
& & & \\
F1w & $R_{9}\ell_{s}^{-2}$ & 
$R_{10} R_{9}({}^{-}\!\!\!\!\ell_{\rm Planck}^{(11)})^{-3}$ & 
M2$(+,-^{8},+^{2})$ \\
& & & \\
\hline
& & & \\
D2 & $R_{9}R_{8}g^{-1}_{A}\ell_{s}^{-3}$ & 
$R_{9}R_{8}({}^{-}\!\!\!\!\ell_{\rm Planck}^{(11)})^{-3}$ &
M2$(+,-^{7},+^{2},-)$\\
& & & \\
\hline
& & & \\
D4 & $R_{9}\ldots R_{6}g^{-1}_{A}\ell_{s}^{-5}$ & 
$R_{10} R_{9}\ldots R_{5}({}^{-}\!\!\!\!\ell_{\rm Planck}^{(11)})^{-6}$ & 
M5$(+,-^{5},+^{5})$ \\
& & & \\
\hline
& & & \\
S5A & $R_{9}\ldots R_{5}g^{-2}_{A}\ell_{s}^{-6}$ & 
$R_{9}\ldots R_{5}({}^{-}\!\!\!\!\ell_{\rm Planck}^{(11)})^{-6}$ & 
M5$(+,-^{4},+^{5},-)$\\
& & & \\
\hline
& & & \\
D6 & $R_{9}\ldots R_{4}g^{-1}_{A}\ell_{s}^{-7}$ & 
$R_{10}^{2}R_{9}\ldots R_{4}({}^{-}\!\!\!\!\ell_{\rm Planck}^{(11)})^{-9}$ & 
KK7M$(+,-^{3},+^{6},-^{\star})$ \\
& & & \\
\hline
& & & \\
KK6A & $R_{9}^{2}R_{8}\ldots R_{4}g_{A}^{-2}\ell_{s}^{-8}$ & 
$R_{10}R_{9}^{2}\ldots R_{4}({}^{-}\!\!\!\!\ell_{\rm Planck}^{(11)})^{-9}$ & 
KK7M$(+,-^{3},+^{5},+^{\star},+)$ \\
& & & \\
\hline
& & & \\
D8 & $R_{9}\ldots R_{2}g^{-1}_{A}\ell_{s}^{-9}$ & 
$R_{10}^{3}R_{9}\ldots R_{4}({}^{-}\!\!\!\!\ell_{\rm Planck}^{(11)})^{-12}$ & 
KK9M$(+,-,+^{8},+^{\star})$ \\
& & & \\
\hline
& & & \\
KK8A & $R_{9}^{3}R_{8}\ldots R_{2}g_{A}^{-3}\ell_{s}^{-11}$ & 
$R_{10}R_{9}^{3}R_{8}\ldots R_{2}
({}^{-}\!\!\!\!\ell_{\rm Planck}^{(11)})^{-12}$ & 
KK9M$(+,-,+^{7},+^{\star},+)$ \\
& & & \\
KK9A & $R_{9}^{3}R_{8}\ldots R_{1}g_{A}^{-4}\ell_{s}^{-12}$ & 
$R_{10}R_{9}^{3}R_{8}\ldots R_{1}
({}^{-}\!\!\!\!\ell_{\rm Planck}^{(11)})^{-12}$ & 
KK9M$(+,+^{8},+^{\star},-)$ \\
& & & \\
\hline\hline
\end{tabular}
\end{center}
\caption[Tabla de masas de los objetos extensos de la Teor'ia de 
Cuerdas tipo~IIA.]
{\small
En esta tabla est'an las masas de los objetos 
extensos de la Teor'ia de Cuerdas tipo~IIA en lenguaje 
10-dimensional (radios de compactificaci'on $R_{i}$, 
constante de acoplo $g_{A}$ 
y longitud de la cuerda $\ell_{s}$) y en lenguaje  11-dimensional 
(de SUGRA $N=1,d=11$)
(radios de compactificaci'on $R_{i}$, constante de Planck reducida 
11-dimensional  ${}^{-}\!\!\!\!\ell_{\rm
  Planck}^{(11)}=\ell_{\rm Planck}^{(11)}/2\pi$). La coordenada que se 
compactifica para relacionar las teor'ias 10- y 11-dimensionales
es, por convenio, $x^{10}$ de forma que el ``radio 11-dimensional''
es $R_{10}= g_{A}\ell_{s} =g_{A}^{2/3}
{}^{-}\!\!\!\!\ell_{\rm Planck}^{(11)}$. Adem'as, las configuraciones 
de los objetos 11-dimensionales que dan lugar a los 
10-dimensionales que dan lugar a los 10-dimensionales se dan en
la siguiente notaci'on notaci'on: la lista de signos 
representa las 11 coordenadas, de $\hat{\hat{x}}{}^{0}$ hasta
$\hat{\hat{x}}{}^{10}$. Un signo $+$ significa que una de las dimensiones
del volumen del mundo ocupa esa direcci'on espacio-temporal. Una estrella
significa que en esa direcci'on el objeto tiene una isometr'ia
especial y la direcci'on correspondiente no se puede
descompactificar.
}
\label{tab-IIAmasses}
\end{table}

\begin{table}
\begin{center}
\begin{tabular}{||c||c||c||c||c||}
\hline\hline
& & & & \\
Objeto IIB  & Masa & \hspace{1cm} & Objeto IIB & Masa\\
\hline\hline
& & & & \\
F1m & $R_{9}^{-1}$ & & KK6A & 
$R_{9}^{2}R_{8}\ldots R_{4}g_{B}^{-2}\ell_{s}^{-8}$ \\
& & & & \\
\hline
& & & & \\
F1w & $R_{9}\ell_{s}^{-2}$ & & D7 &  
$R_{9}\ldots R_{3}g^{-1}_{B}\ell_{s}^{-8}$ \\
& & & & \\
\hline
& & & & \\
D1 & $R_{9}g^{-1}_{B}\ell_{s}^{-2}$ & &  Q7 &
$R_{9}\ldots R_{3}g_{B}^{-3}\ell_{s}^{-8}$ \\
& & & & \\
\hline
& & & & \\
D3 & $R_{9}\ldots R_{7}g^{-1}_{B}\ell_{s}^{-4}$ & &  D9 &  
$R_{9}\ldots R_{1}g^{-1}_{B}\ell_{s}^{-10}$ \\
& & & & \\
\hline
& & & & \\
D5 & $R_{9}\ldots R_{5}g^{-1}_{B}\ell_{s}^{-6}$ & & Q9 &  
$R_{9}\ldots R_{1}g_{B}^{-4}\ell_{s}^{-10}$ \\
& & & & \\
\hline
& & & & \\
S5B &  $R_{9}\ldots R_{5}g_{B}^{-2}\ell_{s}^{-6}$ & & & \\
& & & & \\
\hline\hline
\end{tabular}
\end{center}
\caption[Tabla de masas de los objetos extensos de la Teor'ia 
de Cuerdas tipo~IIB]
{
En esta tabla est'an las masas de los objetos 
extensos de la Teor'ia de Cuerdas tipo~IIB en t'erminos de los 
radios de compactificaci'on $R_{i}$, 
la constante de acoplo de la cuerda $g_{B}$ 
y longitud de la cuerda $\ell_{s}$. Cuando una masa
depende de un  radio con una potencia superior a la unidad,  el
objeto (tipo monopolo de Kaluza-Klein) tiene una isometr'ia 
especial en esa
direcci'on.}
\label{tab-IIBmasses}
\end{table}

%%%%%%%%%%%%%%%%%%%%%%%%%%%%%%%%%%%%%%%%%%%%%%%%%%%%%%%%%%%%%%%%%%%%%%
\subsection{Los objetos extensos de SUGRA $N=1,d=11$}

Hemos dicho repetidas veces que el l'imite de acoplo fuerte de la
Teor'ia de Cuerdas tipo~IIA es SUGRA $N=1,d=11$ y que los objetos
extensos de esa teor'ia de cuerdas provienen de objetos extensos
11-dimensionales. Adem'as de la onda gravitacional, que no es
propiamente un objeto extenso y que est'a igualmente presente en
$d=10$, los objetos extensos de SUGRA $N=1,d=11$ son la M2-brana y la
M5-brana, que se acoplan respectivamente a la 3-forma $\hat{\hat{C}}$
y a su dual la 6-forma $\tilde{\hat{\hat{C}}}$. Cuando una de las
direcciones es compacta, se puede a~nadir a esta lista el monopolo de
Kaluza-Klein (KK7M), que es una soluci'on puramente gravitacional y
por lo tanto presente, como la onda gravitacional, en todas las
dimensiones.

En SUGRA $N=1,d=11$ no hay ning'un campo escalar que pueda
interpretarse como una constante de acoplo y, as'i, la M2 y la M5
tienen un car'acter fundamental. La acci'on de la M2 es 

\begin{equation}
S = -T_{M2}\int d^{3}\xi\ 
\sqrt{|\hat{\hat{g}}_{ij}|}
-{\textstyle\frac{T_{M2}}{3!}}\int d^{3}\xi\ 
\epsilon^{i_{1}\cdots i_{3}} \hat{\hat{C}}_{i_{1}\cdots i_{3}}\, ,
\end{equation}

\noindent y no contiene ning'un otro campo en su volumen del mundo
aparte de los escalares $\hat{\hat{X}}^{\hat{\hat{\mu}}}$. En la
reducci'on dimensional para obtener la D2-brana, el escalar del
volumen del mundo $\hat{\hat{X}}^{10}$ deja de tener significado
espacio-temporal y se puede dualizar en el vector de BI de la D2-brana
\cite{kn:Town8}.

La acci'on de la M5, sin embargo, contiene otros campos: una 2-forma
cuya intensidad de campo es una 3-forma autodual. Por reducci'on
dimensional, esta forma da lugar a la 2-forma con intensidad de campo
autodual de la S5A o a la 2-forma que es el dual del vector de BI de
la D4-brana. 

Para determinar las masas y las tensiones de la M2 y la M5, vamos a
relacionarlas con las de los objetos de la tipo~IIA, utilizando las
relaciones entre las constantes 10- y 11-dimensionales
Ecs.~(\ref{eq:10-11constants}), que reescribimos aqu'i invertidas
por comodidad, llamando $R_{10}$ a lo que all'i llamamos $R_{11}$:

\begin{equation}
\begin{tabular}{|c|}
\hline \\
$
\begin{array}{rcl}
\ell_{s} & = & {}^{-}\!\!\!\!\ell_{\rm  Planck}^{(11)}/R_{10}^{1/2}\, ,\\
& & \\
g_{A} & = & R_{10}^{3/2}/{}^{-}\!\!\!\!\ell_{\rm  Planck}^{(11)}\, ,\\
\end{array}
$
\\ \\ \hline
\end{tabular}
\end{equation}

Podemos empezar con la masa de la cuerda fundamental F1A, que no es
sino la M2 compactificada en el c'irculo 11-dimensional:

\begin{equation}
\label{eq:MM2}
M_{M2}=
M_{F1A}= \frac{R_{9}}{\ell_{s}^{2}}=  
\frac{R_{9}R_{10}}{({}^{-}\!\!\!\!\ell_{\rm
  Planck}^{(11)})^{3}}\, . 
\end{equation}

Cuando la M2 no est'a enrollada en la und'ecima dimensi'on (sino,
por ejemplo, en las $x^{8},x^{9}$) tenemos la D2. Comprob'emoslo:

\begin{equation}
M_{M2} =  \frac{R_{8}R_{9}}{({}^{-}\!\!\!\!\ell_{\rm
  Planck}^{(11)})^{3}} = \frac{R_{8}R_{9}}{g_{A}\ell_{s}^{3}}\, .
\end{equation}

La D4 no es m'as que una M5 enrollada en la und'ecima dimensi'on:

\begin{equation}
M_{M5}=
M_{D4}= \frac{R_{6}\ldots R_{9}}{g_{A}\ell_{s}^{5}}=  
\frac{R_{6}\ldots R_{10}}{({}^{-}\!\!\!\!\ell_{\rm
  Planck}^{(11)})^{6}}\, , 
\end{equation}

\noindent y la S5A una M5 no enrollada ah'i:

\begin{equation}
M_{M5}= \frac{R_{5}\ldots R_{9}}{({}^{-}\!\!\!\!\ell_{\rm
  Planck}^{(11)})^{6}}=\frac{R_{5}\ldots R_{9}}{g_{A}^{2} \ell_{s}^{6}}\, . 
\end{equation}

Para finalizar, podemos ver que la D0-brana no es m'as que un modo de
Kaluza-Klein del gravit'on movi'endose en la und'ecima dimensi'on:

\begin{equation}
M_{D0} = \frac{1}{g_{A}\ell_{s}}=\frac{1}{R_{10}}\, .
\end{equation}

\begin{table}
\begin{center}
\begin{tabular}{||c||c||}
\hline\hline
& \\
Objeto & Masa \\
\hline\hline
& \\
WM & 0 \\
& \\
\hline
& \\
M2 & $R_{10}R_{9}({}^{-}\!\!\!\!\ell_{\rm Planck}^{(11)})^{-3}$ \\
& \\
\hline
& \\
M5 & $ R_{10}\ldots R_{6}
({}^{-}\!\!\!\!\ell_{\rm Planck}^{(11)})^{-6}$ \\
& \\
\hline
& \\
KK7M & $R_{10}^{2} R_{9}\ldots R_{4}
({}^{-}\!\!\!\!\ell_{\rm Planck}^{(11)})^{-9}$ \\
& \\
\hline
& \\
KK9M & $R_{10}^{3}R_{9}\ldots R_{4}
({}^{-}\!\!\!\!\ell_{\rm Planck}^{(11)})^{-12}$ \\
& \\
\hline\hline
\end{tabular}
\end{center}
\caption[Tabla de las masas de los objetos de SUGRA $N=1,d=11$.]
{\small En esta tabla se encuentran las masas de los distintos objetos extensos
de SUGRA $N=1,d=11$ en t'erminos de los radios de compactificaci'on
 $R_{i}$  y de la constante de Planck reducida 11-dimensional
 ${}^{-}\!\!\!\!\ell_{\rm   Planck}^{(11)}
=\ell_{\rm Planck}^{(11)}/2\pi$. Cuando un radio aparece en una potencia
superior a la unidad, el objeto tiene una dimensi'on isom'etrica especial
en esa direcci'on.}
\label{tab-Mmasses}
\end{table}

%%%%%%%%%%%%%%%%%%%%%%%%%%%%%%%%%%%%%%%%%%%%%%%%%%%%%%%%%%%%%%%%%%%%%%
\section{Soluciones gen'ericas y fuentes}
\label{sec-fuentes}

El siguiente paso tras estudiar los estados correspondientes a objetos
extensos en las teor'ias que nos ocupan, es encontrar soluciones
cl'asicas que se puedan asociar a ellos.  Esto se hace en dos etapas:

\begin{enumerate}
\item Se buscan una soluci'on de las acciones de SUGRA
  correspondientes en las que los campos que se acoplan al objeto en
  cuesti'on sean no-triviales (los dem'as deben de tener sus valores
  de vac'io) y tengan adem'as la forma correcta para describir un
  objeto extendido del la dimensi'on correspondiente.
  
  Esto lo vamos a hacer buscando una soluci'on de un modelo
  gen'erico, el {\it modelo $a$} que luego aplicaremos a distintos
  casos.
  
\item Se identifican las constantes de integraci'on de la soluci'on
  de forma que la tensi'on y carga calculadas sobre la soluci'on
  correspondan a las calculadas en la secci'on anterior.
  
  Esto se puede hacer de dos formas: se puede ``compactificar'' la
  soluci'on sobre un $p$-toro en las direcciones correspondientes a
  la $p$-brana y calcular en $d-p$ dimensiones la masa ADM y la carga
  el'ectrica correspondiente en funci'on de las constantes de
  integraci'on y comparar con las de la secci'on anterior. Adem'as,
  en todos los casos de la secci'on anterior se puede buscar la
  fuente: una acci'on de $p$-brana con tensi'on y carga
  determinadas, relacionadas por las ecuaciones de movimiento con las
  constantes de integraci'on.
  
  Aqu'i vamos a utilizar el segundo m'etodo. En la pr'oxima
  secci'on particularizaremos para cada tipo de $p$-brana de la
  Teor'ia de Cuerdas.
\end{enumerate}

%%%%%%%%%%%%%%%%%%%%%%%%%%%%%%%%%%%%%%%%%%%%%%%%%%%%%%%%%%%%%%%%%%%%%%
\subsection{El modelo $a$}

En la Secci'on~\ref{sec-objetosextensos} hemos visto que las
$p$-branas se acoplan gen'ericamente a la m'etrica, a un escalar y a
una $(p+1)$-forma descritas por la acci'on
Ec.~(\ref{eq:acciongenerica}). Es costumbre escribir el escalar $K$
como el exponencial de otro $\varphi$, y sustituir la constante
$\beta$ que controla el acoplo a la forma diferencial por otra $a$
para llegar a la acci'on del {\it modelo $a$}
\cite{kn:HoS,kn:G,kn:GM}

\begin{equation}
\begin{tabular}{|c|}
\hline
\\
$
S = \frac{1}{16\pi G_{N}^{(d)}} {\displaystyle\int} d^{d}x\sqrt{|g|}\, \left[R 
+2(\partial\varphi)^{2}  +\frac{(-1)^{p+1}}{2\cdot (p+2)!}
e^{-2a\varphi}F_{(p+2)}^{2}\right]\, ,  
$
\\
\\
\hline
\end{tabular}
\label{eq:amodelactionpbrane}
\end{equation}

\noindent donde $F_{(p+2)}$ es la intensidad de campo de la 
$(p+1)$-forma $A_{(p+1)}$:

\begin{equation} 
F_{(p+2)}=dA_{(p+1)}\, ,
\hspace{1cm}
F_{(p+2)\mu_{1}\ldots\mu_{p+2}}= (p+2)\partial_{[\mu_{1}}
A_{(p+1)\mu_{2}\ldots\mu_{p+2}]}\, .
\end{equation}

Este modelo contiene muchos casos particulares interesantes: para
$a=0$ el modelo es equivalente al de Einstein y Maxwell (el escalar
est'a desacoplado) y para otros valores, el acoplo del escalar
reproduce el de el escalar de Kaluza y Klein al vector de Kaluza y
Klein (o a su forma dual) o el del dilat'on a formas NSNS o RR.  Es
cierto que en las SUEGRAS que nos interesan aparecen tambi'en
t'erminos de Chern-Simons, pero 'estos no suelen jugar ning'un
papel cuando s'olo hay una $p$-brana presente, aunque s'i lo
hacen cuando hay varias de tipos distintos \cite{kn:Tow7} como veremos
en la Secci'on~\ref{sec-intersections}.

%The equations of motion are

%\begin{equation}
%\left.
%\begin{array}{rcl}
%G_{\mu\nu} +2T^{\varphi}_{\mu\nu} +\frac{(-1)^{p+1}}{2\cdot (p+1)!}
%e^{-2a\varphi} T^{A_{(p+1)}}_{\mu\nu} & = & 0\, ,\\
%& & \\
%\nabla^{2} \varphi +\frac{(-1)^{p+1}}{4\cdot (p+2)!}a 
%e^{-2a\varphi}F_{(p+2)}^{2} & = & 0\, ,\\
%& & \\
%\nabla_{\mu} \left( e^{-2a\varphi} F_{(p+2)}{}^{\mu\nu_{1}\ldots\nu_{p+1}} 
%\right) & = & 0\, ,\\
%\end{array}
%\right\}
%\label{eq:amodelequationspbrane}
%\end{equation}

%\noindent where $T^{A_{(p+1)}}$ is the $(p+1)$-form energy-momentum tensor

%\begin{equation}
%T^{A_{(p+1)}}_{\mu\nu} = F_{(p+2)\mu}{}^{\nu_{1}\ldots\nu_{p+1}}
%F_{(p+2)\nu\nu_{1}\ldots\nu_{p+1}} 
%-{\textstyle\frac{1}{2(p+2)}}g_{\mu\nu}F_{(p+2)}^{2}\, .
%\end{equation} 

%Also in his case, the dilaton equation (for non-constant dilaton)
%can be ontained by using the Bianchi identities for the metric and the
%$(p+1)$-form potential on the other two equations of motion?????????

De acuerdo con la regla general de que las $(p+1)$-formas se acoplan a
objetos $p$-dimensionales, las soluciones est'aticas naturales con la
m'axima simetr'ia (distintas del vac'io) para $p=0$ son
agujeros negros\footnote{Aqu'i usamos este t'ermino en el sentido
  de objeto masivo puntual.} cargados y en general {\it $p$-branas
  negras} cargadas. Como siempre, empezamos haciendo un {\it Ansatz}
adecuado al sistema que la soluci'on pretende describir: un objeto
con $(p+1)$ simetr'ias translacionales (a lo largo de las
direcciones del volumen del mundo de una $p$-brana plana) en las
direcciones $t,\vec{y}_{p}$:

\begin{equation}
  \begin{array}{rcl}
ds^{2} & = & 
f \left[ Wdt^{2}-d\vec{y}^{\, 2}_{p}\right] -g^{-1}\left[ W^{-1}d\rho^{2} 
-\rho^{2}d\Omega_{(\tilde{p}+2)}^{2}\right]\, ,\\
& & \\
A_{t\underline{y}^{1}\ldots\underline{y}^{p}} 
& = & \alpha \left(H^{-1}-1\right)\, , \\
\end{array}
\end{equation}

\noindent donde $f,g,W$ y $H$ son funciones de $\rho$, la coordenada radial en
las direcciones transversas a la $p$-brana y donde 

\begin{equation}
\tilde{p} \equiv d-p-4\, . 
\end{equation}

Para reducir el n'umero de funciones independiente hemos de hacer
algunas hip'otesis m'as: si no hay ``pelo'' escalar, $e^{\varphi}$
ser'a una potencia de $H$. Adem'as sabemos que en el caso del
agujero de Reissner y Nordstr\"om ($d=4,p=a=0$) tambi'en $f$ y $g$
son potencias de $H$ y vamos a suponer que eso es tambi'en cierto
aqu'i de forma que s'olo $H$ y $W$ son funciones independientes,
para las que vamos a suponer la forma

\begin{equation}
H=1 + \frac{h}{\rho^{\tilde{p}+1}}\, ,
\hspace{1cm}
W=1 + \frac{\omega}{\rho^{\tilde{p}+1}}\, .
\end{equation}

Sustituyendo el {\it Ansatz} en las ecuaciones de movimiento, es
inmediato (pero asaz laborioso) encontrar la soluci'on general, que
escribimos de esta forma:

\begin{equation} 
\begin{tabular}{|rcl|} 
\hline & & \\ & & \\ 
$ds^{2}$ & $=$ & $
H^{\frac{2x-2}{p+1}} 
\left[Wdt^{2} -d\vec{y}^{\, 2}_{p} \right]
%$ \\
%& & \\
%& &
%$
-H^{\frac{-(2x-2)}{\tilde{p}+1}} 
\left[W^{-1}d\rho^{2} +\rho^{2}d\Omega_{(\tilde{p} +2)}^{2}\right]\, ,$ \\
& & \\
$e^{-2a\varphi}$ & $=$ & $e^{-2a\varphi_{0}} H^{2x}\, ,
\hspace{.5cm}
A_{t\underline{y}^{1}\ldots \underline{y}^{p}} = 
e^{a\varphi_{0}} \alpha \left(H^{-1}-1\right)\, ,$  \\
& & \\
$H$ & $=$ & $1+ {\displaystyle\frac{h}{\rho^{\tilde{p}+1}}}\, ,
\hspace{.5cm}
W=1+{\displaystyle\frac{\omega}{\rho^{\tilde{p}+1}}}\, ,$ \\
& & \\
$\omega$ & $=$ & $h\left[1- \frac{a^{2}}{4x}\alpha^{2} \right]\, ,$  \\
& & \\
$x$ & $=$ & $\frac{\frac{a^{2}}{2}c}{1+\frac{a^{2}}{2}c}\, ,
\hspace{.5cm}
c=\frac{(p+1)+(\tilde{p}+1)}{(p+1)(\tilde{p}+1)}\, .$
\\ & & \\ \hline
\end{tabular}
\label{eq:HoS}
\end{equation}

Las constantes $d,p,a$ que definen el modelo est'an combinadas en la
constante $x$. La constante $\varphi_{0}$ es el valor del escalar en
el infinito (estas soluciones son asint'oticamente planas en las
direcciones ortogonales al volumen del mundo de la $p$-brana ({\it
  direcciones transversas}) $\rho\rightarrow \infty$) y las constantes
de integraci'on $h,\omega,\alpha$ est'an relacionadas por una
ecuaci'on. La soluci'on est'a escrita de forma que es v'alida para
$a=0$.  Est'a claro tambi'en que en esta soluci'on la carga escalar
no es independiente de la carga ``el'ectrica'' de la $p$-brana, como
quer'iamos.

Como ocurr'ia en la soluci'on de Reissner y
Nordstr\"om, cuando el par'ametro de no-extremalidad $\omega=0$
(l'imite extremo), el ``factor de Schwarzschild'' $W=1$ y $H$
puede ser una funci'on arm'onica arbitraria en el espacio
transverso:

\begin{equation} 
\begin{tabular}{|rcl|} 
\hline & & \\ & & \\ 
$ds^{2}$ & $=$ & $
H^{\frac{2x-2}{p+1}} 
\left(dt^{2} -d\vec{y}^{\, 2}_{p} \right)
%$ \\
%& & \\
%& &
%$
-H^{\frac{-(2x-2)}{\tilde{p}+1}} d\vec{x}_{(\tilde{p}+3)}^{\, 2}\, ,$ \\
& & \\
$e^{-2a\varphi}$ & $=$ & $e^{-2a\varphi_{0}} H^{2x}\, ,
\hspace{.5cm}
A_{t\underline{y}^{1}\ldots \underline{y}^{p}} = 
e^{a\varphi_{0}} \alpha \left(H^{-1}-1\right)\, ,$  \\
& & \\
$\partial_{\underline{m}}\partial_{\underline{m}}H$ & $=$ & $0\, ,$ \\
& & \\
$x$ & $=$ & $\frac{\frac{a^{2}}{2}c}{1+\frac{a^{2}}{2}c}\, ,
\hspace{.5cm}
c=\frac{(p+1)+(\tilde{p}+1)}{(p+1)(\tilde{p}+1)}\, ,
\hspace{.5cm}
\alpha^{2} = \frac{4x}{a^{2}}\, . $
\\ & & \\ \hline
\end{tabular}
\label{eq:HoSextreme}
\end{equation}

Las elecciones de funci'on arm'onica $H$ que m'as nos interesan
son, evidentemente, las de la forma

\begin{equation}
H = 1 +\frac{h}{|\vec{x}_{(\tilde{p}+3)}|^{\tilde{p}+1}}\, ,  
\end{equation}

\noindent para describir una sola $p$-brana extrema situada en 
el origen de coordenadas $\vec{x}_{(\tilde{p}+3)}=0$, 'o

\begin{equation}
H = 1 +\sum_{I=1}{}^{N}\frac{h_{I}}{|\vec{x}_{(\tilde{p}+3)}
-\vec{x}_{(\tilde{p}+3)\, I}|^{\tilde{p}+1}}\, ,  
\end{equation}

\noindent para describir $N$ $p$-branas paralelas localizadas en 
$\vec{x}_{(\tilde{p}+3)}=\vec{x}_{(\tilde{p}+3)\, I}$.

Tambi'en como ocurr'ia en la soluci'on de Reissner y Nordstr\"om
(pero aqu'i s'olo en el caso $a=0$) si tomamos $h=0$, obtenemos
un objeto sin carga y con horizonte: {\it una $p$-brana negra}, con
m'etrica (los otros campos son constantes arbitrarias)

\begin{equation} 
\begin{tabular}{|rcl|} 
\hline & & \\ & & \\ 
$ds^{2}$ & $=$ & $
Wdt^{2} -d\vec{y}^{\, 2}_{p}-
W^{-1}d\rho^{2} +\rho^{2}d\Omega_{(\tilde{p} +2)}^{2}\, ,$ \\
& & \\
$W$ & $=$ & $1+ {\displaystyle\frac{\omega}{\rho^{\tilde{p}+1}}}\, ,$ \\
& & \\ \hline
\end{tabular}
\label{eq:blackpbrane}
\end{equation}

\noindent que generalizan la soluci'on de Schwarzschild $d$-dimensional
($p=0$) \cite{kn:MP,kn:T}. Todas las $p$-branas no-extremas, cargadas o
no, tienen un horizonte de eventos regular y se pueden definir sus
temperaturas y entrop'ias, pero nosotros s'olo nos vamos a
concentrar en los agujeros negros.

%%%%%%%%%%%%%%%%%%%%%%%%%%%%%%%%%%%%%%%%%%%%%%%%%%%%%%%%%%%%%%%%%%%%%%
\subsection{Fuentes}

Nuestra experiencia con el agujero de Reissner y Nordstr\"om nos dice
que podemos encontrar fuentes para las $p$-branas cargadas que son
soluciones del modelo $a$ en el l'imite extremo ($\omega=0$).
Conociendo las fuentes, podemos relacionar la constante de
integraci'on $h$ de la soluci'on Ec.~(\ref{eq:HoSextreme}) con la
tensi'on y la carga que aparecen en la acci'on de la $p$-brana, que
es, evidentemente, nuestro candidato a fuente.

Consideremos, por lo tanto, el siguiente sistema acoplado

\begin{equation}
S = S_{a} +S_{p}\, ,  
\end{equation}

\noindent donde  $S_{a}$ es la acci'on del modelo $a$
Ec.~(\ref{eq:amodelactionpbrane}) y $S_{p}$ es la acci'on de
la $p$-brana cargada

\begin{equation}
\begin{array}{rcl}
S_{p}[X^{\mu},\gamma_{ij}] & = & 
-\frac{T}{2}{\displaystyle\int} d^{p+1}\xi\sqrt{|\gamma|}\,
\left[ e^{-2b\varphi}\, \gamma^{ij}\partial_{i}X^{\mu} \partial_{j}X^{\nu} 
g_{\mu\nu} - (p-1)\right]  \\
& & \\
& & 
-\frac{\mu}{(p+1)!}{\displaystyle\int} 
d^{p+1}\xi\, A_{(p+1)\, \mu_{1}\cdots\mu_{p+1}}
\partial_{i_{1}}X^{\mu_{1}}\ldots\partial_{i_{p+1}}X^{\mu_{p+1}}\, .
\end{array}
\end{equation}

Para resolver las ecuaciones de movimiento, se usa el {\it gauge
  est'atico o f'isico} en el que las primeras $(p+1)$ coordenadas
de la $p$-brana se identifican con las espacio-temporales

\begin{equation}
Y^{i}(\xi)=\xi^{i}\, ,  
\end{equation}

\noindent y el {\it Ansatz}

\begin{equation}
X^{m}(\xi)=0\, ,  
\end{equation}

\noindent para las dem'as. Esto describe una $p$ brana extendida 
en las direcciones $\vec{y}_{(p+1)}$ en reposo en el origen de
coordenadas transversas $\vec{x}_{(\tilde{p}+3)}=\vec{0}$. El {\it
  Ansatz} para los campos del espacio-tiempo es la soluci'on general
extrema del modelo $a$ Ec.~(\ref{eq:HoSextreme}), pero dejando $H$
como una funci'on de $\vec{x}_{(\tilde{p}+3)}$ sin determinar.

El resultado es que todas las ecuaciones se pueden satisfacer si,

\begin{enumerate}
\item  $a$ y $b$ est'an relacionadas por

\begin{equation}
\label{eq:abrel}
a=-(p+1)b\, ,  
\end{equation}

\item la tensi'on y la carga de la $p$-brana est'an relacionadas por

\begin{equation}
\begin{tabular}{|c|}
\hline \\
$\mu = T/\alpha\, ,  $\\ \\ \hline
\end{tabular}  
\end{equation}

\item y $H$ viene dada por

\begin{equation}
H= \epsilon+ \frac{h}{|\vec{x}_{(\tilde{p}+3)}|^{\tilde{p}+1}}\, ,
\end{equation}

\noindent donde $\epsilon$ es una constante arbitraria ($\epsilon=+1$ si 
queremos una soluci'on asint'oticamente plana) y donde $h$ est'a
dada por

\begin{equation}
\begin{tabular}{|c|}
\hline \\
$h= {\displaystyle\frac{16\pi G_{N}^{(d)} T}{(\tilde{p}+1)\alpha^{2} 
\omega_{(\tilde{p}+2)}}}\, ,$\\ \\ \hline
\end{tabular}  
\label{eq:hache}
\end{equation}

\end{enumerate}

La primera condici'on se cumple en todos los casos de inter'es y
est'a en el fondo relacionada con que los objetos extensos han de
corresponder a genuinos objetos supersim'etricos de la SUGRA
correspondiente. La segunda es la condici'on BPS con nuestra
normalizaci'on de los campos. De nuevo est'a relacionada con
supersimetr'ia. La condici'on sobre $H$ selecciona de entre todas
las posibles funciones arm'onicas en el espacio transverso aquella
que tiene un polo del orden adecuado en la posici'on en la que se
encuentra la $p$-brana con el coeficiente que corresponde a su
tensi'on y carga.

Como vemos, la supersimetr'ia juega un papel crucial, incluso
aunque no hay ning'un fermi'on presente en estas f'ormulas.

?`Qu'e pasa con las soluciones no-extremas Ec.~(\ref{eq:HoS})? La
interpretaci'on que se suele hacer de ellas es que describen
$p$-branas excitadas\footnote{Esto quiere decir, en los casos
  est'aticos, que los campos de su volumen del mundo tienen
  configuraciones no-triviales, por ejemplo.} y, por lo tanto, no
supersim'etricas. Su carga debe de ser la de las extremas, pero su
tensi'on es mayor. En las soluciones, la carga depende de $h$ y la
tensi'on de $h$ y $\omega$, por lo que $h$ tendr'a el valor
determinado arriba, pero $\omega$, que parametriza la desviaci'on de
la extremalidad, es arbitraria.

%%%%%%%%%%%%%%%%%%%%%%%%%%%%%%%%%%%%%%%%%%%%%%%%%%%%%%%%%%%%%%%%%%%%%%
\section{Soluciones en $d=11$}

Vamos ahora a utilizar los resultados de la secci'on anterior para
encontrar soluciones que describan los objetos extensos de SUGRA
$N=1,d=11$, que son la M2 y la M5. En esta teor'ia no hay
escalares por lo que estamos en los casos $d=11,a=0,p=2$ y
$d=11,a=0,p=5$. De las soluciones generales Ecs.~(\ref{eq:HoS}) y
(\ref{eq:HoSextreme}) obtenemos inmediatamente los siguientes
resultados:

\subsection{La membrana M2}

En el caso no-extremo, la soluci'on es 

\begin{equation}  
\begin{tabular}{|rcl|}  
\hline & & \\ & & \\ 
$d\hat{\hat{s}}{}^{2}_{E}$ & $=$ & 
$H_{M2}^{-2/3} \left[Wdt^{2}
-d\vec{y}_{2}^{\, 2} \right]
%$ \\ 
%& & \\ 
%& & $
-H_{M2}^{1/3}
\left[W^{-1}d\rho^{2}  +\rho^{2}d\Omega_{(7)}^{2}\right]\, ,$ \\
& & \\ 
$\hat{\hat{C}}_{t\underline{y}^{1}\underline{y}^{2}}$ & $=$ &
$\alpha \left(H_{M2}^{-1}-1\right)\, ,$  \\ 
& & \\ 
$H_{M2}$ & $=$ & $1 +{\displaystyle\frac{h_{M2}}{\rho^{6}}}\, , \hspace{.5cm}
W=1+{\displaystyle\frac{\omega}{\rho^{6}}}\, ,$ \\ 
& & \\ 
$\omega$ & $=$ & $h_{M2} \left[1-\alpha^{2} \right]\, ,$  \\ 
& & \\ \hline 
\end{tabular}
\label{eq:M2} 
\end{equation}

\noindent y en el extremo $\omega=0$ $\alpha=\pm 1$ 

\begin{equation} 
\begin{tabular}{|rcl|} 
\hline & & \\ & & \\ 
$d\hat{\hat{s}}{}^{2}$ & $=$ & 
$H_{M2}^{-2/3}\left[dt^{2} -d\vec{y}_{2}^{\, 2} \right]
-H_{M2}^{1/3}\, d\vec{x}_{8}^{\, 2}\, ,$ \\
& & \\
$\hat{\hat{C}}_{ty_{1}y_{2}}$ & $=$ & 
$\pm \left(H_{M2}^{-1}-1\right)\, ,$  \\
& & \\
$H_{M2}$ & $=$ & $1+ {\displaystyle\frac{h_{M2}}{|\vec{x}_{8}|^{6}}}\, .$ \\
%& & \\
%$\vec{x}_{8}$ & $=$ & $(x^{1},\ldots,x^{8})\, .$ \\
& & \\ \hline
\end{tabular}
\label{eq:extremeM2}
\end{equation}

Para determinar la constante de intergraci'on $h_{M2}$ utilizamos la
Ec.~(\ref{eq:hache})

\begin{equation}
h_{M2}= {\displaystyle\frac{16\pi G_{N}^{(11)} T_{M2}}{6 \omega_{(7)}}}\, ,  
\end{equation}

\noindent la relaci'on entre la tensi'on y la masa de la M2 compactificada
en un 2-toro Ec.~(\ref{eq:MM2})

\begin{equation}
T_{M2}=\frac{M_{M2}}{(2\pi)^{2}R_{9}R_{10}} = \frac{1}{(2\pi)^{2} 
({}^{-}\!\!\!\!\ell_{\rm   Planck}^{(11)})^{3}} 
=\frac{2\pi}{(\ell_{\rm   Planck}^{(11)})^{3}}\, ,
\end{equation}

\noindent y la definici'on de la longitud de Planck
Ec.~(\ref{eq:Plancklength}), y obtenemos

\begin{equation}
\begin{tabular}{|c|} 
\hline \\
$
h_{M2} 
={\displaystyle\frac{(\ell_{\rm   Planck}^{(11)})^{6}}{6\omega_{(7)}}}\, .
$ \\ \\ \hline
\end{tabular}  
\end{equation}

Si queremos que la soluci'on describa $N_{M2}$ M2-branas, simplemente
debemos de reemplazar la funci'on arm'onica por otra con otros
tantos polos, todos ellos con el coeficiente $h_{M2}$. Si las
M2-branas coinciden, habr'a un solo polo con coeficiente
$N_{M2}h_{M2}$. Observaciones enteramente an'alogas a 'estas se
pueden hacer en todos los casos que siguen y las daremos por hechas.

%%%%%%%%%%%%%%%%%%%%%%%%%%%%%%%%%%%%%%%%%%%%%%%%%%%%%%%%%%%%%%%%%%%%%%
\subsection{La 5-brana M5}

En el caso no-extremo tenemos

\begin{equation} 
\begin{tabular}{|rcl|} 
\hline & & \\ & & \\ 
$d\hat{\hat{s}}^{2}$ & $=$ & $H_{M5}^{-1/3}
\left[Wdt^{2} -d\vec{y}_{5}^{\, 2} \right]
%$ \\
%& & \\
%& &
%$
-H_{M5}^{2/3} \left[W^{-1}d\rho^{2} 
+\rho^{2}d\Omega_{(4)}^{2}\right]\, ,$ \\
& & \\
$\tilde{\hat{\hat{C}}}_{t\underline{y}^{1}\ldots\underline{y}^{5}}$ & $=$ & 
$\alpha \left( H_{M5s}^{-1}-1\right)\, ,$  \\
& & \\
%$G$ & $=$ & ???\, ,\\
%& & \\
$H_{M5}$ & $=$ & $1 +{\displaystyle\frac{h_{M5}}{\rho^{3}}}\, ,
\hspace{.5cm}
W=1+{\displaystyle\frac{\omega}{\rho^{3}}}\, ,$ \\
& & \\
$\omega$ & $=$ & $h_{M5} \left[1-\alpha^{2} \right]\, .$  \\
& & \\ \hline
\end{tabular}
\label{eq:M5}
\end{equation}

En el l'imite extremo $\omega=0,\alpha=\pm 1$ tenemos

\begin{equation} 
\begin{tabular}{|rcl|} 
\hline & & \\ & & \\ 
$d\hat{\hat{s}}{}^{2}$ & $=$ & 
$H_{M5}^{-1/3}\left[dt^{2} -d\vec{y}_{5}^{\, 2} \right]
-H_{M5}^{2/3}\, d\vec{x}_{5}^{\, 2}\, ,$ \\
& & \\
$\tilde{\hat{\hat{C}}}_{t\underline{y}^{1}\ldots \underline{y}^{5}}$ & $=$ & 
$\pm \left(H_{M5}^{-1}-1\right)\, ,$  \\
& & \\
%$G_{\underline{i}\underline{j}\underline{k}\underline{l}}$ & $=$ &
%$\pm\epsilon_{\underline{i}\underline{j}\underline{k}\underline{l}\underline{m}}
%\partial{m} \left(H_{M5}^{-1}-1\right) ????\, ,$ \\
%& & \\
$H_{M5}$ & $=$ & $1+ {\displaystyle\frac{h_{M5}}{|\vec{x}_{5}|^{3}}}\, .$ \\
 & & \\ \hline
\end{tabular}
\label{eq:extremeM5}
\end{equation}

\noindent Por el mismo procedimiento que en el caso de la M2-brana 
encontramos 

\begin{equation}
\begin{tabular}{|c|} 
\hline \\
$
h_{M5}= {\displaystyle\frac{(\ell_{\rm Planck}^{(11)})^{3}}{3\omega_{(4)}}}\, .
$ \\ \\ \hline
\end{tabular} 
\end{equation}

%%%%%%%%%%%%%%%%%%%%%%%%%%%%%%%%%%%%%%%%%%%%%%%%%%%%%%%%%%%%%%%%%%%%%%
\section{Soluciones en $d=10$}

La identificaci'on de soluciones de las teor'ias tipo~II con
soluciones del modelo $a$ es un poco m'as complicada. Primero hemos
de identificar cu'ales son los valores de $a$ para cada caso.

La forma gen'erica de las acciones efectivas tipo~II en el sistema de
referencia conforme de la cuerda\footnote{Hacemos 'enfasis en este
  punto a~nadiendo el sub'indice $s$.} es

\begin{equation}
\begin{array}{rcl}
S & = &\frac{1}{16\pi G_{N}^{(d)}} {\displaystyle\int} d^{d}x\sqrt{|g_{s}|}\
\left\{ e^{-2\phi} \left[ R_{s} -4(\partial\phi)^{2} 
+\frac{(-1)^{p_{1}+1}}{2\cdot (p_{1}+2)!} F_{(p_{1}+2)}^{2}\right]
%\right.\\ 
%& & \\
%& & 
%\left.
+\frac{(-1)^{p_{2}+1}}{2\cdot (p_{2}+2)!} F_{(p_{2}+2)}^{2}\right\}\, . \\
\end{array}
\end{equation}

Hay dos tipos de formas diferenciales en esta acci'on: las que se
acoplan al dilat'on con el prefactor $e^{-2\phi}$ ($F_{(p_{1}+2)}$),
que pertenecen al sector NSNS al que se acoplan las $p$-branas
fundamentales y las que no ($F_{(p_{2}+2)}$), que pertenecen al sector
RR, a las que se acoplan las D$p$-branas. Para identificar soluciones
de esta acci'on en la general del modelo $a$, tenemos que reescribir
esta acci'on en la m'etrica de Einstein, identificar al dilat'on como
funci'on de $\varphi$, e identificar el valor de $a$. 'Este es
diferente para las $p$-branas fundamentales y las D$p$-branas, y lo
denotamos por $a_{1}$ y $a_{2}$ respectivamente. Haciendo una
transformaci'on de Weyl a la m'etrica $g_{s}$

\begin{equation}
g_{s\, \mu\nu} = e^{\frac{4}{(d-2)}\phi}\ g_{\mu\nu}\, ,  
\end{equation}

\noindent tenemos

\begin{equation}
\begin{array}{rcl}
  S & = &\frac{1}{16\pi G_{N}^{(d)}} {\displaystyle\int} d^{d}x\sqrt{|g|}\
\left[ R +\frac{4}{(d-2)}(\partial\phi)^{2} 
\right.\\ 
& & \\
& & 
\hspace{3cm}\left.
+\frac{(-1)^{p_{1}+1}}{2\cdot (p_{1}+2)!} e^{-4\frac{(p_{1}+1)}{(d-2)}\phi} 
F_{(p_{1}+2)}^{2}
+\frac{(-1)^{p_{2}+1}}{2\cdot (p_{2}+2)!} 
e^{2\frac{(\tilde{p}_{2} -p_{1})}{(d-2)}\phi} 
F_{(p_{2}+2)}^{2}\right\}\, . \\
\end{array}
\end{equation}

Comparando con la acci'on del modelo $a$
Ec.~(\ref{eq:amodelactionpbrane}) vemos que el dilat'on $\phi$ y el
escalar $\varphi$ est'an relacionados por

\begin{equation}
\phi = {\textstyle\sqrt{\frac{(d-2)}{2}}} \varphi\, ,  
\end{equation}

\noindent y que los valores de la constante $a$ son, para formas NSNS y RR

\begin{eqnarray}
a_{1} & = & \frac{2(p_{1}+1)}{\sqrt{2(d-2)}}\, , 
\hspace{1cm}{\rm (NS-NS)}\\
& & \nonumber \\
a_{2} & = & \frac{-(\tilde{p}_{2}-p_{2})}{\sqrt{2(d-2)}}\, .
\hspace{1cm}{\rm (RR)}
\end{eqnarray}

Antes de ir al sistema de referencia conforme de Einstein,
podr'iamos haber dualizado las intensidades de campo NSNS (para
describir $p$-branas solit'onicas). En ese caso, habr'iamos
encontrado el modelo $a$ con

\begin{equation}
a_{3} = -\frac{2(\tilde{p_{1}}+1)}{\sqrt{2(d-2)}}\, .
\end{equation}

Con estos datos vamos a identificar las soluciones 10-dimensionales.
La 'unica sutileza est'a en la introducci'on del valor del
dilat'on en el infinito, que no se puede deducir correctamente de la
soluci'on gen'erica, sino que hay que introducirlo despu'es
observando c'omo se transforman los distintos campos bajo la
adici'on de una constante al dilat'on.

%%%%%%%%%%%%%%%%%%%%%%%%%%%%%%%%%%%%%%%%%%%%%%%%%%%%%%%%%%%%%%%%%%%%%%
\subsection{La cuerda fundamental F1}

Corresponde al caso $p=1,a=1$ en $d=10$

\begin{equation} 
\begin{tabular}{|rcl|} 
\hline & & \\ & & \\ 
$d\tilde{\hat{s}}^{2}_{E}$ & $=$ & $H_{F1}^{-3/4}
\left[Wdt^{2} -dy^{2} \right]
%$ \\
%& & \\
%& &
%$
-H_{F1}^{1/4} \left[W^{-1}d\rho^{2} 
+\rho^{2}d\Omega_{(7)}^{2}\right]\, ,$ \\
& & \\
$d\hat{s}^{2}_{s}$ & $=$ & $H_{F1}^{-1}
\left[Wdt^{2} -dy^{2} \right]
%$ \\
%& & \\
%& &
%$
-\left[W^{-1}d\rho^{2} +\rho^{2}d\Omega_{(7)}^{2}\right]\, ,$ \\
& & \\
$e^{-2(\hat{\phi}-\hat{\phi}_{0})}$ & $=$ & $ H_{F1}\, ,$\\
& & \\ 
$\hat{B}_{t\underline{y}}$ &  $=$  & $ \alpha \left(H_{F1}^{-1}-1\right)\, ,$  \\
& & \\
$H_{F1}$ & $=$ & $1+ {\displaystyle\frac{h_{F1}}{\rho^{6}}}\, ,
\hspace{.5cm}
W=1+{\displaystyle\frac{\omega}{\rho^{6}}}\, ,$ \\
& & \\
$\omega$ & $=$ & $h_{F1} \left[1-\alpha^{2} \right]\, .$  \\
& & \\ \hline
\end{tabular}
\label{eq:F1}
\end{equation}

En el l'imite extremo $\omega=0,\alpha=\pm 1$ es conocida como la
{\it soluci'on de la cuerda fundamental} \cite{kn:DGHR} y toma la
forma:

\begin{equation} 
\begin{tabular}{|rcl|} 
\hline & & \\ & & \\ 
$d\tilde{\hat{s}}^{2}_{E}$ & $=$ & $H_{F1}^{-3/4}
\left[dt^{2} -dy^{2} \right]-H_{F1}^{1/4}d\vec{x}_{8}^{\, 2}\, ,$ \\
& & \\
$d\hat{s}^{2}_{s}$ & $=$ & $H_{F1}^{-1}
\left[dt^{2} -dy^{\, 2} \right]-d\vec{x}_{8}^{\, 2}\, ,$ \\
& & \\
$e^{-2(\hat{\phi}-\hat{\phi}_{0})}$ & $=$ & $ H_{F1}\, ,$ \\
& & \\
$\hat{B}_{t\underline{y}}$ & $=$ & $\pm \left(H_{F1}^{-1}-1\right)\, ,$  \\
& & \\
$H_{F1}$ & $=$ & $1+ {\displaystyle\frac{h_{F1}}{|\vec{x}_{8}|^{6}}}\, ,$ \\
& & \\ \hline
\end{tabular}
\label{eq:extremeF1}
\end{equation}

Observemos que con $a=p=1$, la fuente del modelo $a$, reexpresada en
el sistema de referencia conforme de la cuerda es realmente la
acci'on de una cuerda fundamental que no se acopla directamente al
dilat'on en ese sistema de referencia (v'ease las
Refs.~\cite{kn:DGHR,kn:DKL}).

El valor de la constante de integraci'on que reproduce correctamente
la tensi'on de la F1 se puede calcular por el mismo m'etodo que
utilizamos con los objetos 11-dimensionales

\begin{equation}
\begin{tabular}{|c|} 
\hline \\
$
h_{F1}={\displaystyle\frac{2^{5} \pi^{6} \ell_{s}^{6}g^{2}}{3\omega_{(7)}}}\, ,
$ \\ \\ \hline
\end{tabular}  
\end{equation}

\noindent pero tambi'en se puede calcular explotando las relaciones
de dualidad con otras soluciones, como veremos en la Secci'on~\ref{sec-relacionesdedualidad}

%%%%%%%%%%%%%%%%%%%%%%%%%%%%%%%%%%%%%%%%%%%%%%%%%%%%%%%%%%%%%%%%%%%%%%
\subsection{La 5-brana solit'onica}

Corresponde al caso $a=-1,p=5$ en diez dimensiones:

\begin{equation} 
\begin{tabular}{|rcl|} 
\hline & & \\ & & \\ 
$d\tilde{\hat{s}}{}^{2}_{E}$ & $=$ & $H_{S5}^{-1/4}
\left[Wdt^{2} -d\vec{y}_{5}^{\, 2} \right]
%$ \\
%& & \\
%& &
%$
-H_{S5}^{3/4} \, \left[W^{-1}d\rho^{2} 
+\rho^{2}d\Omega_{(3)}^{2}\right]\, ,$ \\
& & \\
$d\hat{s}^{2}_{s}$ & $=$ & $\left[Wdt^{2} -d\vec{y}_{5}^{\, 2} \right]
%$ \\
%& & \\
%& &
%$
-H_{S5}\left[W^{-1}d\rho^{2} 
+\rho^{2}d\Omega_{(3)}^{2}\right]\, ,$ \\
& & \\
$e^{-2(\hat{\phi}-\hat{\phi}_{0})}$ & $=$ & $H_{S5}^{-1}\, ,$ \\
& & \\
$ \hat{B}{}^{(6)}{}_{t\underline{y}^{1}\ldots \underline{y}^{5}}$  & $=$  & 
$\alpha e^{-2\hat{\phi}_{0}} \left(H_{S5}^{-1}-1\right)\, ,$  \\
& & \\
$H_{S5}$ & $=$ & $1+ {\displaystyle\frac{h_{ps}}{\rho^{2}}}\, ,
\hspace{.5cm}
W=1+{\displaystyle\frac{\omega}{\rho^{2}}}\, ,$ \\
& & \\
$\omega$ & $=$ & $h_{S5} \left[1-\alpha^{2} \right]\, ,$  \\
& & \\ \hline
\end{tabular}
\label{eq:S5}
\end{equation}

\noindent y, en el l'imite extremo $\omega=0,\alpha=\pm 1$ en el que
es conocida como la {\it soluci'on de la 5-brana solit'onica}
\cite{kn:CHS,kn:CHS2}), toma la forma

\begin{equation} 
\begin{tabular}{|rcl|} 
\hline & & \\ & & \\ 
$d\tilde{\hat{s}}{}^{2}_{E}$ & $=$ & 
$H_{S5}^{-1/4}\left[dt^{2} -d\vec{y}_{5}^{\, 2} \right]
-H_{S5}^{3/4}\, d\vec{x}_{4}^{\, 2}\, ,$ \\
& & \\
$d\hat{s}^{2}_{s}$ & $=$ & $\left[dt^{2} -d\vec{y}_{5}^{\, 2} \right]
-H_{S5}\, d\vec{x}_{4}^{\, 2}\, ,$ \\
& & \\
$e^{-2(\hat{\phi}-\hat{\phi}_{0})}$ & $=$ & $ H_{S5}^{-1}\, ,$ \\
& & \\
$\tilde{\hat{B}}{}^{(6)}{}_{t\underline{y}^{1}\cdots \underline{y}^{5}}$ 
& $=$ & 
$\pm e^{-2\hat{\phi}_{0}}\left(H_{S5}^{-1}-1\right)\, ,$  \\
& & \\
%$\hat{H}_{\underline{i}\underline{j}\underline{k}}$ & $=$ & 
%$\pm e^{2(\hat{\phi}-\hat{\phi}_{0})}
%\epsilon_{\underline{i}\underline{j}\underline{k}\underline{l}}
%\partial_{\underline{l}}\left(H_{S5}^{-1}-1\right)\, ,$ \\
%& & \\
$H_{S5}$ & $=$ & $1+ {\displaystyle\frac{h_{S5}}{|\vec{x}_{4}|^{2}}}\, .$ \\
& & \\ \hline
\end{tabular}
\label{eq:extremeS5}
\end{equation}

El valor de la constante de integraci'on se puede determinar por el
m'etodo que hemos explicado y por relaciones de dualidad y es

\begin{equation}
\begin{tabular}{|c|} 
\hline \\
$
h_{S5} = \ell_{s}^{2}\, .
$ \\ \\ \hline
\end{tabular}  
\end{equation}

%%%%%%%%%%%%%%%%%%%%%%%%%%%%%%%%%%%%%%%%%%%%%%%%%%%%%%%%%%%%%%%%%%%%%%
\subsection{Las D$p$-branas}

Las hay para todos los valores de $p=0,\ldots,9$, y la soluci'on
general para todas ellas en diez dimensiones es

\begin{equation} 
\begin{tabular}{|rcl|} 
\hline & & \\ & & \\ 
$d\tilde{\hat{s}}^{2}_{E}$ & $=$ & $H_{Dp}^{-\frac{(7-p)}{8}}
\left[Wdt^{2} -d\vec{y}_{p}^{\, 2} \right]
%$ \\
%& & \\
%& &
%$
-H_{Dp}^{\frac{(p+1)}{8}}
\left[W^{-1}d\rho^{2} 
+\rho^{2}d\Omega_{(8-p)}^{2}\right]\, ,$ \\
& & \\
$d\hat{s}^{2}_{s}$ & $=$ & $H_{Dp}^{-1/2}
\left[Wdt^{2} -d\vec{y}_{p}^{\, 2} \right]
%$ \\
%& & \\
%& &
%$
-H_{Dp}^{1/2}\left[W^{-1}d\rho^{2} 
+\rho^{2}d\Omega_{(8-p)}^{2}\right]\, ,$ \\
& & \\
$e^{-2(\hat{\phi}-\hat{\phi}_{0})}$ & $=$ & 
$H_{Dp}^{\frac{(p-3)}{2}}\, ,$ \\
& & \\
$ \hat{C}^{(p+1)}{}_{t\underline{y}^{1}\ldots \underline{y}^{p}}$ & $=$ &
$\alpha e^{-\hat{\phi}_{0}} \left( H_{Dp}^{-1}-1\right)\, ,$  \\
& & \\
$H_{Dp}$ & $=$ & $1+ {\displaystyle\frac{h_{Dp}}{\rho^{7-p}}}\, ,
\hspace{.5cm}
W=1+{\displaystyle\frac{\omega}{\rho^{7-p}}}\, ,$ \\
& & \\
$\omega$ & $=$ & $h_{Dp} \left[1-\alpha^{2} \right]\, .$  \\
& & \\ \hline
\end{tabular}
\label{eq:Dp}
\end{equation}

Estrictamente estas soluciones s'olo son v'alidas para $p<7$. Para
$p \geq 7$y, de forma general en todas las soluciones que hemos visto,
cuando las dimensiones transversas son menos de dos, las funciones
$H,W$ han de ser reemplazadas por funciones arm'onicas con un polo en
el espacio bidimensional 

\begin{equation}
H\sim h\log \rho\, ,
\hspace{1cm}  
W\sim \omega\log \rho\, ,
\end{equation}

\noindent o unidimensional transverso.

\begin{equation}
H\sim h \rho\, ,
\hspace{1cm}  
W\sim \omega \rho\, .
\end{equation}

Tambi'en en el caso de la D3-brana la soluci'on no es estrictamente
correcta porque faltan componentes de la 4-forma necesarias para
mantener la autodualidad que hay que calcular.

En el l'imite extremo, tenemos

\begin{equation} 
\begin{tabular}{|rcl|} 
\hline & & \\ & & \\ 
$d\tilde{\hat{s}}^{2}_{E}$ & $=$ & $H_{Dp}^{\frac{p-7}{8}}\left[dt^{2} -d\vec{y}_{p}^{\, 2} \right]
-H_{Dp}^{\frac{p+1}{8}}d\vec{x}_{9-p}^{\, 2}\, ,$ \\
& & \\
$d\hat{s}^{2}_{s}$ & $=$ & $H_{Dp}^{-1/2}\left[dt^{2} -d\vec{y}_{p}^{\, 2} \right]
-H_{Dp}^{1/2}d\vec{x}_{9-p}^{\, 2}\, ,$ \\
& & \\
$e^{-2(\hat{\phi}-\hat{\phi}_{0})}$ & $=$ & $ H_{Dp}^{\frac{(p-3)}{2}}\, ,$ \\
& & \\
$\hat{C}^{(p+1)}{}_{t\underline{y}^{1}\cdots \underline{y}^{p}}$ & $=$ & 
$\pm e^{-\hat{\phi}_{0}}\left(H_{Dp}^{-1}-1\right)\, ,$  \\
& & \\
$H_{Dp}$ & $=$ & $1+ {\displaystyle\frac{h_{Dp}}{|\vec{x}_{9-p}|^{7-p}}}\, .$ \\
& & \\ \hline
\end{tabular}
\label{eq:extremeDp}
\end{equation}

Como en el caso no-extremo, para $p=3$ hay que a~nadir las componentes
que faltan a la 4-forma y hay que sustituir $H_{D7}$ por

\begin{equation}
H_{D7} =   1+ h_{D7}\log{|\vec{x}_{2}|}\, ,
\end{equation}

\noindent y $H_{D8}$ por

\begin{equation}
H_{D8} =   1+ h_{D8}|x|\, ,
\end{equation}

Las constantes de integraci'on vienen dadas por

\begin{equation}
\begin{tabular}{|c|} 
\hline \\
$
h_{Dp} ={\displaystyle\frac{(2\pi\ell_{s})^{(7-p)}g}{(7-p)\omega_{(8-p)}}}\, ,
$ \\ \\ \hline
\end{tabular}  
\end{equation}

\noindent para $p<7$ y  

\begin{equation}
\begin{tabular}{|c|} 
\hline \\
$
h_{D7}=\, , \hspace{.5cm} h_{D8}={\displaystyle\frac{g}{4\pi \ell_{s}}}\, .
$ \\ \\ \hline
\end{tabular}  
\end{equation}

El signo negativo de $h_{D8}$ juega un papel muy importante.

%%%%%%%%%%%%%%%%%%%%%%%%%%%%%%%%%%%%%%%%%%%%%%%%%%%%%%%%%%%%%%%%%%%%%%
\section{Relaciones de dualidad}
\label{sec-relacionesdedualidad}

En la secci'on anterior hemos identificado una serie de candidatos a
ser las soluciones cl'asicas que representan los campos de largo
alcance producidos por los estados de las teor'ias de cuerdas
tipo~II cuya existencia conocemos bien a trav'es del espectro
perturbativo, bien a trav'es del super'algebra y las relaciones de
dualidad.

Para completar la identificaci'on debemos de comprobar que estas
soluciones guardan entre s'i las mismas relaciones de dualidad que
los estados que representan.

%%%%%%%%%%%%%%%%%%%%%%%%%%%%%%%%%%%%%%%%%%%%%%%%%%%%%%%%%%%%%%%%%%%%%%
\subsection{Relaci'on entre soluciones de 11- y 10-dimensionales. 
Compactificaci'on}

Las primeras relaciones que debemos comprobar son las que hay entre
estados/soluciones de SUGRA $N=1,d=11$ y de la Teor'ia de Cuerdas
tipo~IIA.

Hay dos relaciones posibles: o bien el objeto 10-dimensional se
obtiene por compactificaci'on del 11-dimensional en una direcci'on
del volumen del mundo o bien por compactificaci'on de una direcci'on
transversa.

Consideremos el caso m'as sencillo\footnote{Un caso a'un m'as
  sencillo es el de la compactificaci'on de la onda gravitacional
  plana en $d=11$ para dar una onda gravitacional plana o una D0-brana
  en $d=10$. Por falta de espacio no estamos discutiendo aqu'i las
  soluciones puramente gravitacionales como las ondas planas
  gravitacionales y el monopolo de Kaluza-Klein que, sin embargo,
  juegan papeles muy importantes. Las ondas gravitacionales har'an su
  aparici'on en la Secci'on~\ref{sec-intersections}.}: la
compactificaci'on de la M2-brana. Si la M2-brana est'a
compactificada en una direcci'on de su volumen del mundo (dicho de
otro modo: est'a enrollada en la dimensi'on que compactificamos), da
lugar a una F1 en $d=10$ y si es en una direcci'on transversa, da
lugar a una D2-brana.

Para compactificar la M2-brana tenemos que construir una soluci'on
bien definida cuando una coordenada es compacta. Para
compactificaciones en c'irculos o toros estos supone una peque~na
modificaci'on de las soluciones que hemos obtenido anteriormente.

Si una de las direcciones del volumen del mundo de la M2-brana extrema
({\it vgr}.~$y^{2}\equiv z$) es compacta, como, por construcci'on, la
soluci'on no depende de ella, la soluci'on Ec.~(\ref{eq:extremeM2})
sigue siendo completamente v'alida. Tan s'olo necesitamos introducir
el valor del dilat'on en el infinito multiplicando la coordenada
compacta $z$ por 'el:

\begin{equation} 
\left\{
  \begin{array}{rcl}
d\hat{\hat{s}}{}^{2} & = & 
H_{M2}^{-2/3}\left[dt^{2} -dy^{2}-e^{\frac{4}{3}\hat{\phi}_{0}}
dz^{2} \right]-H_{M2}^{1/3}\, d\vec{x}_{8}^{\, 2}\, ,\\
& & \\
\hat{\hat{C}}_{t\underline{y}\underline{z}} & = & 
\pm e^{\frac{2}{3}\hat{\phi}_{0}}\left(H_{M2}^{-1}-1\right)\, ,  \\
& & \\
H_{M2}& = & 1+ {\displaystyle\frac{h_{M2}}{|\vec{x}_{8}|^{6}}}\, . \\
\end{array}
\right.
\label{eq:extremeM2compacwv}
\end{equation}

\noindent Ahora podemos utilizar directamente las
f'ormulas Ecs.~(\ref{eq:11vs10fields}) y (esto es muy importante)
Ecs.~(\ref{eq:rescalings10-11}) para hallar una soluci'on
10-dimensional que tiene la forma de la de la D2-brana extrema
Ec.~(\ref{eq:extremeDp}) pero con la funci'on arm'onica $H_{M2}$ en
vez de $H_{D2}$. S'olo necesitamos, pues, comparar estas funciones,
que son id'enticas excepto por las constantes de integraci'on
$h_{M2}$ y $h_{D2}$. Sin embargo, utilizando las relaciones entre las
constantes 11- y 10-dimensionales Ecs.~(\ref{eq:10-11constants}) para
ver que

\begin{equation}
h_{M2} =\frac{(\ell_{\rm   Planck}^{(11)})^{6}}{6\omega_{(7)}}  
= \frac{(2\pi \ell_{s}g^{1/3})^{6}}{6\omega_{(7)}}= 
 \frac{(2\pi \ell_{s})^{6}g^{2}}{6\omega_{(7)}}
=h_{D2}\, .
\end{equation}

Si es una de las direcciones transversas la que es compacta, entonces
es la funci'on arm'onica $H_{M2}$ la que hay cambiar, puesto que la
que aparece en Ec.~(\ref{eq:extremeM2}) resuelve la ecuaci'on de
Laplace con una fuente puntual en el espacio euclidiano, pero no con
una dimensi'on compacta. El procedimiento corriente para encontrar la
funci'on arm'onica correcta explota el hecho de que las M2-branas
son objetos supersim'etricos (como veremos) y se puede poner a un
n'umero arbitrario de M2-branas paralelas en equilibrio est'atico en
posiciones arbitrarias. Esta libertad est'a asociada a la
arbitrariedad en la elecci'on de la funci'on arm'onica $H_{M2}$ en
ausencia de fuentes. La idea consiste en poner un n'umero infinito de
M2-branas paralelas a intervalos regulares de la coordenada
$x^{10}\equiv z$, siendo el periodo igual a la longitud del
c'irculo en el que vamos a compactificar $z$. Esto corresponde a
elegir una funci'on arm'onica

\begin{equation}
H_{M2} = 1 +h_{M2}\sum_{n=-\infty}^{n=+\infty} 
\frac{1}{(|\vec{x}_{7}|^{2} +(z+2\pi n R_{11})^{2}|)^{3}} \, ,
\end{equation}

\noindent donde esta vez hemos supuesto que $z\in [0,2\pi R_{11}]$ 
por razones t'ecnicas. Esta funci'on es peri'odica en $z$ y tiene
un polo en $\vec{x}_{7}=z=0$, como quer'iamos. 

Sin embargo, ahora no podemos utilizar directamente las f'ormulas de
compactificaci'on usuales porque la soluci'on depende
expl'icitamente de la coordenada peri'odica. Hay dos estrategias
que se pueden seguir: una es sumar la serie anterior para obtener la
funci'on peri'odica, expandirla en serie de Fourier y quedarnos con
el modo cero. Un poco m'as sencillo es simplemente aproximar la suma
por una integral, lo que es v'alido si $|\vec{x}_{7}|>>R_{11}$
\cite{kn:Pee2}. Primero hacemos el cambio de variables

\begin{equation}
u_{n}=\frac{(z-2\pi nR_{11})}{|\vec{x}_{7}|}\, ,
\hspace{1cm}
u_{n}\in[\frac{2\pi n R_{11}}{|\vec{x}_{7}|},
\frac{2\pi (n+1) R_{11}}{|\vec{x}_{7}|}]\, ,  
\end{equation}

\noindent con lo que tenemos 

\begin{equation}
  \begin{array}{rcl}
H_{M2} & = & 1 +{\displaystyle
\frac{h_{M2}}{|\vec{x}_{7}|^{6}}\sum_{n=-\infty}^{n=+\infty} 
\frac{1}{(1+ u_{n}^{2})^{3}}}
\sim 1+ {\displaystyle\frac{h_{M2}}{|\vec{x}_{7}|^{6}} 
\frac{1}{2\pi R_{11}/|\vec{x}_{7}|} \int_{-\infty}^{+\infty} 
\frac{du}{(1+u^{2})^{3}}}\\
& & \\
& = &  1+ {\displaystyle\frac{h_{M2}\omega_{(5)}}{2\pi R_{11}\omega_{(4)}}
\frac{1}{|\vec{x}_{7}|^{5}}}\, .\\ 
\end{array}
\end{equation}

Sustituyendo esta funci'on arm'onica en Ec.~(\ref{eq:extremeM2}) con
$x^{10}=z$, ya podemos utilizar las f'ormulas
Ecs.~(\ref{eq:11vs10fields}) para reducir la soluci'on a diez
dimensiones\footnote{No es necesario hacer los cambios de escala
  Ecs.~(\ref{eq:rescalings10-11}) porque el valor del dilat'on en el
  infinito est'a absorbido en la definici'on de $z$. Para
  recuperarlo simplemente hay que reescalear consistentemente los
  campos 10-dimensionales.} y obtenemos la soluci'on de la D2-brana
extrema Ec.~(\ref{eq:extremeDp}) exactamente puesto que

\begin{equation}
\frac{h_{M2}\omega_{(5)}}{2\pi R_{11}\omega_{(4)}} = h_{D2}\, .  
\end{equation}

De la misma forma podemos comprobar que la M5 se reduce a la D4 y la
S5A. Es 'util la siguiente f'ormula general:

\begin{equation}
H_{p} = 1 +h_{p}\sum_{n=-\infty}^{n=+\infty} 
\frac{1}{(|\vec{x}_{n+1}|^{2} +(z+2\pi n R)^{2}|)^{n/2}}
\sim
 1+ \frac{h_{p} \omega_{(n-1)}}{2\pi R \omega_{(n-2)}} 
\frac{1}{|\vec{x}_{n+1}|^{n-1}}\, ,
\end{equation}

\noindent que nos da la relaci'on entre el coeficiente $h_{p}$ de una 
$p$-brana y el $h_{p}^{\prime}$ de la $p$-brana que se obtiene al
compactificar la otra en una direcci'on transversa de radio $R$:

\begin{equation}
h^{\prime}_{p} = \frac{h_{p} \omega_{(n-1)}}{2\pi R \omega_{(n-2)}} \, . 
\end{equation}

Si repetimos el procedimiento $m$ veces (compactificamos en $T^{m}$),
tenemos

\begin{equation}
\label{eq:esarelacion}
h^{\prime}_{p} = \frac{h_{p} \omega_{(n-1)}}{V^{m}\omega_{(n-m-1)}} \, ,
V^{m}=(2\pi)^{m}R_{1}\ldots R_{m}\, .
\end{equation}

%%%%%%%%%%%%%%%%%%%%%%%%%%%%%%%%%%%%%%%%%%%%%%%%%%%%%%%%%%%%%%%%%%%%%%
\subsection{Dualidad~T entre soluciones 10-dimensionales}

Las relaciones de dualidad~T m'as importantes son las que ligan a
todas las D$p$-branas y consisten en que una D$p$-brana con una
dimensi'on transversa compactificada est'a relacionada con una
D$(p+1)$-brana con una direcci'on del volumen del mundo
compactificada con el radio dual y con la constante de acoplo de la
cuerda dual.

Para comprobar esta relaci'on empezamos por aplicar el algoritmo que
acabamos de explicar para obtener la funci'on arm'onica $H_{Dp}$
peri'odica y su aproximaci'on independiente de la coordenada
transversa $z$. Aplicando las reglas de Buscher tipo~II
Ecs.~(\ref{eq:reglasBuscherIIAtoIIB}) y
(\ref{eq:reglasBuscherIIBtoIIA}), encontramos inmediatamente la
soluci'on de la D$(p+1)$-brana con una funci'on arm'onica con
coeficiente ($p\leq 6$)

\begin{equation}
\frac{h_{Dp}\omega_{(6-p)}}{2\pi R\omega_{(5-p)}}=
\frac{(2\pi \ell_{s})^{7-p}g}{2\pi R} 
\frac{\omega_{(6-p)}}{(7-p)\omega_{(8-p)} \omega_{(6-p)}}\, .
\end{equation}

\noindent Usando en el primer factor las reglas de transformaci'on 
de $g$ y $R$ bajo dualidad~T Ec.~(\ref{eq:tdual}) y usando la
identidad

\begin{equation}
\frac{\omega_{(n-1)}}{n\omega_{(n+1)}\omega_{(n-2)}}=
\frac{1}{(n-1)\omega_{(n)}}\, ,  
\end{equation}

\noindent obtenemos $h_{D(p+1)}(g^{\prime})$, como esper'abamos.

%%%%%%%%%%%%%%%%%%%%%%%%%%%%%%%%%%%%%%%%%%%%%%%%%%%%%%%%%%%%%%%%%%%%%%
\section{Supersimetr'ias residuales}

Para finalizar las comprobaciones sobre nuestra identificaci'on de
soluciones y estados s'olo nos queda ver que las soluciones de
$p$-branas extremas tienen supersimetr'ias residuales (claramente
las no-extremas no las tienen). Para comprobarlo hay que resolver las
ecuaciones de los espinores de Killing correspondientes a las SUGRAS
de las que sean soluciones. Vamos a empezar con las $p$-branas
11-dimensionales.

%%%%%%%%%%%%%%%%%%%%%%%%%%%%%%%%%%%%%%%%%%%%%%%%%%%%%%%%%%%%%%%%%%%%%%
\subsection{Supersimetr'ias residuales de la M2-brana}

Las ecuaciones de los espinores de Killing en SUGRA $N=1,d=11$ son
simplemente
$\delta_{\hat{\hat{\epsilon}}}\hat{\hat{\psi}}_{\hat{\hat{\mu}}}=0$
donde la variaci'on del gravitino viene dada por la
Ec.~(\ref{eq:d11susyrules}). No tenemos m'as que sustituir las
componentes de la conexi'on de esp'in y de la 4-forma de la
soluci'on Ec.~(\ref{eq:extremeM2}). Eligiendo las t'etradas

\begin{equation}
\left\{
  \begin{array}{rcl}
\hat{\hat{e}}_{\underline{i}}{}^{j} & = & H_{M2}^{-1/3}\delta_{i}{}^{j}\, ,\\
& & \\
\hat{\hat{e}}_{\underline{m}}{}^{n} & = & H_{M2}^{1/6}\delta_{m}{}^{n}\, ,\\
  \end{array}
\right.
\end{equation}

\noindent encontramos las componentes no nulas

\begin{equation}
\left\{
  \begin{array}{rcl}
\hat{\hat{\omega}}_{\underline{m}}{}^{nl} & = & -\frac{1}{3}
H_{M2}^{-1}\partial_{\underline{q}}H_{M2}\eta_{m}{}^{[n}\eta^{p]q}\, ,\\
& & \\
\hat{\hat{\omega}}_{\underline{i}}{}^{mj} & = & \frac{2}{3}
H_{M2}^{-3/2}\partial_{\underline{q}}H_{M2}\eta_{i}{}^{[m}\eta^{j]q}\, ,\\
& & \\
\hat{\hat{G}}_{\underline{m}ty^{1}y^{2}} & = & \mp H_{M2}^{-2}
\partial_{\underline{m}} H_{M2}\, ,\\
  \end{array}
\right.
\end{equation}

\noindent y sustituyendo en las ecuaciones de los espinores de Killing, 
llegamos a

\begin{equation}
\left\{
  \begin{array}{rcl}
\delta_{\hat{\hat{\epsilon}}}\hat{\hat{\psi}}_{\underline{i}}
& = & \frac{1}{3}H_{M2}^{-3/2} \partial_{\underline{n}}H_{M2}
\hat{\hat{\Gamma}}_{(i)}{}^{n}
\left(1\mp \frac{i}{2} \epsilon_{(i)jk}
\hat{\hat{\Gamma}}{}^{(i)jk}\right)\hat{\hat{\epsilon}}=0\, ,\\
& & \\
\delta_{\hat{\hat{\epsilon}}}\hat{\hat{\psi}}_{\underline{m}}
& = & 
2\partial_{\underline{m}}\hat{\hat{\epsilon}}
-\frac{1}{6}H_{M2}^{-1}\partial_{\underline{n}} H_{M2}
\left[\hat{\hat{\Gamma}}{}^{mn} 
\mp i\left(\hat{\hat{\Gamma}}{}^{mn} +2\delta^{mn}\right) 
\hat{\hat{\Gamma}}{}^{012}
\right]\hat{\hat{\epsilon}} =0\, .\\
  \end{array}
\right.
\end{equation}

\noindent La primera ecuaci'on se resuelve s'olo si $\hat{\hat{\epsilon}}$
satisface la condici'on algebraica

\begin{equation}
\left(1\mp i\hat{\hat{\Gamma}}{}^{012} \right)\hat{\hat{\epsilon}}=0\, .
\end{equation}

\noindent Utiliz'andola en la segunda ecuaci'on, 'esta se transforma en

\begin{equation}
\delta_{\hat{\hat{\epsilon}}}\hat{\hat{\psi}}_{\underline{m}}
= 2\left(\partial_{\underline{m}} +{\textstyle\frac{1}{6}}
H_{M2}^{-1}\partial_{\underline{m}}H_{M2} \right)\hat{\hat{\epsilon}}=0\, ,
\end{equation}

\noindent cuya soluci'on es $\hat{\hat{\epsilon}}=
H_{M2}^{-1/6}\hat{\hat{\epsilon}}_{0}$, donde
$\hat{\hat{\epsilon}}_{0}$ es un espinor constante. As'i hemos
encontrado los espinores de Killing

\begin{equation}
  \begin{tabular}{|c|}
\hline \\
$\hat{\hat{\epsilon}}=
H_{M2}^{-1/6}\, \hat{\hat{\epsilon}}_{0}\, ,
\hspace{1cm}
\left(1\mp i\hat{\hat{\Gamma}}{}^{012} \right)\hat{\hat{\epsilon}}_{0}=0\, .
$
\\ \\ \hline
  \end{tabular}
\end{equation}

\noindent La condici'on algebraica no es m'as que el proyector
Ec.~(\ref{eq:proyectorpbrana}), como esper'abamos, y, debido a ella,
s'olo la mitad de las componentes de $\hat{\hat{\epsilon}}_{0}$ son
independientes y s'olo $1/2$ de las posibles supersimetr'ias son
preservadas. Lo mismo va a ocurrir en los siguientes soluciones que
describen un 'unico objeto BPS.

Obs'ervese que incluimos los dos posibles signos de la carga.  Para
una sola M2-brana, el signo es irrelevante, pero es a veces crucial
cuando hay otras branas presentes y se estudia la supersimetr'ia
del conjunto.

Obs'ervese tambi'en que el espinor de Killing existe para una
funci'on arm'onica $H_{M2}$ arbitraria. Para ser rigurosos,
deber'iamos de decir que la soluci'on es supersim'etrica si los
espinores de Killing adem'as tienen el comportamiento asint'otico
adecuado: si la soluci'on tiende asint'oticamente al vac'io, los
espinores de Killing han de tender asint'oticamente a los espinores
de Killing del vac'io y adem'as son normalizables. Si $H_{M2}$ es
la funci'on arm'onica para una o varias M2-branas paralelas, estas
condiciones se cumplen.

%%%%%%%%%%%%%%%%%%%%%%%%%%%%%%%%%%%%%%%%%%%%%%%%%%%%%%%%%%%%%%%%%%%%%%
\subsection{Supersimetr'ias residuales de la M5-brana}

El c'alculo de los espinores de Killing de la M5-brana sigue la pauta
del de los de la M2-brana: elegimos una t'etrada

\begin{equation}
\left\{
  \begin{array}{rcl}
\hat{\hat{e}}_{\underline{i}}{}^{j} & = & H_{M5}^{-1/6}\delta_{i}{}^{j}\, ,\\
& & \\
\hat{\hat{e}}_{\underline{m}}{}^{n} & = & H_{M5}^{1/3}\delta_{m}{}^{n}\, ,\\
  \end{array}
\right.
\end{equation}

\noindent y calculamos las componentes no-nulas de la conexi'on 
de esp'in y la intensidad de campo

\begin{equation}
\left\{
  \begin{array}{rcl}
\hat{\hat{\omega}}_{\underline{m}}{}^{nl} & = & -\frac{2}{3}
H_{M5}^{-1}\partial_{\underline{q}}H_{M2}\eta_{m}{}^{[n}\eta^{p]q}\, ,\\
& & \\
\hat{\hat{\omega}}_{\underline{i}}{}^{mj} & = & \frac{1}{3}
H_{M5}^{-3/2}\partial_{\underline{q}}H_{M2}\eta_{i}{}^{[m}\eta^{j]q}\, ,\\
& & \\
\hat{\hat{G}}_{\underline{m}_{1}\cdots \underline{m}_{4}} & = & \pm
\epsilon_{m_{1}\cdots m_{5}}\partial_{\underline{m}_{5}} H_{M5}\, .\\
  \end{array}
\right.
\end{equation}

Al sustituir en las ecuaciones de los espinores de Killing vemos de
nuevo que las ecuaciones correspondientes a las componentes del
gravitino que est'an en las direcciones del volumen del mundo de la
M5-brana se resuelven si el espinor satisface una condici'on
algebraica que, al ser sustituida en las otras ecuaciones las
simplifica. El resultado es

\begin{equation}
  \begin{tabular}{|c|}
\hline \\
$\hat{\hat{\epsilon}}=
H_{M5}^{-1/12}\, \hat{\hat{\epsilon}}_{0}\, ,
\hspace{1cm}
\left(1\mp \hat{\hat{\Gamma}}{}^{012345} \right)\hat{\hat{\epsilon}}_{0}=0\, .
$
\\ \\ \hline
  \end{tabular}
\end{equation}

%%%%%%%%%%%%%%%%%%%%%%%%%%%%%%%%%%%%%%%%%%%%%%%%%%%%%%%%%%%%%%%%%%%%%%
\subsection{Supersimetr'ias residuales de las D$p$-branas}

Teniendo en cuenta que la 2-forma NSNS es cero para estas soluciones y
que la 'unica intensidad de campo que no es cero es
$\hat{G}^{(p+2)}$, las ecuaciones de los espinores de Killing toman la
forma

\begin{equation}
\left\{
\begin{array}{rcl}
\delta_{\hat{\epsilon}}\hat{\psi}_{\hat{\mu}} & = & 
\left\{\partial_{\hat{\mu}} -\frac{1}{4}\not\!\hat{\omega}_{\hat{\mu}}
+\frac{i}{8} e^{\hat{\phi}} \frac{1}{(p+2)!} \not\! \hat{G}^{(p+2)}
\hat{\Gamma}_{\hat{\mu}}\left(-\hat{\Gamma}_{11}\right)^{\frac{p+2}{2}}
\right\}\hat{\epsilon}\, ,\\
& & \\
\delta_{\hat{\epsilon}}\hat{\lambda} & = & 
\left\{\not\partial\hat{\phi}  
-\frac{i}{4} e^{\hat{\phi}} \frac{(p-3)}{(p+2)!} \not\! \hat{G}^{(p+2)}
\left(-\hat{\Gamma}_{11}\right)^{\frac{p+2}{2}}\right\}\hat{\epsilon}\, ,
\end{array}
\right.
\end{equation}

\noindent para la teor'ia tipo~IIA y

\begin{equation}
\left\{
\begin{array}{rcl}
\delta_{\hat{\varepsilon}}\hat{\zeta}_{\hat{\mu}} & = & 
\left\{\partial_{\hat{\mu}} -\frac{1}{4}\not\!\hat{\omega}_{\hat{\mu}}
+\frac{1}{8} e^{\hat{\varphi}} \frac{1}{(p+2)!} \not\! \hat{G}^{(p+2)}
\hat{\Gamma}_{\hat{\mu}}{\cal P}_{\frac{p+3}{2}}
\right\}\hat{\varepsilon}\, ,\\
& & \\
\delta_{\hat{\varepsilon}}\hat{\chi} & = & 
\left\{\not\partial\hat{\varphi}  
+\frac{1}{4} e^{\hat{\varphi}} \frac{(p-3)}{(p+2)!} \not\! \hat{G}^{(p+2)}
{\cal P}_{\frac{p+3}{2}}
\right\}\hat{\varepsilon}\, ,
\end{array}
\right.
\end{equation}

\noindent para la tipo~IIB, donde

\begin{equation}
{\cal P}_{n}=
\left\{
\begin{array}{c}
\sigma^{1}\, ,\,\,\, n\,\, {\rm par}\, ,\\
\\
i\sigma^{2}\, ,\,\,\, n\,\, {\rm impar}\, ,\\
\end{array}
\right.
\end{equation}

\noindent Como en los casos anteriores s'olo necesitamos calcular las
componentes de la conexi'on de esp'in y las intensidades de campo
usando la soluci'on Ec.~(\ref{eq:extremeDp}):

\begin{equation}
\left\{
\begin{array}{rcl}
\not\!\hat{\omega}_{\underline{i}} & = & 
-\frac{1}{2} H_{Dp}^{-3/2} \partial_{\underline{n}}H_{Dp} 
\eta_{ij}\Gamma^{nj}\, ,\\
& & \\
\not\!\hat{\omega}_{\underline{m}} & = & 
\frac{1}{2} H_{Dp}^{-1} \partial_{\underline{n}}H_{Dp} 
\eta_{mq}\hat{\Gamma}^{nq}\, ,\\
& & \\
\not\! \hat{G}^{(p+2)} & = & \mp e^{-\hat{\phi}_{0}} H_{Dp}^{\frac{p}{4}-2} 
\partial_{\underline{m}} H_{Dp} \hat{\Gamma}^{m} \hat{\Gamma}^{01\cdots p}\, .
\end{array}
\right.
\end{equation}

Para hallar una soluci'on primero resolvemos las ecuaciones
correspondientes a las variaciones del dilatino y las componentes del
gravitino en las direcciones del volumen del mundo de las D$p$-branas,
obteniendo como antes una ecuaci'on algebraica que es una ligadura
restringe el n'umero de componentes independientes de los espinores
de Killing. La misma ligadura resuelve todas estas ecuaciones.
Despu'es usamos la ligadura para resolver las ecuaciones
correspondientes a las componentes del gravitino en las direcciones
transversas. El resultado es un sistema de ecuaciones diferenciales
desacopladas que determinan la dependencia del espinor de Killing en
las direcciones transversas a la D$p$-brana $\vec{x}_{(9-p)}$ salvo
por una constante global que corresponde a la normalizaci'on del
espinor. La dependencia en $\vec{x}_{(9-p)}$ est'a concentrada en un
factor que multiplica a un espinor constante $\hat{\epsilon}_{0}$ que
satisface la ligadura algebraica.  El resultado es

\begin{equation}
\begin{tabular}{|crclrcl|}
\hline 
& & & & & & \\
{\bf IIA:} & $\hat{\epsilon}$ & $=$ & 
$H_{Dp}^{-1/8} \hat{\epsilon}_{0}\, ,\hspace{.5cm}$ &
$\left[ 1\mp i\hat{\Gamma}^{01\cdots p} 
\left(-\hat{\Gamma}_{11} \right)^{\frac{p+2}{2}} 
\right]\hat{\epsilon}_{0}$ & $=$ & $0$\, , \\
& & & & & & \\
{\bf IIB:} & $\hat{\varepsilon}$ & $=$ & 
$H_{Dp}^{-1/8} \hat{\varepsilon}_{0}\, ,\hspace{.5cm}$ &
$\left( 1\pm i\hat{\Gamma}^{01\cdots p} {\cal P}_{\frac{p+3}{2}}
\right)\hat{\varepsilon}_{0}$ & $=$ & $0$\, . \\
& & & & & & \\ \hline
\end{tabular}
\end{equation}

%%%%%%%%%%%%%%%%%%%%%%%%%%%%%%%%%%%%%%%%%%%%%%%%%%%%%%%%%%%%%%%%%%%%%%
\subsection{Supersimetr'ias residuales de la F1}

Podemos buscar simult'aneamente los espinores de Killing de la F1 extrema
como soluci'on de las teor'ias tipo~IIA, IIB y heter'otica
escribiendo

\begin{equation}
\left\{
\begin{array}{rcl}
\delta_{\hat{\epsilon}}\hat{\psi}_{\hat{\mu}} & = & 
\left\{\partial_{\hat{\mu}} -\frac{1}{4} \left( \not\!\hat{\omega}_{\hat{\mu}}
+\frac{1}{2}\not\!\! \hat{H}_{\hat{\mu}}{\cal O} \right)
\right\}\hat{\epsilon}\, ,\\
& & \\
\delta_{\hat{\epsilon}}\hat{\lambda} & = & 
\left\{\not\partial\hat{\phi}  
-\frac{1}{12}\not\!\! \hat{H} {\cal O}
\right\}\epsilon\, ,\\
\end{array}
\right.
\end{equation}

\noindent donde  $\epsilon$ es un espinor de Majorana, un par de espinores 
Majorana-Weyl o un 'unico espinor de Majorana y donde ${\cal O}
=\Gamma_{11},\sigma^{3},\mathbb{I}$, respectivamente. Usando la forma
expl'icita de la soluci'on Ecs.~(\ref{eq:extremeF1}), encontramos

\begin{equation}
\begin{array}{rcl}
\not\!\omega_{\underline{i}} & = & 
-H_{F1}^{-3/2} \partial_{\underline{m}} H_{F1} \Gamma_{i}{}^{m}\, , \\
& & \\
\not\!\! H_{\underline{i}} & = & 
\mp 2 H_{F1}^{-1} \partial_{\underline{m}} H_{F1} \Gamma^{01}\, ,\\
& & \\
\not\!\! H_{\underline{m}} & = & \pm \epsilon_{ij} H_{F1}^{-3/2}
\partial_{\underline{m}} H_{F1} \Gamma^{mj}\, ,\\
\end{array}
\end{equation}

\noindent y, siguiendo los mismos pasos que en el caso de las D$p$-branas,
encontramos la soluci'on

\begin{equation}
\begin{tabular}{|rclrcl|}
\hline 
& & & & & \\
$\hat{\epsilon}$ & $=$ & $H_{F1}^{1/4} \hat{\epsilon}_{0}\, ,\hspace{.5cm}$ &
$\left( 1\pm \hat{\Gamma}^{01} {\cal O} \right)\hat{\epsilon}_{0}$ & $=$ & $0$\, .\\
& & & & & \\ \hline
\end{tabular}
\end{equation}

%%%%%%%%%%%%%%%%%%%%%%%%%%%%%%%%%%%%%%%%%%%%%%%%%%%%%%%%%%%%%%%%%%%%%%
\subsection{Supersimetr'ias residuales de la S5}

De nuevo podemos hacer el c'alculo simult'aneo para las tres
teor'ias escribiendo

\begin{equation}
\left\{
\begin{array}{rcl}
\delta_{\hat{\epsilon}}\hat{\psi}_{\hat{\mu}} & = & 
\left\{\partial_{\hat{\mu}} -\frac{1}{4} \left( \not\!\hat{\omega}_{\hat{\mu}}
+\frac{1}{7!} e^{2\hat{\phi}}\not\!\! 
\hat{H}^{(7)\, \hat{a}_{1}\cdots \hat{a}_{7}}
\hat{\Gamma}_{\hat{\mu} \hat{a}_{1}\cdots \hat{a}_{7}}{\cal O} \right)
\right\}\hat{\epsilon}\, ,\\
& & \\
\delta_{\hat{\epsilon}}\hat{\lambda} & = & 
\left\{\not\!\partial\hat{\phi}
+\frac{1}{2}\not\!\! \hat{H}^{(7)} {\cal O}
\right\}\hat{\epsilon}\, ,\\
\end{array}
\right.
\end{equation}

\noindent donde ahora ${\cal O}=\sigma^{3}$ para la~IIB
y $\mathbb{I}$ para las otras dos. El resultado es

\begin{equation}
\begin{tabular}{|rclrcl|}
\hline 
& & & & & \\
$\hat{\epsilon}$ & $=$ & $\hat{\epsilon}_{0}\, ,\hspace{.5cm}$ &
$\left( 1\pm \hat{\Gamma}^{0\cdots 5} {\cal O} \right)\hat{\epsilon}_{0}$ 
& $=$ & $0$\, .\\
& & & & & \\ \hline
\end{tabular}
\end{equation}

%%%%%%%%%%%%%%%%%%%%%%%%%%%%%%%%%%%%%%%%%%%%%%%%%%%%%%%%%%%%%%%%%%%%%%%

%\subsection{Incremento de supersimetr'ia residual en el horizonte}

%??????

\newpage
\chapter{Agujeros negros en Supercuerdas}

%%%%%%%%%%%%%%%%%%%%%%%%%%%%%%%%%%%%%%%%%%%%%%%%%%%%%%%%%%%%%%%%%%%%%%
\section{Introducci'on}

En la 'ultima lecci'on vimos que hay soluciones de las teor'ias
efectivas de las cuerdas (SUEGRAS) que describen con gran precisi'on
los campos de largo alcance de los objetos extensos que aparecen en el
espectro de las teor'ias de cuerdas correspondientes. En general
no son agujeros negros, cuya descripci'on en el marco de la
Teor'ia de Cuerdas es nuestro objetivo. En esta lecci'on vamos a
dar los 'ultimos pasos para encontrar soluciones de tipo agujero
negro extremo\footnote{Los no-extremos se construyen f'acilmente a
  partir de los extremos, pero en general, por falta de espacio, vamos
  a dedicarnos a describir en detalle s'olo los extremos.} que
correspondan a estados de la Teor'ia de Cuerdas. 

Primeramente vamos a probar a construirlos por compactificaci'on
toroidal de las direcciones del volumen del mundo de las soluciones de
$p$-branas que encontramos en le lecci'on anterior. Los agujeros
negros que encontramos tienen horizontes singulares, de volumen cero.
%Vamos a arg\"uir, usando el {\it Principio de correspondencia}, que
%cerca de las singularidades de la geometr'ia cl'asica hay efectos
%cuerd'isticos que han de ser tomados en cuenta y modifican esos
%resultados cl'asicos.

A nosotros nos gustar'ia es obtener un modelo de un agujero negro
extremo cuya geometr'ia cl'asica sea no-singular, cuyo horizonte
tenga un volumen distinto de cero. Para esto hay que construir primero
soluciones que describan m'as de un objeto extenso: intersecciones.
Vamos a estudiar las condiciones bajo las que existen y las soluciones
cl'asicas de las SUEGRAS que las describen.

Inmediatamente despu'es vamos a compactificar una de estas soluciones
que representan intersecciones para obtener en $d=5$ un agujero negro
extremo completamente regular. Realmente es una familia uno de cuyos
miembros es el conocido agujero extremo de Reissner y Nordstr\"om en
$d=5$. 

Tras este primer 'exito, vamos a ver que la Teor'ia de Cuerdas
nos permite calcular el valor de su entrop'ia a partir del
recuento de microestados de la Teor'ia de Campos Conformes de la
cuerda en el vac'io de las $p$-branas correspondientes. Este era
nuestro gran objetivo al comenzar este curso.

Finalmente vamos a enumerar cosas que por falta de espacio se nos han
quedado fuera del tintero.

%%%%%%%%%%%%%%%%%%%%%%%%%%%%%%%%%%%%%%%%%%%%%%%%%%%%%%%%%%%%%%%%%%%%%%
\section{Agujeros de una sola $p$-brana}

Las soluciones naturales de las teor'ias efectivas de cuerdas son
objetos extensos (salvo la D0-brana). El procedimiento m'as sencillo
para generar soluciones de tipo agujero negro es compactificar las
anteriores en las direcciones del volumen del mundo. Recordemos que en
esta operaci'on la funci'on arm'onica no cambia.

Es f'acil ver que la cuerda fundamental F1 Ec.~(\ref{eq:extremeF1})
enrollada en un c'irculo da lugar al agujero negro cargado extremo
en $d=9$

\begin{equation}
\left\{
  \begin{array}{rcl}
d\tilde{s}^{2}_{E} & = & H_{F1}^{-6/7}dt^{2} 
-H_{F1}^{1/7}d\vec{x}_{8}^{\, 2}\, ,\\ 
& & \\
ds^{2}_{s} & = & H_{F1}^{-1}dt^{2} -d\vec{x}_{8}^{\, 2}\, ,\\ 
& & \\
A_{t} & = & \pm \left(H_{F1}^{-1} -1\right)\, ,\\
& & \\
e^{-2(\phi-\phi_{0})} & = & H_{F1}^{1/2}\, ,\\
& & \\
K/K_{0} & = & H_{F1}^{-1/2}\, .\\
  \end{array}
\right.  
\end{equation}

El horizonte est'a en el polo de $H_{F1}$ $\vec{x}_{8}=0$. En ese
punto, el volumen del c'irculo compacto, medido por el {\it
  m'odulo} $K$ va a cero y el dilat'on diverge. Adem'as el 'area
del horizonte es cero.

Una 5-brana solit'onica S5 Ec.~(\ref{eq:extremeS5}) enrollada en un
5-toro da lugar al siguiente agujero negro cargado extremo en cinco
dimensiones:

\begin{equation}
\left\{
  \begin{array}{rcl}
d\tilde{s}^{2}_{E} & = & H_{S5}^{-2/3}dt^{2} 
-H_{S5}^{1/3}d\vec{x}_{4}^{\, 2}\, ,\\ 
& & \\
ds^{2}_{s} & = & dt^{2} -H_{S5}d\vec{x}_{4}^{\, 2}\, ,\\ 
& & \\
A_{t} & = & \pm \left(H_{S5}^{-1} -1\right)\, ,\\
& & \\
e^{-2(\phi-\phi_{0})} & = & H_{S5}^{-1}\, ,\\
& & \\
K/K_{0} & = & 1\, .\\
  \end{array}
\right.  
\end{equation}

En este caso el volumen del 5-toro es finito en el horizonte, y el
dilat'on va a a cero, pero el horizonte sigue teniendo 'area cero en
el sistema de referencia de Einstein (aunque es finito en el de la cuerda).

Las D$p$-branas Ec.~(\ref{eq:extremeDp}) enrolladas en un p-toros dan
lugar a los siguientes agujeros negros cargados extremos en $10-p$
dimensiones:

\begin{equation}
\left\{
  \begin{array}{rcl}
d\tilde{s}^{2}_{E} & = & H_{Dp}^{-\frac{7-p}{8-p}}dt^{2} 
-H_{Dp}^{\frac{1}{(8-p)}}d\vec{x}_{9-p}^{\, 2}\, ,\\ 
& & \\
ds^{2}_{s} & = & H_{Dp}^{-1/2}dt^{2} -H_{Dp}^{1/2}d\vec{x}_{4}^{\, 2}\, ,\\ 
& & \\
A_{t} & = & \pm \left(H_{Dp}^{-1} -1\right)\, ,\\
& & \\
e^{-2(\phi-\phi_{0})} & = & H_{Dp}^{\frac{p-6}{4}}\, ,\\
& & \\
K/K_{0} & = & H_{Dp}^{-p/4}\, .\\
  \end{array}
\right.  
\end{equation}

En todos los casos el volumen del $p$-toro y tambi'en (excepto si $p
\geq 6$) el dilat'on son singulares en el horizonte. El 'area del
horizonte en el sistema de Einstein es siempre cero, aunque para $p=3$
no lo sea en el sistema de referencia conforme de la cuerda.

Podemos compactificar tambi'en dimensiones transversas o una
combinaci'on de ambas, pero en todos los casos vamos a obtener
agujeros negros con horizontes singulares de 'area cero, porque todos
ellos son m'aximamente sim'etricos y sabemos que no hay agujeros
negros regulares m'aximamente supersim'etricos en el sistema de
referencia de Einstein. Esta observaci'on nos da una pista para
construir agujeros regulares: romper m'as supersimetr'ia a
trav'es de intersecciones de $p$-branas.

%%%%%%%%%%%%%%%%%%%%%%%%%%%%%%%%%%%%%%%%%%%%%%%%%%%%%%%%%%%%%%%%%%%%%%%
%\section{El {\it Principio de Correspondencia}}???????

%%%%%%%%%%%%%%%%%%%%%%%%%%%%%%%%%%%%%%%%%%%%%%%%%%%%%%%%%%%%%%%%%%%%%%
\section{Reglas de intersecci'on}
\label{sec-intersections}

Una vez obtenidas las soluciones correspondientes a objetos/estados
``elementales" es l'ogico preguntarse por la existencia de estados
ligados en los que hay presentes varios objetos de tipos distintos, y
por las soluciones que los describen. En las teor'ias
supersim'etricas hay un tipo de estados ligados particularmente
f'acil de describir: aquellos en los que la energ'ia de ligadura
es cero, como cuando hay varios objetos BPS del mismo tipo paralelos y
en equilibrio est'atico formando un estado BPS. Este tipo de estados
juega un papel muy importante en lo que sigue. Si queremos tener
varios objetos supersim'etricos formando un estado supersim'etrico,
podemos empezar por estudiar el super'algebra para ver en qu'e
condiciones es compatible imponer la aniquilaci'on de un mismo estado
por las cargas de supersimetr'ia asociadas a los distintos objetos
(que pueden ser iguales, pero no paralelos) que en general tendr'an
varias direcciones (aparte del tiempo) comunes, es decir: se
intersectar'an.

Como vimos en su momento, la aniquilaci'on de los estados por las
supercargas era completamente equivalente a la acci'on de ciertos
proyectores sobre espinores. Para una $p$-brana extendida en las
direcciones $y^{1}\ldots y^{p}$, que podemos representar as'i
($p=5$)

\begin{equation}
  \begin{array}{r||c|ccccccccc}
5-{\rm brana} & + & + & + & + & + & + & - & - & - & -  \\
  \end{array}
\end{equation}

\noindent  el proyector tiene la forma gen'erica
Ec.~(\ref{eq:proyectorpbrana}) que reescribimos as'i ($Z^{(p)}=\pm
M$)

\begin{equation}
P_{p}\epsilon=\left(1 \pm \Gamma^{01\cdots p}{\cal O}_{p} \right)\epsilon=0\, .
\end{equation}

\noindent La pregunta que nos hemos hecho (?`cu'ales son las
reglas de intersecci'on de los objetos extensos que hemos estudiado?)
se traduce entonces en la compatibilidad de imponer $P_{p}\epsilon=0$
y $P_{p^{\prime}}\epsilon=0$, en el caso de una $p$- y una
$p^{\prime}$-brana. El an'alisis general es complicado por la
presencia de ${\cal O}_{p}$. Si ignoramos su presencia por un momento
(considerando, por ejemplo, objetos iguales pero no paralelos),
podemos ver que

\begin{equation}
[P_{p},P_{p^{\prime}}]=0\, ,  
\end{equation}

\noindent si el n'umero de dimensiones transversas relativas total
(direcciones que para una de ellas son de volumen del mundo y para la
otra no) es $0\,\, {\rm mod}\,\, 4$. Por ejemplo, si tenemos dos S5
esta configuraci'on dar'ia lugar a proyectores compatibles

\begin{equation}
  \begin{array}{r||c|ccccccccc}
S5 & + & + & + & + & + & + & - & - & - & -  \\
S5 & + & + & + & + & - & - & + & + & - & -  \\
  \end{array}
\end{equation}

\noindent porque las direcciones 4 y 5 son del volumen del mundo de la
primera pero transversas a la segunda y las 6 y 7 al rev'es, con lo
que hay 4 dimensiones transversas relativas. Tambi'en 'esta
dar'ia lugar a proyectores compatibles

\begin{equation}
  \begin{array}{r||c|ccccccccc}
S5 & + & + & + & + & + & + & - & - & - & -  \\
S5 & + & + & - & - & - & - & + & + & + & +  \\
  \end{array}
\end{equation}

\noindent porque tiene un total de 8 direcciones transversas relativas.
En la notaci'on de la Ref.~\cite{kn:Pee2}, estas configuraciones son,
respectivamente 

\begin{equation}
{\rm S5}\perp {\rm S5} (3)\, ,
\hspace{1cm}
{\rm S5}\perp {\rm S5} (1)\, .
\end{equation}

En las teor'ias que nos ocupan, ${\cal
  O}_{p}=\mathbb{I},\Gamma_{11},\sigma^{1},i\sigma^{2}$, y depende de
$p\,\, {\rm mod}\,\, 4$ para D$p$-branas, de lo que deducimos que
podemos tener un estado ligado con energ'ia de ligadura cero, de
una D$p$-brana y una D$(p+4)$-brana que se intersectan en $p$
direcciones:

\begin{equation}
{\rm D}_{p}\perp{\rm D}_{(p+4)} (p)\, .  
\end{equation}

\noindent Podemos estudiar casos m'as complicados o 
simplemente generarlos utilizando las reglas de dualidad. Adem'as se
pueden estudiar configuraciones en las que las $p$-branas no son
perpendiculares, sino que forman 'angulos. En todos estos casos, cada
uno de los dos proyectores compatibles reduce el n'umero de
supersimetr'ias residuales a la mitad, de forma que la
configuraci'on total tiene $1/4$ de las supersimetr'ias totales
preservadas.

Antes de generar m'as reglas de intersecci'on por dualidad, es
interesante ver otro m'etodo por el que se pueden deducir, basado en
el estudio de la conservaci'on de la carga asociada a las $p$-branas
en la teor'ia de SUGRA \cite{kn:Tow7,kn:Stro}.  Consideremos una
cuerda fundamental de la teor'ia tipo~IIB F1B (el objeto que se
acopla a la 2-forma NSNS $\hat{\cal B}_{\mu\nu}$). Su carga con
respecto a $\hat{\cal B}_{\mu\nu}$ viene dada, en ausencia de otros
campos, por

\begin{equation}
q_{F1} \sim \int_{S^{7}} e^{-2\varphi} {}^{\star} \hat{\cal H}\, ,
\end{equation}

\noindent donde $S^{7}$ es una 7-esfera que rodea  a la cuerda
en las direcciones transversas. Esta carga es distinta de cero s'olo
si la cuerda es infinita o es cerrada (topol'ogicamente un
c'irculo). La raz'on es que esta integral es invariante bajo
deformaciones continuas de la 7-esfera en las que 'esta no cruce
singularidades, lo que se demuestra usando que, fuera de la cuerda,
que es una fuente localizada del campo $\hat{\cal B}_{\mu\nu}$

\begin{equation}
de^{-2\varphi} {}^{\star} \hat{\cal H}=0\, .
\end{equation}

Si la cuerda tiene extremos libres, la 7-esfera se puede deslizar
continuamente a lo largo de la cuerda hasta ir m'as all'a de los
extremos y entonces contraerla a un punto sin encontrar ninguna
singularidad\footnote{Esto se puede visualizar bien en cuatro
  dimensiones donde las dimensiones transversas a una cuerda son dos
  en vez de ocho, sustituyendo $S^{7}$ por $S^{1}$.}, con lo que la
integral vale cero. 

La situaci'on cambia en presencia de otros campos\footnote{En la
  cuerda heter'otica no hay otros campos que modifiquen la
  situaci'on, pero en las de tipo~II s'i.}. El invariante de
homotop'ia que debemos utilizar para definir la
carga\footnote{Esta definici'on de carga es la de Page
  \cite{kn:Mar}.} es ahora

\begin{equation}
q_{F1} \sim \int_{S^{7}} \left( e^{-2\varphi} {}^{\star} \hat{\cal H}
-{}^{\star}\hat{G}^{(3)}\hat{C}^{(0)}-\hat{G}^{(5)}\hat{C}^{(2)}\right)\, .
\end{equation}

\noindent Consideramos una cuerda semi-infinita con un extremo. A una 
distancia $L$ suficientemente grande del extremo podemos ignorar los
campos distintos de $\hat{\cal B}_{\mu\nu}$, y la carga debe de ser
aproximadamente la misma que en el caso anterior, tanto m'as
aproximada a ella cuanto mayor sea $L$ para un valor fijo del radio
$R$ de la 7-esfera. Si deslizamos la 7-esfera hacia el extremo, los
otros campos deben de empezar a contribuir (si no, estamos en el caso
anterior), pero podemos hacer que la integral siga dando
aproximadamente el mismo valor de $q_{F1}$ haciendo tender $R
\rightarrow 0$ con $R/L$ constante hasta que la 'unica contribuci'on
a la integral provenga del extremo. Ah'i la 7-esfera degenerada se
puede descomponer, por ejemplo, en el producto $S^{5}\times S^{2}$,
suponiendo que sea el tercer t'ermino en el integrando el relevante
en este caso, con lo que toda la contribuci'on a $q_{F1}$ proviene de

\begin{equation}
\int_{S^{5}}\hat{G}^{(5)} \int_{S^{2}}\hat{C}^{(2)}\, .  
\end{equation}

\noindent La primera integral es la carga de una D3-brana 
(por autodualidad de $\hat{G}^{(5)}$), de forma que la cuerda tiene su
extremo en una D3-brana. En cuanto a la segunda integral, si no hay
ninguna D1, podemos suponer que dentro de la D3-brana
$\hat{G}^{(3)}\sim d\hat{C}^{(2)}=0$, y, s'olo localmente,
$\hat{C}^{(2)}=dV^{(1)}$, con lo que

\begin{equation}
q_{F1}\sim \int_{S^{2}}d V^{(1)}\, .  
\end{equation}

La interpretaci'on es clara: una cuerda fundamental puede acabar en
({\it intersectar}) una D3-brana. En el punto del volumen del mundo de
la D3-brana correspondiente a la intersecci'on hay un campo vectorial
excitado de forma que su carga magn'etica es la misma de la cuerda
fundamental. Ese campo es el dual del campo de Born e Infeld que
est'a presente en la acci'on de la D3-brana.

Este argumento parece depender de posibles redefiniciones de campos,
pero en realidad, lo que se ve es que otras variables son m'as
adecuadas para describir otras intersecciones \cite{kn:Tow7}. Por otro
lado, su virtud es que demuestra la relaci'on existente entre la
f'isica en el espacio-tiempo ambiente y en los vol'umenes del
mundo de las $p$-branas. Es posible, por ejemplo, ver las cuerdas
fundamentales ancladas en las D$p$-branas como excitaciones
solit'onicas de los campos de BI llamadas {\it BIones} \cite{kn:CaMa}
etc.

Las sistemas 

\begin{equation}
\label{eq:inter1}
{\rm F}1\perp {\rm D}_{p}(0)\, , 
\end{equation}

\noindent son muy convenientes para
empezar a generar m'as reglas de intersecci'on por dualidad. Todos
ellos est'an relacionados por dualidad~T en direcciones
perpendiculares a la F1. Si tomamos la intersecci'on ${\rm F1B}\perp
{\rm D}_{3}(0)$ y efectuamos una transformaci'on de dualidad~S,
obtenemos una intersecci'on entre D$p$-branas: ${\rm D}_{1}\perp {\rm
  D}_{3}(0)$ y haciendo ahora transformaciones de dualidad~T en
direcciones transversas comunes, obtenemos las intersecciones 

\begin{equation}
\label{eq:inter2}
{\rm  D}_{p}\perp {\rm D}_{p+2}(p-1)\, ,\hspace{.5cm} p\geq 1\, . 
\end{equation}

\noindent Si hacemos
transformaciones de dualidad~T en la direcci'on de la D1 en vez de en
las transversas comunes, tenemos la ${\rm D}_{0}\perp {\rm  D}_{4}(0)$
y haciendo de nuevo dualidad~T en las transversas comunes generamos
las intersecciones

\begin{equation}
\label{eq:inter3}
{\rm  D}_{p}\perp {\rm D}_{p+4}(p)\, . 
\end{equation}

Por dualidad~S de la ${\rm D}_{1}\perp {\rm  D}_{5}(1)$ que est'a en esta 
clase, generamos la 

\begin{equation}
\label{eq:inter4}
{\rm F}1\perp {\rm S5B}(0)\, .
\end{equation}

Si hacemos dualidad~T en una direcci'on del volumen del mundo de la
D3 en la configuraci'on ${\rm D}_{1}\perp {\rm D}_{3}(0)$, en la
familia Ec.~(\ref{eq:inter2}) obtenemos ${\rm D}_{2}\perp {\rm
  D}_{2}(0)$ y haciendo nuevas dualidades~T en las direcciones
transversas comunes, generamos la familia

\begin{equation}
\label{eq:inter5}
{\rm  D}_{p}\perp {\rm D}_{p}(p-2)\, ,\hspace{.5cm} p\geq 2\, . 
\end{equation}

Si hacemos dualidad~S a la ${\rm F1}\perp {\rm D}_{5}(0)$ en la
familia Ec.~(\ref{eq:inter1}), generamos la ${\rm D}_{1}\perp {\rm
  S5B}(0)$ y con dualidades sucesivas en direcciones transversas
relativas a las D$p$-branas, generamos la familia

\begin{equation}
\label{eq:inter6}
{\rm  D}_{p}\perp {\rm S5}(p-1)\, ,\hspace{.5cm} p\geq 1\, ,
\end{equation}

\noindent etc. El resumen de todos estos resultados es:

\begin{equation}
\begin{tabular}{|c|}
\hline \\
\begin{minipage}{10cm}
${\rm F}1 \parallel {\rm S}5$, \hspace{.3cm}
${\rm F}1 \perp {\rm D}_{p} (0)$,

\vspace{.3cm}

${\rm S}5 \perp {\rm S}5 (1)$, \hspace{.3cm}
${\rm S}5 \perp {\rm S}5 (1)$, \hspace{.3cm}
${\rm S}5 \perp {\rm D}_{p} (p-1)\,\,\, (p>1)$,

\vspace{.3cm}

${\rm D}_{p} \perp {\rm D}_{p^{\prime}} (m)\,\,\, p+p^{\prime}=4+2m$, 

\vspace{.3cm}

${\rm W} \parallel {\rm F1}$, \hspace{.3cm}
${\rm W} \parallel {\rm S5}$, \hspace{.3cm}
${\rm W} \parallel {\rm D}_{p}$, \hspace{.3cm}

\vspace{.3cm}

${\rm KK} \perp D_{p} (p-2)$.

\vspace{.3cm}
\end{minipage}
\\ \hline 
\end{tabular}
\end{equation}

En 11 dimensiones, incluyendo intersecciones con ondas gravitacionales
y monopolos de Kaluza-Klein \cite{kn:BdREJvdS}

\begin{equation}
\begin{tabular}{|c|}
\hline \\
\begin{minipage}{13.5cm}
${\rm M}2 \perp {\rm M}2 (0)$, \hspace{.3cm}
${\rm M}2 \perp {\rm M}5 (1)$, \hspace{.3cm}
${\rm M}5 \perp {\rm M}5 (1)$, \hspace{.3cm}
${\rm M}5 \perp {\rm M}5 (3)$,

\vspace{.3cm}

${\rm W} \parallel {\rm M}2$, \hspace{.3cm}
${\rm W} \parallel {\rm M}5$,

\vspace{.3cm}

${\rm KK} \parallel {\rm M}2$, \hspace{.3cm}
${\rm KK} \perp {\rm M}2 (0)$, \hspace{.3cm}
${\rm KK} \parallel {\rm M}5$, \hspace{.3cm}
${\rm KK} \perp {\rm M}5 (1)$, \hspace{.3cm}
${\rm KK} \perp {\rm M}5 (3)$, 

\vspace{.3cm}

${\rm W} \parallel {\rm KK}$, \hspace{.3cm}
${\rm W} \perp {\rm KK}(2)$, \hspace{.3cm}
${\rm W} \perp {\rm KK}(4)$. 
\vspace{.3cm}

\end{minipage}
\\ \hline 
\end{tabular}
\end{equation}

Est'a claro que podemos extender este juego a un n'umero mayor de
branas y que el requisito de que formen un estado ligado con
energ'ia de ligadura cero es que los proyectores asociados
conmuten entre s'i. En el caso de 3 $p$-branas, en general la
supersimetr'ia del sistema se reduce a $1/8$ del total, excepto en
algunos casos en los que la tercera $p$-brana no rompe ninguna
supersimetr'ia adicional. Este caso est'a relacionado con la
creaci'on de $p$-branas cuando otras se cruzan, por ejemplo
\cite{kn:HaWi} cuando una D5-brana y una S5B-brana se cruzan, se crea
una D3-brana asociada a la intersecci'on. El proyector de la D3-brana
no impone ninguna condici'on adicional.

%%%%%%%%%%%%%%%%%%%%%%%%%%%%%%%%%%%%%%%%%%%%%%%%%%%%%%%%%%%%%%%%%%%%%%
\subsection{Soluciones: superposiciones arm'onicas}

Finalmente, veamos qu'e soluciones cl'asicas de las teor'ias
efectivas de cuerdas describen estas configuraciones. Un compendio
'util sobre intersecciones es el de Gauntlett \cite{kn:Ga2}. Las
primeras soluciones fueron identificadas por Papadopoulos y Townsend
en la Ref.~\cite{kn:PT} de entre una familia muy amplia descubierta
por G\"uven en la Ref.~\cite{kn:Gu}. Resulta que estas soluciones, y
todas las correspondientes a sistemas ligados con energ'ia de
ligadura cero se pueden construir por {\it superposici'on arm'onica}
de las soluciones que describen a uno s'olo de los objetos
\cite{kn:Tsey,kn:GKT}, poniendo a cada componente de la m'etrica,
multiplicadas, las funciones arm'onicas que tienen las soluciones
individuales, con las mismas potencias, pero donde las funciones
arm'onicas s'olo dependen ahora de las dimensiones transversas
comunes. Un ejemplo vale aqu'i m'as que mil palabras: si las
soluciones que describen una 'unica cuerda fundamental y una 'unica
D$p$-brana son las Ecs.~(\ref{eq:extremeF1}) y (\ref{eq:extremeDp}),
la soluci'on que describe la intersecci'on de una cuerda fundamental
que est'a en la direcci'on $y$ y una D$p$-brana que est'a en las
direcciones $\vec{z}_{p}\equiv (z^{1},\ldots, z^{p})$ es

\begin{equation} 
\begin{tabular}{|rcl|} 
\hline & & \\ & & \\ 
$d\hat{s}^{2}_{s}$ & $=$ & $H_{Dp}^{-1/2}H_{F1}^{-1}
dt^{2} -H_{Dp}^{+1/2}H_{F1}^{-1}dy^{\, 2}-
H_{Dp}^{-1/2}d\vec{z}_{p}^{\, 2}
-H_{Dp}^{+1/2}d\vec{x}_{8-p}^{\, 2}\, ,$ \\
& & \\
$e^{-2(\hat{\phi}-\hat{\phi}_{0})}$ & $=$ & 
$H_{Dp}^{\frac{(p-3)}{2}}H_{F1} \, ,$ \\
& & \\
$\hat{C}^{(p+1)}{}_{t\underline{z}^{1}\cdots \underline{z}^{p}}$ & $=$ & 
$\pm e^{-\hat{\phi}_{0}}\left(H_{Dp}^{-1}-1\right)\, ,$  \\
& & \\
$\hat{B}_{t\underline{y}}$ & $=$ & $\pm \left(H_{F1}^{-1}-1\right)\, ,$  \\
& & \\
$H_{Dp,F1}$ & $=$ & 
$1+ {\displaystyle\frac{h_{Dp,F1}}{|\vec{x}_{8-p}|^{6-p}}}\, ,$ \\
& & \\ \hline
\end{tabular}
\label{eq:F1Dp}
\end{equation}

Es evidente c'omo extender este ejemplo a otros casos. Lo importante
es que las soluciones as'i obtenidas no son enteramente
satisfactorias porque no nos dicen en qu'e punto la F1 intersecta la
D$p$-brana, y las F1 est'a {\it deslocalizada} sobre el volumen del
mundo de la D$p$-brana. En algunos casos las dependencia de las
funciones arm'onicas se puede extender a algunas de las dimensiones
transversas relativas, pero no hay soluciones que describan
intersecciones completamente localizadas. Una referencia en la que se
discute este problema es \cite{kn:Pee3}.

Que se puedan construir estas soluciones por superposici'on en una
teor'ia tan no-lineal como la gravitaci'on nos da idea de lo
especiales que son, de lo peculiar que es tener objetos cuya
energ'ia de interacci'on es cero y de lo crucial que es la
supersimetr'ia en este juego.

%%%%%%%%%%%%%%%%%%%%%%%%%%%%%%%%%%%%%%%%%%%%%%%%%%%%%%%%%%%%%%%%%%%%%%
\subsection{Intersecciones con ondas gravitacionales y transformaciones
de Lorentz singulares}

Antes de finalizar esta lecci'on vamos a ver un m'etodo de generar
soluciones que describen una onda gravitacional propag'andose en una
direcci'on del volumen del mundo de una $p$-brana cualquiera. El
m'etodo fue ideado originalmente por Aichelburg y Sexl para generar
una onda gravitacional plana de choque a partir de la soluci'on de
Schwarzschild \cite{kn:AS}. Aplicado a las $p$-branas negras cargadas
de las Ecs.~(\ref{eq:HoS}), nos va a dar una onda gravitacional plana
movi'endose en el volumen del mundo de una $p$-brana extrema. Esto
nos va a dar pie para estudiar las ondas planas, que son soluciones
puramente gravitacionales que no pertenecen a las familias de
soluciones que hemos estudiado.

Consideremos una soluci'on de la forma gen'erica

\begin{equation}
\left\{
  \begin{array}{rcl}
ds^{2} & = & H^{\alpha}\left[W dt^{2} -d\vec{y}^{\, 2}_{p-1}-dz^{2} \right]
-H^{\beta} \left[W^{-1}d\rho^{2}+\rho^{2}d\Omega^{2}\right]\, ,\\
& & \\
e^{-2(\phi-\phi_{0})} & = & H^{\gamma}\, ,\\
& & \\
A_{(p+1)\, t\underline{y^{1}}\cdots \underline{y^{p-1}}\underline{z}} & = &
\alpha \left(H^{-1}-1 \right)\, .\\
& & \\
W & = & 1 +{\displaystyle\frac{\omega}{\rho^{n}}}\, ,
\hspace{.5cm}
H=1+{\displaystyle\frac{h}{\rho^{n}}}\, ,\\
  \end{array}
\right.
\end{equation}

\noindent Ni la dimensi'on ni el sistema de referencia conforme en el que 
trabajemos son relevantes. Hacemos la transformaci'on de Lorentz

\begin{equation}
 \left(
   \begin{array}{c}
t \\ z \\
   \end{array}
\right) 
\rightarrow 
 \left(
   \begin{array}{cc}
\cosh\gamma & \sinh\gamma \\ \sinh\gamma & \cosh\gamma \\
   \end{array}
\right) 
 \left(
   \begin{array}{c}
t \\ z \\
   \end{array}
\right)\, , 
\end{equation}

\noindent que deja invariante todo\footnote{Esto es as'i porque 
  la $(p+1)$-forma tiene los dos 'indices $tz$.}  menos este trozo
de la m'etrica:

\begin{equation}
W dt^{2}-dz^{2} \rightarrow dt^{2}-dz^{2} 
+\cosh^{2}\gamma(W-1)(dt +\tanh^{2}\gamma dz)^{2}  \, ,
\end{equation}

\noindent que en el l'imite singular $\gamma\rightarrow \infty$,
$\omega\rightarrow 0$ con $e^{2\gamma}\omega \rightarrow h_{W}$
finito, se convierte en

\begin{equation}
H_{W}^{-1}dt^{2}-H_{W}[dz-(H_{W}^{-1}-1) dt]^{2} \, ,
\hspace{1cm}
H_{W}=1+\frac{h_{W}}{\rho^{n}}\, ,
\end{equation}

\noindent y obtenemos la nueva soluci'on con m'etrica

\begin{equation}
ds^{2} =  H^{\alpha}\left\{
H_{W}^{-1}dt^{2}-H_{W}[dz-(H_{W}^{-1}-1) dt]^{2}
-d\vec{y}^{\, 2}_{p-1}\right\}
-H^{\beta} d\vec{x}^{2}\, ,
\end{equation}

\noindent y los dem'as campos como antes. Esta soluci'on 
tiene dos par'ametros independientes $h$ y $h_{W}$. Si $h=0$ todos
los campos salvo la m'etrica se vuelven triviales y obtenemos una
nueva soluci'on puramente gravitatoria: {\it la onda gravitacional
  plana}, v'alida en cualquier dimensi'on y para cualquier funci'on
arm'onica $H_{W}$ de las coordenadas $\vec{x}_{n}$, aunque nosotros
escogemos aqu'i una determinada ($d>4$):

\begin{equation}
\begin{tabular}{|rcl|} 
\hline  & & \\  
$ds^{2}$  & $=$ & $
H_{W}^{-1}dt^{2}-H_{W}[dz+\alpha(H_{W}^{-1}-1) dt]^{2}
-d\vec{x}_{d-2}^{2}\, ,$ \\
& & \\
$H_{W}$ & $=$  & $1+{\displaystyle\frac{h_{W}}{|\vec{x}_{d-2}|^{d-4}}}\, ,
\hspace{.5cm}\alpha=\pm 1\, ,$
\\ & & \\
\hline
\end{tabular}
\end{equation}

\noindent donde hemos incluido el par'ametro $\alpha$: 
$\alpha=+1 (-1)$ cuando la onda se propaga en la direcci'on positiva
(negativa) del eje $z$.

La soluci'on general claramente representa la onda propag'andose en
la direcci'on $-z$ del volumen del mundo de la $p$-brana y obedece la
regla de superposici'on arm'onica.

Para determinar la constante de integraci'on $h_{W}$ en t'erminos de
las constantes f'isicas podemos utilizar de nuevo el m'etodo de la
fuente, que aqu'i ser'a la acci'on de una part'icula sin masa. El
resultado es, primeramente, que $H_{W}$ debe de ser reemplazada por

\begin{equation}
\begin{tabular}{|c|} 
\hline  \\  
$H_{W}=1+{\displaystyle\frac{h_{W}}{|\vec{x}_{d-2}|^{d-4}}}
 \delta (u_{\alpha})\, .
\hspace{.5cm}
u_{\alpha}=\frac{1}{\sqrt{2}}(t-\alpha z)\, .$
\\  \\
\hline
\end{tabular}  
\end{equation}

\noindent (El factor $\delta (u_{\alpha})$ aparece tambi'en al tomar el l'imite 
$\gamma \rightarrow 0 $ con cuidado \cite{kn:AS}.) Segundo, 

\begin{equation}
\begin{tabular}{|c|} 
\hline  \\  
$h_{W}= -\alpha {\displaystyle
\frac{\sqrt{2} |p^{z}| 8\pi G^{(d)}_{N}}{(d-4)\omega_{(d-3)}}}\, ,$
\\  \\
\hline
\end{tabular}  
\end{equation}

\noindent donde $p^{z}$ es el momento que transporta la onda. 'Este 
es un par'ametro continuo (a diferencia de la masa ya la carga de las
$p$-branas BPS), excepto si la direcci'on $z$ es compacta, en cuyo
caso est'a cuantizado en m'ultiplos enteros de $1/R_{z}$. Si
queremos obtener la soluci'on compactificada, tenemos que tener en
cuenta que a funci'on arm'onica $H_{W}$ depende de $z$ a trav'es de
$u$ y tenemos que hacer una expansi'on en serie de Fourier y
quedarnos con el modo cero

\begin{equation}
\delta(u_{\alpha}) \sim -\alpha\frac{\sqrt{2}}{2\pi R_{z}}\, ,
\end{equation}

\noindent lo que, combinado con la cuantizaci'on del momento nos da

\begin{equation}
\label{eq:esaotrarelacion}
\begin{tabular}{|c|} 
\hline  \\  
$h_{W}= {\displaystyle
\frac{ |N| 8G^{(d)}_{N}}{R_{z}^{2}(d-4)\omega_{(d-3)}}}\, .$
\\  \\
\hline
\end{tabular}  
\end{equation}

%%%%%%%%%%%%%%%%%%%%%%%%%%%%%%%%%%%%%%%%%%%%%%%%%%%%%%%%%%%%%%%%%%%%%%
\subsection{Agujeros negros a partir de intersecciones}

Ahora estamos en condiciones de construir nuevas soluciones de tipo
agujero negro a partir de intersecciones de $p$-branas.  Las m'as
sencillas son las de tipo ${\rm D}_{p} \parallel {\rm D}_{p+4}$. No
queremos utilizar branas con $p>6$ porque si no no obtendr'iamos
soluciones asint'oticamente planas, y esto s'olo nos deja tres
posibilidades: $D0 \parallel D4$, $D1\parallel D5$ y $D2 \parallel D6$
que est'an relacionadas por dualidad~T y son adecuadas para agujeros
negros en $d=6,5,4$ respectivamente, o dimensiones m'as bajas
compactificando dimensiones transversas. Sin embargo, sabemos que los
agujeros negros regulares tienen m'as de dos cargas (y est'an
compuestas por m'as de dos objetos) por lo que vamos a tener que
a~nadirlos. Lo m'as sencillo es a~nadir una onda gravitacional en
una de las direcciones del volumen del mundo comunes. Esto s'olo es
posible en los dos 'ultimos casos. A los dos primeros se les puede
a~nadir una cuerda fundamental.

Si queremos obtener un agujero negro en $d=4$ adem'as vamos a
necesitar un cuarto objeto que suele ser un monopolo de Kaluza-Klein o
una S5.  Por simplicidad vamos a considerar el caso en $d=5$
$D1\parallel D5$ al que vamos a a~nadir una onda gravitacional en la
direcci'on de la cuerda. Esta configuraci'on fue estudiada por
primera vez por Callan y Maldacena en la Ref.~\cite{kn:CaMa2}, como
alternativa m'as simple a la construcci'on original de Strominger y
Vafa \cite{kn:SV}, y su versi'on dual $W\parallel D2\parallel D6$ fue
considerada poco despu'es por Maldacena y Strominger en la
Ref.~\cite{kn:MaSt}.  En $d=4$ se usaron las configuraciones
$W\parallel D2\parallel D6$ m'as una S5 en el mismo art'iculo y
las configuraciones $D0\parallel D4$ y $D1 \parallel D5$ con F1 y
monopolo de Kaluza-Klein por Johnson, Khuri y Myers en
Ref.~\cite{kn:JKM}. Nosotros no hemos mencionado el momento angular,
pero es posible tener tambi'en agujeros negros supersim'etricos con
momento angular en $d=5$ y su construcci'on a partir de
configuraciones de objetos extensos de Teor'ia de Cuerdas se hizo
en la Ref.~\cite{kn:BMPV}.

%%%%%%%%%%%%%%%%%%%%%%%%%%%%%%%%%%%%%%%%%%%%%%%%%%%%%%%%%%%%%%%%%%%%%%
\section{El agujero extremo ${\rm W} \parallel {\rm D1} \parallel {\rm D5}$ 
  en $d=5$}

La soluci'on de la que vamos a partir se construye usando el
principio de superposici'on arm'onica. La configuraci'on de
$N_{D1}$ D1-branas, $N_{D5}$ D5-branas y una onda gravitacional con
n'umero de momento $N_{W}$ es

\begin{equation}
  \begin{array}{r||c|ccccccccc}
{\rm D}1 & + &      +      & \sim & \sim & \sim & \sim & - & - & - & -  \\
{\rm D}5 & + &      +      &  +   &  +   &  +   &  +   & - & - & - & -  \\
{\rm W}  & + & \rightarrow & \sim & \sim & \sim & \sim & - & - & - & -  \\
  \end{array}
\end{equation}

\noindent donde un signo $+$ indica que esa direcci'on es del 
volumen del mundo de la brana (isom'etrica, pues), un signo menos que
es una direcci'on transversa en la que la soluci'on conserva la
dependencia, un $\sim$ que es una direcci'on transversa pero que
hemos compactificado, eliminando primero la dependencia en esa
coordenada por el procedimiento explicado en el cap'itulo
anterior, y el signo $\rightarrow$ denota la direcci'on de
propagaci'on de la onda, y tambi'en vamos a eliminar la dependencia
en la misma. Las direcciones espaciales con $+,\sim$ 'o $\rightarrow$
van a estar compactificadas en un 5-toro $T^{5}=S^{1}\times T^{4}$ de
volumen $V^{5}=2\pi R V^{4}$ donde $R$ es el radio de la
coordenada $y^{1}$ y $V^{4}=(2\pi)^{4} R_{2}\ldots R_{4}$ el del
$T^{4}$ en el que est'an compactificadas las coordenadas
$y^{2},\ldots, y^{4}$ y las funciones arm'onicas van a depender
'unicamente en las 4 coordenadas transversas comunes $\vec{x}_{4}$.

El principio de superposici'on arm'onica nos da la siguiente
soluci'on 10-dimensional, en el sistema de referencia de la cuerda:

\begin{equation}
\left\{
  \begin{array}{rcl}
d\hat{s}_{s}^{2} & = & 
H_{D1}^{-1/2}H_{D5}^{-1/2}
\left\{ H_{W}^{-1}dt^{2} 
-H_{W} \left[dy^{1}+\alpha_{W} (H_{W}^{-1}-1) dt\right]^{2} 
\right\} \\
& & \\
& & 
\hspace{5cm}
-H_{D1}^{1/2}H_{D5}^{-1/2}d\vec{y}^{\, 2}_{4}
-H_{D1}^{1/2}H_{D5}^{1/2} d\vec{x}_{4}^{\, 2}\, , \\
& & \\
e^{-2(\hat{\phi}-\hat{\phi}_{0})} & = & 
H_{D5}/H_{D1}\, ,\\
& & \\
\hat{C}^{(2)}{}_{t\underline{y}^{1}} 
& = & \alpha_{D1} \left(H_{D1}^{-1} -1\right)\, ,\\
& & \\
\hat{C}^{(6)}{}_{t\underline{y}^{1}\cdots \underline{y}^{5}} 
& = & \alpha_{D5} \left(H_{D5}^{-1} -1 \right)\, ,\\
  \end{array}
\right.
\end{equation}

\noindent donde $\alpha_{D1,D5,W}^{2}=1$
son los signos de las cargas y el momento y donde las funciones
arm'onicas est'an dadas por

\begin{equation}
H_{i}=1+\frac{r_{i}^{2}}{|\vec{x}_{4}|^{2}}\, ,
\hspace{1cm}
i=D1,D5,W\, ,
\end{equation}

\noindent donde, para la D5-brana, ninguna de cuyas dimensiones transversas
ha sido compactificada

\begin{equation}
r_{D5}^{2}= N_{D5}h_{D5}= N_{D5}\ell_{s}^{2}g\, ,  
\end{equation}

\noindent para la D1-brana, cuatro de cuyas dimensiones transversas han sido 
compactificadas en un 4-toro, usando la relaci'on
Ec.~(\ref{eq:esarelacion})

\begin{equation}
r_{D1}^{2}= N_{D1}h_{D1} \frac{\omega_{(5)}}{V^{4}\omega_{(1)}} = 
\frac{N_{D1} \ell_{s}^{6} g}{V}\, , 
\hspace{1cm}
V\equiv R_{2}\ldots R_{5}\, ,    
\end{equation}

\noindent y para la onda gravitacional, que se propaga en una 
direcci'on compacta y tiene 4 direcciones compactas transversas,
usando primero Ec.~(\ref{eq:esaotrarelacion}), donde $G_{N}^{(10)}$
viene dada por Ec.~(\ref{eq:GN10A}), y luego Ec.~(\ref{eq:esarelacion})
 
\begin{equation}
r_{W}^{2}= h_{W} \frac{\omega_{(5)}}{V^{4}\omega_{(1)}} =
\frac{N_{W}\ell_{s}^{8} g^{2}}{R^{2}V}\, .
\end{equation}

Los tres $N_{D1},N_{D5},N_{W}$ son enteros positivos.

Podemos compactificar inmediatamente la soluci'on a 5 dimensiones,
obteniendo una soluci'on con tres cargas relativas a tres campos
vectoriales, y con tres m'odulos: el dilat'on $\phi$, $K_{V}$ que
mide el volumen local del 4-toro y $K_{R}$ que mide la longitud del
c'irculo en el que se enrolla la D-cuerda:

\begin{equation}
  \begin{tabular}{|c|}
\hline \\
$
  \begin{array}{rcl}
d\tilde{s}_{E}^{2} & = & \left(H_{D1}H_{D5}H_{W} \right)^{-2/3}dt^{2}
-\left(H_{D1}H_{D5}H_{W} \right)^{1/3}d\vec{x}_{4}^{\, 2}\, ,\\
& & \\
ds^{2}_{s} & = & \left(H_{D1}H_{D5}\right)^{-1/2}H_{W}^{-1}dt^{2} 
-\left(H_{D1}H_{D5}\right)^{1/2} d\vec{x}_{4}^{\, 2}\, ,\\
& & \\
A^{(D1,D5,W)}{}_{t} & = & \alpha_{D1,D5,W} 
\left(H_{D1,D5,W}^{-1} -1 \right)\, ,\\
& & \\
K_{V}/K_{V0} & = & H_{D1}/H_{D5}\, ,
\hspace{.5cm}
e^{-2(\phi-\phi_{0})}= K_{R}/K_{R0} =  \left(H_{D1}H_{D5}\right)^{-1/4}
H_{W}^{1/2}\, .\\
  \end{array} 
$
\\ \\ \hline
\end{tabular}
\end{equation}

Los tres m'odulos son finitos en el horizonte de este agujero negro
extremo.  Adem'as, el volumen del horizonte (en el sistema de
referencia conforme de Einstein) es tambi'en finito, como en el
agujero de Reissner y Nordstr\"om extremo

\begin{equation}
  \begin{array}{rcl}
A & = & \omega_{(3)} \left(\lim_{|\vec{x}_{4}|\rightarrow 0} |\vec{x}_{4}|^{6} \ H_{D1}H_{D5}H_{W} \right)^{1/2} = 
2\pi^{2} \left( r_{D1}r_{D5}r_{W}\right)^{1/2} \\
& & \\
& = & 2\pi^{2} {\displaystyle\sqrt{N_{D1}N_{D5}N_{W}}\,  
\frac{\ell_{s}^{8}g^{2}}{RV}}\, .\\
\end{array}
\end{equation}

La entrop'ia en cualquier dimensi'on viene dada por un cuarto del
volumen del horizonte medida en unidades de Planck \cite{kn:MP}:

\begin{equation}
S= \frac{A}{4 G_{N}^{(5)}}\, ,
\hspace{1cm}
G_{N}^{(5)} = \frac{G_{N}^{(10)}}{(2\pi)^{5}RV}= 
\frac{\pi}{4}\frac{\ell_{s}^{8}g^{2}}{RV}\, , 
\end{equation}

\noindent lo que implica la bell'isima f'ormula

\begin{equation}
  \begin{tabular}{|c|}
\hline \\
$ 
S= 2\pi {\displaystyle\sqrt{N_{D1}N_{D5}N_{W}}}\, .
$
\\ \\ \hline
\end{tabular}
\end{equation}

%la f'ormula invariante bajo dualidad??????????

Al expresar todos los par'ametros de la teor'ia y la soluci'on
en funci'on de constantes de la Teor'ia de Cuerdas hemos obtenido
un valor para la entrop'ia que no depende de ninguno de los
m'odulos $g,R,V$ ni de la longitud de la cuerda $\ell_{s}$, y s'olo
depende de n'umeros enteros, lo que nos hace concebir esperanzas de
que este valor se pueda explicar a trav'es del recuento de estados.
Observemos que la masa del agujero negro s'i depende de los
m'odulos:

\begin{equation}
\begin{tabular}{|c|}
\hline \\
$ 
M =  {\displaystyle\frac{N_{D1}R}{g\ell_{s}} +\frac{N_{D5}RV}{g \ell_{s}^{6}}
+\frac{N_{W}}{R}}\, .
$
\\ \\ \hline
\end{tabular}
\end{equation}

Si $r_{D1}=r_{D5}=r_{W}$ entonces todos los m'odulos toman un valor
constante\footnote{En rigor, dado que estas constantes dependen de los
  n'umeros enteros $N_{D1},N_{D5},N_{W}$, esto no es posible salvo
  para valores espaciales de los m'odulos $g,R,V$. Sin embargo, si
  los enteros $N_{D1},N_{D5},N_{W}$ son suficientemente grandes,
  podemos estar arbitrariamente cerca de la igualdad.}, y la m'etrica
es (no hay distinci'on entre los sistemas de la cuerda y Einstein)

\begin{equation}
ds^{2}= H^{-2}dt^{2}-Hd\vec{x}_{4}^{\, 2}\, ,  
\hspace{1cm}
H=H_{D1}=H_{D5}=H_{W}\, ,
\end{equation}

\noindent justamente la del agujero negro extremo de 
Reissner y Nordstr\"om en $d=5$. Esta m'etrica est'an entre las
soluciones extremas del modelo $a$ con $a=p=0,d=5$
Ec.~(\ref{eq:HoSextreme}). La 'unica diferencia es que ah'i
est'a cargada s'olo con respecto a un vector y aqu'i con respecto
a 3. Esto es necesario para que podamos considerar la soluci'on como
una soluci'on de la teor'ia efectiva de cuerdas~IIB, que tiene
muchos m'as campos cuyas ecuaciones de movimiento se han de
satisfacer incluso aunque tomen su valor de vac'io.

Hemos as'i conseguido no s'olo obtener una soluci'on de agujero
negro extremo con entrop'ia distinta de cero en una teor'ia de
SUGRA, sino que hemos identificado todos sus componentes elementales:
D1- y D5-branas m'as una onda gravitacional. El siguiente paso es
intentar deducir de esta composici'on macrosc'opica, el n'umero de
microestados del sistema y obtener de 'el la entrop'ia.

%%%%%%%%%%%%%%%%%%%%%%%%%%%%%%%%%%%%%%%%%%%%%%%%%%%%%%%%%%%%%%%%%%%%%%
\section{Microestados y entrop'ia de 
${\rm W} \parallel {\rm D1} \parallel {\rm D5}$}

Desde el punto de vista de la Teor'ia de Cuerdas microsc'opica,
el agujero negro que hemos obtenido es un vac'io en el que se debe
cuantizar la cuerda, teniendo en cuenta las condiciones de contorno
que impone la presencia de D-branas. Pero no sabemos cuantizar la
cuerda en 'el porque en general la constante de acoplo de la cuerda
va a ser grande y s'olo sabemos cuantizar perturbativamente las
cuerdas. 

En este punto hacemos uso de la independencia de la entrop'ia de
la constante de acoplo, una independencia asociada a la
supersimetr'ia. Si tomamos el l'imite $g\rightarrow 0$, puesto
que todos los $r_{i}$s llevan potencias positivas de $g$, la
soluci'on tiende al espacio plano. En este l'imite de espacio
plano s'i sabemos cuantizar las cuerdas e identificar los
microestados. El resultado, del que derivaremos la entrop'ia
ser'a v'alido para $g$ grande, en el que recuperamos el agujero
negro extremo. 'Este es uno de los puntos cruciales de este
c'alculo. Otra raz'on por la que el c'alculo es v'alido es porque
los 'unicos microestados que contribuyen a la entrop'ia son
tambi'en BPS. Su n'umero est'a determinado por la
supersimetr'ia (es decir, por la cinem'atica, no la din'amica) y
es completamente independiente de $g$, como de hecho demuestra la
expresi'on que hemos obtenido para $S$.

As'i pues, debemos de identificar la teor'ia de cuerdas
definida en el vac'io de D1- y D5-branas que adem'as tiene
momento en la direcci'on de la D1. En este punto echamos de menos el
no haber podido dar paralelamente un curso de Teor'ia de Cuerdas
basado en el enfoque tradicional de la f'isica bidimensional en
vez de en la f'isica espacio-temporal que nosotros hemos seguido,
pero es imposible explicar ambos enfoques en un tiempo reducido.
Intentaremos explicar al menos la esencia de esta parte del c'alculo.

Las teor'ias de cuerdas son casos especiales de Teor'ias de
Campos Conformes bidimensionales \cite{kn:Gin}. Estas teor'ias
est'an caracterizadas por su {\it carga central} $c$. Adem'as tienen
un espectro infinito de estados cuya degeneraci'on crece con la
energ'ia. El comportamiento asint'otico de la degeneraci'on de
estados viene dado por la {\it f'ormula de Cardy}

\begin{equation}
\rho(E) \sim e^{\sqrt{\pi (c-24 E_{0})E L/3}}\, ,
\end{equation}

\noindent donde $E_{0}$ es la m'inima energ'ia  y $L$ la longitud
de la coordenada espacial, que aqu'i es $2\pi R$. De todo el
espectro de la teor'ia s'olo van a contribuir
significativamente\footnote{Hay contribuciones de otros estados a esta
  f'ormula de la entrop'ia, que s'olo debe de ser considerada
  aproximada \cite{kn:CdWM}. } a $\rho(E)$ modos correspondientes a
cuerdas abiertas con un extremo en una D1 y otro en una D5, y con
momento en la direcci'on $y^{1}$. El n'umero de estos modos es
proporcional al producto $N_{D1}N_{D5}$ puesto que pueden empezar o
acabar en D1- y D5-branas distintas., y por lo tanto esperamos que
$c$, que es proporcional al n'umero de grados de libertad de la
teor'ia conforme, sea proporcional a este producto. Una
evaluaci'on precisa que no podemos reproducir aqu'i da
precisamente $c= 6 N_{D1}N_{2}$. Por otro lado, la energ'ia de
estos modos es igual a su momento $N_{W}/R$. Sustituyendo en la
f'ormula de Cardy todos estos datos, tenemos

\begin{equation}
\rho(E)= e^{2\pi \sqrt{N_{D1}N_{D5}N_{W}}}\, ,  
\end{equation}

\noindent lo que nos lleva inmediatamente al valor de la entrop'ia
que calculamos m'as arriba.

%%%%%%%%%%%%%%%%%%%%%%%%%%%%%%%%%%%%%%%%%%%%%%%%%%%%%%%%%%%%%%%%%%%%%%
\section{Comentarios finales}

En el breve tiempo de que hemos dispuesto hemos intentado describir,
desde sus fundamentos, los modelos de agujero negro de la Teor'ia
de Cuerdas. Hemos llegado a explicar c'omo una configuraci'on de
objetos extensos de la teor'ia, en acoplo fuerte, da origen a una
m'etrica de agujero negro cargado extremo con horizonte regular y
c'omo el conocer esa configuraci'on nos permite hacer un c'alculo
estad'istico de la entrop'ia.

Hay muchas cosas m'as que desear'iamos haber podido contar: los
modelos de agujeros extremos en $d=4$ \cite{kn:MaSt,kn:JKM}, los de
agujeros negros cuasi-extremos en $d=5$ \cite{kn:CaMa2,kn:HoMaSt},
$d=4$ \cite{kn:HLM} y con rotaci'on \cite{kn:BLMPSV} y c'omo estos
modelos explican la radiaci'on de Hawking
\cite{kn:DaMat,kn:MaSt2,kn:DaGMat}, el {\it Principio de
  correspondencia} \cite{kn:Sen6,kn:HoPo}, la relaci'on de estos
c'alculos de la entrop'ia con el agujero negro en tres
dimensiones de Ba~nados, Teitelboim y Zanelli
\cite{kn:BTZ2,kn:Hy,kn:SfSk,kn:Car2}, y la relaci'on de todo esto con
la dualidad entre SUGRA y Teor'ias de Campos Conformes
\cite{kn:AGMOO} a trav'es de la geometr'ia del horizonte.
Esperamos que al lector interesado estas lecciones le sirvan como un
buen punto de apoyo para adentrarse en este interesante mundo de
ideas.

\newpage
\appendix
\chapter{Convenios y f'ormulas}
\label{sec-conventions}

%%%%%%%%%%%%%%%%%%%%%%%%%%%%%%%%%%%%%%%%%%%%%%%%%%%%%%%%%%%%%%%%%%%%%%
\section{Convenios de geometr'ia diferencial}

Utilizamos letras griegas $\mu,\nu,\rho,\ldots$ como 'indices
tensoriales en la base de coordenadas ({\it 'indices curvos}) y
letras latinas $a,b,c\ldots$ como 'indices tensoriales en la base
del espacio tangente asociada a una t'etrada ({\it 'indices
  Lorentz} o {\it planos}). Usamos dobles gorros sobre objetos
once-dimensionales, un gorro para objetos diez-dimensionales y ninguno
para objetos en menos dimensiones.  Simetrizamos y antisimetrizamos
con peso uno (dividiendo por $n!$). A veces utilizamos el siguiente
convenio: los 'indices que no escribimos expl'icitamente
est'an completamente antisimetrizados en el orden correspondiente.
Esto es similar a la notaci'on de formas diferenciales, pero los
factores num'ericos difieren.

Nuestra signatura es $(+-\cdots -)$. $\eta$ es la m'etrica de
Minkowski y la m'etrica en general es $g$ ($\hat{\hat{g}}$ es la
m'etrica 11-dimensional, $\hat{g}$ es la m'etrica 10-dimensional en
el sistema de referencia conforme de la cuerda). Los 'indices
planos y curvos est'an relacionados por las t'etradas
$e_{a}{}^{\mu}$ y sus inversos $e_{\mu}{}^{a}$, que satisfacen

\begin{equation}
e_{a}{}^{\mu}e_{b}{}^{\nu}g_{\mu\nu}=\eta_{ab}\, ,
\hspace{1cm}
e_{\mu}{}^{a}e_{\nu}{}^{b}\eta_{ab}=g_{\mu\nu}\, .  
\end{equation}

$\nabla$ es la derivada covariantes total (con respecto a
reparametrizaciones y transformaciones Lorentz locales) y ${\cal D}$
es la derivada covariante Lorentz. Est'an definidas por

\begin{equation}
\end{equation}

\noindent y sobre tensores del espacio tangente y espinores  ($\psi$) por

\begin{equation}
\begin{array}{rcl}
\nabla_{\mu}\xi^{\nu} & =  & \partial_{\mu}\xi^{\nu}
+\Gamma_{\mu\rho}{}^{\nu}\xi^{\rho}\, , \\
& & \\
{\cal D}_{\mu}\xi^{a}  & =  &  \partial_{\mu}\xi^{a} 
+\omega_{\mu b}{}^{a} \xi^{b}\, , \\
& & \\
\nabla_{\mu}\psi 
& = & \partial_{\mu} \psi -{\textstyle\frac{1}{4}}
\omega_{\mu}{}^{ab}\Gamma_{ab}\psi\, ,\\
\end{array}
\end{equation}

\noindent donde $\Gamma_{ab}$ es el producto antisim'etrico de dos 
matrices gamma. Los tensores de curvatura y torsi'on correspondientes
est'an definidos a trav'es de las identidades de Ricci

\begin{equation}
  \begin{array}{rcl}
\left[ \nabla_{\mu} , \nabla_{\nu} \right]\ \xi^{\rho} & = &
R_{\mu\nu\sigma}{}^{\rho}(\Gamma)\, \xi^{\sigma} 
+T_{\mu\nu}{}^{\sigma}\nabla_{\sigma}  \xi^{\rho}\, , \\
& & \\
\left[ {\cal D}_{\mu} , {\cal D}_{\nu} \right]\, \xi^{a} & = &
R_{\mu\nu b}{}^{a} (\omega)\xi^{b}\, ,\\
\end{array}
\end{equation}

\noindent y las curvaturas est'an dadas en funci'on de las conexiones por

\begin{equation}
\label{eq:curvatures}
\begin{array}{rcl}
R_{\mu\nu\rho}{}^{\sigma}(\Gamma) & = & 
2\partial_{[\mu}\Gamma_{\nu]\rho}{}^{\sigma}
+ 2\Gamma_{[\mu|\lambda}{}^{\sigma} \Gamma_{\nu]\rho}{}^{\lambda}\, ,\\
& & \\
R_{\mu\nu a}{}^{b} (\omega) & = & 2\partial_{[\mu}\, \omega_{\nu] a}{}^{b} 
-2\omega_{[\mu| a}{}^{c}\,\omega_{|\nu]c}{}^{b}\, .\\
\end{array}  
\end{equation}

Imponiendo  el postulado de las t'etradas

\begin{equation}
\nabla_{\mu}e_{a}{}^{\mu}=0\, ,  
\end{equation}

\noindent las dos conexiones est'an relacionadas por

\begin{equation}
\omega_{\mu a}{}^{b} = \Gamma_{\mu a}{}^{b} 
+e_{a}{}^{\nu}\partial_{\mu}e_{\nu}{}^{b}\, ,
\end{equation}

\noindent y las curvaturas de ambas conexiones est'an a su vez
 relacionadas por

\begin{eqnarray}
R_{\mu\nu\rho}{}^{\sigma}(\Gamma) = e_{\rho}{}^{a} e^{\sigma}{}_{b}
R_{\mu\nu a}{}^{b}(\omega)\, .
\end{eqnarray}

Si imponemos el postulado m'etrico

\begin{equation}
\nabla_{\mu}g_{\rho\sigma}=0  \, ,
\end{equation}

\noindent entonces la conexi'on se puede escribir siempre as'i:

\begin{equation}
\Gamma_{\mu\nu}{}^{\rho} =
\biggl\{\!
\begin{array}{c}
\mbox{\scriptsize $\rho$} \\
\mbox{\scriptsize $\mu\, \nu$} \\
\end{array}\!\biggr\}
+K_{\mu\nu}{}^{\rho}
=\Gamma_{\mu\nu}{}^{\rho} (g) +K_{\mu\nu}{}^{\rho}\, ,
\end{equation}

\noindent donde

\begin{equation}
\biggl\{\!
\begin{array}{c}
\mbox{\scriptsize $\rho$} \\
\mbox{\scriptsize $\mu\, \nu$} \\
\end{array}\!
\biggr\}
={\textstyle\frac{1}{2}}g^{\rho\sigma}
\left\{\partial_{\mu}g_{\nu\sigma} +\partial_{\nu}g_{\mu\sigma}
-\partial_{\sigma}g_{\mu\nu} \right\}\, .
\end{equation}

\noindent son los {\it s'imbolos de Christoffel},  y $K$,
es el {\it tensor de contorsi'on}, que depende del {\it tensor de
  torsi'on} $T$ as'i:

\begin{equation}
K_{\mu\nu}{}^{\rho} = {\textstyle\frac{1}{2} }g^{\rho\sigma}
\left\{ T_{\mu\sigma\nu} +T_{\nu\sigma\mu} -T_{\mu\nu\sigma}\right\}\, .
\end{equation}

Si, adem'as del postulado m'etrico imponemos el postulado de las t'etradas,
entonces la relaci'on entre $\Gamma$ y $\omega$ implica

\begin{equation}
\label{eq:spincon}
\omega_{abc} = \omega_{abc}(e) + K_{abc}\, ,
\hspace{1cm}
\omega_{abc}(e) = -\Omega_{abc}+\Omega_{bca} -\Omega_{cab}\, ,
\hspace{1cm}
\Omega_{ab}{}^{c} = 
e_{a}{}^{\mu}e_{b}{}^{\nu} \partial_{[\mu}e^{c}{}_{\nu]}\, .
\end{equation}

$\omega(e)$ es la conexi'on de esp'in relacionada con la
conexi'on de Levi-Civit\`a $\Gamma(g)$ por el postulado de las
t'etradas. 

%%%%%%%%%%%%%%%%%%%%%%%%%%%%%%%%%%%%%%%%%%%%%%%%%%%%%%%%%%%%%%%%%%%%%%

\section{Matrices gamma y espinores}

%%%%%%%%%%%%%%%%%%%%%%%%%%%%%%%%%%%%%%%%%%%%%%%%%%%%%%%%%%%%%%%%%%%%%%%%%%

\subsection{$d=11$}

Nuestras matrices gamma 11-dimensionales satisfacen

\begin{equation}
\left\{ \hat{\hat{\Gamma}}{}^{\hat{\hat{a}}},
\hat{\hat{\Gamma}}{}^{\hat{\hat{b}}}\right\} =
+2\hat{\hat{\eta}}{}^{\hat{\hat{a}}\hat{\hat{b}}}\, ,
\end{equation}

\noindent con la und'ecima matriz gamma
$\hat{\hat{\Gamma}}{}^{10}$ relacionada con las otras por

\begin{equation}
\hat{\hat{\Gamma}}{}^{\hat{\hat{10}}} = i\hat{\hat{\Gamma}}{}^{\hat{\hat{0}}}
\ldots\hat{\hat{\Gamma}}{}^{\hat{\hat{9}}}\equiv-i\hat{\Gamma}_{\hat{11}}\, ,  
\end{equation}

\noindent donde $\hat{\Gamma}_{\hat{11}}$ ser'a la matriz de quiralidad
en 10 dimensiones. Son puramente imaginarias (es decir, est'an en una
representaci'on de Majorana)

\begin{equation}
\hat{\hat{\Gamma}}{}^{\hat{\hat{a}}\, \star} 
= -\hat{\hat{\Gamma}}{}^{\hat{\hat{a}}}\, ,
\end{equation}

\noindent y son todas, salvo
$\hat{\hat{\Gamma}}{}^{\hat{\hat{0}}}$, antiherm'iticas :

\begin{equation}
\begin{array}{rcl}
\hat{\hat{\Gamma}}{}^{\hat{\hat{0}}\, \dagger} & = & 
+\hat{\hat{\Gamma}}{}^{\hat{\hat{0}}}\, . \\
& & \\
\hat{\hat{\Gamma}}{}^{\hat{\hat{\imath}}\, \dagger} & = & 
-\hat{\hat{\Gamma}}{}^{\hat{\hat{\imath}}}\, ,
\hspace{.5cm}
\hat{\hat{\imath}}= 1,\ldots,10\, .\\
\end{array}
\end{equation}

Las propiedades de hermiticidad combinadas con su el hecho de que son
imaginarias puras, implican que todas ellas son sim'etricas excepto
$\hat{\hat{\Gamma}}{}^{\hat{\hat{0}}}$, que es antisim'etrica:

\begin{equation}
\begin{array}{rcl}
\hat{\hat{\Gamma}}{}^{\hat{\hat{0}}\, T} & = & 
-\hat{\hat{\Gamma}}{}^{\hat{\hat{0}}}\, . \\
& & \\
\hat{\hat{\Gamma}}{}^{\hat{\hat{\imath}}\, T} & = & 
+\hat{\hat{\Gamma}}{}^{\hat{\hat{\imath}}}\, ,
\hspace{.5cm}
\hat{\hat{\imath}}= 1,\ldots,10\, .\\
\end{array}
\end{equation}

$\hat{\hat{\Gamma}}{}^{\hat{\hat{0}}}$ tiene la propiedad

\begin{equation}
\hat{\hat{\Gamma}}{}^{\hat{\hat{0}}}\, \hat{\hat{\Gamma}}{}^{\hat{\hat{a}}}\,
\hat{\hat{\Gamma}}{}^{\hat{\hat{0}}}  
=\hat{\hat{\Gamma}}{}^{\hat{\hat{a}}\, \dagger}\, .
\end{equation}

\noindent por lo que podemos escoger como matriz de conjugaci'on de Dirac 
$\hat{\hat{\cal D}}$ la matriz antisim'etrica y real

\begin{equation}
 \hat{\hat{\cal D}} =i\hat{\hat{\Gamma}}{}^{0} \, ,
\end{equation}

\noindent que satisface

\begin{equation}
\label{eq:gammaherm}
\hat{\hat{\cal D}}\, 
\hat{\hat{\Gamma}}{}^{\hat{\hat{a}}_{1}\cdots \hat{\hat{a}}_{n}} \,
\hat{\hat{\cal D}}{}^{-1}
=(-1)^{\left[n/2 \right]} 
\left(\hat{\hat{\Gamma}}{}^{\hat{\hat{a}}_{1}\cdots \hat{\hat{a}}_{n}}
\right)^{\dagger}\, .
\end{equation}

Nuestra matriz de conjugaci'on de carga es igual a la de 
conjugaci'on de Dirac

\begin{equation}
\hat{\hat{\cal C}}=\hat{\hat{\cal D}}=i\hat{\hat{\Gamma}}{}^{0}\, ,  
\end{equation}

\noindent y, por lo tanto

\begin{equation}
\hat{\hat{\cal C}}{}^{\, T}=  \hat{\hat{\cal C}}{}^{\, \dagger}=  
\hat{\hat{\cal C}}{}^{-1}= - \hat{\hat{\cal C}}\, ,
\end{equation}

\noindent y

\begin{equation}
\hat{\hat{\cal C}}\, \hat{\hat{\Gamma}}{}^{\hat{\hat{a}}}\
\hat{\hat{\cal C}}{}^{-1}= -\hat{\hat{\Gamma}}{}^{\hat{\hat{a}}\, T}\, .
\end{equation}

\noindent Esta 'ultima propiedad implica

\begin{equation}
\label{eq:gammatrans}
\hat{\hat{\cal C}}\,
\hat{\hat{\Gamma}}{}^{\hat{\hat{a}}_{1}\cdots \hat{\hat{a}}_{n}}\,
\hat{\hat{\cal C}}{}^{-1}= 
(-1)^{n+\left[n/2\right]}\left(
\hat{\hat{\Gamma}}{}^{\hat{\hat{a}}_{1}\cdots 
\hat{\hat{a}}_{n}}\right){}^{T}\, .  
\end{equation}

La definici'on habitual de conjugado de Dirac 

\begin{equation}
\bar{\hat{\hat{\lambda}}} 
=\hat{\hat{\lambda}}{}^{\dagger}\hat{\cal{\cal D}}\, ,
\end{equation}

\noindent y conjugado de Majorana

\begin{equation}
\hat{\hat{\lambda}}{}^{c}= \hat{\hat{\lambda}}{}^{T} \hat{\hat{\cal C}}\, ,
\end{equation}

\noindent y nuestra
elecci'on de matrices de conjugaci'on de Dirac y carga
$\hat{\hat{\cal C}}=\hat{\hat{\cal D}}$ implican que la condici'on de
Majorana

\begin{equation}
\bar{\hat{\hat{\lambda}}}=\hat{\hat{\lambda}}{}^{c}\, ,  
\end{equation}

\noindent  es equivalente a requerir que todas las componentes
de un espinor de Majorana sean reales. Usando (\ref{eq:gammatrans}) y
la definici'on de espinor de Majorana ({\it anticonmutante}), tenemos

\begin{equation}
\label{eq:spintrans}
\overline{\hat{\hat{\epsilon}}}\,
\hat{\hat{\Gamma}}{}^{\hat{\hat{a}}_{1}\cdots \hat{\hat{a}}_{n}}\,
\hat{\hat{\psi}} =  
(-1)^{n+\left[n/2\right]}\, 
\bar{\hat{\hat{\psi}}} \,
\hat{\hat{\Gamma}}{}^{\hat{\hat{a}}_{1}\cdots \hat{\hat{a}}_{n}}\,
\hat{\hat{\epsilon}}\, ,
\end{equation}

\noindent con lo que esta combinaci'on bilineal es sim'etrica para
 $n=0,3,4,7,8$ y antisim'etrica para $n=1,2,5,6,9,10$.

Por otro lado, tomando el conjugado herm'itico de la misma combinaci'on
\footnote{Usamos el convenio $(ab)^{\star}=
  +a^{\star}b^{\star}$ para variables anticonmutantes.} y usando
(\ref{eq:gammaherm}) tenemos

\begin{equation}
\left(\bar{\hat{\hat{\epsilon}}}\, 
\hat{\hat{\Gamma}}{}^{\hat{\hat{a}}_{1}\cdots \hat{\hat{a}}_{n}}\,
\hat{\hat{\psi}}\right)^{\dagger} =  
(-1)^{\left[n/2\right]} 
\bar{\hat{\hat{\psi}}}\, 
\hat{\hat{\Gamma}}{}^{\hat{\hat{a}}_{1}\cdots \hat{\hat{a}}_{n}}\,
\hat{\hat{\epsilon}}\, ,
\end{equation}

\noindent lo que implica, al comparar con Ec.~(\ref{eq:spintrans}) 
que pata $n$ par es real e imaginario para $n$ impar.

Finalmente, tenemos la identidad

\begin{equation}
\label{eq:dualgamma11}
\hat{\hat{\Gamma}}{}^{\hat{\hat{a}}_{1}\cdots\hat{\hat{a}}_{n}}
=i\frac{(-1)^{[n/2]+1}}{(11-n)!}
\hat{\hat{\epsilon}}{}^{\hat{\hat{a}}_{1}\cdots\hat{\hat{a}}_{n}
\hat{\hat{b}}_{1}\cdots\hat{\hat{b}}_{11-n}}
\hat{\hat{\Gamma}}_{\hat{\hat{b}}_{1}\cdots\hat{\hat{b}}_{11-n}}\, .
\end{equation}

%%%%%%%%%%%%%%%%%%%%%%%%%%%%%%%%%%%%%%%%%%%%%%%%%%%%%%%%%%%%%%%%%%%%%%%%%%

\subsection{$d=10$}
\label{sec-d10gammas}

La representaci'on de Majorana de las matrices gamma 11-dimensionales
se puede construir a partir de una representaci'on de Majorana
10-dimensional (son matrices de la misma dimensi'on)

\begin{equation}
\label{eq:11vs10gamma}
\left\{
\begin{array}{rcl}
\hat{\hat{\Gamma}}{}^{\hat{a}} & = &
\hat{\Gamma}^{\hat{a}}\, ,\,\,\, \hat{a}=0,\ldots,9\, ,\\
& & \\
\hat{\hat{\Gamma}}{}^{10} & = &
+i\hat{\Gamma}^{0}\ldots\hat{\Gamma}^{9}\, .\\
\end{array}
\right.
\end{equation}

Los espinores 10-dimensionales son id'enticos a los 11-dimensionales
y las mismas definiciones e identidades son v'alidas para ellos, pero
en 10 dimensiones tambi'en se pueden definir espinores de Weyl, que
poseen propiedades adicionales. Los espinores de Weyl se definen con
la matriz de quiralidad $\hat{\Gamma}_{11}$

\begin{equation}
\hat{\Gamma}_{11} = -\hat{\Gamma}^{0}\ldots\hat{\Gamma}^{9}
=i\hat{\hat{\Gamma}}{}^{10}\, ,
\end{equation}

\noindent que es herm'itica y satisface
$(\hat{\Gamma}_{11})^{2}=+1$. Espinores de Weyl de quiralidad positiva
$\hat{\psi}^{(+)}$ y negativa $\hat{\psi}^{(-)}$ se definen por

\begin{equation}
\hat{\Gamma}_{11}\hat{\psi}^{(\pm)}=\pm \hat{\psi}^{(\pm)}\, .
\end{equation}

Adem'as, en 10 dimensiones se pueden definir espinores de
Majorana-Weyl. Es 'util trabajar en una representaci'on de
Majorana-Weyl de las matrices gamma en la que 'estas son imaginarias
y $\hat{\Gamma}_{11}$ toma la forma

\begin{equation}
\hat{\Gamma}_{11}=\mathbb{I}_{16\times 16}\otimes \sigma^{3}=
\left( 
\begin{array}{cc}
\mathbb{I}_{16\times 16} & 0 \\
& \\
0 & -\mathbb{I}_{16\times 16} \\
\end{array}
\right)\, .
\end{equation}

En la representaci'on de Majorana-Weyl, cada espinor de Majorana
(real) se puede construir como suma directa de un espinor de
quiralidad positiva y otro de quiralidad negativa de 16 componentes:

\begin{equation}
\hat{\psi}=
\left(
\begin{array}{c}
\hat{\psi}^{(+)} \\
\\
\hat{\psi}^{(-)} \\
\end{array}
\right)\, .
\end{equation}

Finalmente, tenemos la identidad

\begin{equation}
\label{eq:dualgamma10}
\Gamma_{11}\hat{\Gamma}^{\hat{a}_{1}\cdots \hat{a}_{n}}=
\frac{(-1)^{[(10-n)/2]+1}}{(10-n)!}
\hat{\epsilon}^{\hat{a}_{1}\cdots \hat{a}_{n}
\hat{b}_{1}\cdots \hat{b}_{10-n}}
\hat{\Gamma}_{\hat{b}_{1}\cdots \hat{b}_{10-n}}\, .
\end{equation}

%%%%%%%%%%%%%%%%%%%%%%%%%%%%%%%%%%%%%%%%%%%%%%%%%%%%%%%%%%%%%%%%%%%%%%
\subsection{$d=9$}
\label{sec-d9gammas}

La representaci'on 10-dimensional de Majorana-Weyl, se puede construir
a partir de una representaci'on puramente real (Majorana) de las
matrices gamma 9-dimensionales:

\begin{equation}
\left\{
\begin{array}{rcl}
\hat{\Gamma}^{a} & = & \Gamma^{a}\otimes \sigma^{2}\, ,
\ \ a=0,\ldots,8\, ,\\
& & \\
\hat{\Gamma}^{9} & = & \mathbb{I}_{16\times 16} 
\otimes i\sigma^{1}\, ,\\
\end{array}
\right.
\end{equation}

\noindent donde $\Gamma^{8}$ satisface

\begin{equation}
\Gamma^{8}=\Gamma^{0}\cdots\Gamma^{7}\, .
\end{equation}

Como antes, $\Gamma^{8}$ ser'a proporcional a la matriz de quiralidad
8-dimensional

\begin{equation}
\Gamma_{(8)\, 9}=i\Gamma^{8}=i\Gamma^{0}\cdots\Gamma^{7}\, .
\end{equation}

Se puede comprobar expl'icitamente que con estas definiciones, la
  representaci'on 10-dimensional es quiral y que $\hat{\Gamma}_{11}=
  \mathbb{I}_{16\times 16}\otimes \sigma^{3}$.

%%%%%%%%%%%%%%%%%%%%%%%%%%%%%%%%%%%%%%%%%%%%%%%%%%%%%%%%%%%%%%%%%%%%%%

\subsection{$d=4$}

En 4 dimensiones podemos utilizar representaciones Majorana o Weyl,
pero no Majorana-Weyl. Nosotros utilizamos una representaci'on
puramente imaginaria (Majorana). La matriz de quiralidad es

\begin{equation}
\gamma_{5}=-i\gamma^{0}\gamma^{1}\gamma^{2}\gamma^{3}
=\frac{i}{4!}\epsilon_{abcd}\gamma^{abcd}\, ,
\end{equation}

\noindent herm'itica e imaginaria (y, por lo tanto, antisim'etrica).
En esta representaci'on podemos usar como matrices de conjugaci'on
de Dirac y de conjugaci'on de carga ${\cal D}={\cal
  C}_{-}=i\gamma^{0}$, que es real y antisim'etrica.  La condici'on
de Majorana es que los espinores de Majorana son reales
$\psi=\psi^{*}$.

Finalmente, tambi'en tenemos la identidad

\begin{equation}
\label{eq:dualgammaidentityind4}
\gamma^{a_{1}\cdots a_{n}} =\frac{(-1)^{\left[n/2\right]}i}{(4-n)!}
\epsilon^{a_{1}\cdots a_{n}b_{1}\cdots b_{4-n}} \gamma_{b_{1}\cdots b_{4-n}}
\gamma_{5}\, .
\end{equation}

%%%%%%%%%%%%%%%%%%%%%%%%%%%%%%%%%%%%%%%%%%%%%%%%%%%%%%%%%%%%%%%%%%%%%%
\section{Geometr'ia extr'inseca}
\label{sec-extrinsic}

Sea una hipersuperficie $\Sigma$ inmersa en un espacio-tiempo
$d$-dimensional con m'etrica $g_{\mu\nu}$ y sea  
$n^{\mu}$ su vector unitario normal:

\begin{equation}
n^{\mu}n_{\mu}=\varepsilon\, ,
\hspace{1cm}
\left\{
  \begin{array}{l}
\varepsilon=+1\, ,\,\,\,\Sigma\,\,\, {\rm tipo\,\,\, espacio}\, ,\\
\\
\varepsilon=-1\, ,\,\,\,\Sigma\,\,\, {\rm tipo\,\,\, luz}\, .\\
  \end{array}
\right.  
\end{equation}

La m'etrica inducida en $\Sigma$ es

\begin{equation}
h_{\mu\nu}=g_{\mu\nu}-\varepsilon n_{\mu}n_{\nu}\, .  
\end{equation}

$h_{\mu\nu}$ tiene car'acter $(d-1)$-dimensional pero est'a escrita
en forma $d$-dimensional y es evidentemente singular y no se puede
invertir. Sus 'indices se suben y bajan con $g$. Obs'ervese que
$h_{\mu\nu}n^{\nu}=0$ y por ello $h$ se puede usar para proyectar
tensores sobre $\Sigma$.

Una forma de medir cuan curvada est'a $\Sigma$ en el espacio-tiempo
ser'ia medir la variaci'on de su vector unitario normal sobre
ella. Matem'aticamente esto se expresa as'i:

\begin{equation}
{\cal K}_{\mu\nu}\equiv h_{\mu}{}^{\alpha} h_{\nu}{}^{\beta}
\nabla_{(\alpha}n_{\beta)}\, ,
\end{equation}

\noindent donde  ${\cal K}_{\mu\nu}$ recibe el nombre de
{\it curvatura extr'inseca} o {\it segunda forma  fundamental}.

Podemos considerar un campo de vectores unitarios $n^{\mu}(x)$
definido sobre todo el espacio-tiempo. Este campo determina una
familia de hipersuperficies. Podemos entonces calcular la derivada de
Lie de las m'etricas inducidas sobre las mismas en la direcci'on de
los vectores unitarios. El resultado es que esta derivada es dos veces
la curvatura extr'inseca

\begin{equation}
{\cal K}_{\mu\nu} ={\textstyle\frac{1}{2}}\pounds_{n}h_{\mu\nu}\, .  
\end{equation}

La traza de la curvatura extr'inseca se denota por ${\cal K}$ y
est'a dada por

\begin{equation}
\label{eq:traceextrinsiccurvature}
{\cal K}= h^{\mu\nu}{\cal K}_{\mu\nu} = h^{\mu\nu}\nabla_{\mu}n_{\nu}\, .
\end{equation}

%%%%%%%%%%%%%%%%%%%%%%%%%%%%%%%%%%%%%%%%%%%%%%%%%%%%%%%%%%%%%%%%%%%%%%
\section{$n$-Esferas}
\label{sec-sph}

La $n$-dimensional esfera de radio unidad $S^{n}$ puede definirse como
la hipersuperficie de radio constante $r=1$ en el espacio
euclidiano $(n+1)$-dimensional. $r$ es una de las coordenadas
esf'ericas $(n+1)$-dimensionales
$\{r,\varphi,\theta_{1},\ldots,\theta_{n-1}\}$ definidas en funci'on
de las cartesianas $\{x^{1},\ldots ,x^{n}\}$ por las relaciones

\begin{equation}
\left\{
\begin{array}{rcl}
x^{1} & = & \rho_{n-1} \sin\varphi\, , \\
& & \\
x^{2} & = & \rho_{n-1} \cos\varphi\, ,  \\
& & \\
x^{3} & = & \rho_{n-2} \cos\theta_{1}\, ,\\
& & \\
\vdots & & \vdots \\
& & \\
x^{k} & = & \rho_{n-k+1} \cos \theta_{k-2}\, ,
\hspace{1cm}
3 \leq k \leq n+1\, , \\
\end{array}
\right.
\end{equation}

\noindent donde

\begin{equation}
\left\{
\begin{array}{rcl}
\rho_{l} & = & [(x^{1})^{2}+\ldots +(x^{n+1-l})^{2}]^{1/2} = 
r \prod_{m=1}^{l} \sin \theta_{n-m}\, ,  \\
& & \\
\rho_{0} & = & r = [(x^{1})^{2} + \ldots +(x^{n+1})^{2}]^{1/2}\, , \\
\end{array}
\right.
\end{equation}

La forma de volumen en $S^{n}$ es

\begin{equation}
d\Omega^{n} \equiv d\varphi
\prod_{i=1}^{n-1} \sin^{i}\theta_{i} d\theta_{i}\, ,
\end{equation}

En coordenadas cartesianas del espacio ambiente $(n+1)$-dimensional
toma la forma

\begin{equation}
\label{eq:snvolume}
d\Omega^{n} = \frac{1}{n!r^{n+1}} \epsilon_{\mu_{1}\ldots\mu_{n+1}}
x^{\mu_{n+1}} dx^{\mu_{1}}\ldots dx^{\mu_{n}}\, .
\end{equation}

Otras identidades 'utiles son

\begin{equation}
\left\{
\begin{array}{rcl}
d^{n+1}x & = & r^{n}drd\Omega^{n}\, ,\\
& & \\
r^{n} d\Omega^{n} & = & d^{n}y\sqrt{|g|}\, , \\
\end{array}
\right.
\end{equation}

\noindent donde las $y$s son coordenadas en $S^{n}$.

El volumen de la $n$-esfera de radio unidad $S^{n}$ viene dado por la
integral de la forma de volumen sobre toda la esfera:

\begin{equation}
\omega_{(n)} = \int_{S^{n}}d \Omega^{n} =
\frac{2\pi^{\frac{n+1}{2}}}{\Gamma(\frac{n+1}{2})}\, .
\end{equation}

Usando

\begin{equation}
\Gamma(x+1) = x\Gamma(x)\, ,
\hspace{.5cm}
\Gamma (0) =1\, ,
\hspace{.5cm}
\Gamma (1/2) = \pi^{1/2}\, ,
\end{equation}

\noindent obtenemos $\omega_{(1)}=2\pi\, ,\omega_{(2)}=4\pi\, , 
\omega_{(3)}=2\pi^{2}$ etc.

La m'etrica inducida en $S^{n}$ en coordenadas esf'ericas es
$d\Omega^{2}_{(n)}$ y su relaci'on con la m'etrica del espacio
euclidiano ambiente es:

\begin{equation}
\begin{array}{rcl}
d\vec{x}^{\, 2} & = & d\rho_{0}^{2} +\rho_{0}^{2} d\theta_{n-1}^{2}+\ldots
+\rho_{n-2}^{2}d\theta_{1}^{2}+\rho_{n-1}^{2}d\varphi^{2} \\
& & \\
& = & dr^{2} +r^{2} 
\left\{d\theta^{2}_{n-1} +\sin^{2}\theta_{n-1} 
\left[ d\theta^{2}_{n-2} +\sin^{2}\theta_{n-2} 
\left( d\theta^{2}_{n-3} +\sin^{2}\theta_{n-3} 
\left(\cdots \right.\right.\right.\right.\\
& & \\
& & 
\left.\left.
\cdots
\sin^{2}\theta_{2} 
\left( d\theta^{2}_{1} +\sin^{2}\theta_{1}d\varphi^{2}\right)
\cdots\right]
\right\} \\
& & \\
& = & dr^{2} +r^{2}d\Omega_{(n)}^{2}\, .\\
\end{array}
\end{equation}

\newpage
%%%%%%%%%%%%%%%%%%%%%%%%%%%%%%%%%%%%%%%%%%%%%%%%%%%%%%%%%%%%%%%%%
%%%%%%%%%%%%%%%%%%%%%%%%%%%%%%%%%%%%%%%%%%%%%%%%%%%%%%%%%%%%%%%%%
%%%%%%%%%%%%%%%%%%%%%%%%%%%%%%%%%%%%%%%%%%%%%%%%%%%%%%%%%%%%%%%%%
%%%%%%%%%%%%%%%%%%%%%%%%%%%%%%%%%%%%%%%%%%%%%%%%%%%%%%%%%%%%%%%%%
%%%%%%%%%%%%%%%%%%%%%%%%%%%%%%%%%%%%%%%%%%%%%%%%%%%%%%%%%%%%%%%%%
%%%%%%%%%%%%%%%%%%%%%%%%%%%%%%%%%%%%%%%%%%%%%%%%%%%%%%%%%%%%%%%%%


\small

\begin{thebibliography}{260}

%%%%%%%%%%%%%%%%%%%%%%%%%%%%%%%%%%%%%%%%%%%%%%%%%%%%%%%%%%%%%%%%%%%%%%
%intro%%%%%%%%%%%%%%%%%%%%%%%%%%%%%%%%%%%%%%%%%%%%%%%%%%%%%%%%%%%%%%%%
%%%%%%%%%%%%%%%%%%%%%%%%%%%%%%%%%%%%%%%%%%%%%%%%%%%%%%%%%%%%%%%%%%%%%%
%reviews on string theory BHs

\bibitem{kn:Pee2} A.W.~Peet,
                  {\sl TASI Lectures on Black Holes in String Theory},
                  [{\tt hep-th/0008241}].               

\bibitem{kn:M} J.M.~Maldacena,
               Ph.D.~Thesis, Princeton University,
               {\sl Black Holes in String Theory},
               [{\tt hep-th/9607235}].

\bibitem{kn:DM} S.R.~Das and S.D.~Mathur,
                {\sl The Quantum Physics of Black Holes: 
                Results from String Theory},
                {\it Ann.~Rev. Nucl.~Part.~Sci.}~\textbf{50} (2000) 153-206.
                [{\tt gr-qc/0105063}].

\bibitem{kn:Ho2} G.T.~Horowitz,
                {\sl The Origin of Black Hole Entropy in String
                Theory},
                Contribution to the Proceedings of the Pacific
                Conference on Gravitation and Cosmology, Seoul, Korea,
                February 1-6 1996.
                [{\tt gr-qc/9604051}].

\bibitem{kn:Ho3} G.T.~Horowitz,
                 {\sl Quantum States of Black Holes},
                 Presented at {\it Symposium on Black Holes and 
                 Relativistic Stars}, (dedicated to memory of 
                 S.~Chandrasekhar), Chicago, IL, 14-15 December 1996. 
                 [{\tt gr-qc/9704072}].

        
\bibitem{kn:M2} J.M.~Maldacena,
                {\sl Black Holes And D-Branes},
                Lectures given at {\it 33rd Karpacz Winter School 
                of Theoretical Physics: Duality - Strings and Fields},
                Karpacz, Poland, 13-22 February 1997. 
                [{\tt hep-th/9705078}].

\bibitem{kn:Pee} A.W.~Peet,
                 {\sl The Bekenstein Formula and String Theory 
                 (N-Brane Theory)},
                 {\it Class.~Quantum Grav.}~\textbf{15} (1998) 3291.
                 [{\tt hep-th/9712253}].

\bibitem{kn:Ske} K.~Skenderis, 
                 {\sl  Black Holes and Branes in String Theory},
                 Talk given at the International School of Nuclear Physics: 
                 36th Course: {\it From the Planck Length to the Hubble Radius}, 
                 Erice, 29 August-7 September 1998.
                 \textit{Lect. Notes Phys.}~\textbf{541} (2000) 325-364.
                 [{\tt hep-th/9901050}].

\bibitem{kn:BaKw} B.~E.~Baaquie and L.~C.~Kwek,
                  {\sl Superstrings, Gauge Fields and Black Holes}
                  {\it Int.~J.~Mod.~Phys.}~\textbf{A16} (2001) 2605.
                  [{\tt hep-th/0002165}].

\bibitem{kn:Moh2} T.~Mohaupt,
                  {\sl Black Holes in Supergravity and String Theory}, 
                  {\it Class.~Quantum Grav.}~\textbf{17} (2000) 3429.
                  [{\tt hep-th/0004098}].

\bibitem{kn:Wad} S.~Wadia,
                 {\sl Status of Microscopic Modeling of Black Holes 
                 by D1-D5 System},
                 Talk given at 9th Marcel Grossmann Meeting on 
                 {\sl Recent Developments in Theoretical and Experimental 
                 General Relativity, Gravitation and Relativistic Field 
                 Theories} (MG 9), Rome, Italy, 2-9 Jul 2000
                 [{\tt hep-th/0011286}].

\bibitem{kn:My} R.C.~Myers,
                {\sl Black Holes and String Theory},
                Summary of Lectures given at {\it Fourth Mexican School
                on Gravitation and Mathematical Physics},
                [{\tt gr-qc/0107034}].
                
%%%%%%%%%%%%%%%%%%%%%%%%%%%%%%%%%%%%%%%%%%%%%%%%%%%%%%%%%%%%%%%%%%%%%%
%lecci'on 1%%%%%%%%%%%%%%%%%%%%%%%%%%%%%%%%%%%%%%%%%%%%%%%%%%%%%%%%%%%
%%%%%%%%%%%%%%%%%%%%%%%%%%%%%%%%%%%%%%%%%%%%%%%%%%%%%%%%%%%%%%%%%%%%%%

\bibitem{kn:GH} G.W.~Gibbons and S.W.~Hawking,
                {\sl Action Integrals and Partition Functions in
                Quantum Gravity},
                {\it Phys.~Rev.}~\textbf{D15} (1977) 2752.
                
\bibitem{kn:GH5} G.W.~Gibbons and S.W.~Hawking (Eds.),
                 {\sl Euclidean Quantum Gravity},
                 World Scientific, Singapore (1993).

\bibitem{kn:Ver} E.~Verdager,
                 {\sl Termodin'amica de Agujeros Negros},
                 lecciones dadas en la {\it IV Escuela La Hechicera de 
                 Relatividad, Campos y Astrof'isica}, M'erida, Venezuela, 
                 1998

\bibitem{kn:BiDa} N.D.~Birrell and P.C.W.~Davies,
                  {\sl Quantum Fields in Curved Space},
                  Cambridge University Press, Cambridge U.K.~(1989). 

\bibitem{kn:W10} R.M.~Wald,
                 {\sl Quantum Field Theory in Curved Spacetime 
                 and Black Hole Thermodynamics},
                 The University of Chicago Press, Chicago (1994).

\bibitem{kn:W8} R.M.~Wald,
                {\sl The Thermodynamics of Black Holes},
                {\it Living Rev.~Rel.}~\textbf{4} (2001) 6.
                {\tt http://www.livingreviews.org/Articles/Volume4/2001-6wald}
                [{\tt gr-qc/9912119}].

\bibitem{kn:FN} I. Novikov and V.P.~Frolov,
                {\sl Physics of Black Holes},
                Kluwer Academic Publishers, The Netherlands, 1989.

\bibitem{kn:Tow5} P.K.~Townsend,
                  {\sl Black Holes},
                  Lecture notes 
                  [{\tt gr-qc/9707012}].

\bibitem{kn:H} D.~Hilbert,
               {\sl Die Grundlagen der Physik},
               {\it Konigl.~Gesell.~d.~Wiss.~G\"ottingen,
               Nachhr.~Math.-Phys.~Kl} 295-407.

\bibitem{kn:W} R.~Wald,
               {\sl General Relativity},
               The University of Chicago Press, Chicago, 1984.

\bibitem{kn:Schw} K.~Schwarzschild,
                  {\sl \"Uber das Gravitationsfled eines Massenpunktes
                  nach der Einsteinschen Theorie},
                  {\it Sitzunsberichte der Deutsch Akademie der 
                  Wissenschaften zu Berlin, Klasse f\"ur Mathematik, 
                  Physik und Technik}, \textbf{1916} 189-196.

\bibitem{kn:Birk5} G.D.~Birkhoff,
                   {\sl Relativity and Modern Physics},
                   Harvard University Press, Cambridge, 
                   Massachusets (1923).

\bibitem{kn:MTW} C.W.~Misner, K.S.~Thorne and J.A.~Wheeler,
                 {\sl Gravitation},
                 W.H.~Freeman and Co., New York, 1973.

\bibitem{kn:CW} J.~Ciufolini and J.A.~Wheeler,
                {\sl Gravitation and Inertia},
                Princeton University Press,
                Princeton, New Jersey (1995).

\bibitem{kn:chan} S.~Chandrasekhar,
                  {\sl The Mathemetical Theory of Black Holes},
                  Clarendon Press, Oxford (1983).

\bibitem{kn:ADM} R.~Arnowitt, S.~Deser and C.~Misner,
                 in {\sl Gravitation: An Introduction to Current 
                 Research}, Ed.~L.~Witten, Wiley, New York (1962).

\bibitem{kn:HoRo} G.T.~Horowitz and S.F.~Ross,
                  {\sl Naked Black Holes},
                  {\it Phys.~Rev.}~\textbf{D56} (1997) 2180-2187.
                  [{\tt hep-th/9704058}].

\bibitem{kn:HoRo2} G.T.~Horowitz and S.F.~Ross,
                   {\sl Properties of Naked Black Holes},
                   {\it Phys.~Rev.}~\textbf{D57} (1998) 1098-1107.
                   [{\tt hep-th/9709050}].

\bibitem{kn:Edd} A.S.~Eddington,
                 {\it Nature}~\textbf{113} (1924) 192.

\bibitem{kn:Fink} D.~Finkelstein,
                  {\sl Past-Future Asymmetry of the Gravitational Field 
                  of a Point Particle},
                  {\it Phys.~Rev.}~\textbf{110} (1958) 965-967.

\bibitem{kn:Kr} M.~Kruskal,
                {\sl Maximal Extension of the Schwarzschild Metric},
                {\it Phys.~Rev.}~\textbf{119} (1960) 1743-1745.

\bibitem{kn:Sz} G.~Szekeres,
                {\sl On the Singularities of a Riemannian Manifold},
                {\it Pbl.~Mat.~Debrecen}~\textbf{7} (1960) 285-301.

\bibitem{kn:Th} K.S.~Thorne,
                {\sl Black Holes \& Time Warps},
                W.W. Norton and Co., New York, 1994.

%%%%%%%%%%%%%%%%%%%%%%%%%%%%%%%%%%%%%%%%%%%%%%%%%%%%%%%%%%%%%%%%%%%%%%
%no-hair

%perturbaciones

\bibitem{kn:Pr1} R.H.~Price,
                 {\sl Non-spherical Perturbations of Gravitational Collapse.
                 I. Scalar and Gravitational Perturbations},
                 {\it Phys.~Rev.}~\textbf{D5} (1972) 2419-2438.

\bibitem{kn:Pr2} R.H.~Price,
                 {\sl Non-spherical Perturbations of Gravitational Collapse.
                 II. Integrer-Spin, Zero-REst-Mass-Fields},
                 {\it Phys.~Rev.}~\textbf{D5} (1972) 2439-2454.

%general

\bibitem{kn:He} M.~Heusler,
                {\sl Black Holes Uniqueness Theorems},
                Cambridge University Press, Cambridge (U.K.) 1996.

\bibitem{kn:He3} M.~Heusler,
                 {\sl Stationary Black Holes: Uniqueness and Beyond},
                 {\it Living~Rev.~Rel.}~\textbf{1} (1998) 6.\\
                 {\tt http://www.livingreviews.org/Articles/Volume1/1998-6heusler}

%Sch

\bibitem{kn:I} W.~Israel,
               {\sl Event Horizons in Static Vacuum Space-Times},
               {\it Phys.~Rev.}~\textbf{164} (1967) 1776-1779.
%RN

\bibitem{kn:I2} W.~Israel,
                {\it Commun.~Math.~Phys.}~\textbf{8} (1968) 245.
%K
\bibitem{kn:Ca2} B.~Carter,
                 {\sl Axisymmetric Black Hole Has Only 
                 Two Degrees of Freedom},
                 {\it Phys.~Rev.~Lett.}~\textbf{26} (1971) 331-333.

\bibitem{kn:W3} R.M.~Wald,
                {\sl Final States of Gravitational Collapse},
                {\it Phys.~Rev.~Lett.}~\textbf{26} (1971) 1653-1655.
%scal
\bibitem{kn:Chas} J.E.~Chase,
                  {\it Commun.~Math.~Phys.}~\textbf{19} (1970) 276.

\bibitem{kn:B4} J.D.~Bekenstein
                {\sl Novel ``No-Scalar-Hair'' Theorem for Black Holes},
                {\it Phys.~Rev.}~\textbf{D51}  (1995)  R6608-R6611.

\bibitem{kn:MB} A.E.~Mayo and J.D.~Bekenstein,
                {\sl No Hair for Spherical Black Holes: Charged and 
                Nonminimally Coupled Scalar Field With Selfinteraction},
                {\it Phys.~Rev.}~\textbf{D54} (1996) 5059-5069.
                [{\tt gr-qc/9602057}].

\bibitem{kn:SZ} D.~Sudarsky and T.~Zannias,
                {\sl Spherical Black Holes Cannot Support Scalar Hair},
                {\it Phys.~Rev.}~\textbf{D58} (1998) 087502.
                [{\tt gr-qc/9712083}].

\bibitem{kn:JNW}  A.I.~Janis, E.T. Newman and J.~Winicour,
                  {\sl Reality of the Schwarzschild Singularity},
                  {\it Phys.~Rev.~Lett.}~\textbf{20} (1968) 878.

\bibitem{kn:ALC} A.G.~Agnese and M.~La Camera,
                 {\sl Gravitation without Black Holes},
                 {\it Phys.~Rev.}~\textbf{D31} (1985) 1280-1286.

\bibitem{kn:B2} J.D.~Bekenstein,
                {\sl Black Hole Hair: 25 Years After},
                Talk given at the {\it 2nd International Sakharov
                Conference on Physics}, Moscow, Russia, 20-23 May 1996.
                [{\tt gr-qc/9605059}].

\bibitem{kn:He2} M.~Heusler,
                 {\sl No-Hair Theorems and Black Holes with Hair},
                 {\it Helv.~Phys.~Acta}~\textbf{69} (1996) 501.
                 [{\tt gr-qc/9610019}].

\bibitem{kn:RW} R.~Ruffini and J.A.~Wheeler,
                {\sl Introducing the Black Hole},
                {\it Physics Today}, \textbf{24} (1971) 30-36.

\bibitem{kn:O} T.~Ort'in,
               {\sl Time-Symmetric Initial-Data Sets in 4-D Dilaton
               Gravity},
               {\it Phys.~Rev.}~\textbf{D52} (1995) 3392-3405.
               [{\tt hep-th/9501094}].

%%%%%%%%%%%%%%%%%%%%%%%%%%%%%%%%%%%%%%%%%%%%%%%%%%%%%%%%%%%%%%%%%%%%%%
%Censura topologica

\bibitem{kn:Haw7} S.W.~Hawking,
                  {\it Comm.~Math.~Phys.}~\textbf{25} (1972) 152.

\bibitem{kn:FSW} J.L.~Friedman, K.~Schleich and D.M.~Witt,
                 {\sl Topological Censorship}
                 {\it Phys.~Rev.~Lett.}~\textbf{71} (1993) 1486-1489.
                 {\sl Erratum} {\it ibid.}~\textbf{75} (1995) 1872.
                 [{\tt gr-qc/9305017}].

%%%%%%%%%%%%%%%%%%%%%%%%%%%%%%%%%%%%%%%%%%%%%%%%%%%%%%%%%%%%%%%%%%%%%%
%Agujeros negros topologicos

\bibitem{kn:Va} L.~Vanzo,
                {\sl Black Holes with Unusual Topology},
                {\it Phys.~Rev.}~\textbf{D56} (1997) 6475-6483.
                [{\tt gr-qc/9705004}].

%%%%%%%%%%%%%%%%%%%%%%%%%%%%%%%%%%%%%%%%%%%%%%%%%%%%%%%%%%%%%%%%%%%%%%
%termodinamica de BHs

\bibitem{kn:Haw3} S.W.~Hawking,
                  {\sl Black Holes and Thermodynamics},
                  {\it Phys.~Rev.}~\textbf{D13} (1976) 191-197.

\bibitem{kn:Haw} S.W.~Hawking,
                 {\sl The Event Horizon},
                 in {\it Black Holes}, Gordon and Breach, New York,
                 1973.

\bibitem{kn:Ca} B.~Carter,
                {\sl Properties of the Kerr Metric},
                in {\it Black Holes}, Gordon and Breach, New York, 1973.

\bibitem{kn:BCH} J.M.~Bardeen, B.~ Carter and S.W.~Hawking,
                 {\sl The Four Laws of Black Hole Mechanics},
                 {\it Commun.~Math.~Phys.}~\textbf{31} (1973) 161-170.

\bibitem{kn:Ch} D.~Christodoulou,
                Ph.D.~Thesis, Princeton University, (unpublished).

\bibitem{kn:B} J.~Bekenstein,
                Ph.D.~Thesis, Princeton University, (unpublished).

\bibitem{kn:B5} J.D.~Bekenstein,
                {\sl Black Holes and the Second Law},
                {\it Lett. Nuovo Cim.}~\textbf{4} (1972) 737.

\bibitem{kn:B6} J.D.~Bekenstein,
                {\sl Black Holes and Entropy},
                {\it Phys.~Rev.}~\textbf{D9} (1973) 2333.

\bibitem{kn:B7} J.D.~Bekenstein, 
                {\sl Generalized Second Law of Thermodynamics 
                in Black Hole Physics},
                {\it Phys.~Rev.}~\textbf{D9} (1974) 3292.

\bibitem{kn:Sm} L.~Smarr,
                {\sl Mass Formula for Kerr Black Holes},
                {\it Phys.~Rev.~Lett.}~\textbf{30} (1973) 71-73.
                Erratum {\i ibid.}~\textbf{30} (1973) 521.

\bibitem{kn:W2} R.M.~Wald,
                {\sl The First Law of Black Hole Mechanics},
                Proceedings of {\sl College Park 1993, 
                Directions in general relativity}, Volume 1, 
                pages 358-366.
               [{\tt gr-qc/9305022}].

\bibitem{kn:HS} M.~Heusler and N.~Straumann,
                {\sl The First Law of Black Hole Physics For 
                a Class of Nonlinear Matter Models},
                {\it Class.~Quantum.~Grav.}~\textbf{10} (1993) 1299.

\bibitem{kn:GKK} G.W.~Gibbons, R.~Kallosh and B.~Kol,
                 {\sl Moduli, Scalar Charges and the First Law of
                 Black Hole Thermodynamics},
                 {\it Phys.~Rev.~Lett.}~\textbf{77}(1996) 4992-4995.
                 [{\tt hep-th/9607108}].

\bibitem{kn:W11} R.M.~Wald,
%                {\sl ?????},
                 {\it Ann.~Phys.}~\textbf{82} (1974) 548.

\bibitem{kn:Haw2} S.W.~Hawking,
                 {\sl Particle Creation by Black Holes},
                 {\it Commun.~Math.~Phys.}~\textbf{43} (1975) 199-220.
                 Reprinted in Ref.~\cite{kn:GH5}.

\bibitem{kn:Haw5} S.W.Hawking,
                  in {\it General Relativity, An Einstein Centenary
                  Survey}, eds.~S.W.~Hawking and W.~Israel
                  (Cambridge, 1979), Chapter 15.


\bibitem{kn:Reiss} H.~Reissner,
                   {\it Ann.~Phys.}~\textbf{50} (1916) 106.


\bibitem{kn:Pen} R.~Penrose,
                 {\sl Structure of Space-Time},
                 in {\it Battelle Rencontres}, C.M.~Will, J.A.~Wheeler Eds.
                 Benjamin, N.Y.
  
\bibitem{kn:No8} G. Nordstr\"om,
                 {\it Proc.~Kon. Ned.~Akad.~Wet.}~\textbf{20} (1918) 1238.

\bibitem{kn:Ma} S.D.~Majumdar,
                {\sl A Class of Exact Solutions of Einstein's 
                Field Equations},
                {\it Phys.~Rev.}~\textbf{72} (1947) 390-398.

\bibitem{kn:Pa} A.~Papapetrou,
               {\sl A Static Solution of the Equations of the Gravitational 
               Field for an Arbitrary Charge-Distribution},
               {\it Proc.~Roy.~Irish.~Acad.}~\textbf{A51} (1947) 191.

\bibitem{kn:HaHa} J.B.~Hartle and S.W.~Hawking,
                  {\sl Solutions of the Einstein-Maxwell Equations with
                  Many Black Holes}
                  {\it Commun.~Math.~Phys.}~\textbf{26}, (1972) 87.

\bibitem{kn:Rob} I.~Robinson,
                 {\sl A Solution of the Maxwell-Einstein Equations},
                 {\it Bull.~Acad.~Polon.~Sci.}~\textbf{7} (1959) 351.

\bibitem{kn:Bert} B.~Bertotti,
                  {\sl Uniform Electromagnetic Field in the Theory of GR},
                  {\it Phys.~Rev.}~\textbf{116} (1959) 1331.

\bibitem{kn:GWG} G.W.~Gibbons, {\sl Aspects of Supergravity Theories},
                 (three lectures) in: {\sl Supersymmetry, Supergravity
                 and Related Topics},  eds.~F.~del 'Aguila, J.~de
                 Azc'arraga and L.~Ib'a~nez, World Scientific,
                 Singapore, 1985, page 147.

\bibitem{kn:W12} R.M.Wald,
                 {\sl The ``Nernst Theorem'' and Black Hole Thermodynamics},
                 {\it Phys.~Rev.}~\textbf{D56} (!997) 6467-6474.
                 [{\tt gr-qc/9704008}].

\bibitem{kn:PSSTW} J.~Preskill, P.~Schwarz, A.~Shapere, S.~Trivedi
                   and F.~Wilczek,
                   {\sl Limitations on the Statistical Description
                   of Black Holes},
                   {\it Mod.~Phys.~Lett.}~\textbf{A6} (1991) 2353-2361.

\bibitem{kn:GK}  G.W.~Gibbons and R.E.~Kallosh,
                 {\sl Topology, Entropy and Witten Index of Dilaton
                 Black Holes},
                 {\it Phys.~Rev.}~\textbf{D51} (1995) 2839-2862.
                 [{\tt hep-th/9407118}].

\bibitem{kn:Te3} C.~Teitelboim,
                 {\sl Action and Entropy of Extreme and Non-Extreme
                 Black Holes},
                 {\it Phys.~Rev.}~\textbf{D51} (1995) 4315-4318.
                 \textbf{Erratum} {\it ibid.}~\textbf{D52} (1995) 6201.
                 [{\tt hep-th/9410103}].

\bibitem{kn:HaHo} S.W.~Hawking and G.T.~Horowitz,
                  {\sl Gravitational Hamiltonian, Action, Entropy
                  and Surface Terms},
                  {\it Class.~Quantum Grav.}~\textbf{13} (1996) 1487-1498.
                  [{\tt gr-qc/9501014}].         


\bibitem{kn:HT} C.M.~Hull and P.K.~Townsend,
                {\sl Unity of Superstring Dualities},
                {\it Nucl.~Phys.}~\textbf{B438} (1995) 109-137.
                [{\tt hep-th/9410167}].

\bibitem{kn:HR} S.W.~Hawking and S.F.~Ross,
                {\sl Duality Between Electric and Magnetic Black Holes},
                {\it Phys.~Rev.}~\textbf{D52} (1995) 5865-5876. 
                [{\tt hep-th/9504019}].

\bibitem{kn:DHT} S.~Deser, M.~Hennaux and C.~Teitelboim,
                 {\sl Electric-Magnetic Black-Hole Duality},
                 {\it Phys.~Rev.}~\textbf{D55} (1997) 826-828.
                 [{\tt hep-th/9607182}].

\bibitem{kn:De} S.~Deser,
                {\sl Black-Hole Electromagnetic Duality},
                Lectures given at 7th Mexican School of Particles and
                Fields and 1st Latin American Symposium on
                High-Energy Physics (VII-EMPC and I-SILAFAE -
                Dedicated to Memory of Juan Jose Giambiagi), M'erida,
                Yucat'an, M'exico, 30 October to 6 November 1996. 
                [{\tt hep-th/9701157}].

%%%%%%%%%%%%%%%%%%%%%%%%%%%%%%%%%%%%%%%%%%%%%%%%%%%%%%%%%%%%%%%%%%%%%%
%leccion 2
%%%%%%%%%%%%%%%%%%%%%%%%%%%%%%%%%%%%%%%%%%%%%%%%%%%%%%%%%%%%%%%%%%%%%%

\bibitem{kn:dSG} V.~de Sabbata and M.~Gasperini,
                 {\sl Introduction to Gravitation},
                 World Scientific, Singapore (1985).



%%%%%%%%%%%%%%%%%%%%%%%%%%%%%%%%%%%%%%%%%%%%%%%%%%%%%%%%%%%%%%%%%%%%%%
%Literatura basica de supersimetria y supergravedad

\bibitem{kn:vN} P.~van Nieuwenhuizen,
                {\sl Supergravity},
                {\it Phys.~Rept.}~\textbf{68} (1981) 189-398.

\bibitem{kn:WB} J.~Wess and J.~Bagger,
                {\sl Supersymmetry and Supergravity},
                Princeton University Press, Princeton (1992).

\bibitem{kn:Wes2} P.C.~West,
                  {\sl An Introduction to Supersymmetry and Supergravity},
                  Extended second edition.
                  World Scientific, Singapore (1990).                  

\bibitem{kn:Fr} P.G.O.~Freund,
                {\it Introduction to Supersymmetry},
                Cambridge University Press (1986).

\bibitem{kn:vP} A.~van Proeyen,
                {\sl Tools for Supersymmetry},
                Lectures given at {\it Spring School on Quantum Field Theory:
                Supersymmetry and Superstrings},
                Calimanesti, Romania, 24-30  April 1998.
                {\tt hep-th/9910030}.

\bibitem{kn:Bi2} A.~Bilal,
                 {\sl Introduction to Supersymmetry},
                 Lectures given at the summer scholl {\it Gif 2000},
                 at Paris.
                 {\tt hep-th/0101055}.

\bibitem{kn:SaSe} {\it Supergravities in Diverse Dimensions}
                  Vols.~1 and 2, eds.~A.~Salam and E.~Sezgin,
                  North Holland/World Scientific
                  Amsterdam/Singapore 1989.

%%%%%%%%%%%%%%%%%%%%%%%%%%%%%%%%%%%%%%%%%%%%%%%%%%%%%%%%%%%%%%%%%%%%%%

\bibitem{kn:HLS} R.~Haag, J.T.~Lopuszanski and M.~Sohnius.
                 {\sl All Possible Generators of Supersymmetries 
                 of the S~Matrix},
                 {\it Nucl.~Phys.}~\textbf{B88} (1975) 257.

\bibitem{kn:CDAF} L.~Castellani, R.~D'Auria and P.~Fr'e,
                  {\sl Supergravity and Superstrings, 
                  A Geometric Perspective},
                  3 Vols. World Scientific, Singapore (1991).
 
\bibitem{kn:vHvP} J.W.~van Holten and A.~van Proeyen,
                  {\sl $N=1$ Supersymmetry algebras in $D=2$,
                  $D=3$, $D=4$ Mod 8},
                  {\it J.~Phys.}~\textbf{A15} (1982) 3763.

%%%%%%%%%%%%%%%%%%%%%%%%%%%%%%%%%%%%%%%%%%%%%%%%%%%%%%%%%%%%%%%%%%%%%%
%cotas de Bogomol'nyi

\bibitem{kn:FSZ} S.~Ferrara, C.A.~Savoy and B.~Zumino,
                 {\sl General Massive Multiplets in Extended
                 Supersymmetry},
                 {\it Phys.~Lett.}~\textbf{100B} (1981) 393.

\bibitem{kn:DT2} S.~Deser and C.~Teitelboim,  
                 {\sl Supergravity has Positive Energy}
                 {\it Phys.~Rev.~Lett.}~\textbf{39} (1977) 249.

\bibitem{kn:WO} E.~Witten and D.~Olive,
                {\sl Supersymmetry Algebras that Include Topological
                Charges},
                {\it Phys.~Lett.}~\textbf{78B} (1978) 97.

%%%%%%%%%%%%%%%%%%%%%%%%%%%%%%%%%%%%%%%%%%%%%%%%%%%%%%%%%%%%%%%%%%%%%%
%positividad de la energia en RG y SUGRA


\bibitem{kn:Gri} M.~Grisaru, 
                 {\sl Positivity of the Energy in Einstein's Theory}
                 {\it Phys.~Lett.}~\textbf{73B} (1978) 207.

\bibitem{kn:SY} R.~Schoen and S.-T.~Yau,
                {\sl On the Proof of the Positive mass Conjecture in 
                General Relativity},
                {\it Comm.~Math.~Phys.}~\textbf{65} (1979) 45-76.

\bibitem{kn:Wi5} E.~Witten, 
                 {\sl A New Proof of the Positive Energy Theorem},      
                 {\it Comm.~Math.~Phys.}~\textbf{80}, (1981) 381-402.

\bibitem{kn:Ne} J.M. Nester,
                {\sl A new Gravitational Energy Expression with a
                Simple Positivity Proof},
                {\it Phys.~Lett.}~\textbf{83A}, (1981) 241.

\bibitem{kn:IN} W.~Israel and J.M. Nester,
                {\sl  Positivity of the Bondi Gravitational Mass},
                {\it Phys.~Lett.}~\textbf{85A}, (1981) 259.


\bibitem{kn:GHu} G.W.~Gibbons and C.M.~Hull,
                 {\sl A Bogomol'nyi bound for General Relativity
                 and Solitons in $N=2$ Supergravity},
                 {\it Phys.~Lett.}~\textbf{109B} (1982) 190.

\bibitem{kn:ILPT} J.M.~Izquierdo, N.D.~Lambert, G.~Papadopoulos 
                  and  P.K.~Townsend,
                  {\sl Dyonic Membranes},
                  {\it Nucl.~Phys.}~\textbf{B460} (1996) 560-578.
                  [{\tt hep-th/9508177}].

\bibitem{kn:Hu5} C.M.~Hull,
                 {\sl The Positivity of Gravitational Energy and
                 Global Supersymmetry},
                 {\it Comm.~Math.~Phys.}~\textbf{90} (1983) 545-561. 

%%%%%%%%%%%%%%%%%%%%%%%%%%%%%%%%%%%%%%%%%%%%%%%%%%%%%%%%%%%%%%%%%%%%%%
%N=2

\bibitem{kn:FPVN} S.~Ferrara and P.~van Nieuwenhuizen,
                  {\sl Consistent Supergravity with Complex 
                  Spin 3/2 Gauge Fields},
                  {\it Phys.~Rev.~Lett.}~\textbf{37} (1976) 1669-1671. 

\bibitem{kn:To} K.P.~Tod,
               {\sl All Metrics Admitting Supercovariantly
               Constant Spinors},
               {\it Phys.~Lett.}~\textbf{121B}, (1981) 241.

\bibitem{kn:IW} W.~Israel and G.A.~Wilson,
                {\sl A Class of Stationary Electromagnetic Vacuum Fields},
                {\it J.~Math.~Phys.}~\textbf{13}, (1972) 865.

\bibitem{kn:Pe} Z.~Perj'es,
                {\sl Solutions of the Coupled Einstein-Maxwell Equations
                Representing the Fields of Spinning Sources},
                {\it Phys.~Rev.~Lett.}~\textbf{27} (1971) 1668.


%%%%%%%%%%%%%%%%%%%%%%%%%%%%%%%%%%%%%%%%%%%%%%%%%%%%%%%%%%%%%%%%%%%%%%
%N=4

\bibitem{kn:CSF} E.~Cremmer, J.~Scherk and S.~Ferrara,
                 {\sl $SU(4)$ Invariant Supergravity Theory},
                 {\it Phys.~Lett.}~{74B} (1978) 64.

\bibitem{kn:G} G.W.~Gibbons, 
               {\sl Antigravitating Black Hole Solitons
               with Scalar Hair in $N=4$ Supergravity}, 
               {\it Nucl.~Phys.}~\textbf{B207}, (1982) 337.

\bibitem{kn:GM}  G.W.~Gibbons and K.~Maeda, 
                 {\sl Black Holes and Membranes in Higher Dimensional 
                 Theories with Dilaton Fields}, 
                 {\it Nucl.~Phys.}~\textbf{B298}, (1988) 741.

\bibitem{kn:STW} A.~Shapere, S.~Trivedi and F.~Wilczek, 
                 {\sl Dual Dilaton Dyons}, 
                 {\it Mod.~Phys.~Lett.}~\textbf{A6}, (1991) 2677.

\bibitem{kn:KLOPP} R.~Kallosh, A.~Linde, T.~Ort'in, A.~Peet and
                   A.~Van Proeyen, 
                   {\sl Supersymmetry as a Cosmic Censor}, 
                   {\it Phys.~Rev.} \textbf{D46} (1992) 5278-5302.
                   [{\tt hep-th/9205027}].

\bibitem{kn:O1} T.~Ort'in, 
                {\sl Electric-Magnetic Duality and Supersymmetry 
                in Stringy Black Holes},
                {\it Phys.~Rev.}~\textbf{D47} (1993) 3136-3143.
                [{\tt hep-th/9208078}].

\bibitem{kn:KO} R.~Kallosh and T.~Ort'in,
                {\sl Charge Quantization of Axion--Dilaton Black Holes},
                {\it Phys.~Rev.}~\textbf{D48} (1993) 742--747.
                [{\tt hep-th/9302109}].

\bibitem{kn:KKOT} R.~Kallosh, D.~Kastor, T.~Ort'in and T.~Torma,
                  {\sl Supersymmetry and Stationary Solutions in
                  Dilaton-Axion Gravity}, 
                  {\it Phys.~Rev.}~\textbf{D50} (1994) 6374.
                  [{\tt hep-th/9406059}].

\bibitem{kn:To2} K.P.~Tod,
                 {\sl More on Supercovariantly Constant Spinors},
                 {\it Class.~Quantum Grav.}~\textbf{12} (1995) 1801-1820.

\bibitem{kn:BKO3} E.~Bergshoeff, R.~Kallosh and T.~Ort'in,
                  {\sl Stationary Axion/Dilaton Solutions and
                  Supersymmetry},
                  {\it Nucl.~Phys.}~\textbf{B478} (1996) 156-180.
                  [{\tt hep-th/9605059}].

\bibitem{kn:L-TO} E.~Lozano-Tellechea and T.~Ort'in,
                  {\sl The General, Duality-Invariant Family of Non-BPS
                  Black-Hole Solutions of  $N=4$, $d=4$ Supergravity},
                  {\it Nucl.~Phys.}~\textbf{B569} (2000) 435-450.
                 [{\tt hep-th/9910020}].

\bibitem{kn:FKS} S.~Ferrara, R.~Kallosh and A.~Strominger, 
                 {\sl $N=2$ Extremal Black Holes}, 
                 {\it Phys.~Rev.}~\textbf{D52} (1995) 5412-5416.
                 [{\tt hep-th/9508072}].

\bibitem{kn:BLS} K.~Behrndt, D.~L\"ust and  W.A.~Sabra,
                 {\sl Stationary Solutions of $N=2$ Supergravity},
                 {\it Nucl.~Phys.}~\textbf{B510} (1998) 264.
                 [{\tt  hep-th/9705169}].

\bibitem{kn:Moh} T.~Mohaupt,
                 {\sl Black Hole Entropy, Special Geometry and Strings},
                 [{\tt hep-th/0007195}].

\bibitem{kn:DAF} R.~D'Auria and P. Fr'e,
                 {\sl  BPS Black Holes in Supergravity: 
                 Duality Groups, P-Branes, 
                 Central Charges and The Entropy}
                 Lecture notes for the {\it 8th Graduate School in 
                 Contemporary Relativity and Gravitational Physics: 
                 The Physics of Black Holes} (SIGRAV 98),
                 Villa Olmo, Italy, 20-25 Apr 1998.
                 [{\tt hep-th/9812160}].

%%%%%%%%%%%%%%%%%%%%%%%%%%%%%%%%%%%%%%%%%%%%%%%%%%%%%%%%%%%%%%%%%%%%%%
%N=8

%%%%%%%%%%%%%%%%%%%%%%%%%%%%%%%%%%%%%%%%%%%%%%%%%%%%%%%%%%%%%%%%%%%%%%
%composite BHS

\bibitem{kn:KK} R.~Kallosh and B.~Kol,
                {\sl $E_{7}$ Symmetric Area of the Black Hole Horizon},
                {\it Phys.~Rev.}~\textbf{D53} (1996) 5344-5348. 
                [{\tt hep-th/9602014}].

\bibitem{kn:KhO1} R.R.~Khuri and T.~Ort'in,
                  {\sl Supersymmetric Black Holes in $N=8$ Supergravity},
                  {\it Nucl.~Phys.}~\textbf{B467}), (1996) 355-382.
                  [{\tt hep-th/9512177}].

\bibitem{kn:Ra} J.~Rahmfeld, 
                {\sl Extremal Black Holes as Bound States},
                {\it Phys.~Lett.}~\textbf{B372} (196) 198-203. 
                [{\tt hep-th/9512089}].

\bibitem{kn:O3} T.~Ort'in,
                {\sl Massless String Theory Black Holes as
                Black Diholes and Quadruholes},
                {\it Phys.~Rev.~Lett.}~\textbf{76} (1996) 3890.
                [{\tt hep-th/9602067}].

\bibitem{kn:O7} T.~Ort'in,
                {\sl Extremality Versus Supersymmetry in Stringy
                Black Holes},
                {\it Phys.~Lett.}~\textbf{B422}: (1998) 93-100.
                [{\tt hep-th/9612142}].


%%%%%%%%%%%%%%%%%%%%%%%%%%%%%%%%%%%%%%%%%%%%%%%%%%%%%%%%%%%%%%%%%%%%%%
%%%%%%%%%%%%%%%%%%%%%%%%%%%%%%%%%%%%%%%%%%%%%%%%%%%%%%%%%%%%%%%%%%%%%%
%%%%%%%%%%%%%%%%%%%%%%%%%%%%%%%%%%%%%%%%%%%%%%%%%%%%%%%%%%%%%%%%%%%%%%
%3

%%%%%%%%%%%%%%%%%%%%%%%%%%%%%%%%%%%%%%%%%%%%%%%%%%%%%%%%%%%%%%%%%%%%%%

\bibitem{kn:Wi} E.~Witten,
                {\sl String Theory Dynamics in Various Dimensions},
                {\it Nucl.~Phys.}~\textbf{B443} (1995) 85-126.
                [{\tt hep-th/9503124}].

\bibitem{kn:DHS} M.~Dine, P.~Huet and N.~Seiberg,
                 {\sl Large and Small Radius in String Theory},
                 {\it Nucl.~Phys.}~\textbf{B322} (1989) 301.

\bibitem{kn:DLP} J.~Dai, R.G.~Leigh and J.~Polchinski,
                 {\sl New Connections between String Theories},
                 {\it Mod.~Phys.~Lett.}~\textbf{A4}, No.~21 (1989) 2073.

\bibitem{kn:BHO} E.~Bergshoeff, C.M.~Hull and T.~Ort'in,
                 {\sl Duality in the Type~II Superstring Effective
                 Action},
                 {\it Nucl.~Phys.}~\textbf{B451} (1995) 547-578.
                 [{\tt hep-th/9504081}].

\bibitem{kn:BRGPT} E.~Bergshoeff, M.~de Roo, M.B.~Green,
                   G.~Papadopoulos and P.K.~Townsend,
                   {\sl Duality of Type~II 7-Branes and 8-Branes},
                   {\it Nucl.~Phys.}~\textbf{B470} (1996) 113-135.
                   [{\tt hep-th/9601150}].

\bibitem{kn:GPR} A.~Giveon, M.~Porrati and E.~Rabinovici,
                 {\sl Target Space Duality in String Theory},
                 {\it Phys.~Rep.}~\textbf{244} (1994) 77-202.
                 [{\tt hep-th/9401139}].


\bibitem{kn:Bu} T.~Buscher,
                {\sl Quantum Corrections and Extended Supersymmetry
                in New Sigma Models},
                {\it Phys.~Lett.}~\textbf{159B} (1985) 127;
                {\sl A Symmetry of the String Background Field
                Equations},
                {\it ibid} \textbf{194B} (1987) 59;
                {\sl Path Integral derivation of Quantum Duality in
                Non-Linear Sigma Models},
                {\it ibid} \textbf{201B} (1988) 466.

\bibitem{kn:HorWit} P.~Horava and E.~Witten,
                    {\sl Heterotic and Type I String Dynamics 
                    from Eleven Dimensions},
                    {\it Nucl.~Phys.}~\textbf{B460} (1996) 506-524.
                    [{\tt hep-th/9510209}].

\bibitem{kn:Da2} A.~Dabholkar,
                 {\sl Lectures on Orientifolds and Duality},
                 in {\it High Energy Physics and Cosmology 1997},
                 Proceedings Trieste 1997, The ICTP Series in Theoretical 
                 Physics, Vol.~14, eds.~E.~Gava et al.,  
                 World Scientific, Singapore (1998), p.~128.
                 [{\tt hep-th/9804208}].


\bibitem{kn:Hu6} C.M.~Hull,
                 {\sl  String-String Duality in Ten-Dimensions},
                 {\it Phys.~Lett.}~\textbf{B357} (1995) 545-551.
                 [{\tt hep-th/9506194}].

\bibitem{kn:Da3} A.~Dabholkar,
                 {\sl  Ten-Dimensional Heterotic String as a Soliton},
                 {\it Phys.~Lett.}~\textbf{B357} (1995) 307-312.
                 [{\tt hep-th/9506160}].

\bibitem{kn:PW} J.~Polchinski and E.~Witten,
                {\sl Evidence for Heterotic - Type I String Duality},
                {\it Nucl.~Phys.}~\textbf{B460} (1996) 525.
                [{\tt hep-th/9510169}].

\bibitem{kn:Hu7} C.M.~Hull,
                 {\sl The Nonperturbative SO(32) Heterotic String},
                 {\it Phys.~Lett.}~\textbf{B462} (1999) 271-276.
                 [{\tt hep-th/9812210}].

\bibitem{kn:BEHvdSHL}  E.~Bergshoeff, E.~Eyras, R.~Halbersma, 
                       J.P.~van der Schaar, C.M.~Hull and Y.~Lozano,
                       {\sl Space-Time Filling Branes and Strings with 
                       Sixteen Supercharges},
                       {\it Nucl.~Phys.}~\textbf{B564} (2000) 29-59.
                       [{\tt hep-th/9812224}].

%%%%%%%%%%%%%%%%%%%%%%%%%%%%%%%%%%%%%%%%%%%%%%%%%%%%%%%%%%%%%%%%%%%%%%
%Literatura basica de cuerdas





\bibitem{kn:GSW} M.B.~Green, J.H.~Schwarz and E.~Witten,
                 {\sl Superstring Theory},
                 (two volumes) 
                 Cambridge University Press, Cambridge (U.K.) (1987).

\bibitem{kn:LT} D.~L\"ust, S.~Theisen,
                {\sl Lectures on String Theory},
                Springer Verlag, berlin 1989,
                (Lecture notes in physics, 346).

\bibitem{kn:P3} J.~Polchinski,
                {\sl String Theory},
                Vols.~1 and 2, Cambridge University Press (1998).

\bibitem{kn:Ki2} E.~Kiritsis, 
                 {\sl Introduction to Superstring Theory},
                 [{\tt hep-th/9709062}].
                 (To be published in book form by Leuven University Press).

\bibitem{kn:EM} E.~'Alvarez and P.~Meessen,
                {\sl String Primer},
                {\it JHEP}~\textbf{9902} (1999) 015.
                [{\tt hep-th/9810240}].

\bibitem{kn:J} C.V.~Johnson,
               {\sl D-Brane Primer},
               lectures given at ICTP, TASI, and BUSSTEPP
               [{\tt hep-th/0007170}].


%%%%%%%%%%%%%%%%%%%%%%%%%%%%%%%%%%%%%%%%%%%%%%%%%%%%%%%%%%%%%%%%%%%%%%

\bibitem{kn:CJS} E.~Cremmer, B.~Julia and J.~Scherk,
                 {\sl Supergravity Theory in 11 Dimensions},
                 {\it Phys.~Lett.}~\textbf{76B} (1978) 409.

\bibitem{kn:A} O.~Aharony,
               {\sl String Theory Dualities From M Theory},
               {\it Nucl.~Phys.}~\textbf{B476} (1996) 470-483.
               [{\tt hep-th/9604103}].

\bibitem{kn:BRO} E.~Bergshoeff, M.~de Roo and T.~Ort'in,
                 {\sl The Eleven-Dimensional Five-Brane},
                 {\it Phys.~Lett.}~\textbf{B386} (1996) 85-90.
                 [{\tt hep-th/9606118}].

\bibitem{kn:BLO} E.~Bergshoeff, Y.~Lozano and T.~Ort'in,
                 {\sl Massive Branes},
                 {\it Nucl.~Phys.}~\textbf{B518} (1998) 363.
                 [{\tt hep-th/9712115}].

\bibitem{kn:Ro2} L.J.~Romans,
                 {\sl Massive $N=2a$ Supergravity in Ten Dimensions},
                 {\it Phys.~Lett.}~\textbf{169B} (1986) 374.

\bibitem{kn:Hu3} C.M.~Hull, 
                 {\sl Exact PP-Wave Solutions of Eleven-Dimensional 
                 Supergravity},
                 {\it Phys.~Lett.}~\textbf{139B} (1984) 39.

\bibitem{kn:SS} J.~Scherk and J.H.~Schwarz,
                {\sl How to Get Masses from extra Dimensions},
                {\it Nucl.~Phys.}~\textbf{B153} (1979) 61-88.

\bibitem{kn:HW} P.~Howe and P.C.~West,
                {\sl The Complete $N=2,d=10$ Supergravity},
                {\it Nucl.~Phys.}~\textbf{B238} (1984) 181-219.

\bibitem{kn:L-TO2} E.~Lozano-Tellechea and T.~Ort'in,
                   {\sl 7-Branes and Higher Kaluza-Klein Branes},
                   {\it Nucl.~Phys.}~\textbf{B607} (2001) 213-236.
                   [{\tt hep-th/0012051}].

\bibitem{kn:JHS} J.H.~Schwarz,
                 {\sl Covariant Field Equations of Chiral $N=2,D=10$
                 Supergravity},
                 {\it Nucl.~Phys.}~\textbf{B226} (1983) 269-288.

\bibitem{kn:BBO} E.~Bergshoeff, H.-J.~Boonstra and T.~Ort'in,
                 {\sl $S$~Duality and Dyonic $p$-Brane Solutions
                 in Type~II String Theory},
                 {\it Phys.~Rev.}~\textbf{D53} 7206-7212.
%                {\tt hep-th/9508091}.

\bibitem{kn:MO} P.~Meessen and T.~Ort'in,
                {\sl An $Sl(2,\mathbb{Z})$ Multiplet of Nine-Dimensional 
                Type~II Supergravity Theories},
                {\it Nucl.~Phys.}~\textbf{B541}  (1999) 195-245.
                [{\tt hep-th/9806120}].





%%%%%%%%%%%%%%%%%%%%%%%%%%%%%%%%%%%%%%%%%%%%%%%%%%%%%%%%%%%%%%%%%%%%%%
%%%%%%%%%%%%%%%%%%%%%%%%%%%%%%%%%%%%%%%%%%%%%%%%%%%%%%%%%%%%%%%%%%%%%%
%%%%%%%%%%%%%%%%%%%%%%%%%%%%%%%%%%%%%%%%%%%%%%%%%%%%%%%%%%%%%%%%%%%%%%
%4

%%%%%%%%%%%%%%%%%%%%%%%%%%%%%%%%%%%%%%%%%%%%%%%%%%%%%%%%%%%%%%%%%%%%%%
%Reviews de branas y soluciones similares de supergravities y cuerdas


\bibitem{kn:Du4} M.J.~Duff,
                 {\sl  Supermembranes: the First Fifteen Weeks}
                 {\it Class.~Quantum Grav.}~\textbf{5} (1988) 189-205.

\bibitem{kn:Tow2} P.K.~Townsend,
                  {\sl Three Lectures on Supermembranes},
                  Proceedings of the 1988 Trieste School. 

\bibitem{kn:Tow3} P.K.~Townsend,
                  {\sl Three Lectures on Supersymmetry and 
                  Extended Objects},
                  Proceedings of the 13th GIFT Seminar on
                  Theoretical Physics: Recent Problems in
                  Mathematical Physics, {\sl Integrable Systems, 
                  Quantum Groups and Quantum Field Theory} Salamanca, 
                  Spain, 15-27 June 1992. L.A.~Ibort and M.A.~Rodr'iguez
                  Editors, Kluwer, 1993.

\bibitem{kn:DKL} M.J.~Duff, R.R.~Khuri and J.X.~Lu,
                 {\sl String Solitons},
                 {\sl Phys.~Rep.}~\textbf{259} (1995) 213-326.
                 [{\tt hep-th/9412184}].

\bibitem{kn:CvSo} M.~Cveti\v{c} and H.H.~Soleng,
                  {\sl Supergravity Domain Walls},
                  {\it Phys.~Rept.}~\textbf{282} (1997) 159.
                  [{\tt hep-th/9604090}].

\bibitem{kn:Sc}  J.H.~Schwarz, {\sl Lectures on Superstring and 
                 M~Theory Dualities}, 
                 given at the ICTP {\sl Spring School and Workshop 
                 on String Theory, Gauge Theory and Quantum Gravity}, 
                 Trieste, Italy, 18-29 March 1996 and at the 
                 Theoretical Advanced 
                 Study Institute in Elementary Particle Physics (TASI 96), 
                 {\sl Fields, Strings, and Duality}, Boulder, 
                 Colorado (U.S.A.), 2-28 June 1996.
                 [{\tt hep-th/9607201}].

\bibitem{kn:Tow} P.K.~Townsend,
                 {\sl Four Lectures on M~Theory},
                 given at the {\sl ICTP Summer School in High-energy 
                 Physics and Cosmology}, Trieste, Italy, 
                 10 June to 26 July 1996.  
                 [{\tt hep-th/9612121}].
                 
\bibitem{kn:Ste} K.S.~Stelle,
                 {\sl Lectures on Supergravity $p$-Branes},
                 given at the {\sl ICTP Summer School in High-energy 
                 Physics and Cosmology}, Trieste, Italy, 
                 10 June to 26 July 1996.                  
                 [{\tt hep-th/9701088}].

\bibitem{kn:Y} D.~Youm,
               {\sl Black Holes and Solitons in String Theory},
               {\it Phys.~Rept.}~\textbf{316} (1999) 1-232.
               [{\tt hep-th/9710046}].

\bibitem{kn:Ca3} B.~Carter,
                 {\sl Essentials of Classical Brane Dynamics},
                 Proceedings of {\it Peyresq 5 meeting, ``Quantum Spacetime, Brane 
                 Cosmology and Stochastic Effective Theories''},
                 [{\tt gr-qc/0012036}].

%%%%%%%%%%%%%%%%%%%%%%%%%%%%%%%%%%%%%%%%%%%%%%%%%%%%%%%%%%%%%%%%%%%%%%

\bibitem{kn:Na} Y. Nambu,
                Lectures at the 1970 Copenhagen Symposium.

\bibitem{kn:Go} T.~Goto,
                {\sl Relativistic Quantum Mechanics of One-Dimensional 
                Mechanical Continuum and Subsidiary Condition 
                of Dual resonance Model},  
                {\it Prog.~Theor.~Phys.}~\textbf{46} (1971) 1560-1569.

\bibitem{kn:BDH} L.~Brink, P~Di Vecchia and P.~Howe,
                 {\sl A Locally Supersymmetric and Reparametrization 
                 Invariant Action for the Spinning String},
                 {\it Phys.~Lett.}~\textbf{65B} (1976) 471-474.

\bibitem{kn:Mar} D.~Marolf,
                 {\sl Chern-Simons Terms and the Three Notions of Charge},
                 {\it Proceedings of the E.S.~Fradkin Memorial Conference}.
                 [{\tt hep-th/0006117}].

\bibitem{kn:BdRO} E.~Bergshoeff, M.~de Roo and T.~Ort'in,
                  {\sl The Eleven-Dimensional Five-Brane},
                  {\it Phys.~Lett.}~\textbf{B386} (1996) 85-90.
                  [{\tt hep-th/9606118}].

\bibitem{kn:BLNPST} I.~Bandos, K.~Lechner, A. Nurmagambetov, P.~Pasti, 
                    D.~Sorokin and M.~Tonin,
                    {\sl Covariant Action for the Super-Five-Brane 
                    of M-Theory},
                    {\it Phys.~Rev.~Lett.}~\textbf{78} (1997) 4332.
                    [{\tt hep-th/9701149}].

\bibitem{kn:Town8} P.K.~Townsend,
                   {\sl D-Branes from M-Branes},
                   {\it Phys.~Lett.}~\textbf{B373} (1996) 68-75.
                   [{\tt hep-th/9512062}].

\bibitem{kn:HoS} G.T.~Horowitz and A.~Strominger,
                 {\sl Black Strings and p-Branes},
                 {\it Nucl.~Phys.}~\textbf{B360} (1991) 197.

\bibitem{kn:MP} R.C.~Myers and M.J.~Perry,
                {\sl Black Holes in Higher Dimensional Space-Times},
                {\it Ann.~Phys.}~\textbf{172} (1986) 304.

\bibitem{kn:T} F.~Tangherlini,
%               {\sl ?????},
               {\it Nuovo Cimento} \textbf{77} (1963) 636.

\bibitem{kn:P} J.~Polchinski,
               {\sl Dirichlet Branes and Ramond--Ramond
               Charges},
               {\it Phys.~Rev.~Lett.}~\textbf{75} (1995) 4724-4727.
               [{\tt  hep-th/9510017}].

\bibitem{kn:Ba} C.~Bachas,
                {\sl (Half) a Lecture on D~branes},
                given at the {\sl Workshop on Gauge Theories, 
                Applied Supersymmetry and Quantum Gravity}, 
                Imperial College, London, July 1996 and at the 
                {\sl Institut d'Et'e},
                Ecole Normale Sup'erieure, Paris, August 1996.
                Ecole Polytechnique Report CPTH-PC491-0197 and
                [{\tt hep-th/9701019}].

\bibitem{kn:Ba2} C.~Bachas,
                 {\sl Lectures on D-branes},
                 Based on lectures given in 1997 at the Isaac Newton 
                 Institute, Cambridge, the Trieste Spring School 
                 on String Theory, and at the 31rst International 
                 Symposium Ahrenshoop in Buckow.
                 [{\tt hep-th/9806199}].


\bibitem{kn:Dou} M.R.~Douglas,
                 {\sl Superstring Dualities, Dirichlet Branes and the
                 Small Scale Structure of Space},
                 Talk given at NATO Advanced Study Institute: 
                 Les Houches Summer School on Theoretical Physics, 
                 Session 64: {\sl Quantum Symmetries}, 
                 Les Houches, France, 1 August to 8 September 1995.
                 Rutgers U.~Report RU-96-91 and 
                 {\tt hep-th/9610041}.

\bibitem{kn:P2} J.~Polchinski,
                {\sl TASI Lectures on D~branes},
                given at the Theoretical Advanced Study Institute in 
                Elementary Particle Physics (TASI 96), 
                {\sl Fields, Strings, and Duality}, Boulder, 
                Colorado (U.S.A.), 2-28 June 1996.
                ITP Report NSF-ITP-96-145 and
                [{\tt hep-th/9611050}].

\bibitem{kn:Van} I.V.~Vancea,
                 {\sl Introductory Lectures to D-branes},
                 lectures delivered at {\it Jorge Andre Swieca School
                 on Particles and Fields}, Campos do Jordao, Brazil, 2001.
                 [{\tt hep-th/0109029}].

\bibitem{kn:For} S.~F\"orste, 
                 {\sl Strings, Branes and Extra Dimensions},
                 [{\tt hep-th/0110055}].

\bibitem{kn:KS} P.~Koerber and A.~Sevrin,
                {\sl Testing the $\alpha^{\prime\, 3}$ Term in 
                the Non-Abelian Open Superstring  Effective Action},
                {\it JHEP}~\textbf{0109} (2001) 009.
                [{\tt hep-th/0109030}].  


\bibitem{kn:DGHR} A.~Dabholkar, G.W.~Gibbons, J.~Harvey and
                  F.~Ruiz-Ruiz,
                  {\sl Superstrings and Solitons},
                  {\it Nucl.~Phys.}~\textbf{B340} (1990) 33.

\bibitem{kn:CHS} C.G.~Callan, J.A.~Harvey and A.~Strominger,
                 {\sl World-Sheet Approach to Heterotic Instantons
                 and Solitons},
                 {\it Nucl.~Phys.}~\textbf{B359} (1991) 611-634.

\bibitem{kn:CHS2} C.G.~Callan, J.A.~Harvey and A.~Strominger,
                  {\sl Supersymmetric String Solitons},
                  in the proceedings of
                  {\it String Theory and Quantum Gravity '91},
                  Trieste 1991.
                  [{\tt  hep-th/911203}].

%%%%%%%%%%%%%%%%%%%%%%%%%%%%%%%%%%%%%%%%%%%%%%%%%%%%%%%%%%%%%%%%%%%%%%
%intersecciones

%en general

\bibitem{kn:Tow7} P.K.~Townsend,
                  {\sl Brane Surgery},
                  {\it Nucl.~Phys.~Proc.~Suppl.}~\textbf{B475} (1996).
                  [{\tt hep-th/9609217}].

\bibitem{kn:Stro} A.~Strominger,
                  {\sl Open $p$-Branes},
                  {\it Phys.~Lett.}~\textbf{B383} (1996) 44.
                  [{\tt hep-th/9512059}].

\bibitem{kn:CaMa} C.G.~Callan Jr.~and J.M.~Maldacena,
                  {\sl Brane Dynamics from the Born-Infeld Action},
                  {\it Nucl.~Phys.}~\textbf{B513} (1998) 198-212.
                  [{\tt hep-th/9708147}].

\bibitem{kn:HaWi} A.~Hanany and E.~Witten,
                  {\sl Tyle~IIB Superstrings, BPS Monopoles and 3-Dimensional 
                  Gauge Dynamics},
                  {\it Nucl.~Phys.}~\textbf{B492} (1997) 152-190.
                  [{\tt hep-th/9611230}].

\bibitem{kn:BdREJvdS} E.~Bergshoeff, M.~de Roo, E.~Eyras, B.~Janssen and
                      J.P.~van der Schaar,
                      {\sl Intersections Involving Monopoles and Waves in 
                      eleven Dimensions},
                      {\it Class.~Quantum Grav.}~\textbf{14} (1997) 2757.
                      [{\tt hep-th/9704120}].
%soluciones

\bibitem{kn:Ga2} J.P.~Gauntlett,
                 {\sl Intersecting Branes},
                 lectures given at the APCTP Winter School 
                 {\sl Dualities of Gauge and String Theories}, 
                 Korea, February 1997.
                 [{\tt hep-th/9705011}].

\bibitem{kn:Gu} R.~G\"{u}ven,
                {\sl Black $p$-Brane Solutions of $D=11$ Supergravity 
                Theory},
                {\it Phys.~Lett.}~\textbf{276B} (1992) 49.

\bibitem{kn:PT} G.~Papadopoulos and P.K.~Townsend,
                {\sl Intersecting M-Branes},
                {\it Phys.~Lett.}~\textbf{B380} (1996) 273-279.
                [{\tt hep-th/9603087}].

\bibitem{kn:Tsey} A.A.~Tseytlin,
                  {\sl Harmonic Superpositions of M~Branes},
                  {\it Nucl.~Phys.}~\textbf{B475} (1996) 149.
                  [{\tt hep-th/9604035}].

\bibitem{kn:GKT} J.P.~Gauntlett, D.A.~Kastor and J.~Traschen,
                 {\sl Overlapping Branes in M~Theory},
                 {\it Nucl.~Phys.}~\textbf{B478} (1996) 544.
                 [{\tt hep-th/9604179}].

\bibitem{kn:Pee3} A.W.~Peet,
                  {\sl Baldness/Delocalization in Intersecting Brane Systems},
                  {\it Class.~Quantum Grav.}~\textbf{17} (2000) 1235.
                  [{\tt hep-th/9910098}].

\bibitem{kn:AS} P.~Aichelburg and R.~Sexl, 
                {\sl On the Gravitational Field of a Massless Particle},
                {\it Gen.~Relativ.~Gravit.}~\textbf{2} (1971) 303.


%%%%%%%%%%%%%%%%%%%%%%%%%%%%%%%%%%%%%%%%%%%%%%%%%%%%%%%%%%%%%%%%%%%%%%
%BHs and strings

%correspondence (not complementarity!!)

\bibitem{kn:Sen6} A.~Sen,
                  {\sl Extreme Black Holes and Elementary Strings},
                  {\it Mod.~Phys.~Lett.}~\textbf{A10} (1995) 2081.
                  [{\tt hep-th/9504147}].
 
\bibitem{kn:HoPo} G.T.~Horowitz and J.~Polchinski,
                  {\sl A Correspondence Principle for Black Holes and Strings},
                  {\it Phys.~Rev.}~\textbf{D55} (1997) 6189.
                  [{\tt hep-th/9612146}].

%1 entropy



\bibitem{kn:SV} A.~Strominger and C.~Vafa,
                {\sl Microscopic Origin of the Bekenstein-Hawking Entropy},
                {\it Phys.~Lett.}~\textbf{B370} (1996) 99.
                [{\tt hep-th/9601029}].

\bibitem{kn:CaMa2} C.G.~Callan Jr.~and J.M.~Maldacena,
                   {\sl D-Brane Approach to Black Holes Quantum Mechanics},
                   {\it Nucl.~Phys.}~\textbf{B472} (1996) 591-608.
                   [{\tt hep-th/9602043}].

\bibitem{kn:BMPV} J.C.~Beckenridge, R.C.~Myers, A.W.~Peet and C.~Vafa,
                  {\sl D-Branes and Spinning Black Holes},
                  {\it Phys~.Lett.}~\textbf{B391} (1997) 93-98.
                  [{\tt hep-th/9602065}].

%\bibitem{kn:HoS2} G.~T.~Horowitz and A.~Strominger,
%                  {\sl Counting States of Near-Extremal Black Holes},
%                  {\it Phys.~Rev.~Lett.}~\textbf{77} (1996) 2368.
%                  [{\tt hep-th/9602051}].

\bibitem{kn:MaSt} J.M.~Maldacena and A.~Strominger,
                  {\sl Statistical Entropy of Four-Dimensional Extremal 
                  Black Holes},
                  {\it Phys.~Rev.~Lett.}~\textbf{77} (1996) 428.
                  [{\tt hep-th/9603060}].

\bibitem{kn:JKM} C.V.~Johnson, R.R.~Khuri and R.C.~Myers,
                 {\sl Entropy of 4-d Extremal Black Holes},
                 {\it Phys.~Lett.}~\textbf{B378} (1996) 78.
                 [{\tt hep-th/9603061}].

\bibitem{kn:HoMaSt} G.T.~Horowitz, J.M.~Maldacena and A.~Strominger,
                    {\sl Nonextremal Black Hole Microstates and U~Duality},
                    {\it Phys.~Lett.}~\textbf{B383} (1996) 151.
                    [{\tt hep-th/9603109}].
                    
\bibitem{kn:BLMPSV} J.C.~Beckenridge, D.A.~Lowe, R.C.~Myers, A.W.~Peet, 
                    A.~Strominger and C.~Vafa,
                    {\sl Macroscopic and Microscopic Entropy of Near-Extremal
                    Spinning Black Holes},
                    {\it Phys.~Lett.}~\textbf{B381} (1996) 423.
                    [{\tt hep-th/9603078}].

\bibitem{kn:HLM} G.T.~Horowitz, D.A.~Lowe and J.M.~Maldacena,
                 {\sl Statistical Entropy of Nonextremal Four-Dimensional 
                 Black Holes and U~Duality},
                 {\it Phys.~Rev.~Lett.}~\textbf{77} (1996) 430.
                 [{\tt hep-th/9603195}].


%2 Hawking radiation

\bibitem{kn:DaMat} S.R.~Das and S.D.~Mathur,
                   {\sl Comparing Decay Rates for Black Holes and D-Branes},
                   {\it Nucl.~Phys.}~\textbf{B478} (1996) 561.
                   [{\tt hep-th/9606185}].

\bibitem{kn:MaSt2} J.M.~Maldacena and A.~Strominger,
                   {\sl Black Hole Greybody Factors and D-Brane Spectroscopy},
                   {\it Phys.~Rev.}~\textbf{D55} (1997) 861.
                   [{\tt hep-th/9609026}].

\bibitem{kn:DaGMat} S.R.~Das, G.W.~Gibbons and S.D.~Mathur,
                    {\sl Universality of Low Energy Absoption Cross Sections 
                    for Black Holes},
                    {\it Phys.~Rev.~Lett.}~\textbf{78} (1997) 417.
                    [{\tt hep-th/9609052}].

%%%%%%%%%%%%%%%%%%%%%%%%%%%%%%%%%%%%%%%%%%%%%%%%%%%%%%%%%%%%%%%%%%%%%%
%relacion con 2+1 BH

\bibitem{kn:BTZ2} M.~Ba~nados, C.~Teitelboim and J.~Zanelli,
                  {\sl Black Hole in Three-Dimensional Spacetime}
                  {\it Phys.~Rev.~Lett.}~\textbf{69} (1992) 1849-1851.
                  [{\tt hep-th/9204099}].

\bibitem{kn:Hy} S.~Hyun,
                {\sl U-Duality Between Three and Higher Dimensional 
                Black Holes},
                {\tt hep-th/9704055}.

\bibitem{kn:SfSk} K.~Sfetsos and K.~Skenderis,
                  {\sl Microscopic Derivation of the Bekenstein-Hawking Entropy 
                  Formula for  Non-Extremal Black Holes},
                  {\it Nucl.~Phys.}~\textbf{B517} (1998) 179.
                  [{\tt hep-th/9711138}].

\bibitem{kn:Car2} S.~Carlip,
                  {\sl Black Hole Entropy from Horizon Conformal Field Theory},
                  {\it Nucl.~Phys.~Proc.~Suppl.}~\textbf{88} (2000) 10-16.
                  [{\tt gr-qc/9912118}].

\bibitem{kn:AGMOO} O.~Aharony, S.~S.~Gubser, J.~Maldacena, H.~Ooguri 
                   and Y.~Oz,
                   {\sl Large N Field Theories, String Theory and Gravity},
                   {\it Phys.~Rept.}~\textbf{323} (2000) 183.
                   [{\tt hep-th/9905111}].                       

\bibitem{kn:Gin} P.~Ginsparg,
                 {\sl Applied Conformal Field Theory},
                 Les Houches, Session XLIX {\it Fields, Strings and Critical
                 Phenomena}, Eds.~E.~Br'ezin and J.~Zinn-Justin, Elsevier 
                 (1989).

\bibitem{kn:CdWM} G.L.~Cardoso, B.~de Wit and T.~Mohaupt,
                  {\sl Area Law Corrections form State Counting and Supergravity},
                  Proceedings of {\it Strings'99},
                  {\it Class.~Quantum Grav.}~\textbf{17} (2000) 1007.
                  [{\tt hep-th/9910179}].




%%%%%%%%%%%%%%%%%%%%%%%%%%%%%%%%%%%%%%%%%%%%%%%%%%%%%%%%%%%%%%%%%%%%%%
%otros


%%%%%%%%%%%%%%%%%%%%%%%%%%%%%%%%%%%%%%%%%%%%%%%%%%%%%%%%%%%%%%%%%%%%%%


%\bibitem{kn:PRC} T.~Ort'in,
%                 {\sl Gravity and Strings},
%                 (en preparaci'on).

%%%%%%%%%%%%%%%%%%%%%%%%%%%%%%%%%%%%%%%%%%%%%%%%%%%%%%%%%%%%%%%%%%%%%%



\end{thebibliography}
\end{document}